\documentclass[11pt]{article}

\usepackage{amsmath}
\usepackage{amssymb}
\usepackage{graphicx}
\usepackage{natbib}
\usepackage{bm}
\usepackage{bbm}
\usepackage{threeparttable}
\usepackage{xcolor}
\usepackage{soul}
\usepackage{amsthm}
\usepackage{amsfonts}
\usepackage{enumitem}   
\usepackage{setspace}
\usepackage{caption}
\usepackage{subcaption}
\usepackage{hyperref}
\hypersetup{hidelinks}
\hypersetup{
  colorlinks   = true, %Colours links instead of ugly boxes
  urlcolor     = blue, %Colour for external hyperlinks
  linkcolor    = blue, %Colour of internal links
  citecolor   = blue %Colour of citations
}

\usepackage{multibib}
\newcites{sm}{Additional References}

\theoremstyle{plain}
\newtheorem{prop}{Proposition}
\newtheorem{theo}{Theorem}

\newtheorem{lemma}{Lemma}

\theoremstyle{remark}
\newtheorem{remark}{Remark}

\newcommand{\by}{\bm{y}}
\newcommand{\la}{\lambda}
\newcommand{\ga}{\gamma}
\newcommand{\bphi}{\bm{\phi}}
\newcommand{\btheta}{\bm{\theta}}
\newcommand{\iH}{\mathcal{H}}
\newcommand{\tN}{t_{N(t^-)}}
\newcommand{\Ntl}{N(t^-)}
\newcommand{\R}{\mathbb{R}}
\newcommand{\tx}{\tilde{x}}

\newcommand{\bw}{\bm{w}} 
\newcommand{\bth}{\bm{\theta}} 
\newcommand{\sF}{\mathcal{F}}
\newcommand{\x}{\bm{x}}
\newcommand{\bbeta}{\bm{\beta}}
\newcommand{\eps}{\epsilon}
\newcommand{\RomanNumeralCaps}[1]{\MakeUppercase{\romannumeral #1}}

\newcommand{\bZ}{\bm{Z}}
\newcommand{\bz}{\bm{z}}
\newcommand{\br}{\bm{r}}

\graphicspath{{figs/}}

\newtheorem{definition}{Definition}

\theoremstyle{remark}

\begin{document}

\def\spacingset#1{\renewcommand{\baselinestretch}%
{#1}\small\normalsize} \spacingset{1.2}
%%%%%%%%%%%%%%%%%%%%%%%%%%%%%%%%%%%%%%%%%%%%%%%%%%%%%%%%%%%%%%%%%%%%%%%%%%%%%%
\title{\bf Mixture Modeling for Temporal Point Processes with Memory}
\author{
Xiaotian Zheng, Athanasios Kottas, and Bruno Sans\'o\thanks{
Xiaotian Zheng (xzheng@uga.edu) is an Assistant Professor in the Department of 
Statistics, University of Georgia, USA. Athanasios Kottas
and Bruno Sans\'o  are Professors in the Department of Statistics, University of California, Santa Cruz, USA. }
}
\maketitle

\bigskip
\begin{abstract}
We propose a constructive approach to building temporal point processes that incorporate 
dependence on their history. The dependence is modeled through the conditional density of 
the duration, i.e., the interval between successive event times, using a mixture of 
first-order conditional densities for each one of a specific number of lagged durations. 
Such a formulation for the conditional duration density accommodates high-order
dynamics, and it thus enables flexible modeling for point processes 
with memory. The implied conditional intensity function admits a representation as a 
local mixture of first-order hazard functions. By specifying appropriate families of 
distributions for the first-order conditional densities, with different shapes for the 
associated hazard functions, we can obtain either self-exciting or self-regulating 
point processes. From the perspective of duration processes, we 
develop a method to specify a stationary marginal density. The resulting model, 
interpreted as a dependent renewal process, introduces high-order Markov dependence 
among identically distributed durations. Furthermore, we provide extensions to cluster 
point processes. These can describe duration clustering behaviors attributed to different 
factors, thus expanding the scope of the modeling framework to a wider range of applications. 
Regarding implementation, we develop a Bayesian approach to inference, 
model checking, and prediction.
We investigate point process model properties analytically, and illustrate the methodology 
with both synthetic and real data examples.
\end{abstract}


\vfill

\newpage
\spacingset{1.55}

\section{Introduction}

Temporal point processes are stochastic models for sequences of 
random events that occur in continuous time, with irregular durations, i.e., 
intervals between successive arrival times. Throughout this article,
event time and arrival time will be used interchangeably for 
the occurrence time of an event. Data corresponding to point patterns 
are common in a wide range of applications, 
such as earthquake occurrences \citep{ogata1988statistical}, 
recurrent events \citep{cook2007statistical}, 
financial high frequency trading and orders \citep{hautsch2011econometrics}, 
and neural spike trains \citep{tang2021multivariate},
to name a few. For many point patterns,
it is believed that occurrence of a future event depends on the past.
This motivates the use of point processes with memory, for example,
the Hawkes process \citep{hawkes1971point} with full memory, 
or renewal processes with lagged dependence. 
The goal of this article is to propose a modeling framework for point 
processes with high-order memory, 
replacing the independent durations of renewal processes
with high-order dependent ones,
and including the ability to model
duration-clustering behaviors present in applications such as health care 
\citep{yang2018clustering}, climatology \citep{cowpertwait2001renewal}, and
finance \citep{o1997market}.

As such, this article explores construction of point processes based on
models for the durations.
For point processes with memory, the collection of dependent
durations form a discrete-time stochastic process,
and thus a time series model for durations induces
conditional densities on the arrival times. Hereafter,
we refer to these conditional densities as conditional arrival densities, and 
notice that they uniquely determine the distribution of the resulting 
point process \citep{daley2003introduction}.
A common approach to model point process dependence is to specify the conditional 
intensity of the process, namely, the instantaneous event 
rate conditional on the process history (e.g., the Hawkes process).
In fact, a point process can be equivalently characterized by its conditional
intensity or the conditional arrival densities. 
The latter approach benefits from the vast literature on conditional density 
modeling. Density-based modeling naturally leads to a well-defined point process,
with its conditional intensity derived through a normalization
of the conditional arrival densities against the associated
survival functions \citep{daley2003introduction}. 
Constructing point processes using duration models, usually coupled with a limited 
memory assumption, can be computationally attractive for inference, as this approach 
facilitates evaluation of the resulting likelihood. In Section \ref{sec:mtdpp},
we provide further discussion of the duration-based approach that induces conditional 
arrival densities, as well as its connection to the conditional intensity approach.

Statistical models for 
duration process dynamics date back at least to \cite{wold1948stationary}
who proposed a first-order Markov chain with an additive model representation. 
Subsequent developments \citep{jacobs1977mixed,gaver1980first} investigate 
specific families for the duration process stationary marginal distribution. 
Since durations are positive-valued, a structure with an additive error process 
is in general restrictive. 
A popular class of models in finance is built from the autoregressive 
conditional duration (ACD) structure \citep{engle1998autoregressive}.
The ACD model assumes independent and identically distributed (i.i.d.) 
multiplicative errors for the durations, with each multiplicative factor 
modeled as a linear function of the past factors and durations. Extensions of 
this class of models provide additional flexibility through the multiplicative 
factor specification or the error distribution choice.
We refer to \cite{pacurar2008autoregressive} and \cite{bhogal2019conditional} for 
comprehensive reviews. For these models, the conditional intensity function is 
obtained by scaling the baseline hazard function with multiplicative factors.
The baseline hazard corresponds to the error distribution, typically chosen within 
a parametric family. A restriction of ACD models is their limited 
capacity to handle non-linear dynamics. %
Moreover, the distributional properties of likelihood estimators
are sensitive to the tail behavior of the durations
\citep{cavaliere2024tail}.
Regarding computation, the ACD model structure complicates inference when 
high-order memory is necessary, e.g., estimating the correlated multiplicative 
factors may require approximations \citep{strickland2006bayesian}.

A different approach to modeling duration dependence involves mixture transition 
distribution (MTD) models \citep{le1996modeling}, which describe the 
transition density of a time series as a weighted combination of first-order 
conditional densities for each one of a specified number of lags. 
\cite{hassan2006modeling} propose a bivariate MTD model for the joint conditional 
distribution of the duration and a continuous mark, i.e., a random variable
associated with the point events. \cite{hassan2013modelling} extend 
the model to include a discrete mark. However, these approaches 
do not investigate point process properties, such as stationarity, 
and require certain families of distributions for 
the duration and mark, which can be practically restrictive. 
\cite{hassan2006modeling} point out 
the difficulties of finding suitable parameterizations
to ensure model stability and prediction capability.

In this article, we introduce a class of temporal point processes 
that builds on the idea of 
describing duration process dynamics with MTD models. 
To use traditional high-order autoregressive models, a transformation of the 
durations or their conditional means is typically needed to handle the dependent, 
positive-valued durations.
This introduces the challenge of inference
under a constrained, possibly high-dimensional parameter space. 
For example, coefficients may need to be 
restricted to avoid negative-valued durations, and implementing 
stationarity conditions in practice
can be difficult, especially under the assumption of high-order
dependence.
The aforementioned work that uses 
MTD models attempts to handle the former issue, albeit under restrictive 
structures. A key contribution of the present article is the development of 
an MTD point process (MTDPP) constructive framework that provides flexible 
modeling of high-order
dynamics for the duration process, without parameter
constraints. The framework allows 
for various types of practically
relevant point patterns, such as those with self-excitation or self-regulation
effects. In addition, it provides an efficient inferential approach, as
the MTDPP likelihood evaluation grows linearly with the number of events. 
Thus, our proposed method is computationally scalable,
especially for large point patterns with high-order memory.

Within the MTDPP framework, we provide easily-implemented conditions to construct 
point processes that correspond to pre-specified families of
marginal distributions for the durations.
In addition, we obtain a limit result for the mean value 
function, analogous to that for renewal processes. 
The resulting class of models has identically distributed, 
dependent durations and can be interpreted as a 
class of renewal processes that incorporate high-order dependence
among durations.
To the best of our knowledge, the proposed model is the first to enable 
simultaneous modeling of high-order dependence and stationary durations, with 
computationally efficient inference.

Moreover, we develop an extension to handle duration clustering, 
based on a two-component mixture for the conditional duration density. 
In this setting, one component of the 
mixture corresponds to an independent durations model that 
accounts for external factors. 
The other component is an MTDPP that models self-excitation. Point patterns of this type 
can be found, for instance, in hospital emergency department visits of patients, where 
long durations may be observed between clusters of multiple visits in short 
bursts \citep{yang2018clustering}, and in financial markets where fluctuation can be 
caused by either external or internal processes \citep{filimonov2012quantifying}. 
The model extension accounts for the possibility of two different factors that may 
drive the point process dynamics.

The rest of the article is organized as follows. Section \ref{sec:mtdpp} introduces
the MTDPP framework, including study of model properties, approaches to constructing 
various types of MTDPP models, and the extension to cluster point processes. 
(Technical details and proofs of the theoretical results can be found 
in the Supplementary Material.)
Section \ref{sec:inference} develops the Bayesian 
model formulation, Markov chain Monte Carlo (MCMC) inference, an approach 
for predicting future events, and a model validation 
method. In Section \ref{sec:data}, we illustrate the proposed methodology with 
synthetic and real data examples. Finally, Section \ref{summary} concludes with 
a summary and discussion.

%
%--------------------------------------------------------------------------
%

\section{Temporal MTD point processes}
\label{sec:mtdpp}

We consider a temporal point process $N(t)$ defined on the positive half-line $\R^+$,
where $N(t) = \sum_{i\geq 1}\mathbbm{1}(\{t_i\leq t\})$ is a right-continuous,
integer-valued function, $t_1, t_2,\ldots\in\R^+$ denote the event times, 
and $\mathbbm{1}(A)$ is the indicator function for set $A$.
A temporal point process is usually modeled by its conditional intensity, 
$\la^*(t) \equiv \la(t\,|\,\iH_t) = \lim_{dt\rightarrow 0}E[dN(t)\,|\,\iH_t]/dt$, 
where $dN(t) = N(t+dt) - N(t)$, and 
$\iH_t$ is the process history up to but not including $t$.
The point process is said to have memory if $\la^*(t)$ depends on the process history.

The likelihood for a point process realization is typically written in terms
of $\la^*(t)$ which characterizes the point process \citep{daley2003introduction}.
An alternative way to characterize the point process probability structure
is to use the collection of conditional arrival densities, denoted as
$p^*_i(t) \equiv p_i(t\,|\,\iH_{t})$, supported on $(t_{i-1},\infty)$, 
with associated conditional survival functions $S^*_i(t) =$
$1 - \int_{t_{i-1}}^tp^*_i(u)du$.
When $i = 1$, $p_1^*(t) \equiv p_1(t)$ and $S_1^*(t) = 1 - \int_0^tp^*_1(u)du$,
where $p_1$ is the marginal density of the first event time.
The likelihood for point pattern $0 < t_1 < \ldots < t_n < T$ is given by 
\begin{equation}\label{eq:likhod_2}
p(t_1,\dots,t_n) \, = \,
\left\{ \prod_{i=1}^np^*_i(t_i) \right\} \,
\left\{ 1 - \int_{t_n}^Tp^*_{n+1}(u) \, du \right\} .
\end{equation}
The last component of \eqref{eq:likhod_2} defines the likelihood normalizing term,
i.e., the probability of no events occurring in the interval $(t_n, T]$. 
Since the normalizing term corresponds to a conditional %c.d.f.,
cumulative distribution function (c.d.f.),
it may be available in closed-form for particular model formulations 
for the conditional arrival densities.

Using the collection of conditional densities $p_i^*$ and survival functions $S_i^*$, 
we can define the hazard functions as $\la_i^*(t) =$ $p^*_i(t)/S^*_i(t)$, for 
$i = 1,\dots, n$. The hazard function is naturally interpreted as the conditional 
instantaneous event rate. Consequently, given the set of arrival times, we can write 
the conditional intensity of the process as $\la^*(t) =$
$\la_i^*(t),\; t_{i-1}<t\leq t_i,\;1 \leq i \leq n$.
Since $p^*_i(t) = \la^*_i(t)\exp(-\int_{t_{i-1}}^t\la^*_i(u) \, du)$, 
we can use \eqref{eq:likhod_2} to 
rewrite the likelihood in terms of $\la^*(t)$, 
$p(t_1,\dots,t_n) = $
$\left\{ \prod_{i=1}^n\la^*(t_i) \right\}
\exp\left(-\int_0^T\la^*(t) \, dt\right)$.
Point process modeling via the collection of 
$\{p_i^*(t)\}$ leads to locally integrable $\la^*(t)$.
In contrast, when $\la^*(t)$ is directly specified, 
its local integrability needs to be verified; moreover, 
likelihood evaluation involves integrating $\la^*(t)$, which 
is typically analytically intractable.

\subsection{Conditional duration density}
\label{sec:mtdpp_def}

Here, we introduce an approach to obtain 
conditional arrival densities by 
specifying the conditional densities of durations.
Consider an ordered sequence of arrival times $0 = t_0 < t_1 < \ldots < t _n < T$,
and denote the durations by $x_i = t_i - t_{i-1}$, for $i = 1,\dots, n$. 
The density of $x_i$ conditional on the 
past durations
is modeled as a weighted combination of first-order transition densities, 
each of which depends on a specific past duration, i.e.,
$f(x_i\,|\,x_{i-1},\dots,x_1) = \sum_{l=1}^L w_l \, f_l(x_i\,|\,x_{i-l})$,
where $w_l \geq 0$, for all $l$, and $\sum_{l=1}^Lw_l = 1$.
Transforming the conditional density of $x_i$ to that for 
$t_i = t_{i-1} + x_i$, for every $i$, creates conditional arrival densities 
that uniquely determine the point process.
The construction above is valid for durations $x_i$ with $i > L$. 
The formal MTDPP definition is given as follows.

\vspace{5pt}

\begin{definition}\label{def:mtdpp}
Let $N(t)$ be a temporal point process 
with event arrival times $t_1,t_2,\ldots\in\R^+$.
Denote by $f^*(t - \tN) \equiv f(t-\tN\,|\,\iH_{t})$ the 
conditional duration density. Then, $N(t)$ is said to be
an \textit{MTD point process} if
(i) for $\Ntl \geq L$, 
the conditional duration density 
\begin{equation}\label{eq:mtdpp2}
f^*(t - \tN) =  \sum_{l=1}^{L} w_l \, f_{l}(t - \tN\,|\,t_{\Ntl-l+1}- t_{\Ntl-l});
\end{equation}
(ii) for $1\leq \Ntl\leq L-1$, the conditional duration density 
\begin{equation}\label{eq:mtdpp1}
\begin{aligned}
f^*(t - \tN)  = 
\!\!\!\!\sum_{l=1}^{\Ntl-1}\!\!\!\! w_l \, f_{l}(t - \tN\,|\,
t_{\Ntl-l+1}- t_{\Ntl-l}) + 
\left( 1 - \!\!\!\!\! \sum_{r=1}^{\Ntl-1}\!\!\!\! w_r \right)
f_{\Ntl}(t-\tN\,|\,t_1);
\end{aligned}
\end{equation}
(iii) for the initial event, where $\Ntl = 0$, $t_1\sim f_0(t_1)$.
In both \eqref{eq:mtdpp2} and \eqref{eq:mtdpp1}, the weights $w_l\geq 0$, 
for $l = 1,\dots, L$, with $\sum_{l=1}^Lw_l = 1$.
The marginal density $f_0$ and the conditional densities
$f_l$, $l = 1,\dots, L$, are supported on the set of non-negative real numbers.
\end{definition}

\vspace{5pt}

\begin{remark}
The marginal density $f_0$ and the conditional density $f^*(t-\tN)$
define the conditional arrival densities 
$\{p^*_i\}$ for point pattern $\{t_i\}_{i=1}^n$, 
by taking $p^*_1(t) \equiv f_0(t)$ 
and $p_i^*(t) \equiv f^*(t - t_{i-1})$, $t > t_{i-1}$, 
for $i = 2,\dots, n$.
Thus, specification of densities $f_0(t)$ and $f^*(t-\tN)$ suffices to 
characterize the probability structure of the resulting MTDPP.
\end{remark}

\begin{remark}
The two different expressions \eqref{eq:mtdpp2} and \eqref{eq:mtdpp1} 
for the conditional duration density allow us to study stationarity 
conditions for the MTDPP (Section \ref{sec:properties}). 
For brevity, we will use \eqref{eq:mtdpp2} 
to discuss model properties throughout the rest of the article.
For inference on the unknown weights
$\{w_1,\dots,w_L\}$ and the parameters of component densities $\{f_1,\dots,f_L\}$, 
Equation \eqref{eq:mtdpp2} is the relevant expression,
since we work with a conditional likelihood. 
Moreover, the mixture model structure enables an efficient computational 
scheme for high-order dynamics
(Section \ref{sec:inference}),
without constraints on the parameter space.
\end{remark}

The specification of the conditional density $f^*(t-\tN)$ involves 
the first-order conditional density $f_l$, for $l = 1,\dots,L$.
Following \cite{zheng2021construction}, we build $f_l$ 
from a bivariate positive-valued random vector $(U_l,V_l)$ with joint density
$f_{U_l,V_l}$ and marginals $f_{U_l}$ and $f_{V_l}$, by taking $f_l\equiv f_{U_l|V_l}$ 
as the conditional density of $U_l$ given $V_l$. In general, there are two
strategies to define 
the joint density $f_{U_l,V_l}$,
one through specific marginal densities, and the other through a pair of 
compatible conditional densities \citep{arnold1999conditional}. The two
conditional densities $f_{U_l|V_l}$ and $f_{V_l|U_l}$ are said to be compatible 
if there exists a bivariate density with its conditionals given by $f_{U_l|V_l}$ 
and $f_{V_l|U_l}$. We note that each strategy has its own benefits depending on 
the modeling objective. In Section \ref{sec:mtdpp-constrction}, we illustrate 
construction 
of the conditional densities $f_l$ with various examples for different goals.

An important result of using the MTD model for the conditional duration 
density is a mixture formulation for the implied conditional intensity 
$\la^*(t) \equiv h^*(t-\tN) = f^*(t-\tN)/S^*(t-\tN)$, where $h^*(t-\tN)$ and 
$S^*(t-\tN)$ are the hazard and survival function, respectively, associated 
with $f^*(t-\tN)$. 
Similarly, for the $l$th component, 
we have that $h_l(u\,|\,v) = f_l(u\,|\,v)/S_l(u\,|\,v)$,
where $h_l$ and $S_l$ are, respectively, the hazard and survival function associated
with $f_l$. 
We can write the conditional intensity $\la^*(t)$ as
\begin{equation}\label{eq:cond_intens}
\la^*(t) \, = \, \sum_{l=1}^L w_l^*(t) \, h_l(t-\tN\,|\,t_{\Ntl-l+1} - t_{\Ntl-l}),
\end{equation}
with weights
$w^*_l(t) =$ $w_l \, S_{l}(t - \tN\,|\,t_{\Ntl-l+1} - t_{\Ntl-l})/S^*(t-\tN)$, where
$S^*(t-\tN) = $ $\sum_{l=1}^L w_l \, S_{l}(t -\tN\,|\,t_{\Ntl-l+1} - t_{\Ntl-l})$.
Note that $w^*_l(t)\geq 0$ and $\sum_{l=1}^Lw^*_l(t)=1$ for all $t$.
The time-dependent weights, $w^*_l(t)$, provide local adjustment, and thus the
flexibility to accommodate a wide range of conditional intensity shapes.

In addition to model flexibility, the mixture formulation of $\la^*(t)$ guides 
modeling choice. Each mixture component $h_l$ is a first-order hazard function.
For example, if we select $f_l$ such that $h_l\leq B_l$, for constant $B_l > 0$,
and for all $l$, then $\la^*(t) \leq \sum_{l=1}^{L} w_l^*(t) \, B_l$, for every $t$.
Moreover, choosing $f_l$ such that $h_l$ has certain shapes results in 
particular types of point processes. A point process is said
to be self-exciting if a new arrival causes the conditional intensity to jump,
and is called self-regulating (or self-correcting) if a new arrival causes the 
conditional intensity to drop. 
If $h_l$ monotonically decreases, for all $l$, the resulting MTDPP is self-exciting; 
see Section \ref{sec:mtdpp-constrction} for details.

\subsection{Model properties}
\label{sec:properties}

We first investigate stationarity, in particular,  
conditions for first-order strict stationarity, such that the 
MTDPP has a stationary marginal density, $f_X$, for the duration process.
The constructive approach to build $f_l$ as the conditional density 
of $U_l$ given $V_l$
based on random vector $(U_l,V_l)$ allows us to obtain a stationary marginal 
density $f_X$, using the approach in \cite{zheng2021construction}. 
We summarize the conditions in the following proposition.

\begin{prop}\label{prop:sta_mtdpp}
Consider an MTD point process $N(t)$ with event arrival times 
$0 < t_1 < t_2 < \dots$, where $t_i\in\R^+$, $i\geq 1$.
Let $\{X_i: i\geq 1\}$ be the duration process, 
such that $t_1 = x_1$, and $t_i = t_{i-1} + x_i$, for $i\geq 2$.
The duration process has a stationary marginal density $f_X$ if: 
(i) $t_1\sim f_0(t_1)\equiv f_X(t_1)$ for $\Ntl=0$;
(ii) the density $f_l$ in $\eqref{eq:mtdpp2}$ and \eqref{eq:mtdpp1} is 
taken to be the conditional density $f_{U_l|V_l}$ of a bivariate positive-valued 
random vector $(U_l,V_l)$ with marginal densities $f_{U_l}$ and $f_{V_l}$, 
such that $f_{U_l}(x) = f_{V_l}(x) = f_X(x)$, for all $x$ and for all $l$.
\end{prop}

We refer to the class of MTDPPs that satisfies the conditions in 
Proposition \ref{prop:sta_mtdpp} as stationary MTDPPs. 
Compared to renewal processes that have i.i.d. durations,
stationary MTDPPs can be interpreted as dependent renewal 
processes, where the durations are identically distributed but
dependent up to $L$-order. In fact, the independence assumption of
classical renewal processes is often restrictive \citep{coen2019modelling},
and thus incorporating high-order dependence among durations may be necessary. 
An example that involves the analysis of the recurrence interval distribution 
is presented in Section \ref{sec:recur}.

In renewal theory, the rate of renewals in the long run corresponds to 
$\lim_{t\rightarrow\infty}N(t)/t$,
which equals $1/\mu$, provided $\mu$ is finite. Here, $\mu$ 
denotes the mean under the stationary marginal duration distribution. We obtain an 
analogous result for stationary MTDPPs.

\begin{theo}\label{theo:limit}
Consider a stationary MTD point process $N(t)$ built from 
Proposition \ref{prop:sta_mtdpp}, with the additional structure for 
condition (ii) that the $(U_l,V_l)$ are identically distributed as 
$(U,V)$, for $l = 1,\dots,L$, with $f_{U\,|\,V}(u\,|\,v)$ strictly positive 
and continuous for all $u,v$. 
Suppose that there exists a function $\tau(x)\geq 1$, for $x\geq0$, which 
is everywhere finite, and a compact set $C \subset [0,\infty)$, such that 
for some $\beta< 1$, $b<\infty$, $E[\tau(U)\,|\,V = v] \leq$
$\beta \tau(v) + b\mathbbm{1}_C(v)$ is satisfied. Then, 
as $t\rightarrow\infty$, $N(t)/t\rightarrow 1/\mu$ almost surely,
provided $\mu<\infty$.
\end{theo}

Note that implementing condition (ii) of Proposition \ref{prop:sta_mtdpp} 
(i.e., $f_{U_l}(x) = f_{V_l}(x) = f_X(x)$, for all $l$)
often results in 
$f_{U_l\,|\,V_l}(u\,|\,v)=f_{U\,|\,V}(u\,|\,v)$, for all $l$;
see Section \ref{sec:mtdpp-constrction}.
Theorem \ref{theo:limit} thus provides practical sufficient conditions 
for constructing MTDPPs that ensure the almost sure convergence of $N(t)/t$,
mainly through verifying an inequality for the conditional expectation
of $U$ given $V$. We refer to the Supplementary Material for 
two examples that illustrate the application of Theorem 1, 
based on models introduced in Section \ref{sec:mtdpp-constrction}.

Similar to the classical renewal theorem, Theorem \ref{theo:limit} 
provides information about the average renewal rate, the difference being 
that MTDPPs allow dependence among waiting times between renewals.
Of interest is also the asymptotic behavior of the mean-value 
function, $m(t) = E[N(t)]$, i.e., $\lim_{t\rightarrow\infty}m(t)/t$.
In general, function $m(t)$ is not analytically available for MTDPPs. 
However, an upper bound for 
$\lim_{t\rightarrow\infty}m(t)/t$ can be obtained for MTDPPs with 
bounded component hazard functions (see Supplementary Material).

Finally, note that 
the structured mixture formulation of the MTDPP conditional duration density
distinguishes it from standard finite mixture models. The mixture components
of the conditional duration density are ordered by lagged durations, as lag $l$ 
enters into the $l$-th component, for $l = 1,\dots, L$. Such a formulation results 
in likelihood asymmetry and indicates a single labeling of the components. Thus, 
identifiability for MTDPP models is generally not as major a challenge as for 
traditional finite mixture models. Study of identifiability can be conducted on 
a case-by-case basis; we refer to the Supplementary Material for specific results 
based on the models introduced in the next section.

\subsection{Construction of MTD point processes}
\label{sec:mtdpp-constrction}

We provide guidance to construct MTDPPs, focusing on the conditional density
$f_l$. As discussed in Section \ref{sec:mtdpp_def}, we derive $f_l$
from a bivariate density $f_{U_l,V_l}$, which can be specified through 
compatible conditionals $f_{U_l|V_l}$ and $f_{V_l|U_l}$, 
or through marginals $f_{U_l}$ and $f_{V_l}$. The former is 
particularly useful when the objective is to construct self-exciting or
self-regulating MTDPPs, by choosing $f_{U_l|V_l}$ such that its associated hazard 
function is monotonically decreasing or increasing, respectively. 
We illustrate this approach in Example 1.

In light of Proposition \ref{prop:sta_mtdpp}, the strategy of constructing MTDPPs 
through pre-specified marginals is natural for modeling dependent renewal processes.
This strategy is also useful when interest lies in the shape 
of the marginal hazard function. For example, \cite{grammig2000non} 
point out that it may be more appropriate to consider non-monotonic hazard 
functions for modeling financial duration processes. We implement this MTDPP 
construction approach using bivariate copula functions for $f_{U_l,V_l}$,
illustrated in Example 2.

\paragraph{\normalfont\textbf{Example 1: Self-exciting and self-regulating MTDPPs}}\mbox{}

\vspace{5pt}

\noindent
We build a self-exciting MTDPP based on a new class
of bivariate distributions (Proposition \ref{prop:bi-slo}), which are derived from 
the pair of Lomax conditionals in \cite{arnold1999conditional}. 
The Lomax distribution is a shifted version of the Pareto Type I distribution, such 
that the support is $\mathbb{R}^{+}$. 
The density function is given by 
$a b^{-1}(1 + u b^{-1})^{-(a+1)}$, for $u > 0$, 
with a monotonically decreasing hazard function $a(b + u)^{-1}$,
where $a>0$ is the shape parameter 
and $b>0$ the scale parameter. Hereafter, we use $P(\cdot \,|\, b, a)$ to denote,
depending on the context, either the density function or the distribution for a 
Lomax random variable (we follow the same notation for other distributions).

\begin{prop}\label{prop:bi-slo}
Consider a positive-valued random vector $(X,Y)$ with 
bivariate Lomax density 
$f_{X,Y}(x,y)\propto (\la_0 + \la_1x + \la_2y)^{-(\alpha+1)}$. Let $(U,V) = 
(\alpha X, \alpha Y)$. Then, the bivariate random vector $(U,V)$ has 
conditionals $f_{U|V}(u|v) = P(u\,|\,\la_1^{-1}(\alpha\la_0+\la_2v), \alpha)$ 
and $f_{V|U}(v|u) = P(v\,|\,\la_2^{-1}(\alpha\la_0+\la_1u), \alpha)$, and 
marginals $f_{U}(u) = P(u\,|\,\la_1^{-1}\alpha\la_0, \alpha-1)$
and $f_{V}(v) = P(v\,|\,\la_2^{-1}\alpha\la_0, \alpha-1)$,
where $\la_0>0$, $\la_1>0$, $\la_2>0$, and $\alpha > 1$.
\end{prop}

Since $(X,Y)$ is scaled by $\alpha$,
we refer to the distribution of $(U,V)$ as the bivariate 
scaled-Lomax distribution. The difference 
with the original Lomax distribution is
that the shape parameter of the scaled-Lomax distribution is part of the 
scale parameter. 
Similar to the Lomax distribution, the scaled-Lomax distribution
has a monotonically decreasing hazard function, and thus both distributions
can be used to construct self-exciting MTDPPs.

We start with the bivariate scaled-Lomax densities 
$f_{U_l,V_l}$ with parameters $\alpha_l, \la_{0l}, \la_{1l}, \la_{2l}$. 
We simplify the parameterization by setting $\la_l = \la_{1l} = \la_{2l}$, 
and letting $\phi_l = \la_{0l}/\la_l$, which yields 
$f_{U_l}(u) = P(u\,|\,\alpha_l\phi_l, \alpha_l-1)$, 
$f_{V_l}(v) = P(v\,|\,\alpha_l\phi_l, \alpha_l-1)$, and
$f_{U_l|V_l}(u|v) =$ $P(u\,|\,\alpha_l\phi_l + v, \alpha_l)$,
where $\phi_l > 0$ and $\alpha_l>1$, for all $l$.
Taking $f_l \equiv f_{U_l|V_l}$, we obtain the conditional duration density,
\begin{equation}\label{eq:slmtdpp}
f^*(t-t_{\Ntl}) \, = \, \sum_{l=1}^L w_l \,
P(t - \tN\,|\,\alpha_l\phi_l + t_{\Ntl-l+1} - t_{\Ntl-l}, \alpha_l).
\end{equation}
Then, specifying $f_0$ for the initial event,
we complete the construction for the scaled-Lomax MTDPP, which 
is a self-exciting point process.

If we let $f_0(t) =$ $P(t\,|\,\alpha_1\phi_1, \alpha_1-1)$, and 
set $\alpha_l = \alpha$ and 
$\phi_l = \phi$, for $l = 1,\dots, L$, then both conditions in Proposition \ref{prop:sta_mtdpp}
are satisfied, and the model admits a stationary duration density
$f_X(x) = P(x\,|\,\alpha\phi,\alpha-1)$.
Furthermore, this model
satisfies the conditions in Theorem \ref{theo:limit}, provided 
$\alpha > 2$ (see Supplementary Material).
The next result describes the limiting behavior of the stationary scaled-Lomax 
MTDPP conditional duration distribution.

\begin{prop}\label{prop:limitexp}
Consider the stationary scaled-Lomax MTDPP with marginal duration density
$P(\alpha\phi,\alpha-1)$. As $\alpha \rightarrow \infty$, the 
conditional duration distribution converges in distribution to the 
exponential distribution with rate parameter $\phi^{-1}$.
\end{prop}

According to \eqref{eq:cond_intens}, the conditional intensity of the 
scaled-Lomax MTDPP can be expressed as $\la^*(t) = $
$\sum_{l=1}^Lw^*_l(t) \,
\{ \phi_l + \alpha_l^{-1}(t - \tN + t_{\Ntl-l+1} - t_{\Ntl-l}) \}^{-1}$.
For each $l$, the $l$th component of the conditional intensity is bounded above by 
$\phi_l^{-1}$. Thus, $\la^*(t) \leq \sum_{l=1}^L w^*_l(t) \, \phi_l^{-1}$, for 
any $t$, and $\lim_{t\rightarrow\infty}m(t)/t\leq \sum_{l=1}^L w_l \, \phi_l^{-1}$;
see the Supplementary Material for further details.

Finally, we note that if we remove $\alpha$ from the scale parameter 
component in \eqref{eq:slmtdpp}, i.e.,
$f_l(u\,|\,v) = P(u\,|\,\phi_l + v, \alpha_l)$,
then $f_l$ corresponds to the bivariate
Lomax distribution of \cite{arnold1999conditional}. 
If, furthermore, we take $f_0(t) = P(t\,|\,\phi_1,\alpha_1-1)$,
the resulting point
process is referred to as the Lomax MTDPP, which is also
a self-exciting point process.
A self-regulating MTDPP can be constructed 
through compatible conditionals associated with monotonically increasing hazard functions,
such as gamma conditionals; see \cite{arnold1999conditional} for relevant bivariate 
distributions.

\paragraph{\normalfont\textbf{Example 2: Dependent renewal MTDPPs}}\mbox{}

\vspace{5pt}

\noindent
Motivated by Proposition \ref{prop:sta_mtdpp}, we can select a stationary 
density $f_X$, and take $f_{U_l}(x) = f_{V_l}(x) = f_X(x)$, 
for every $x$ and for all $l$. Given the desired
marginals, what remains is to specify the joint density $f_{U_l,V_l}$ to 
obtain $f_{U_l|V_l}$. In this example, we introduce the idea of specifying a
bivariate copula function $C:[0,1]^2\rightarrow[0,1]$ to build $f_{U_l,V_l}$, 
which provides a general scheme to construct MTDPPs given a stationary 
marginal $f_X$.

Let $F_{U_l,V_l}$ be the joint c.d.f. of the 
random vector $(U_l,V_l)$, and denote by $F_{U_l},F_{V_l}$ 
the corresponding marginal c.d.f.s. Given $F_{U_l}$ and $F_{V_l}$,
there exists a unique copula $C_l$ such that
$F_{U_l,V_l}(u,v) = C_l(F_{U_l}(u), F_{V_l}(v))$, and the joint density
$f_{U_l,V_l}$ is given by $c_{l}(u,v)f_{U_l}(u)f_{V_l}(v)$,
where $c_l(u,v) = 
\partial^{2} C(F_{U_l}(u),F_{V_l}(v))/(\partial F_{U_l}\partial F_{V_l})$ 
is the copula density \citep{sklar1959fonctions}. Hence, based on a marginal 
duration density $f_X$ and a copula $C_l$, we have
$f_l(u) \equiv f_{U_l|V_l}(u\,|\,v) = c_l(u,v)f_X(u)$. 
The conditional duration density of the resulting MTDPP is 
\begin{equation}\label{eq:copula_mtdpp}
f^*(t-\tN) \, = \, \sum_{l=1}^L w_l \, c_l(t-\tN, t_{\Ntl-l+1}-t_{\Ntl-l})
\, f_X(t-\tN).
\end{equation}
We refer to this class of models as copula MTDPPs. Their conditional intensity 
in \eqref{eq:cond_intens} involves hazard function components
$h_l(u\,|\,v) =f_l(u\,|\,v)/S_l(u\,|\,v)$, where 
$S_l(u\,|\,v) = 1 - \partial C_l(F_{U_l}(u), F_{V_l}(v))/\partial F_{V_l}$. 
A closed-form expression for $h_l$ relies on the specific copula function
(e.g., a Gaussian copula leads to an analytically intractable $h_l$).

For certain copulas, the conditional and marginal densities belong to the same 
family of distributions. As a particular example, consider the three-parameter 
Burr density,  $\mathrm{Burr}(x\,|\,\ga,\la,\psi) =$
$\psi\ga x^{\ga-1}\la^{-\ga} \{ 1 + (x/\la)^{\ga} \}^{-(\psi+1)}$, for $x > 0$, 
with shape parameters $\ga>0$, $\psi >0$, and scale parameter $\la > 0$.
The corresponding hazard function is monotonically decreasing when $0< \ga \leq 1$, 
and hump-shaped when $\ga > 1$.
In the Supplementary Material, we derive a bivariate Burr distribution built from 
Burr marginals and a heavy right tail copula, such that the conditionals are also 
Burr distributions.

To construct a class of Burr MTDPPs, for each $l$, we specify $f_{U_l,V_l}$ with
the bivariate Burr density such that the marginals are $f_{U_l}(x) =$ $f_{V_l}(x) =$ 
$f_X(x) =$ $\mathrm{Burr}(x\,|\,\ga,\la,\kappa-1)$, where $\kappa > 1$.
Then, for all $l$, the conditional density,  
$f_{U_l|V_l}(u\,|\,v) = \mathrm{Burr}(u\,|\,\ga,\tilde\la(v), \kappa)$,
where $\tilde\la(v) = (\la^{\ga} + v^{\ga})^{1/\ga}$. Hence, the 
conditional duration density of the Burr MTDPP is 
\begin{equation}\label{eq:bmtdpp}
f^*(t-\tN) \, = \, \sum_{l=1}^L w_l \, \mathrm{Burr}(t-\tN\,|\,\ga,
\tilde\la(t_{\Ntl-l+1}-t_{\Ntl-l}), \kappa).
\end{equation}
If we let $f_0(t) = \mathrm{Burr}(t\,|\,\ga,\la,\kappa-1)$, the Burr MTDPP 
has stationary marginal $f_X(x) = \mathrm{Burr}(x\,|\,\ga,\la,\kappa-1)$.
Moreover, as shown in the Supplementary Material, when 
$\kappa > \max\{2,1+1/\gamma\}$, the stationary Burr MTDPP
satisfies the conditions in Theorem \ref{theo:limit}.

The stationary Burr MTDPP model includes as a special case (with $\ga = 1$)
the Lomax MTDPP with marginal $P(x\,|\,\la, \kappa - 1)$. 
Moreover, when $\kappa = 2$, it reduces to a model with log-logistic 
stationary marginal density $LL(x\,|\,\ga,\la) =
\ga x^{\ga-1}\la^{-\ga} \{ 1 + (x/\la)^{\ga} \}^{-2}$, for $x>0$.

\subsection{Extension to MTD cluster point processes}
\label{sec: mtdcpp}

In practice, there may exist different factors that drive duration 
process dynamics. As an example from hydrology, durations of dry 
spells can be classified into two types, corresponding to cyclonic and 
anticyclonic weather \citep{cowpertwait2001renewal}. A point process model 
for such data should be able to account for the two weather types, as the 
lengths of the dry spells may be distinctly different. 
Here, we extend MTDPPs to MTD cluster point processes (MTDCPPs), 
based on a two-component mixture model.

\begin{definition}\label{def:mtdcpp}
Let $N(t)$ be a temporal point process defined on $\R^+$ with 
event arrival times $t_1,t_2,\ldots\in\R^+$.
Denote by $f_C^*(t-\tN)$ the conditional duration density.
Then, $N(t)$ is said to be an \textit{MTD cluster point process} if 
(i) $t_1\sim f_I(t_1)$ for $\Ntl = 0$; 
(ii) for $\Ntl\geq 1$, the conditional duration density is given by
\vspace{-12pt}
\begin{equation}\label{eq:mtdcpp}
\begin{aligned}
f_C^*(t - \tN) \, = \, \pi_0 \, f_I(t-\tN) + (1-\pi_0) \, f^*(t-\tN),
\end{aligned}
\end{equation}
where $0\leq\pi_0\leq 1$, $f_I(t)$ is a density on $\R^+$,
and $f^*(t - \tN)$ is the conditional duration density of a
self-exciting MTD point process.
\end{definition}

Similar to the MTDPP, we use densities $f_I$ and $f_C^*$ to define the
conditional arrival densities $p^*_i$ of event time $t_i$, for an observed point
pattern $\{t_i\}_{i=1}^n$, by taking $p^*_1(t) = f_I(t)$ and 
$p_i^*(t) = f_C^*(t - t_{i-1})$, $t > t_{i-1}$, for $i = 2,\dots, n$.
Equation \eqref{eq:mtdcpp} specifies the generating mechanism of a 
point pattern for two distinct types of intervals: $f_I$ generates an interval
independent of previous ones (e.g., a typical seasonal interval 
between rainfall or flood occurrences), 
while $f^*$ models dependency patterns among 
shorter durations triggered by preceding events,
forming clusters (e.g., rainfall or flood clusters).
An example involving market endogeneity for
studying financial market microstructure, including a comparison with 
an alternative model, is given in 
Section \ref{sec:data-fx}. When $\pi_0 = 1$, the MTDCPP 
reduces to a renewal process of single-type intervals;
furthermore, if $f_I$ corresponds to the exponential distribution, 
the model reduces to a homogeneous Poisson process. When $\pi_0 = 0$, 
the model becomes a self-exciting MTDPP.

Let $h_I$ and $S_I$ be the hazard and survival functions associated with $f_I$. 
The conditional intensity 
of the MTDCPP, denoted as $\la_C^*(t)$, extends the mixture form 
in \eqref{eq:cond_intens} as follows:
\begin{equation}\label{eq:mtdcpp_cond_intens}
\begin{aligned}
\la_C^*(t) 
& = \pi_0(t) \, h_I(t-\tN) \, + \, 
\sum_{l=1}^L \pi_l(t) \, h_l(t-\tN\,|\,t_{\Ntl-l+1}-t_{\Ntl-l}),
\end{aligned}
\end{equation}
where 
$\pi_0(t) = \pi_0 \, S_I(t-\tN)/S_C^*(t-\tN),\;$ 
$\pi_l(t) = (1-\pi_0) \, w_l \, S_l(t-\tN\,|\,t_{\Ntl-l+1}-t_{\Ntl-l})/S_C^*(t-\tN)$,
for $l = 1,\dots, L$, $S_C^*(t-\tN) = \pi_0S_I(t-\tN) + (1-\pi_0)S^*(t-\tN)$,
and we have that $\pi_l(t)\geq 0$, for $l = 0,\dots, L$, and 
$\sum_{l=0}^L \pi_l(t) = 1$, for all $t$.

Compared to the MTDPP conditional intensity function, 
the MTDCPP conditional intensity has an extra term contributed 
by component $f_I$, with appropriately renormalized time-dependent 
weights. If we take an exponential density with rate parameter 
$\mu$ for $f_I$, and a Lomax MTDPP for $f^{*}$, 
the resulting model is referred to as the Lomax MTDCPP. Note that we 
consider the Lomax instead of the 
scaled-Lomax MTDPP to avoid potential identifiability issues, 
indicated by Proposition \ref{prop:limitexp}.

%
%--------------------------------------------------------------------------
%

\section{Bayesian implementation}
\label{sec:inference}

\subsection{Conditional likelihood and prior specification}

Let $0 = t_0 < t_1 < \ldots < t_n < T$ be the observed point pattern, 
with durations $x_i = t_i - t_{i-1}$, for $i = 1,\dots,n$.
We outline the approach to posterior inference for MTDPP models based on a 
conditional likelihood. The Supplementary Material includes the corresponding 
details for MTDCPP models.

The point process likelihood can be expressed equivalently using 
event times $\{t_i\}_{i=1}^n$ or durations
$\{x_i\}_{i=1}^n$. For brevity, we use the latter,
and take $\tx_{n+1} = T - t_n$.
Combining \eqref{eq:likhod_2} and \eqref{eq:mtdpp2}, 
the likelihood conditional on $(x_1,\dots,x_L)$ is 
$$
p(x_1,\dots,x_n,\tx_{n+1};\btheta, \bm w) \propto 
\left( \sum_{l=1}^L w_l \, S_l(\tx_{n+1}\,|\,x_{n+1-l}, \btheta_l) \right) \,
\prod_{i=L+1}^n
\left\{ \sum_{l=1}^L w_l \, f_l(x_i\,|\,x_{i-l},\btheta_l) \right\}
$$
where $f_l(x_i\,|\,x_{i-l},\btheta_l)$ corresponds to the conditional density
$f_l$ in \eqref{eq:mtdpp2} with parameters $\btheta_l$,
$S_l(\tx_{n+1}\,|\,x_{n+1-l},\btheta_l) =
1 - \int_0^{\tx_{n+1}}f_l(u\,|\,x_{n+1-l},\btheta_l) \, du$, 
$\btheta = \{\btheta_l\}_{l=1}^L$, and $\bm w = (w_1,\dots, w_L)^\top$.
The Bayesian model formulation involves priors for $\btheta$ and $\bm w$, where 
the prior for $\btheta$ depends on the choice of the component densities 
$f_l$, $l = 1,\ldots, L$.

We take the weights $w_l$ as increments of a c.d.f. $G$, i.e.,
$w_l = G(l/L) - G((l-1)/L)$, for $l = 1,\dots,L$, where 
$G$ has support on the unit interval. Flexible estimation of the weights 
depends on the shape of $G$. Thus, we consider a Dirichlet process (DP) prior
for $G$, denoted as $\text{DP}(\alpha_0, G_0)$, where 
$G_0 = \text{Beta}(a_0, b_0)$ is the baseline c.d.f., and $\alpha_0 > 0$ is the 
precision parameter. 
Based on its original definition \citep{ferguson1973bayesian}, the DP implies a Dirichlet 
distribution, $\mathrm{Dir}(\bm w\,|\,\alpha_0a_1,\dots,\alpha_0a_L)$, for the vector of 
weights $\bm w$, where $a_l = G_0(l/L) - G_0((l-1)/L)$, for $l=1,\dots,L$.
The prior expectation is $E(\bm{w}) = (a_1,\dots,a_L)^\top$. We denote this prior for 
the weights as $\mathrm{CDP}(\cdot\,|\,\alpha_0, a_0, b_0)$.

As discussed in Section \ref{sec:properties}, the mixture components of the MTDPP 
conditional duration density are ordered by lagged durations. The structured mixture model 
formulation motivates the approach to define the prior for the weights through 
distribution $G$. Note that, although the DP-based model for the weights gives rise 
to a Dirichlet distribution, the approach differs from directly assigning a Dirichlet 
prior distribution. The DP prior supports general distributional shapes for $G$, and 
it thus allows for flexible estimation of the weights, as demonstrated with 
the data example of Section \ref{sec:recur}, as well as for incorporating prior 
information (e.g., directionality), as outlined below.

As it is natural to assume that near lagged durations contribute more than distant 
ones, our default choice for $G_0$ is $\mathrm{Beta}(1, b_0)$, with $b_0 > 1$.
Such a choice yields a decreasing density for $G_0$, and thus given the regular 
cutoff points, the weights exhibit a decreasing pattern in prior expectation. 
Given $L$, a larger $b_0$ leads to a greater penalization of the weights 
for distant lags towards zero.
The DP precision parameter $\alpha_0$ represents the degree of prior belief;
as $\alpha_0$ increases, DP realizations for $G$ are less variable around $G_0$. 
Our default choice is $\alpha_0 = 5$, which suggests a moderate prior belief in 
a decreasing pattern for the weights, 
while allowing for a certain amount of variation. The Supplementary Material
includes results from prior sensitivity analysis for the mixture weights, 
using a simulation study.

The (almost sure) discreteness of the DP prior for $G$ induces sparsity 
in the weights.
This supports the strategy of fitting an over-specified mixture model, viewing 
$L$ as an upper bound on the number of effective components \citep{gelman2013bayesian}.
We select $L$ conservatively 
such that the weights of the last few lags are close to zero a posteriori, i.e., 
the nearer lags adequately account for process dependence.
In practice, the autocorrelation function (ACF) and partial autocorrelation 
function (PACF) of the observed duration time series can be used to guide the 
choice of $L$, with a sensitivity analysis to ensure that the selected $L$ is
a reasonable upper bound. Results from this strategy, as implemented for the 
data example of Section \ref{sec:recur}, are provided in the Supplementary Material.

\subsection{Posterior simulation}
\label{sec: mcmc}

We outline an MCMC method (Metropolis-within-Gibbs) for posterior simulation.
Similar to finite mixtures, we augment the model with configuration 
variables $\ell_i$, taking values in $\{1, \dots, L\}$, with discrete distribution 
$\sum_{l=1}^L w_l \, \delta_l(\ell_i)$, where $\delta_l(\ell_i) = 1$
if $\ell_i = l$ and $0$ otherwise, for $i = L+1,\dots, n+1$. 
The posterior distribution for the augmented model is 
\vspace{-15pt}
$$
\begin{aligned}
& p(\btheta, \bm w, \ell_{L+1},\dots,\ell_{n+1}\,|\,x_1,\dots,x_n,\tx_{n+1}) 
\propto  \mathrm{Dir}(\bm w\,|\,\alpha_0a_1,\dots,\alpha_0a_L) \times
\prod_{l=1}^Lp(\btheta_l)  \\
& \times 
\left( S_{\ell_{n+1}}(\tx_{n+1}\,|\,x_{n+1-\ell_{n+1}},\btheta_{\ell_{n+1}})
\sum_{l=1}^L w_l \, \delta_l(\ell_{n+1}) \right) \,
\prod_{i=L+1}^n \left\{ f_{\ell_i}(x_i \,|\,x_{i-\ell_i},\btheta_{\ell_i})
\sum_{l=1}^L w_l \, \delta_l(\ell_i) \right\}.
\end{aligned}
$$
The posterior full conditional distribution 
of $\ell_i$ is a discrete distribution on $\{ 1,...,L \}$ with probabilities 
proportional to $w_l \, f_l(x_i\,|\,x_{i-l},\btheta_l)$, for $i = L+1,\dots,n$,
and with probabilities proportional to 
$w_l \, S_l(\tx_{n+1}\,|\,x_{n+1-l},\btheta_l)$, for $i = n+1$. 
Given the configuration variables, we update the weights $\bm{w}$ with a 
Dirichlet posterior full conditional distribution 
with parameter vector $(\alpha_0 a_1 + M_1, \dots, \alpha_0 a_L + M_L)^\top$,
where $M_l = |\{i:\ell_i = l,\,L+1\leq i\leq n+1\}|$, for
$l = 1,...,L$, and $|\{\cdot\}|$ returns the size of set $\{\cdot\}$.
The updates for parameters $\btheta_l$ depend on the component densities $f_l$,
$l = 1,\dots, L$. In the Supplementary Material, we provide details of the MCMC 
algorithms for specific models implemented in Section \ref{sec:data}.

\subsection{Computational complexity}

We provide rough estimates 
of the floating point operation (flop) counts per iteration
of the Metropolis-within-Gibbs sampler for the MTDPP model, expressed
in terms of sample size $n$ and order $L$, where $L\ll n$. 
Updating $\bm w$ involves $O(n)$ flops for counting $\{M_l\}$, and %typically 
$O(L)$ flops for sampling from a Dirichlet distribution.
Sampling each $\ell_i$ requires evaluating a mixture 
of $L$ components, resulting in a total of $O(nL)$ flops. 
Subsequently, sampling each element of the parameter vector $\btheta_l$ requires 
$O(n)$ flops, for $l = 1,\dots, L$.

The total flop count is thus of the
order $O(Ln)$, similar to
that of standard ACD models that are also duration-based.
The latter require $O(L'n)$ flops for likelihood evaluation, where $L'$ is
the total number of lagged durations and lagged conditional means,
although computational burden increases
when conditional means are modeled as latent variables 
\citep[e.g., stochastic ACD;][]{strickland2006bayesian}.
In general, both duration-based models offer scalability over a standard
Hawkes process model, for which likelihood evaluation involves $O(n^2)$ flops.
Empirical comparison of the computation time of our models with 
alternative models can be found in the Supplementary Material.

\subsection{Inference, model checking, and prediction}
\label{sec:mc}

Using the MCMC algorithm, we obtain posterior samples that provide full 
inference for any functional of the point process. For example, given the
posterior draws for the model parameters, we obtain posterior 
realizations for the conditional intensity function by evaluating \eqref{eq:cond_intens} 
or \eqref{eq:mtdcpp_cond_intens} over a grid of time points.
Similarly, for stationary MTDPPs, we can obtain point and interval estimates
for the marginal duration density.

For model assessment, we use 
the time-rescaling theorem \citep{daley2003introduction}, 
according to which $\{\Lambda^*(t_i)\}_{i=1}^n$ is a realization from a unit rate Poisson 
process, where $\Lambda^*(t) =
\int_0^t\la^*(u)du$ is the conditional cumulative intensity,
and $\{0 < t_1 < \dots < t_n  < T\}$ is the observed point pattern. If the 
model is correctly specified, $U^*_i =$ $1 - \exp\{-(\Lambda^*(t_i) - \Lambda^*(t_{i-1}))\}$,
for $i = 1,\dots,n$, are independent uniform random variables on $(0,1)$. Thus, the model 
can be assessed graphically using quantile-quantile plots for the estimated $U^*_i$.

For MTDPP models, 
$\Lambda^*(t_i) =$ $\sum_{j=1}^i\int_{t_{j-1}}^{t_j} h^*(u-t_{j-1})du$, and thus
$\Lambda^*(t_i) - \Lambda^*(t_{i-1}) =$ $\int_{t_{i-1}}^{t_i} h^*(u-t_{i-1})du$. 
Using the relationship between the conditional survival and cumultive intensity functions, 
we have $S^{*}(t - t_{i-1}) =$ $\exp( - \int_{t_{i-1}}^{t} h^*(u-t_{i-1})du )$, for
$t_{i-1} < t \leq t_{i}$. Therefore, $S^{*}(t_{i} - t_{i-1}) =$ 
$\exp\{-(\Lambda^*(t_i) - \Lambda^*(t_{i-1}))\}$, which allows us to obtain posterior 
samples for the $U^*_i$ from $U^*_i =$ $1 - S^{*}(t_{i} - t_{i-1}) =$
$1 - \sum_{l=1}^L w_l \, S_{l}(t_{i} - t_{i-1}  \,|\, t_{i-l} - t_{i-l-1},\btheta_l)$.
Replacing survival function $S^{*}$ with $S^{*}_{C}$, the approach can also be 
used for MTDCPPs.

Finally, we consider prediction for future events. Let $D_n$ denote the observed 
point pattern $\{ 0 = t_0 < t_1 < \ldots < t_n < T \}$, with corresponding observed 
durations $x_i = t_i - t_{i-1}$, for $i = 1,\dots,n$. Note that $D_n$ includes the 
constraint that the next (unobserved) event time $t_{n+1} > T$, i.e., that the next 
(unobserved) duration $x_{n+1} > T - t_{n}$. We can predict $t_{n+1}$ via prediction 
of $x_{n+1}$, incorporating the condition that $x_{n+1} > T - t_n$. 
The posterior predictive density for the next duration can be written as 
$$
p(x_{n+1} \,|\, D_n) = \int\int \left\{ 
\sum_{l=1}^L w_l^*(T) \, \tilde f_l(x_{n+1} \,|\, x_{n+1-l},\btheta_l) \right\}
p(\btheta,\bm w\,|\,D_n) \, d\btheta\,d\bm w,
$$
where the weights $w_l^*(T) =$ $w_l \, S_l(T-t_n\,|\,x_{n+1-l},\btheta_l)/
\{ \sum_{l=1}^L w_l \, S_l(T-t_n\,|\,x_{n+1-l},\btheta_l) \}$, and
$\tilde f_l(x_{n+1} \,|\,x_{n+1-l},\btheta_l)$
$= f_l(x_{n+1}\,|\,x_{n+1-l},\btheta_l)/S_l(T-t_n\,|\,x_{n+1-l},\btheta_l)$, 
for $x_{n+1} \in (T-t_n,\infty)$, is the $l$-th component density truncated 
below at $T - t_n$.
The Supplementary Material includes details for the derivation, and the 
extension to $k$-step-ahead predictions, for $k \geq 2$. Also provided in 
the Supplementary Material are details on prediction for MTDCPP models.

%
%-------------------------------------------------------------------------
%

\section{Data illustrations}
\label{sec:data}

We illustrate the scope of the modeling framework through one synthetic and two 
real data examples. In the simulation example, we explore inference for conditional 
intensities and duration hazard functions of 
different shapes, using the Burr MTDPP that allows for monotonic and 
non-monotonic hazard functions. The goal of the first real data example is 
to demonstrate the practical utility of stationary MTDPPs for scenarios 
where the duration-independence assumption of renewal processes needs to be 
relaxed. The second real data example examines the capacity of MTDCPPs 
to detect and quantify duration clustering behaviors; this was also evaluated through 
a simulation study, the details of which can be found in the Supplementary Material.
Also available in the Supplementary Material are additional simulation examples,
model comparison results,
prior sensitivity analysis for the mixture weights, and graphical model assessment 
results obtained using the approach of Section \ref{sec:mc}. The model assessment 
results indicate good model fit for all data examples.

We implemented all MCMC algorithms in the R programming language, 
with C++ code integrated to update latent variables.
MCMC convergence diagnostics and computing times 
are available in the Supplementary Material.
As examples of computing times, fitting the Burr MTDPP ($L=3$) to 572 
observations took 78 seconds for 55000 iterations, while fitting the MTDCPP 
($L=15$) to 3961 event times took 17 minutes for 155000 iterations, both on 
a Linux server with 512 GB of RAM and two Intel Xeon Gold 6348 processors.

\subsection{Simulation study}
\label{sec: sim}

\begin{figure*}[t!]
    \centering
    \captionsetup[subfigure]{justification=centering, font=footnotesize}
    \begin{subfigure}[b]{0.3\textwidth}
         \centering
         \includegraphics[width=\textwidth]{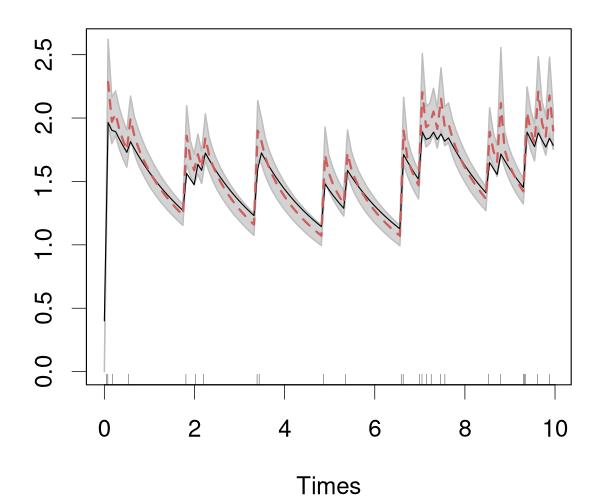}
    \end{subfigure}     
    \begin{subfigure}[b]{0.3\textwidth}
         \centering
         \includegraphics[width=\textwidth]{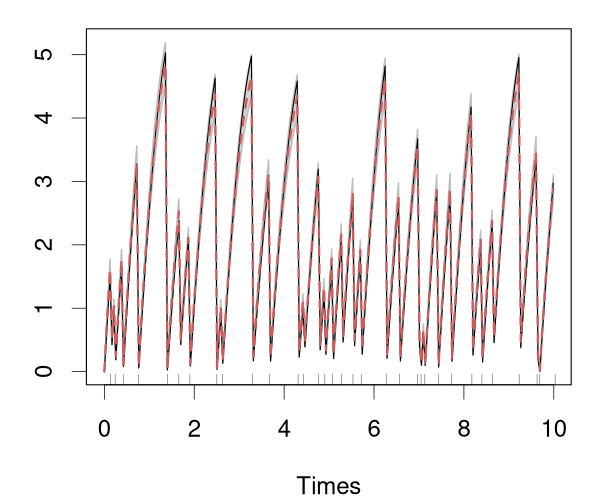}
    \end{subfigure}
    \begin{subfigure}[b]{0.3\textwidth}
         \centering
         \includegraphics[width=\textwidth]{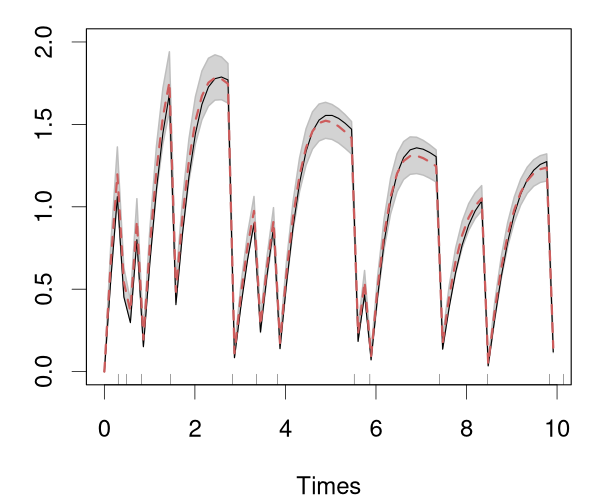}
    \end{subfigure} 
    \begin{subfigure}[b]{0.3\textwidth}
         \centering
         \includegraphics[width=\textwidth]{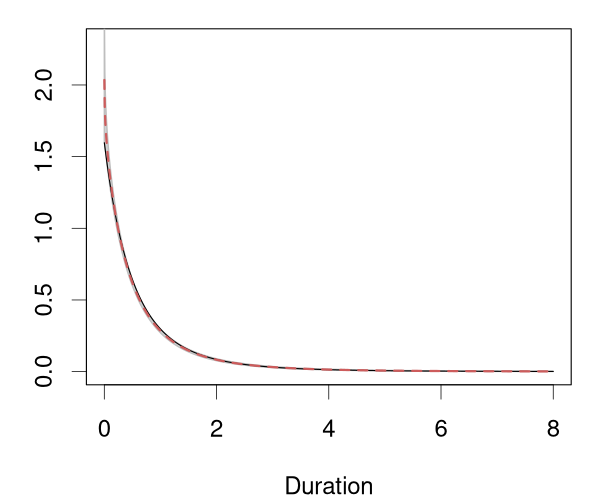}
    \end{subfigure}   
    \begin{subfigure}[b]{0.3\textwidth}
         \centering
         \includegraphics[width=\textwidth]{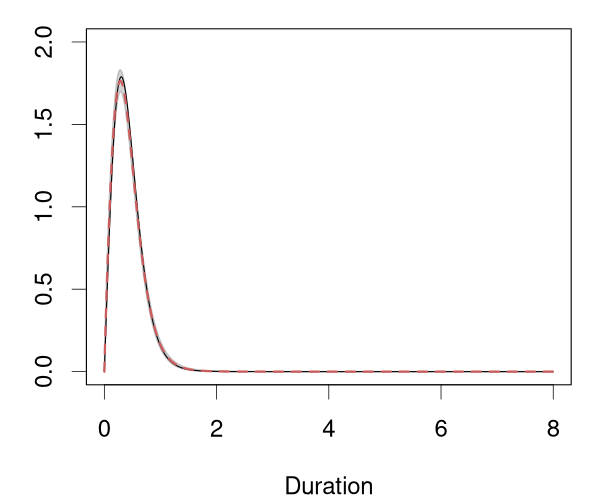}
    \end{subfigure}
    \begin{subfigure}[b]{0.3\textwidth}
         \centering
         \includegraphics[width=\textwidth]{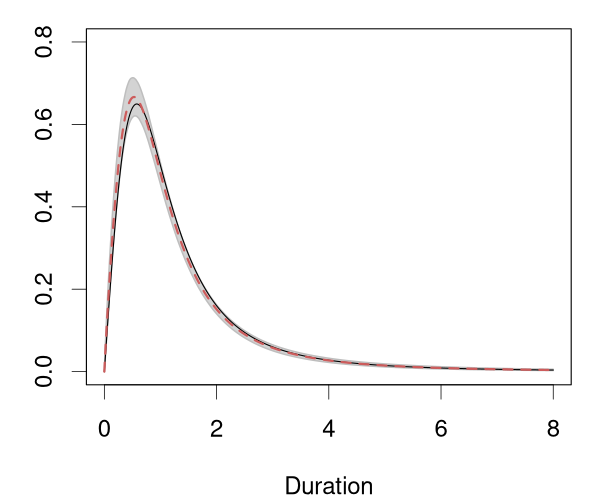}
    \end{subfigure}    
    \begin{subfigure}[b]{0.3\textwidth}
         \centering
         \includegraphics[width=\textwidth]{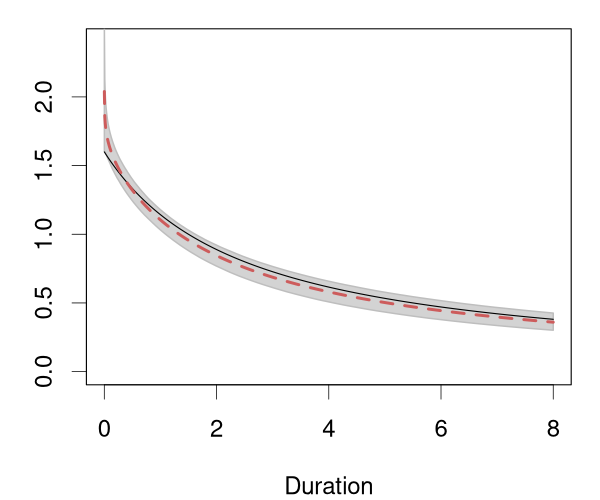}
           \caption{scaled-Lomax}
    \end{subfigure}   
    \begin{subfigure}[b]{0.3\textwidth}
         \centering
         \includegraphics[width=\textwidth]{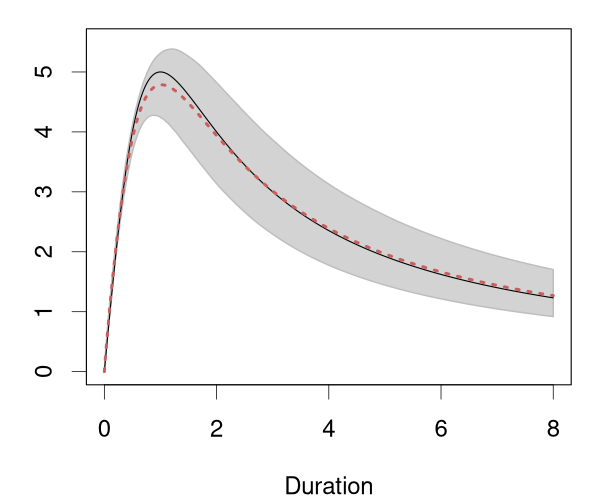}
           \caption{Burr}
    \end{subfigure}
    \begin{subfigure}[b]{0.3\textwidth}
         \centering
         \includegraphics[width=\textwidth]{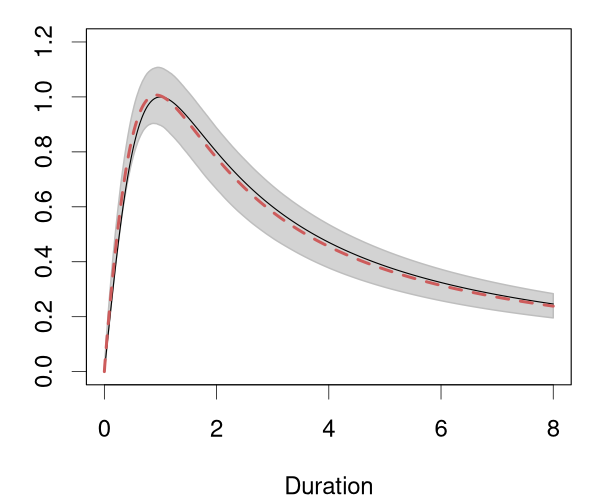}
          \caption{Log-logistic}
    \end{subfigure}    
    \caption{\footnotesize
    Synthetic data example. 
    The first, second, and third rows plot  
    posterior means (red dashed lines) and $95\%$ credible interval
    estimates (grey bands) for the
    conditional intensity, marginal duration density, and marginal duration 
    hazard function. The black solid lines correspond to the true functions. 
    }
    \label{fig: mtdpp-sim1}
\end{figure*}

We generated data from three stationary MTDPP models (discussed in Section 
\ref{sec:mtdpp-constrction}) with scaled-Lomax, Burr, and log-logistic 
marginal duration distributions. The respective parameters were set at 
$(\phi,\alpha) = (0.5, 5)$, $(\la,\ga,\kappa) = (1, 2, 6)$,
and $(\la,\ga) = (1,2)$, such that the hazard function for the durations is 
decreasing for the scaled-Lomax MTDPP, and hump-shaped for the other two models; 
see Figure \ref{fig: mtdpp-sim1}. The model order was $L = 3$ for all simulations, 
with decaying weights $\bm w = (0.5, 0.3, 0.2)$. For each simulated point pattern, 
we chose the observation window to obtain around 2000 event times.

We applied the Burr MTDPP model in \eqref{eq:bmtdpp}, with $L = 3$, to the three 
synthetic data sets. Recall that the hazard function of the marginal 
$\mathrm{Burr}(\ga,\la,\kappa-1)$ duration distribution is decreasing when $\ga\leq 1$,
and hump-shaped when $\ga > 1$. 
We thus assigned a $\mathrm{Ga}(1,1)$ prior to $\ga$, where 
$\mathrm{Ga}(a,b)$ denotes the gamma distribution with mean $a/b$. Moreover, the $m$th 
moment of the $\mathrm{Burr}(\ga,\la,\kappa-1)$ distribution exists if 
$\ga (\kappa-1) > m$. Independently of $\ga$, we placed a truncated gamma prior, 
$\mathrm{Ga}(6,1)\mathbbm{1}(\kappa > 1)$, on $\kappa$. Since 
$\text{E}(\kappa) = 6.004$, the prior choice for $\ga$ and $\kappa$ implies that, 
in prior expectation, the first five moments of the marginal duration distribution exist. 
The scale parameter $\la$ was assigned a $\mathrm{Ga}(1,1)$ prior, and the
vector of weights a CDP$(5, 1, 2)$ prior.

Figure \ref{fig: mtdpp-sim1} plots point and interval estimates for the point process 
conditional intensity, as well as for the duration process marginal density and 
its associated hazard function. 
Results for each synthetic data set were based on $5000$ posterior 
samples collected after appropriate burn-in and thinning.
Note that, although the true data generating mechanisms correspond to MTDPPs, the 
Burr MTDPP is a mis-specified model for two of the simulated data sets. 
However, the model is able to distinguish between 
monotonic and non-monotonic hazard functions for the marginal duration distribution. 
Overall, based on a single process realization, the Burr MTDPP model provides
reasonably accurate estimates for different point process functionals, with 
uncertainty bands that effectively contain the true functions.

\subsection{IVT recurrence interval analysis}
\label{sec:recur}

Integrated water vapor transport (IVT) is a vector representing the total 
amount of water vapor being transported in an atmospheric column. 
Atmospheric rivers (ARs), which are corridors of enhanced IVT, 
play a vital role in transporting moisture into western North America.
Identifying and tracking ARs is central to understanding high-impact weather 
events, such as extreme precipitation and flooding.
\cite{rutz2019atmospheric} review several of the AR detection algorithms, 
most of which use IVT thresholds as input. 
Appropriately thresholding
the IVT is important to improve AR detection; e.g., 
\cite{barata2022fast} provide a time-varying quantile estimate of the IVT
using a dynamic statistical model.

In this example, we take on a different perspective to study the 
IVT, based on the general idea that strong ARs tend to associate
with extreme IVT magnitudes.
We obtain a collection of recurrent events for which the IVT magnitude 
exceeds a given threshold;
the durations between these consecutive 
events are referred to as recurrence intervals.
Modeling extreme events using a point process approach is motivated by the 
asymptotic behavior of threshold exceedances; 
for a large threshold, the exccedances and the associated event times 
can be considered as a marked Poisson process 
\citep[e.g.,][]{kottas2007bayesian}.
On the other hand, the Poisson process assumption 
may be too restrictive, as well as unsuitable for applications where the inferential 
interest lies in the stationary distribution of the durations between event times.
Studying the recurrence interval distribution is important 
in many areas, such as study of earthquakes above a certain magnitude,
and extreme returns. Depending on the correlation structure of
the original time series, the recurrence interval distribution may exhibit
different types of tail behavior (e.g., power law).
Furthermore, the recurrence intervals can be dependent \citep{santhanam2008return}.
A generalization of the renewal process is needed in order to
capture the dependence among durations.

\begin{figure*}[t!]
    \centering
    \captionsetup[subfigure]{justification=centering, font=scriptsize}
    \begin{subfigure}[b]{0.3\textwidth}
         \centering
         \includegraphics[width=\textwidth]{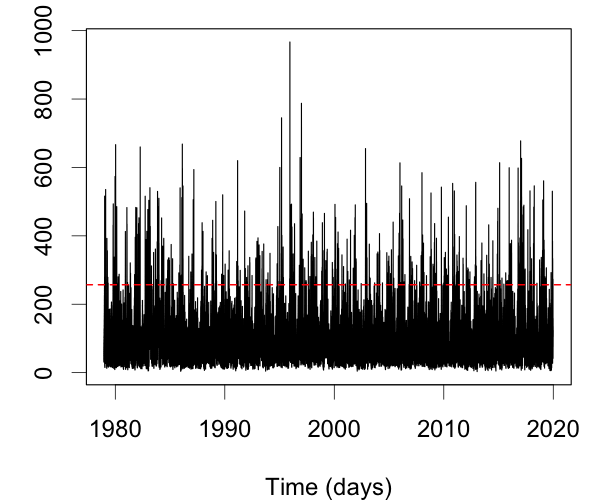}
         \caption{Average daily IVT magnitude}
    \end{subfigure}
    \begin{subfigure}[b]{0.3\textwidth}
         \centering
         \includegraphics[width=\textwidth]{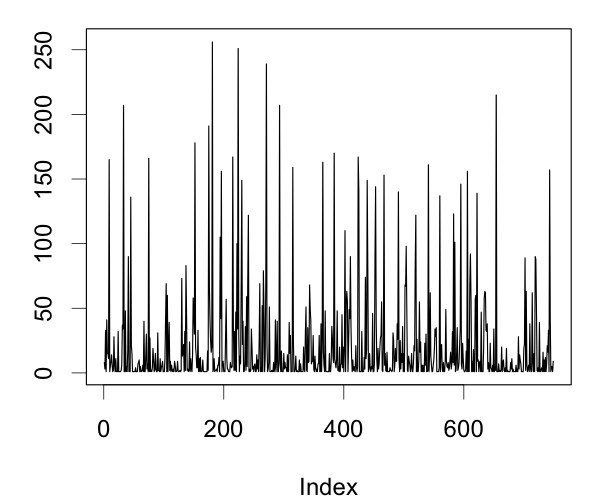}
         \caption{Durations}
    \end{subfigure}
    \begin{subfigure}[b]{0.3\textwidth}
         \centering
         \includegraphics[width=\textwidth]{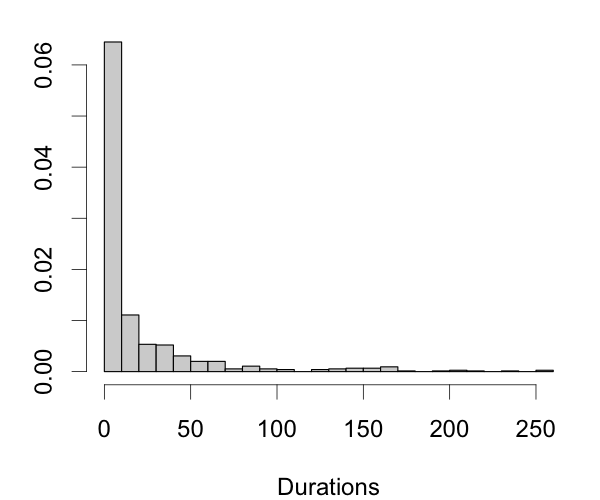}
         \caption{Histogram of the durations}
    \end{subfigure}\\
    \vspace{10pt}
     \begin{subfigure}[b]{0.3\textwidth}
         \centering
         \includegraphics[width=\textwidth]{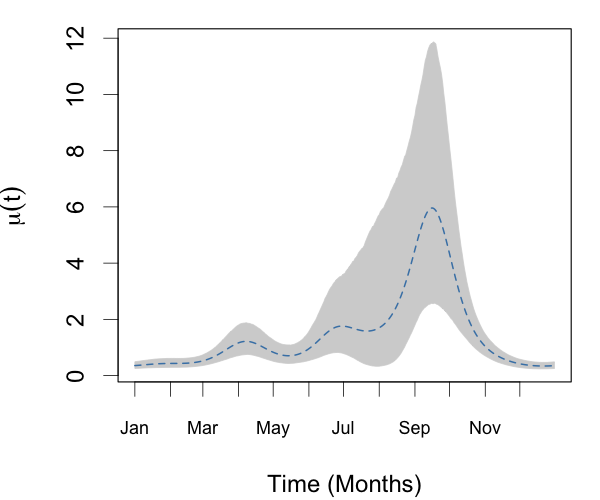}
         \caption{Harmonic function $\mu(t)$}
    \end{subfigure}
    \begin{subfigure}[b]{0.3\textwidth}
         \centering
         \includegraphics[width=\textwidth]{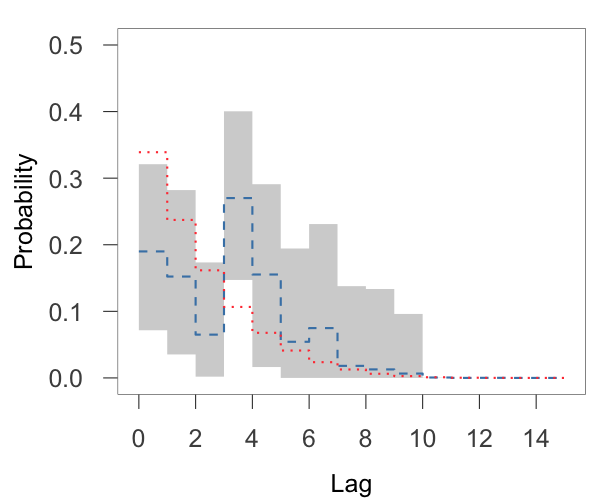}
         \caption{Estimated weights}
    \end{subfigure}
    \begin{subfigure}[b]{0.3\textwidth}
         \centering
         \includegraphics[width=\textwidth]{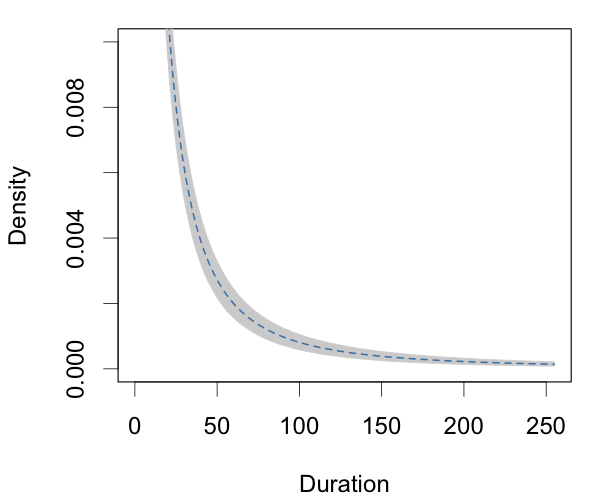}
         \caption{Stationary marginal estimate}
    \end{subfigure} 
    \caption{\footnotesize
    Recurrence interval analysis. Panel (a) shows the average 
    daily IVT magnitude with the $0.95$ quantile (red line) fixed over time.
    Panels (b) and (c) plot the recurrence intervals and their histogram
    respectively. Panel (d) shows the harmonic function $\mu(t)$ for an
    one-year window. Panel (e) shows inference results for the weights.
    Panel (f) is the close view of the stationary marginal estimate,
    specifically at the tail of the marginal density. 
    In Panels (d), (e), and (f), the blue dashed line
    and the light grey polygon correspond to the posterior mean and the pointwise
    $95\%$ credible interval estimates, respectively. The red dotted line
    in panel (e) is the prior mean.
    }
    \label{fig:ivt}
\end{figure*}

Here, we demonstrate the potential of MTDPPs for the aforementioned dual goal:
model the stationary recurrence interval distribution; and, capture 
the recurrence intervals dependence.
The data set involves a time series of average daily IVT magnitude.
The time series has 14965 observations, spanning from January 1, 1979 to December 31, 2019, 
with all February 29s omitted, corresponding to the city of Santa Cruz in California.
The data are publicly available in the R 
package \textit{exdqlm} \citep{barata2022fast}. Using the $0.95$ quantile threshold, 
we obtained 749 events of IVT exceedances. 
The histogram of the durations (Figure \ref{fig:ivt}(c)) 
suggests a heavy right tail for the recurrence interval distribution.

We consider the scaled-Lomax MTDPP. 
As previously discussed in Section \ref{sec:mtdpp-constrction}, the model has 
a stationary scaled-Lomax marginal distribution $P(\alpha\phi, \alpha-1)$
for the recurrence intervals, and the conditional duration 
distribution converges to the exponential distribution
with rate parameter $\phi^{-1}$, as $\alpha\rightarrow\infty$. 
Let $\{t_i\}$ and $\{x_i\}$ be the observed event 
times and durations, respectively. To account for potential 
seasonality, we use the following 
multiplicative model, $x_i = \mu(t_i)z_i$, with
$\log\mu(t_i) = \sum_{j=1}^J\{\beta_{1j}\sin(j\omega t_i)+
\beta_{2j}\cos(j\omega t_i)\},$
where $\omega = 2\pi/T_0$, and $T_0 = 365$ is the period for daily data.
We assume the stationary scaled-Lomax MTDPP model for $\{z_i\}$, 
such that the conditional duration density is
$f^*(x_i) =$ $\mu(t_i)^{-1}\sum_{l=1}^Lw_lP\left(\mu(t_i)^{-1}x_i\,|\,
\alpha\phi+\mu(t_{i-l})^{-1}x_{i-l},\alpha\right)$.
We took $J = 5$, and assigned mean-zero, dispersed normal priors 
to the regression parameter vector.
The shape and scale parameters $\alpha$ and $\phi$ received 
$\mathrm{Ga}(6,1)\mathbbm{1}(\alpha>1)$ and $\mathrm{Ga}(1,1)$ priors, respectively. 
We chose model order $L = 15$; this was based on the ACF and PACF plots 
of the original data and the detrended data based on a harmonic regression, with a sensitivity 
analysis for $L$ (details can be found in the Supplementary Material).
For the weights, we considered a $\mathrm{CDP}(5,1,6)$ prior, which implies
a decreasing trend in prior expectation (Figure \ref{fig:ivt}(e)).

We report the results based on the multiplicative model fitted to the
original data. Posterior inference used $5000$ 
samples obtained after appropriate burn-in and thinning.
The posterior mean and $95\%$ credible interval estimates of the harmonic component
coefficients imply the presence of annual and semiannual seasonality. 
The posterior estimates of the corresponding coefficients 
$(\beta_{11},\beta_{21},\beta_{22})$ are 
$-0.58\,(-0.86, -0.29)$, $-0.68\,(-1.06, -0.33)$, and $-0.53\,(-0.82, -0.23)$.
Figure \ref{fig:ivt}(d) shows the function $\mu(t)$ evaluated at a grid over a period of one year.
Smaller durations between high IVT magnitudes tend to appear from November to March, 
corresponding to high atmospheric river frequency during that period. In fact, this 
time interval corresponds to the usual flooding period in California (e.g., the most 
recent floods in California were caused by multiple atmospheric rivers between December 
2022 and March 2023). Figure \ref{fig:ivt}(e) shows the estimated weights. Lags one, 
two, four and five are the most influential, which suggests serial dependence 
in the durations.
The posterior mean and $95\%$ credible interval estimates of $\alpha$ and $\phi$
were $2.01\,(1.72, 2.35)$ and $4.92\,(3.35, 6.92)$, respectively. 
Inference for $\alpha$ suggests that, even after adjusting 
for seasonality, the distribution of the recurrence intervals is heavy-tailed.
Figure \ref{fig:ivt}(f) shows a marginal density tail that decays 
very slowly, in particular when compared to the histogram of the observed durations 
in Figure \ref{fig:ivt}(c), where the seasonality is not accounted for.

We also assessed our model in comparison to a renewal process %(RP) 
model with scaled-Lomax marginal distribution (see the Supplementary Material). 
The MTDPP outperforms the renewal process in terms of both goodness-of-fit 
and prediction, thus demonstrating the benefits of incorporating duration 
dependence in the particular recurrence interval analysis.

%
%--------------------------------------------------------------------------
%

\subsection{Mid-price changes of the AUD/USD exchange rate}
\label{sec:data-fx}

Financial markets involve complex human activities, with both external and 
internal factors driving market dynamics. 
It is suggested that, for high-frequency financial data, 
price dynamics is more endogenous, driven largely by internal factors within 
the market itself \citep{filimonov2012quantifying}. 
To understand financial market microstructure, it is important to quantify 
the level of endogeneity, measured as the proportion of price
movements due to internal rather than external processes. 
Here, we explore modeling for 
endogeneity quantification from the duration clustering perspective using the 
MTDCPP, where each price move is considered as an event.

We analyze the price movements of the AUD/USD foreign exchange rate.
A price movement is recorded when a mid-price change occurs,
where mid-price is defined as the average of the best bid and ask prices
\citep{filimonov2012quantifying}. 
The data set consists of 121 non-overlapping point patterns, with total number of 
events ranging from 108 to 3961. Each point pattern corresponds to an one-hour 
time window of the trading week from 20:00 Greenwich Mean Time (GMT) July 19 
to 21:00 GMT July 24 in 2015. Analyzing sequences of point patterns within 
small time windows avoids to some extent the issue of nonstationarity, such as 
diurnal pattern. We refer to \cite{chen2018direct} for more details about the data,
which are available in R package \textit{RHawkes} \citep{RHawkes}.

We considered the Lomax MTDCPP, that is, model \eqref{eq:mtdcpp} with $f_I$ 
given by an exponential density with rate parameter $\mu$, and $f^{*}$ corresponding 
to the stationary Lomax MTDPP. 
In particular, $f^*$ is regarded as the driver of internal factors (e.g., market
participants' anticipations and reactions to market prices), 
while external information is driven by $f_I$. 
Thus, the probability $(1-\pi_0)$ 
can be used to quantify market endogeneity.

We applied the model to each of the 121 point patterns
and, for illustrative purposes, considered the 
same model specification for all point patterns.
We used a $\mathrm{Beta}(5,5)$ 
prior for $\pi_0$. The prior assigns small probabilities to values of $\pi_0$ 
around $0$ or $1$, which correspond to the less likely scenarios where the market 
is driven by only an internal or an external process, 
as suggested by previous studies \citep{filimonov2012quantifying,wheatley2016hawkes,chen2018direct}.
For component-density parameters, we used a $\mathrm{Ga}(1,1)$ prior for $\mu$,
and $\mathrm{Ga}(\alpha\,|\,6,1)\mathbbm{1}(\alpha>1)$ and 
$\mathrm{Ga}(\phi\,|\,1, 1)$ priors for the shape and scale parameter of the 
Lomax model, respectively. Based on the PACF of the observed 
durations (see the Supplementary Material for details), 
we chose model order $L = 15$ for all point patterns, and the mixture 
weights were assigned a CDP$(5, 1, 6)$ prior.

\begin{figure*}[t!]
    \centering
    \captionsetup[subfigure]{justification=centering, font=scriptsize}
    \begin{subfigure}[b]{0.35\textwidth}
         \centering
         \includegraphics[width=\textwidth]{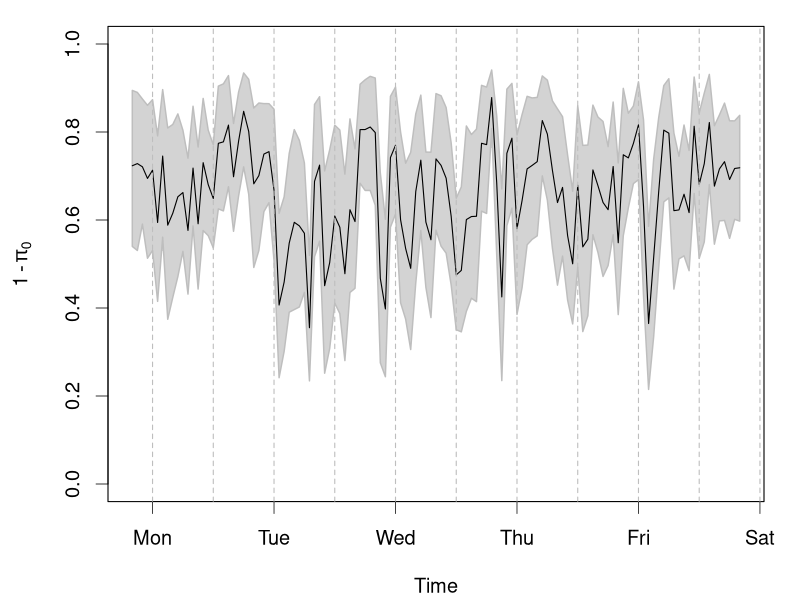}
    \end{subfigure}    
    \hspace{20pt}
    \begin{subfigure}[b]{0.35\textwidth}
         \centering
         \includegraphics[width=\textwidth]{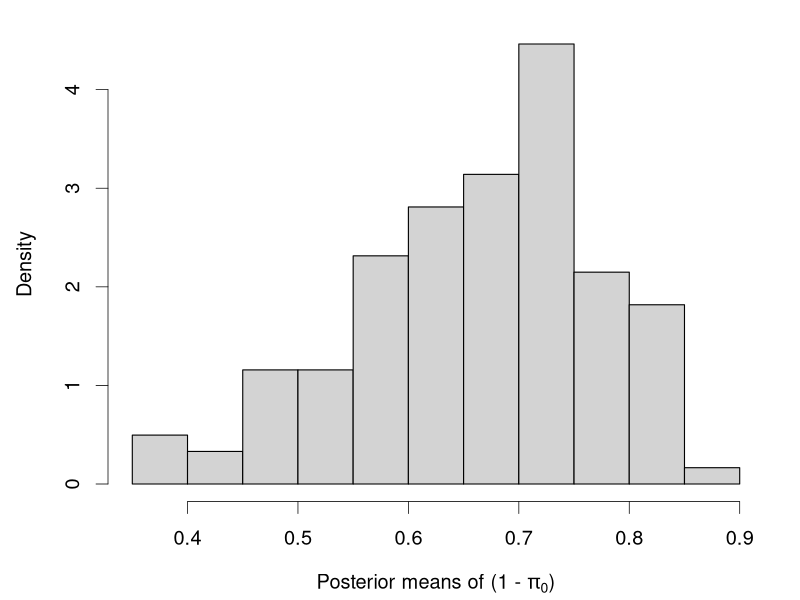}
    \end{subfigure}    
    \caption{\footnotesize
    AUD/USD foreign exchange market endogeneity analysis.    
    Left panel: Time series of the posterior means (solid lines) and 
    pointwise $95\%$ credible intervals (grey polygons) for 
    the level of endogeneity $1-\pi_0$.
    Vertical dashed lines correspond to midnight and midday GMT. 
    Right panel: Histogram of the posterior means of 
    $1-\pi_0$ for the $121$ one-hour time windows.
    }
    \label{fig:params}
\end{figure*}

Results from each point pattern were based on $10000$ posterior 
samples collected after appropriate burn-in and thinning.
Here, we focus on inference for the level of endogeneity $(1-\pi_0)$.
The Supplementary Material includes results for the other model 
parameters. In particular, the estimates of the mean waiting time
$1/\mu$ for external factors exhibit diurnal patterns, particularly 
around midnight and midday GMT, across the 121 point patterns.

The time series of posterior means and interval estimates of $(1-\pi_0)$ 
(for the $121$ one-hour time windows) shows that the level 
of endogeneity fluctuated heavily over the trading week
(Figure \ref{fig:params}, left panel). The histogram of the posterior means 
is skewed to the left (Figure \ref{fig:params}, right panel), 
with median $0.68$ and quartiles $(0.59, 0.74)$, suggesting that the market 
dynamics were mostly driven by internal processes.

A similar conclusion was drawn by \cite{chen2018direct}, where a renewal 
Hawkes (RHawkes) process \citep{wheatley2016hawkes} was applied to the same 
data; the mean and quartiles of their estimates for 
the level of endogeneity over the trading week were, respectively, 
$0.66$ and $(0.53, 0.80)$.
The RHawkes process extends the Hawkes process to 
capture dependence between clusters, by replacing the immigrant Poisson process 
with a renewal process. Both the Hawkes and RHawkes process models include a 
branching ratio parameter, which can be used to quantify the level of endogeneity 
\citep{filimonov2012quantifying, chen2018direct}.
Under a different stochastic model structure, the MTDCPP was able to quantify 
the extent to which the observed dynamics are caused by internal factors versus 
external influences.
Moreover, the MTDCPP demonstrated superior out-of-sample predictive 
performance compared to the RHawkes process, while requiring less computation 
time for large point patterns (see the Supplementary Material).

We note that the MTDCPP does not require stationarity assumptions. 
In contrast, to use the branching ratio as an estimator for the level 
of endogeneity, stationarity is essential for the Hawkes and RHawkes 
processes. However, 
market activities are commonly nonstationary \citep{filimonov2012quantifying}. 
The lack of stationarity
is typically attributed to seasonal trends, which can be addressed by splitting the 
time window into small intervals, as shown in this example. Still, one has to balance
the size of the intervals and the number of the events within the interval 
to ensure reliable estimates are produced.
Moreover, even after removing seasonality, stationarity is not necessarily guaranteed. 
Therefore, MTDCPP models may be useful in applications where stationarity assumptions
are not plausible.

%
%--------------------------------------------------------------------------
%

\section{Summary and discussion}
\label{summary}

We have developed a new class 
of stochastic models for temporal point patterns with self-excitation or 
self-regulation effects, identically distributed but dependent durations, or 
clustered durations.
The modeling framework allows for different approaches to building the point 
process: through marginal duration distributions, when the inferential goal 
pertains to the intervals between event times; or through conditional hazard functions,
when interest lies in the point process dependence structure on its history.
Both strategies connect naturally to existing point process models. The former is 
analogous to renewal process modeling, while the latter 
involves the same motivation 
of Hawkes processes.
We have presented several examples of implementing these strategies.
The Burr model, which allows for 
point process functionals with flexible shapes, can be considered as a default 
choice for general purposes, while the scaled-Lomax model, which provides 
flexible tail behavior, may be considered for 
scientific applications where such property is relevant.

Our framework builds from a structured mixture model for the point process conditional 
duration density. The resulting point process has restricted memory, i.e., its evolution 
depends on recent events. This assumption is generally suitable for relatively 
large point patterns. For scenarios where one anticipates more extensive history 
dependence, a large value for the order of the mixture model can be used. The 
nonparametric prior for the weights allows efficient inference with a large order. 
On the other hand, there are applications where data correspond to many processes 
that exhibit a relatively small number of point events, such as the analysis of 
recurrent event gap times for multiple patients in medical studies. For such data, 
a small order is more appropriate. In fact, even the special case where the conditional 
duration density depends on the most recent lag provides a meaningful 
generalization of renewal processes commonly used for this type of analysis.

In many applications, point patterns include information on marks, that is, random 
variables associated with each point event, such that the data generating mechanism 
corresponds to a marked point process. Consider, for instance, continuous marks, $\by$. 
The marked point process intensity can be developed from $\la^*(t,\by) =$
$\la_g^*(t) \, m^{*}_{t}(\by)$, where $\la_g^*(t)$ is the conditional intensity for 
the event times (referred to as the ground process intensity), and $m^{*}_{t}(\by)$
is the time-dependent mark distribution \citep{daley2003introduction}. The proposed
framework can be utilized for marked point processes by combining an MTDPP or MTDCPP 
model for the ground process with a model for the mark distribution.

\section*{Acknowledgements}
This research was supported in part by the National Science Foundation under 
award SES 1950902. 
The authors thank the Associate Editor and the 
reviewers for valuable comments.

\section*{Supplementary Material}
The Supplementary Material includes proofs for the theoretical results, details
for the bivariate Burr distribution, 
additional details for the MCMC algorithms and predictions, 
additional data examples and results, MCMC diagnostics,
and model checking results.

\bibliographystyle{jasa3}
\bibliography{ref}

\begin{thebibliography}{35}
\newcommand{\enquote}[1]{``#1''}
\expandafter\ifx\csname natexlab\endcsname\relax\def\natexlab#1{#1}\fi
\expandafter\ifx\csname url\endcsname\relax
  \def\url#1{{\tt #1}}\fi
\expandafter\ifx\csname urlprefix\endcsname\relax\def\urlprefix{URL }\fi

\bibitem[\protect\citeauthoryear{Arnold, Castillo, and Sarabia}{Arnold
  et~al.}{1999}]{arnold1999conditional}
Arnold, B.~C., Castillo, E., and Sarabia, J.~M. (1999), {\em Conditional
  Specification of Statistical Models\/}, New York: Springer.

\bibitem[\protect\citeauthoryear{Barata, Prado, and Sans{\'o}}{Barata
  et~al.}{2022}]{barata2022fast}
Barata, R., Prado, R., and Sans{\'o}, B. (2022), \enquote{Fast inference for
  time-varying quantiles via flexible dynamic models with application to the
  characterization of atmospheric rivers,} {\em The Annals of Applied
  Statistics\/}, 16, 247--271.

\bibitem[\protect\citeauthoryear{Bhogal and Thekke~Variyam}{Bhogal and
  Thekke~Variyam}{2019}]{bhogal2019conditional}
Bhogal, S.~K. and Thekke~Variyam, R. (2019), \enquote{Conditional duration
  models for high-frequency data: a review on recent developments,} {\em
  Journal of Economic Surveys\/}, 33, 252--273.

\bibitem[\protect\citeauthoryear{Cavaliere, Mikosch, Rahbek, and
  Vilandt}{Cavaliere et~al.}{2024}]{cavaliere2024tail}
Cavaliere, G., Mikosch, T., Rahbek, A., and Vilandt, F. (2024), \enquote{Tail
  behavior of ACD models and consequences for likelihood-based estimation,}
  {\em Journal of Econometrics\/}, 238, 105613.

\bibitem[\protect\citeauthoryear{Chen and Stindl}{Chen and
  Stindl}{2018}]{chen2018direct}
Chen, F. and Stindl, T. (2018), \enquote{Direct likelihood evaluation for the
  renewal Hawkes process,} {\em Journal of Computational and Graphical
  Statistics\/}, 27, 119--131.

\bibitem[\protect\citeauthoryear{Chen and Stindl}{Chen and
  Stindl}{2022}]{RHawkes}
--- (2022), {\em RHawkes: Renewal Hawkes Process\/}. R package version 1.0.

\bibitem[\protect\citeauthoryear{Coen, Guti{\'e}rrez, and Mena}{Coen
  et~al.}{2019}]{coen2019modelling}
Coen, A., Guti{\'e}rrez, L., and Mena, R.~H. (2019), \enquote{Modelling
  failures times with dependent renewal type models via exchangeability,} {\em
  Statistics\/}, 53, 1112--1130.

\bibitem[\protect\citeauthoryear{Cook, Lawless et~al.}{Cook
  et~al.}{2007}]{cook2007statistical}
Cook, R.~J., Lawless, J.~F., et~al. (2007), {\em The Statistical Analysis of
  Recurrent Events\/}, New York: Springer.

\bibitem[\protect\citeauthoryear{Cowpertwait}{Cowpertwait}{2001}]{cowpertwait2001renewal}
Cowpertwait, P.~S. (2001), \enquote{A renewal cluster model for the
  inter-arrival times of rainfall events,} {\em International Journal of
  Climatology\/}, 21, 49--61.

\bibitem[\protect\citeauthoryear{Daley and Vere-Jones}{Daley and
  Vere-Jones}{2003}]{daley2003introduction}
Daley, D.~J. and Vere-Jones, D. (2003), {\em An Introduction to the Theory of
  Point Processes: Volume I: Elementary Theory and Methods\/}, New York:
  Springer.

\bibitem[\protect\citeauthoryear{Engle and Russell}{Engle and
  Russell}{1998}]{engle1998autoregressive}
Engle, R.~F. and Russell, J.~R. (1998), \enquote{Autoregressive conditional
  duration: a new model for irregularly spaced transaction data,} {\em
  Econometrica\/}, 1127--1162.

\bibitem[\protect\citeauthoryear{Ferguson}{Ferguson}{1973}]{ferguson1973bayesian}
Ferguson, T.~S. (1973), \enquote{A {B}ayesian analysis of some nonparametric
  problems,} {\em The Annals of Statistics\/}, 209--230.

\bibitem[\protect\citeauthoryear{Filimonov and Sornette}{Filimonov and
  Sornette}{2012}]{filimonov2012quantifying}
Filimonov, V. and Sornette, D. (2012), \enquote{Quantifying reflexivity in
  financial markets: Toward a prediction of flash crashes,} {\em Physical
  Review E\/}, 85, 056108.

\bibitem[\protect\citeauthoryear{Gaver and Lewis}{Gaver and
  Lewis}{1980}]{gaver1980first}
Gaver, D.~P. and Lewis, P. (1980), \enquote{First-order autoregressive gamma
  sequences and point processes,} {\em Advances in Applied Probability\/}, 12,
  727--745.

\bibitem[\protect\citeauthoryear{Gelman, Stern, Carlin, Dunson, Vehtari, and
  Rubin}{Gelman et~al.}{2013}]{gelman2013bayesian}
Gelman, A., Stern, H.~S., Carlin, J.~B., Dunson, D.~B., Vehtari, A., and Rubin,
  D.~B. (2013), {\em Bayesian Data analysis\/}, New York: Chapman and Hall/CRC,
  third edition.

\bibitem[\protect\citeauthoryear{Grammig and Maurer}{Grammig and
  Maurer}{2000}]{grammig2000non}
Grammig, J. and Maurer, K.-O. (2000), \enquote{Non-monotonic hazard functions
  and the autoregressive conditional duration model,} {\em The Econometrics
  Journal\/}, 3, 16--38.

\bibitem[\protect\citeauthoryear{Hassan and El-Bassiouni}{Hassan and
  El-Bassiouni}{2013}]{hassan2013modelling}
Hassan, M.~Y. and El-Bassiouni, M.~Y. (2013), \enquote{Modelling Poisson marked
  point processes using bivariate mixture transition distributions,} {\em
  Journal of Statistical Computation and Simulation\/}, 83, 1440--1452.

\bibitem[\protect\citeauthoryear{Hassan and Lii}{Hassan and
  Lii}{2006}]{hassan2006modeling}
Hassan, M.~Y. and Lii, K.-S. (2006), \enquote{Modeling marked point processes
  via bivariate mixture transition distribution models,} {\em Journal of the
  American Statistical Association\/}, 101, 1241--1252.

\bibitem[\protect\citeauthoryear{Hautsch}{Hautsch}{2011}]{hautsch2011econometrics}
Hautsch, N. (2011), {\em Econometrics of Financial High-Frequency Data\/}, New
  York: Springer Science \& Business Media.

\bibitem[\protect\citeauthoryear{Hawkes}{Hawkes}{1971}]{hawkes1971point}
Hawkes, A.~G. (1971), \enquote{Point spectra of some mutually exciting point
  processes,} {\em Journal of the Royal Statistical Society: Series B
  (Methodological)\/}, 33, 438--443.

\bibitem[\protect\citeauthoryear{Jacobs and Lewis}{Jacobs and
  Lewis}{1977}]{jacobs1977mixed}
Jacobs, P. and Lewis, P. (1977), \enquote{A mixed autoregressive-moving average
  exponential sequence and point process (EARMA 1, 1),} {\em Advances in
  Applied Probability\/}, 9, 87--104.

\bibitem[\protect\citeauthoryear{Kottas and Sans{\'o}}{Kottas and
  Sans{\'o}}{2007}]{kottas2007bayesian}
Kottas, A. and Sans{\'o}, B. (2007), \enquote{Bayesian mixture modeling for
  spatial Poisson process intensities, with applications to extreme value
  analysis,} {\em Journal of Statistical Planning and Inference\/}, 137,
  3151--3163.

\bibitem[\protect\citeauthoryear{Le, Martin, and Raftery}{Le
  et~al.}{1996}]{le1996modeling}
Le, N.~D., Martin, R.~D., and Raftery, A.~E. (1996), \enquote{Modeling flat
  stretches, bursts outliers in time series using mixture transition
  distribution models,} {\em Journal of the American Statistical
  Association\/}, 91, 1504--1515.

\bibitem[\protect\citeauthoryear{Ogata}{Ogata}{1988}]{ogata1988statistical}
Ogata, Y. (1988), \enquote{Statistical models for earthquake occurrences and
  residual analysis for point processes,} {\em Journal of the American
  Statistical Association\/}, 83, 9--27.

\bibitem[\protect\citeauthoryear{O'Hara}{O'Hara}{1995}]{o1997market}
O'Hara, M. (1995), {\em Market Microstructure Theory\/}, London: Blackwell.

\bibitem[\protect\citeauthoryear{Pacurar}{Pacurar}{2008}]{pacurar2008autoregressive}
Pacurar, M. (2008), \enquote{Autoregressive conditional duration models in
  finance: a survey of the theoretical and empirical literature,} {\em Journal
  of Economic Surveys\/}, 22, 711--751.

\bibitem[\protect\citeauthoryear{Rutz, Shields, Lora, Payne, Guan, Ullrich,
  O’brien, Leung, Ralph, Wehner et~al.}{Rutz
  et~al.}{2019}]{rutz2019atmospheric}
Rutz, J.~J., Shields, C.~A., Lora, J.~M., Payne, A.~E., Guan, B., Ullrich, P.,
  O’brien, T., Leung, L.~R., Ralph, F.~M., Wehner, M., et~al. (2019),
  \enquote{The atmospheric river tracking method intercomparison project
  (ARTMIP): quantifying uncertainties in atmospheric river climatology,} {\em
  Journal of Geophysical Research: Atmospheres\/}, 124, 13777--13802.

\bibitem[\protect\citeauthoryear{Santhanam and Kantz}{Santhanam and
  Kantz}{2008}]{santhanam2008return}
Santhanam, M. and Kantz, H. (2008), \enquote{Return interval distribution of
  extreme events and long-term memory,} {\em Physical Review E\/}, 78, 051113.

\bibitem[\protect\citeauthoryear{Sklar}{Sklar}{1959}]{sklar1959fonctions}
Sklar, M. (1959), \enquote{Fonctions de repartition an dimensions et leurs
  marges,} {\em Publications de l’Institut de Statistique de L’Universit\'e
  de Paris\/}, 8, 229--231.

\bibitem[\protect\citeauthoryear{Strickland, Forbes, and Martin}{Strickland
  et~al.}{2006}]{strickland2006bayesian}
Strickland, C.~M., Forbes, C.~S., and Martin, G.~M. (2006), \enquote{Bayesian
  analysis of the stochastic conditional duration model,} {\em Computational
  Statistics \& Data Analysis\/}, 50, 2247--2267.

\bibitem[\protect\citeauthoryear{Tang and Li}{Tang and
  Li}{2021}]{tang2021multivariate}
Tang, X. and Li, L. (2021), \enquote{Multivariate temporal point process
  regression,} {\em Journal of the American Statistical Association\/}, 1--16.

\bibitem[\protect\citeauthoryear{Wheatley, Filimonov, and Sornette}{Wheatley
  et~al.}{2016}]{wheatley2016hawkes}
Wheatley, S., Filimonov, V., and Sornette, D. (2016), \enquote{The Hawkes
  process with renewal immigration \& its estimation with an EM algorithm,}
  {\em Computational Statistics \& Data Analysis\/}, 94, 120--135.

\bibitem[\protect\citeauthoryear{Wold}{Wold}{1948}]{wold1948stationary}
Wold, H. (1948), \enquote{On stationary point processes and Markov chains,}
  {\em Scandinavian Actuarial Journal\/}, 1948, 229--240.

\bibitem[\protect\citeauthoryear{Yang, Delcher, Shenkman, and Ranka}{Yang
  et~al.}{2018}]{yang2018clustering}
Yang, C., Delcher, C., Shenkman, E., and Ranka, S. (2018), \enquote{Clustering
  inter-arrival time of health care encounters for high utilizers,} in {\em
  2018 IEEE 20th International Conference on e-Health Networking, Applications
  and Services (Healthcom)\/}, IEEE.

\bibitem[\protect\citeauthoryear{Zheng, Kottas, and Sansó}{Zheng
  et~al.}{2022}]{zheng2021construction}
Zheng, X., Kottas, A., and Sansó, B. (2022), \enquote{On construction and
  estimation of stationary mixture transition distribution models,} {\em
  Journal of Computational and Graphical Statistics\/}, 31, 283--293.

\end{thebibliography}


\begin{thebibliography}{28}
\newcommand{\enquote}[1]{``#1''}
\expandafter\ifx\csname natexlab\endcsname\relax\def\natexlab#1{#1}\fi
\expandafter\ifx\csname url\endcsname\relax
  \def\url#1{{\tt #1}}\fi
\expandafter\ifx\csname urlprefix\endcsname\relax\def\urlprefix{URL }\fi

\bibitem[\protect\citeauthoryear{Ahmad}{Ahmad}{1994}]{ahmad1994small}
Ahmad, K. (1994), \enquote{Small sample results for a nonlinear discriminant
  function estimated from a mixture of two Burr type XII distributions,} {\em
  Computers \& Mathematics with Applications\/}, 28, 13--20.

\bibitem[\protect\citeauthoryear{Ahmad}{Ahmad}{1988}]{ahmad1988identifiability}
Ahmad, K.~E. (1988), \enquote{Identifiability of finite mixtures using a new
  transform,} {\em Annals of the Institute of Statistical Mathematics\/}, 40,
  261--265.

\bibitem[\protect\citeauthoryear{Al-Moisheer et~al.}{Al-Moisheer
  et~al.}{2016}]{al2016mixture}
Al-Moisheer, A. et~al. (2016), \enquote{A mixture of two Burr Type III
  distributions: Identifiability and estimation under type II censoring,} {\em
  Mathematical Problems in Engineering\/}, 2016.

\bibitem[\protect\citeauthoryear{Belfrage}{Belfrage}{2022}]{belfrage2022acdm}
Belfrage, M. (2022), \enquote{ACDm: Tools for autoregressive conditional
  duration models,} .

\bibitem[\protect\citeauthoryear{Carrasco and Chen}{Carrasco and
  Chen}{2002}]{carrasco2002mixing}
Carrasco, M. and Chen, X. (2002), \enquote{Mixing and moment properties of
  various GARCH and stochastic volatility models,} {\em Econometric Theory\/},
  18, 17--39.

\bibitem[\protect\citeauthoryear{Chandra}{Chandra}{1977}]{chandra1977mixtures}
Chandra, S. (1977), \enquote{On the mixtures of probability distributions,}
  {\em Scandinavian Journal of Statistics\/}, 105--112.

\bibitem[\protect\citeauthoryear{Crawford}{Crawford}{1994}]{crawford1994application}
Crawford, S.~L. (1994), \enquote{An application of the Laplace method to finite
  mixture distributions,} {\em Journal of the American Statistical
  Association\/}, 89, 259--267.

\bibitem[\protect\citeauthoryear{DiCiccio and Efron}{DiCiccio and
  Efron}{1996}]{diciccio1996bootstrap}
DiCiccio, T.~J. and Efron, B. (1996), \enquote{Bootstrap confidence intervals,}
  {\em Statistical Science\/}, 11, 189--228.

\bibitem[\protect\citeauthoryear{Folland}{Folland}{1999}]{folland1999real}
Folland, G.~B. (1999), {\em Real Analysis: Modern Techniques and Their
  Applications\/}, John Wiley \& Sons.

\bibitem[\protect\citeauthoryear{Frees and Valdez}{Frees and
  Valdez}{1998}]{frees1998understanding}
Frees, E.~W. and Valdez, E.~A. (1998), \enquote{Understanding relationships
  using copulas,} {\em North American Actuarial Journal\/}, 2, 1--25.

\bibitem[\protect\citeauthoryear{Fr{\"u}hwirth-Schnatter}{Fr{\"u}hwirth-Schnatter}{2006}]{fruhwirth2006finite}
Fr{\"u}hwirth-Schnatter, S. (2006), {\em Finite Mixture and Markov Switching
  Models\/}, New York: Springer.

\bibitem[\protect\citeauthoryear{Gneiting and Raftery}{Gneiting and
  Raftery}{2007}]{gneiting2007strictly}
Gneiting, T. and Raftery, A.~E. (2007), \enquote{Strictly proper scoring rules,
  prediction, and estimation,} {\em Journal of the American Statistical
  Association\/}, 102, 359--378.

\bibitem[\protect\citeauthoryear{Kalliovirta, Meitz, and Saikkonen}{Kalliovirta
  et~al.}{2015}]{kalliovirta2015gaussian}
Kalliovirta, L., Meitz, M., and Saikkonen, P. (2015), \enquote{A Gaussian
  mixture autoregressive model for univariate time series,} {\em Journal of
  Time Series Analysis\/}, 36, 247--266.

\bibitem[\protect\citeauthoryear{Kent}{Kent}{1983}]{kent1983identifiability}
Kent, J.~T. (1983), \enquote{Identifiability of finite mixtures for directional
  data,} {\em The Annals of Statistics\/}, 984--988.

\bibitem[\protect\citeauthoryear{McLachlan, Lee, and Rathnayake}{McLachlan
  et~al.}{2019}]{mclachlan2019finite}
McLachlan, G.~J., Lee, S.~X., and Rathnayake, S.~I. (2019), \enquote{Finite
  mixture models,} {\em Annual Review of Statistics and its Application\/}, 6,
  355--378.

\bibitem[\protect\citeauthoryear{Mengersen and Tweedie}{Mengersen and
  Tweedie}{1996}]{mengersen1996rates}
Mengersen, K.~L. and Tweedie, R.~L. (1996), \enquote{Rates of convergence of
  the Hastings and Metropolis algorithms,} {\em The annals of Statistics\/},
  24, 101--121.

\bibitem[\protect\citeauthoryear{Meyn and Tweedie}{Meyn and
  Tweedie}{2012}]{meyn2012markov}
Meyn, S.~P. and Tweedie, R.~L. (2012), {\em Markov Chains and Stochastic
  Stability (2nd ed.)\/}, London: Springer.

\bibitem[\protect\citeauthoryear{Raftery}{Raftery}{1985}]{raftery1985model}
Raftery, A.~E. (1985), \enquote{A model for high-order Markov chains,} {\em
  Journal of the Royal Statistical Society Series B: Statistical
  Methodology\/}, 47, 528--539.

\bibitem[\protect\citeauthoryear{Resnick}{Resnick}{2013}]{resnick2013adventures}
Resnick, S.~I. (2013), {\em Adventures in Stochastic Processes\/}, Boston:
  Birkhäuser.

\bibitem[\protect\citeauthoryear{Schwarz}{Schwarz}{1978}]{schwarz1978estimating}
Schwarz, G. (1978), \enquote{Estimating the dimension of a model,} {\em The
  Annals of Statistics\/}, 461--464.

\bibitem[\protect\citeauthoryear{Spiegelhalter, Best, Carlin, and Van
  Der~Linde}{Spiegelhalter et~al.}{2002}]{spiegelhalter2002bayesian}
Spiegelhalter, D.~J., Best, N.~G., Carlin, B.~P., and Van Der~Linde, A. (2002),
  \enquote{Bayesian measures of model complexity and fit,} {\em Journal of the
  Royal Statistical Society Series B: Statistical Methodology\/}, 64, 583--639.

\bibitem[\protect\citeauthoryear{Tadikamalla}{Tadikamalla}{1980}]{tadikamalla1980look}
Tadikamalla, P.~R. (1980), \enquote{A look at the Burr and related
  distributions,} {\em International Statistical Review/Revue Internationale de
  Statistique\/}, 337--344.

\bibitem[\protect\citeauthoryear{Teicher}{Teicher}{1961}]{teicher1961identifiability}
Teicher, H. (1961), \enquote{Identifiability of mixtures,} {\em The Annals of
  Mathematical Statistics\/}, 32, 244--248.

\bibitem[\protect\citeauthoryear{Teicher}{Teicher}{1963}]{teicher1963identifiability}
--- (1963), \enquote{Identifiability of finite mixtures,} {\em The Annals of
  Mathematical Statistics\/}, 1265--1269.

\bibitem[\protect\citeauthoryear{Tierney}{Tierney}{1994}]{tierney1994markov}
Tierney, L. (1994), \enquote{Markov chains for exploring posterior
  distributions,} {\em the Annals of Statistics\/}, 1701--1728.

\bibitem[\protect\citeauthoryear{Titterington, Smith, and Makov}{Titterington
  et~al.}{1985}]{titterington1985statistical}
Titterington, D.~M., Smith, A.~F., and Makov, U.~E. (1985), {\em Statistical
  Analysis of Finite Mixture Distributions\/}, Wiley, New York.

\bibitem[\protect\citeauthoryear{Venter}{Venter}{2002}]{venter2002tails}
Venter, G.~G. (2002), \enquote{Tails of copulas,} in {\em Proceedings of the
  Casualty Actuarial Society\/}, volume~89.

\bibitem[\protect\citeauthoryear{Yakowitz and Spragins}{Yakowitz and
  Spragins}{1968}]{yakowitz1968identifiability}
Yakowitz, S.~J. and Spragins, J.~D. (1968), \enquote{On the identifiability of
  finite mixtures,} {\em The Annals of Mathematical Statistics\/}, 39,
  209--214.

\end{thebibliography}

% \newpage
\clearpage\pagebreak\newpage
%%%%%%%%%% Merge with supplemental materials %%%%%%%%%%

\spacingset{1.2} 

\begin{center}
\Large\bf Supplementary Material for 
       ``Mixture Modeling for Temporal Point Processes with Memory"
\end{center}
\setcounter{section}{0}
\setcounter{equation}{0}
\setcounter{figure}{0}
\setcounter{table}{0}

\renewcommand{\thesection}{S\arabic{section}}
\renewcommand{\thefigure}{S\arabic{figure}}
\renewcommand{\thetable}{S\arabic{table}}
\renewcommand{\thedefinition}{S\arabic{definition}}

\vspace{50pt}

\spacingset{1.55} 

\section{Theoretical results}

\subsection{Mean long-run rate}

We provide an upper bound for 
the rate, $\lim_{t\rightarrow\infty}E[N(t)]/t$,
for MTDPPs with bounded component hazard functions,
following the notation developed in the main paper.

\begin{prop}\label{prop:mt_ub}
Consider an MTD point process $N(t)$ with conditional intensity given by 
$\la^*(t) \, = \, \sum_{l=1}^L w_l^*(t) \, h_l(t-\tN\,|\,t_{\Ntl-l+1} - t_{\Ntl-l})$,
such that, for all $l$, the component hazard functions 
satisfy $h_l\leq B_l$. Then, 
$\lim_{t\rightarrow\infty}E[N(t)]/t \, \leq \, \sum_{l=1}^L w_l \, B_l$.
\end{prop}

\begin{proof}[\textbf{Proof}]
The definition of 
$\la^*(t)$ yields that $m(t) = E[N(t)] = E[\int_0^t\la^*(u)du]$.
Since our interest is in $\lim_{t\rightarrow\infty}m(t)/t$,
consider time $t$ large enough such that $N(t) > L$.

Recall that $\la^*(t)\equiv h^*(t-\tN) = f^*(t-\tN)/S^*(t-\tN)$,
where $h^*(t-\tN)$ and $S^*(t-\tN)$ are the hazard and survival functions, 
respectively, associated with $f^*(t-\tN)$.
Let $t_0 = 0$. We have that

\begin{equation}\label{eq:la_decomp}
\begin{aligned}
\int_0^t\la^*(u)du 
& = \sum_{i=1}^{\Ntl}\int_{t_{i-1}}^{t_i}h^*(u-t_{i-1})du + \int_{\tN}^th^*(u-\tN)du\\
& = \sum_{i=1}^{\Ntl}(-\log\{S^*(t_i-t_{i-1})\}) - \log\{S^*(t - \tN)\}.
\end{aligned}
\end{equation}
For $i = 1,\dots,\Ntl$, by Jensen's inequality, we have that
\begin{equation}\label{eq:jensen}
\begin{aligned}
-\log\{S^*(t_i-t_{i-1})\} & = -\log\left\{\sum_{l=1}^Lw_lS_l(t_i-t_{i-1}\,|\,t_{i-l}-t_{i-1-l})\right\}\\
& \leq \sum_{l=1}^Lw_l\left(-\log\{S_l(t_i-t_{i-1}\,|\,t_{i-l}-t_{i-1-l})\}\right)\\
& = \sum_{l=1}^Lw_l\int_{t_{i-1}}^{t_i}h_l(u-t_{i-1}\,|\,t_{i-l}-t_{i-1-l})du 
= \tilde\Lambda^*(t_i-t_{i-1}),
\end{aligned}
\end{equation}
where $\tilde\Lambda^*(a-t_k) = \sum_{l=1}^Lw_l
\int_{t_k}^{a}h_l(u-t_k\,|\,t_{k-l+1}-t_{k-l})du$.
Similarly, applying Jensen's inequality, we obtain 
$- \log\{S^*(t - \tN)\}\leq\tilde\Lambda^*(t-\tN)$, 
and combining \eqref{eq:la_decomp} and \eqref{eq:jensen},
we have that
$
\int_0^t\la^*(u)du \,\leq \,
\sum_{i=1}^{\Ntl}
\tilde\Lambda^*(t_i-t_{i-1})\,+\, \tilde\Lambda^*(t-\tN).
$

If $h_l \leq B_l$ for all $l$, 
then $\tilde\Lambda^*(t_i-t_{i-1})\leq \sum_{l=1}^Lw_l(t_i-t_{i-1})B_l$,
for $i = 1,\dots,\Ntl$, and $\tilde\Lambda^*(t-\tN)\leq 
\sum_{l=1}^Lw_l(t-\tN)B_l$.
Then we have that 
\begin{equation}
\begin{aligned}
    \int_0^t\la^*(u)du \, & \leq \,
    \sum_{i=1}^{\Ntl}\sum_{l=1}^Lw_l(t_i-t_{i-1})B_l
    + \sum_{l=1}^Lw_l(t-\tN)B_l\\
    & = \tN\sum_{l=1}^Lw_lB_l + (t-\tN)\sum_{l=1}^Lw_lB_l 
    = t\sum_{l=1}^Lw_lB_l.
\end{aligned}
\end{equation}
Hence, the function $m(t)\leq t\sum_{l=1}^Lw_lB_l$.
It follows that $\lim_{t\rightarrow\infty}m(t)/t \leq \sum_{l=1}^Lw_lB_l$.
\end{proof}

Proposition \ref{prop:mt_ub} implies that the 
mean long-run rate is no larger than a convex 
combination of the hazard rates upper bounds.
As an example, consider the scaled-Lomax MTDPP 
in which the $l$th hazard function is bounded above by 
$\phi_l^{-1}$, $l = 1,\dots, L$. 
Thus, by Proposition \ref{prop:mt_ub}, 
we have that $\lim_{t\rightarrow\infty}m(t)/t\leq \sum_{l=1}^L w_l \, \phi_l^{-1}$.

\subsection{Theorem 1}

We first introduce the notation. 
Let $\{X_i\}\equiv\{X_i: i\geq 1\}$ be the MTD duration process of order $L$.
The transition density of $X_i$ given $(X_{i-1}=x_{i-1},
\dots,X_{i-L}=x_{i-L})$, for $i > L$, is 
\begin{equation}\label{eq: mtd}
\begin{aligned}
p(x_i\,|\,x_{i-1},\dots,x_{i-L}) = \sum_{l=1}^Lw_lf_l(x_i\,|\,x_{i-l}).
\end{aligned}
\end{equation}
Under the assumptions of Theorem 1, 
the conditional density $f_l(u\,|\,v)\equiv f_{U\,|\,V}(u\,|\,v)$,
for $l = 1,\dots, L$, with $f_{U\,|\,V}(u\,|\,v)$
strictly positive and continuous for all $u,v$. 
Define the transition kernel $G(x, A) = \int_Af_{U\,|\,V}(y\,|\, x)dy$
for $x\in\mathcal{X}$ and 
$A\in\mathcal{B}(\mathcal{X})$, where 
$\mathcal{B}(\mathcal{X})$ is the Borel $\sigma$-algebra. 
Under the assumptions of Theorem 1, 
$G$ admits an invariant distribution with density $f_X$,
such that $f_X(u) = \int f_{U\,|\,V}(u\,|\,v)f_X(v)dv$.

Consider the Markov chain $\{\bZ_i\}\equiv\{\bZ_i: i\geq L\}$,
with $\bZ_i = (X_i,\dots, X_{i-L+1})^\top\in\mathcal{S}$
and $\mathcal{S} \equiv \mathcal{X}^L$ equipped with the 
Borel $\sigma$-algebra $\mathcal{B}(\mathcal{S}$). 
Let $\mathrm{Pr}(\bZ_i\in S\,|\,\bZ_{i-1} = \br) = P(\br, S)$
be the transition probability of $\bZ_i\in S$ given 
$\bZ_{i-1} = \br$, where $P(\br, S)$ is the transition kernel of $\{\bZ_i\}$,
with $\br\in\mathcal{S}$ and $S\in\mathcal{B}(\mathcal{S})$.
Following the argument of Theorem 1 in \citesm{raftery1985model}, 
we have 
\begin{equation}\label{eq: Z_kernel}
\begin{aligned}
P(\br, S) = \int \mathbbm{1}_S(\by)\left[\sum_{l=1}^Lw_lG(r_l, dy_0)\right]
\prod_{j=1}^{L-1}\delta_{r_j}(dy_j),
\end{aligned}
\end{equation}
with vectors $\by = (y_0,y_1,\dots,y_{L-1})^\top$
and $\br =  (r_1,\dots,r_L)^\top$, and 
where $\delta_{x}$ is the delta measure on $\mathcal{X}$ such that
for any $A\in\mathcal{B}(\mathcal{X})$, $\delta_x(A) = 1$
if $x\in A$ and $\delta_x(A) = 0$ otherwise.

Following \citesm{meyn2012markov}, for any measurable function $h$, we write
\begin{equation}
    Ph(\br) = \int P(\br, d\by)h(\by),
\end{equation}
where $h$ is either bounded or nonnegative.

\subsubsection{Lemmas}

We present two lemmas before proving Theorem 1.
Hereafter, we refer to \citesm{meyn2012markov} as MT.

\begin{lemma}\label{lemma: le1}
The chain $\{\bZ_i\}$ with transition kernel in \eqref{eq: Z_kernel} is 
irreducible and aperiodic.
\end{lemma}

\begin{proof}
We follow the argument of Theorem 1 in \citesm{kalliovirta2015gaussian}.
Note that the density of the $L$-step transition kernel 
$P^{(L)}(\bz_L, S)$
given initial state $\bz_L = (x_1,\dots,x_L)^\top$ is 
\begin{equation}\label{eq: L-step}
p(\bz_{2L}\,|\,\bz_L) = \prod_{i=L+1}^{2L}\sum_{l=1}^Lw_lf_{U\,|\,V}(x_i\,|\,x_{i-l}),
\end{equation}
with $\bz_{2L}=(x_{2L},\dots,x_{L+1})^\top$.
Since $p(\bz_{2L}\,|\,\bz_L) > 0$ for all $\bz_L, \bz_{2L}\in\mathcal{S}$,
from every $\bz_L$, the chain $\{\bZ_i\}$ can reach any subset of 
$\mathcal{S}$ with positive Lebesgue measure in $L$ steps, and thus 
the chain $\{\bZ_i\}$ is irreducible and aperiodic 
(MT, Chapters 4.2 and 5.4).
\end{proof}

From Lemma \ref{lemma: le1}, the chain $\{\bZ_i\}$ is $\varphi$-irreducible, 
where $\varphi$ is the Lebesgue measure.
By Proposition 4.2.2 in MT, 
there exists a maximal irreducibility measure $\psi$ that is a
probability measure,
such that $\{\bZ_i\}$ is $\psi$-irreducible.

To prove Theorem 1, we will first establish that 
the chain $\{\bZ_i\}$ is geometrically ergodic,
which requires the concept of petite sets (MT, Chapter 5.5).
The following lemma shows that $\{\bZ_i\}$ is a T-chain, which 
allows us to work with compact sets, since by Theorem 6.2.5 in MT,
all compact sets are petite for a T-chain.

In particular, let $\mathcal{C}(\mathcal{S})$ denote the class of bounded 
continuous functions from $\mathcal{S}$ to $\R$.
It is known (MT, Chapter 6) 
that a Markov chain is weak Feller if and only if 
the transition kernel $P$ maps $\mathcal{C}(\mathcal{S})$ to $\mathcal{C}(\mathcal{S})$,
and that a $\psi$-irreducible Markov chain is a T-chain if it is weak Feller and the support of $\psi$ has 
non-empty interior.

\begin{lemma}
The chain $\{\bZ_i\}$ with transition kernel in \eqref{eq: Z_kernel} 
is a T-chain.    
\end{lemma}

\begin{proof}
Take $h\in \mathcal{C}(\mathcal{S})$, and we have
\begin{equation}
\begin{aligned}
Ph(\br) 
& = \int P(\br, d\by)h(\by)\\
& = \int \left[\sum_{l=1}^Lw_lG(r_l, dy_0)\right]
\prod_{j=1}^{L-1}\delta_{r_j}(dy_j)\,h(y_0,y_1,\dots,y_{L-1})\\
&= \sum_{l=1}^Lw_l\int h(y_0,r_1,\dots,r_{L-1})f_{U\,|\,V}(y_0\,|\,r_l)dy_0.    
\end{aligned}
\end{equation}
The boundedness of $Ph(\br)$ 
follows from the fact that $h$ is bounded and $f_{U\,|\,V}$ integrates to one.
Note that $h$ is bounded and continuous in $\br$, and that 
$f_{U\,|\,V}(u\,|\,v)$ is 
continuous in $v$ for each $u$. It follows from the generalized Dominated
Convergence Theorem \citepsm[][Chapter 2.3]{folland1999real} that 
$\int h(y_0,r_1,\dots,r_{L-1})f_{U\,|\,V}(y_0\,|\,r_l)dy_0$ is a continuous
function of $\br\in\mathcal{S}$, for $l = 1,\dots, L$,
and thus $Ph(\br)$ is continuous. 
By Proposition 6.1.1 in MT, the chain $\{\bZ_i\}$ is weak Feller.
Since $\{\bZ_i\}$ can reach any subset of $\mathcal{S}$ with positive 
Lebesgue measure in $L$ steps, the support of 
$\psi$ has non-empty interior. It follows from Theorem 6.2.9 in MT that 
$\{\bZ_i\}$ is a T-chain.
\end{proof}

\subsubsection{Proof of Theorem 1}
We now give a proof of Theorem 1. 
Specifically, we first use Theorem 15.0.1 in MT to establish
the geometric ergodicity of $\{\bZ_i\}$ for any initial 
condition $\bz_L$. This requires 
that $\{\bZ_i\}$ is irreducible and aperiodic, and that there exists 
an everywhere finite function $\tau'\geq1$, 
and a petite set $C'$, such that for some $\beta'<1$, $b'<\infty$, 
the following `geometric drift condition' is satisfied:
\begin{equation}\label{eq: V4}
P\tau'(\br) \leq \beta' \tau'(\br) + b'\mathbbm{1}_{C}(\br).
\end{equation}
For drift conditions to test various forms
of stability, we refer to Appendix B in MT for details;
see also \citesm{mengersen1996rates} and \citesm{carrasco2002mixing}
in the context of MCMC and generalized random coefficient models,
respectively.

\begin{proof}
    
By Lemma 1, the chain $\{\bZ_i\}$ is irreducible and aperiodic.
For the geometric drift condition, 
we consider the following test function,
\begin{equation}
\tau'(y_0,\dots,y_{L-1}) = \rho_0\tau(y_0) + \dots + \rho_{L-1}\tau(y_{L-1}),
\end{equation}
with finite $\rho_j>0$, for $j = 0, 1,\dots,L-1$, to be specified later.
Since by assumption 
$\tau$ is everywhere finite, $\tau'$ is everywhere finite.

The geometric drift condition for the chain $\{\bZ_i\}$ is then given by
\begin{equation}
\begin{aligned}
P\tau'(\br) & = \int P(\br,d\by) \tau'(\by)\\
& = \int\left[\sum_{l=1}^Lw_lG(r_l, dy_0)\right]\prod_{j=1}^{L-1}\delta_{r_j}(dy_j)
    \left(\rho_0\tau(y_0) + \sum_{j=1}^{L-1}\rho_j\tau(y_j)\right)\\
& = \rho_0\sum_{l=1}^Lw_lE(\tau(U)\,|\,V = r_l) + \sum_{j=1}^{L-1}\rho_j\tau(r_j).
\end{aligned}
\end{equation}
Under the assumptions in Theorem 1, there exists a compact set $C$,
such that for some $\beta<1$, $b<\infty$, 
$E(\tau(U)\,|\,V = x) \leq \beta \tau(x) + b\mathbbm{1}_{C}(x)$. It follows that
\begin{equation}
\begin{aligned}
P\tau'(\br) & \leq \rho_0\sum_{l=1}^Lw_l[\beta \tau(r_l) + b\mathbbm 1_{C}(r_l)] + 
\sum_{j=1}^{L-1}\rho_j\tau(r_j)\\
& = \sum_{l=1}^{L-1}(\rho_0\beta w_l + \rho_l)\tau(r_l) + \rho_0\beta w_L\tau(r_L) + 
b\rho_0\sum_{l=1}^Lw_l\mathbbm{1}_{C}(r_l).\\
\end{aligned}
\end{equation}
To find $\{\rho_j\}$ that satisfies the inequality in \eqref{eq: V4}, we set $\rho_0 = 1$ and 
choose $\rho_j$, $j = 1,\dots,L-1$, such that for some $\beta'< 1$,
\begin{equation}
\begin{cases}
    \beta w_l + \rho_l \leq \beta'\rho_{l-1} < \rho_{l-1},\; l = 1,\dots, L-1,\\
    \beta w_L \leq \beta'\rho_{L-1} < \rho_{L-1}.
\end{cases}
\end{equation}

We start by setting $\rho_{L-1} = \beta w_L + \epsilon_{L-1}$ for some $\epsilon_{L-1} > 0$. 
We then define recursively, for $l=L-2,\dots, 1$,
$\rho_{l} = \beta w_{l+1} + \rho_{l+1} + \epsilon_l$,
for some $\epsilon_l>0$. It remains to verify $\beta w_1 + \rho_1 < 
\rho_0 = 1$. Note that
\begin{equation}
\begin{aligned}
\beta w_1 + \rho_1 
& = \beta w_1 + (\beta w_2 + \rho_2 + \epsilon_1)\\
& = \vdots\\
& = \beta (w_1 +\cdots+w_L) + \epsilon_1 + \cdots + \epsilon_{L-1}\\
& = \beta + \epsilon_1 + \cdots + \epsilon_{L-1} < 1,
\end{aligned}
\end{equation}
provided that we choose $\epsilon_1 + \cdots + \epsilon_{L-1} < 1 - \beta$.
It follows that
\begin{equation}\label{eq: drift}
\begin{aligned}
P\tau'(\br)
&\leq \beta' \tau'(\bm r) + b\rho_0\sum_{l=1}^Lw_l\mathbbm{1}_{C}(r_l)\\
&\leq \beta' \tau'(\bm r) + b'\mathbbm{1}_{C'}(\bm r),
\end{aligned}
\end{equation}
for some $\beta'<1$, where $b' = b\rho_0$ and $C' = C^L$ is compact.
By Lemma 2, $\{\bZ_i\}$ is a T-chain, 
and thus $C'$ is a petite set.

Using Theorem 15.0.1 in MT, $\{\bZ_i\}$ is geometrically ergodic 
with an invariant distribution $\pi$, in the sense that there exists constants 
$\rho>1$, $R<\infty$, such that, for all initial condition $\bz_L$,
$\sum_{n=1}^{\infty}\rho^n||P^{(n)}(\bz_L,\cdot) - \pi||_{\tau'}\leq R\tau'(x)$.
Here, the $\tau'$-norm is defined as $||\la||_{\tau'} =$
$\sup_{g:|g|\leq\tau'}|\int g(y)\la(dy)|$, for any signed measure $\la$.

Since $\{\bZ_i\}$ is geometrically ergodic, by Theorem 3 in 
\citesm{tierney1994markov}, for any initial distribution,
$n^{-1}\sum_{i=1}^n\bZ_i\rightarrow \bm\mu_Z\;a.s.$,
provided that $\bm\mu_Z = \int \bz\pi(d\bz)< \infty$.
Note that under the assumptions in Theorem 1,  
each coordinate of the invariant distribution $\pi$ has marginal density $f_X$.
Then we have $\bm\mu_Z = (\mu,\dots,\mu)^\top$,
and $n^{-1}\sum_{i=1}^nX_i\rightarrow \mu\;a.s.$,
where $\mu = \int xf_X(x)dx < \infty$.

Let $T_{N(t)} = \sum_{i=1}^{N(t)}X_i$ be the 
last arrival time prior to $t$ or the arrival time at $t$. 
We follow Chapter 3.3 in \citesm{resnick2013adventures}. 
As $t\rightarrow\infty$,
$T_{N(t)}/N(t)=\sum_{i=1}^{N(t)}X_i/N(t)\rightarrow\mu\;a.s.$,
since as $t\rightarrow\infty$,
$N(t)\rightarrow\infty\;a.s.$.
Note that $T_{N(t)} \leq t < T_{N(t) + 1}$, and that
$T_{N(t)}/N(t) \leq t/N(t) < T_{N(t) + 1}/N(t)$. 
 Observing that 
$T_{N(t)+1}/N(t) = \{T_{N(t)+1}/(N(t)+1)\}\{(N(t)+1)/N(t)\}$,
where the first term $T_{N(t)+1}/(N(t)+1)\rightarrow\mu\;a.s.$, and
the second term $(N(t)+1)/N(t)\rightarrow 1$, we can conclude
that $N(t)/t\rightarrow1/\mu\;a.s.$.

\end{proof}

\subsubsection{Examples}

\paragraph{Scaled-Lomax MTDPP} The density of the transition kernel 
$G$ is
\begin{equation}\label{eq: lomax-dens}
f_{U\,|\, V}(u\,|\,v) = \frac{\alpha}{\alpha\phi+v}
\left(1 + \frac{u}{\alpha\phi+v}\right)^{-(\alpha+1)},
\end{equation}
and the marginal densities are
\begin{equation}
f_X(x) = f_U(x) = f_V(x) = \frac{\alpha-1}{\alpha\phi}
\left(1 + \frac{x}{\alpha\phi}\right)^{-(\alpha)}.
\end{equation}
Note that $\mu = \int xf_X(x)dx = \alpha\phi/(\alpha-2)$ exists if
$\alpha > 2$.
Similarly, $E(U\,|\,V = v) = (\alpha\phi+v)/(\alpha-1)$ exists if $\alpha>1$.

Consider a test function $\tau(x) = 1 + x$ for $x\geq0$, and assume $\alpha>1$. Then
\begin{equation}
\begin{aligned}
E(\tau(U)\,|\,V =v)  = 1 + \frac{\alpha\phi+v}{\alpha-1}
= \beta^*\tau(v) + b^*,
\end{aligned}
\end{equation}
where $\beta^*=1/(\alpha-1)<1$ provided that $\alpha>2$,
and $b^* = 1 + \frac{\alpha\phi-1}{\alpha-1}$.

Set $\tilde\beta = (1-\beta^*)/2$ and define the compact set $C$ as 
$C = \{x: \tau(x)\leq b^*/\tilde\beta\}$.
It follows that, for $x\in C$, 
$E(\tau(U)\,|\,V =v)\leq \beta^* b^*/\tilde\beta + b^* < \infty$.
For $x\notin C$ (i.e. $\tau(x)>b^*/\tilde\beta)$, we have that 
\begin{equation}
E(\tau(U)\,|\,V =v) \leq \beta^*\tau(v) + \tilde\beta\tau(v) = \beta\tau(v),
\end{equation}
where $\beta = \beta^*+\tilde\beta = (1+\beta^*)/2<1$.

Let $b = \beta^* b^*/\tilde\beta + b^*$. The following condition,
$E(\tau(U)\,|\,V =v) \leq \beta\tau(v) + b\mathbbm{1}_C(v)$,
is satisfied. Thus, by Theorem 1, $N(t)/t\rightarrow1/\mu\;a.s.$,
provided that $\alpha > 2$.

\paragraph{Burr MTDPP} The density of the transition kernel 
$G$ is
\begin{equation}\label{eq: burr-dens}
f_{U\,|\, V}(u\,|\,v) = 
\kappa\ga u^{\ga-1}\frac{1}{\tilde\la(v)^{\ga}}
\left\{ 1 + \left(\frac{u}{\tilde\la(v)}\right)^{\ga} \right\}^{-(\kappa+1)},
\end{equation}
where $\tilde\la(v) = (\la^{\ga} + v^{\ga})^{1/\ga}$.
The marginal densities are
\begin{equation}
f_X(x) = f_U(x) = f_V(x) = 
(\kappa-1)\ga x^{\ga-1}\frac{1}{\la^{\ga}}
\left\{ 1 + \left(\frac{x}{\la}\right)^{\ga} \right\}^{-(\kappa)}.
\end{equation}
Note that $E(X^a) = \int x^af_X(x)dx = 
(\kappa-1)\lambda^{a}B(\kappa - 1 - a/\gamma, 1 + a/\gamma)$ exists
if $\kappa > 1 + a/\gamma$.
Similarly, $E(U^a\,|\,V = v) = \kappa\tilde\lambda(v) B(\kappa-a/\gamma,1+a/\gamma)$ 
exists if $\kappa > a/\gamma$. The marginal mean $\mu = E(X)$ exists
if $\kappa > 1+ 1/\gamma$.

Consider a test function $\tau(x) = 1 + x^\gamma$ for $x\geq0$,
and assume $\kappa > \gamma/\gamma = 1$. Then
\begin{equation}
\begin{aligned}
E(\tau(U)\,|\,V =v)  =  1 + \kappa(\la^{\ga} + v^{\ga})B(\kappa - 1, 2) 
= \beta^*\tau(v) + b^*,
\end{aligned}
\end{equation}
where $\beta^*=\kappa B(\kappa - 1, 2) = 1/(\kappa-1) < 1$
provided that $\kappa>2$,
and $b^* = 1 + \kappa(\la^\gamma-1)B(\kappa-1,2) < \infty$.

Similar to the previous example, 
we set $\tilde\beta = (1-\beta^*)/2$ and define a compact set $C$ as
$C = \{x: \tau(x)\leq b^*/\tilde\beta\}$.
Then we have that 
$E(\tau(U)\,|\,V =v) \leq \beta\tau(v) + b\mathbbm{1}_C(v)$,
is satisfied, 
where $\beta = \beta^*+\tilde\beta$ and $b = \beta^* b^*/\tilde\beta + b^*$.
Thus, by Theorem 1, $N(t)/t\rightarrow1/\mu\;a.s.$,
provided that $\kappa > \max\{2, 1+1/\gamma\}$.

\subsection{Proofs of propositions}

In this section, we provide proofs of the propositions in the main paper,
following the notation developed in the main paper.

\vspace{5pt}

\begin{proof}[\textbf{Proof of Proposition 1}]
By Definition 1 in the main paper,
the conditional density of duration $X_i$ 
is $f^*(x_i)\equiv f^*(t_i-t_{i-1})$, 
where $x_i = t_i - t_{i-1}$, for $i\geq 2$. 
Then, according to (3) and (4) in the main paper, we have 
$f^*(x_i) = \sum_{l=1}^L w_l \, f_{l}(x_i\,|\, x_{i-l}),\;i > L$,
and 
$f^*(x_i)= \sum_{l=1}^{i-2} w_l \, f_{l}(x_i\,|\, x_{i-l}) + 
(1 - \sum_{r=1}^{i-2}w_r)f_{i-1}(x_i\,|\,x_1),\;i = 2,\dots, L$.

Let $f_X$ be a marginal density of interest, and 
denote by $g_i(x_i)$ the marginal density of $X_i$, for $i\geq 1$.
By condition (i) in Proposition 1, $g_1(x_1)\equiv f_X(x_1)$,
and let $p(x_1)\equiv g_1(x)$ denote the density of $X_1$.
Using both conditions (i) and (ii) in the proposition,
we have, for $i\geq 2$, 
\begin{equation}\label{eq:sta}
    \begin{aligned}
    g_i(x_i) = \int f^*(x_i)\,p(x_1,\dots,x_{i-1})\,dx_1\cdots dx_{i-1}
    = f_X(x_i),
\end{aligned}
\end{equation}
where $p(x_1,\dots,x_{i-1})$ is the joint density
for random vector $(X_1,\dots, X_{i-1})$.
The second equality in \eqref{eq:sta} 
follows from the argument of Proposition 1 in \cite{zheng2021construction}.

\end{proof}

\begin{proof}[\textbf{Proof of Proposition 2}]
Let $(U,V) = (\alpha X, \alpha Y)$, where the joint density of $(X,Y)$ is 
$f_{X,Y}(x,y) \propto (\la_0 + \la_1x + \la_2y)^{-(\alpha+1)}$, 
which corresponds to the bivariate Lomax distribution of 
\cite{arnold1999conditional}.
By change of variable, we obtain the joint density of $(U,V)$, namely, 
\begin{equation}
f_{U,V}(u,v)\propto(\la_0a + \la_1u/\alpha + \la_2v/\alpha)^{-(\alpha+1)},
\end{equation}
with normalizing constant 
\begin{equation}
C = \int_0^{\infty}\int_0^{\infty}(\la_0 + \la_1u/\alpha + \la_2v/\alpha)^{-(\alpha+1)}dudv
= \alpha\la_0^{-(\alpha-1)}\{(\alpha-1)\la_1\la_2\}^{-1}.
\end{equation}
The marginal density of $U$ is
\begin{equation}
\begin{aligned}
f_U(u)  & = C^{-1}\int_0^{\infty}\alpha^{-2}(\la_0 + \la_1u/\alpha +
\la_2v/\alpha)^{-(\alpha+1)}dv\\
& = (\alpha-1)(\la_0\alpha)^{-1}\la_1\{1 + (\la_0\alpha)^{-1}\la_1u\}^{-\alpha}.
\end{aligned}
\end{equation}
Since $u$ and $v$ are symmetric in the joint density
$f_{U,V}(u,v)$, the marginal density 
\begin{equation}
f_V(v)= (\alpha-1)(\la_0\alpha)^{-1}\la_2\{1 + (\la_0\alpha)^{-1}\la_2v\}^{-\alpha}.
\end{equation}
It follows that the conditional density,
\begin{equation}
f_{U|V}(u\,|\,v) = f_{U,V}(u,v)/f_V(v) = 
\alpha\la_1(\alpha\la_0+\la_2v)^{-1}\{1+\la_1u(\alpha\la_0+\la_2v)^{-1}\}^{-(\alpha+1)}.
\end{equation}
Similarly, we have 
$f_{V|U}(v\,|\,u) =
\alpha\la_2(\alpha\la_0+\la_1u)^{-1}\{1+\la_2v(\alpha\la_0+\la_1u)^{-1}\}^{-(\alpha+1)}$.
\end{proof}

\vspace{5pt}

\begin{proof}[\textbf{Proof of Proposition 3}]
The survival function of the conditional duration distribution can be 
expressed as 
\begin{equation}
S^*(t - \tN) = \sum_{l=1}^{t_L} w_l^* \, 
\left(1 + \frac{t - \tN}{\alpha\phi +t_{\Ntl-l+1}-t_{\Ntl-l}}\right)^{-\alpha},
\end{equation}
where $t_L = \min\{\Ntl, L\}$.
In particular, 
for $\Ntl\geq L$, $w_l^* = w_l$, for $l = 1,\dots, L$.
When $1\leq \Ntl < L$, $w_l^* = 1,\dots,t_L-1$,
and $w_{t_L}^* = 1-\sum_{r=1}^{t_L-1}w_r$. It follows that
the weights $w_l^*$ satisfy 
$\sum_{l=1}^{t_L}w_l^* = 1$ for $N(t)\geq 1$.

Then, for $N(t)\geq 1$, we have that 
\begin{equation}\label{eq:lomax_mtdpp_surv}
\begin{aligned}
&S^*(t - \tN)\\ 
= &\sum_{l=1}^{t_L}w_l^*\left\{\left(1 + \frac{t - \tN}{\alpha\phi + 
t_{\Ntl-l+1}-t_{\Ntl-l}}\right)^{-(\alpha\phi + 
t_{\Ntl-l+1}-t_{\Ntl-l})}\right\}^{1/\phi}\\
&\;\;\;\;\;\;\;\;\;\;\;\;\times\left(1 + \frac{t-\tN}{\alpha\phi +
t_{\Ntl-l+1}-t_{\Ntl-l}}\right)^{(t_{\Ntl-l+1}-t_{\Ntl-l})/\phi}.
\end{aligned}
\end{equation}
As $\alpha\rightarrow\infty$,
the limits of the first term and the second term 
in the $l$th mixture component of \eqref{eq:lomax_mtdpp_surv}
are $\exp(-(t-\tN)\phi^{-1})$ and $1$, respectively.
More specifically, the limit of the first term is obtained by using the results
that (i) $\lim_{n\rightarrow\infty}(1+x/n)^n = \exp(x)$;
(ii) $\lim_{n\rightarrow\infty}g_1(n)/g_2(n) = 
\lim_{n\rightarrow\infty}g_1(n)/\lim_{n\rightarrow\infty}g_2(n)$,
provided that both $\lim_{n\rightarrow\infty}g_1(n)$
and $\lim_{n\rightarrow\infty}g_2(n)$ exist, and 
$\lim_{n\rightarrow\infty}g_2(n)\neq 0$.

Since $\sum_{l=1}^{t_L}w_l^* = 1$, 
it follows that, as $\alpha\rightarrow\infty$, the survival function of the 
conditional duration distribution converges to $\exp(-(t-\tN)\phi^{-1})$, 
which is the survival function of an exponential distribution with rate parameter $\phi^{-1}$.

\end{proof}

\subsection{Identifiability}

Identifiability of a standard finite mixture model,
commonly referred to as generic identifiability (e.g.,
\citealtsm{fruhwirth2006finite}), 
has been well addressed in the literature (see, e.g., \citealtsm{teicher1961identifiability,teicher1963identifiability,yakowitz1968identifiability,chandra1977mixtures,kent1983identifiability,crawford1994application}). 
Generally, a regular finite mixture model is said to be 
identifiable if no two sets of parameter values, up to permutation
of the components, produce the same distribution or density 
\citepsm{mclachlan2019finite}. 
We refer to Chapter 3 in \citesm{titterington1985statistical}
for a discussion of the generic identifiability and
relevant theoretical results.

We present below a definition of the identifiability
for MTDPPs.

\begin{definition}\label{def: mtdpp_iden}
\textit{Given a realization of durations, $(x_1,\dots, x_n)$, 
consider an MTDPP model with conditional duration density 
given in (4) of the main paper, with parameters
$\{\bw,\bth\}\in\bm\Psi$, the associated parameter space,
where $\bw = (w_1,\dots, w_L)^\top$ is the vector of weights and 
$\bth = \{\bth_1,\dots,\bth_L\}$ denotes the component parameters.
The MTDPP is said to be identifiable if for any 
two sets of parameters $\{\bw,\bth\},\{\bw',\bth'\}\in\bm\Psi$,
\begin{equation}\label{eq: id_def}
\sum_{l=1}^Lw_lf_l(x_i\,|\,x_{i-l}, \bth_l) = 
\sum_{l=1}^{L'}w_l'f_l(x_i\,|\,x_{i-l},\bth'_l)
\end{equation}
for each $i = L+1,\dots, n$ and for all possible values of $x_i$, 
implies that
$L = L'$, $w_l = w_l'$, and $\bth_l = \bth'_l$, $l = 1,\dots, L$.}
\end{definition}

Definition \ref{def: mtdpp_iden} suggests that we can use 
established results for standard finite mixture models 
to verify the identifiability of an MTDPP, by treating
the conditional density in \eqref{eq: id_def}
for each $i$ as a standard finite mixture model.
A similar idea has been used in \cite{hassan2006modeling} to verify 
the identifiability of their proposed bivariate MTD models. 
As examples, we demonstrate the identifiability of the Burr
MTDPP, the Lomax MTDPP, and the scaled-Lomax MTDPP, which
are illustrated in Section 2.3 of the main paper.

\paragraph{Burr MTDPP} 
To verify the identifiability of the class of Burr MTDPPs,
it suffices to show that a finite mixture of the corresponding
Burr distributions is identifiable.
Identifiability for finite mixtures of Burr distributions has been 
studied in \citesm{ahmad1994small} and \citesm{al2016mixture}. 
In particular, \citesm{ahmad1994small} shows that, 
for the two-parameter Burr-Type-\RomanNumeralCaps{7}
distribution with c.d.f. $F(x) = 1 - (1 + x^\gamma)^{-\kappa}$, 
the corresponding finite mixture model is identifiable with a common shape 
parameter $\gamma$, using Theorem 1 in \citesm{teicher1963identifiability}. 
More recently, \citesm{al2016mixture} shows that the finite mixtures of 
two-parameter Burr-Type-\RomanNumeralCaps{3}
distributions with c.d.f. $F(x) = (1 + x^{-\gamma})^{-\kappa}$ is identifiable, using Theorem 2.4 in \citesm{chandra1977mixtures}.

Here, we show that the finite mixtures of three-parameter Burr-Type-\RomanNumeralCaps{7}
distributions with c.d.f. $F(x) = 1 - (1 + (x/\la)^\gamma)^{-\kappa}$ is identifiable,
using Theorem 2.4 in \citesm{chandra1977mixtures}.

\begin{proof}
Let $\sF=\{F(x;\kappa,\gamma,\lambda) = 1 - (1 + (x/\la)^\gamma)^{-\kappa};\kappa>0,\gamma>0,\lambda>0\}$
be the family of three-parameter Burr-Type-\RomanNumeralCaps{7} distributions.
Consider the following transformation,
\begin{equation}
\phi_i(t) = E(X^t) = \lambda_i^{t}\kappa_i\,B(\kappa_i - t/\gamma_i, 
1 + t/\gamma_i),\;t < \kappa_i\gamma_i,
\end{equation}
where $B(a, b) = \Gamma(a)\Gamma(b)/\Gamma(a+b)$ is the beta function, and
$E(X^t)$, for a random variable $X$, is taken with respect to 
a Burr distribution with c.d.f. $F_i(x)\in\sF$, where 
$F_i(x) \equiv F(x;\kappa_i,\gamma_i,\lambda_i) 
= 1 - (1 + (x/\la_i)^{\gamma_i})^{-\kappa_i}$.

We order the family lexicographically by:
$F_1(x) < F_2(x)$ if $\kappa_1 < \kappa_2$, or 
if $\kappa_1 = \kappa_2$ but $\gamma_1 < \gamma_2$,
or if $\kappa_1 = \kappa_2, \gamma_1 = \gamma_2$ but $\lambda_1 > \lambda_2$.
Then we have that $D_{\kappa_1\gamma_1}\subseteq D_{\kappa_2\gamma_2}$,
where $D_{\kappa_i\gamma_i} = (-\infty, \kappa_i\gamma_i)$, 
$i = 1,2$.
Take $t_1 = \kappa_1\gamma_1$ and note that $t_1$ in the closure of $D_{\kappa_1\gamma_1}$.
Then we have that
\begin{equation}
\begin{aligned}
\lim_{t\rightarrow t_1}\phi_1(t) & = 
\lim_{t\rightarrow \kappa_1\gamma_1}
\lambda_1^{t}\kappa_1B(\kappa_1 - t/\gamma_1, 
1 + t/\gamma_1)\\
& = \lim_{t\rightarrow \kappa_1\gamma_1}
\lambda_1^{t}\,\Gamma(\kappa_1-t/\gamma_1)\,
\Gamma(1 + t/\gamma_1) = \infty,
\end{aligned}
\end{equation}
since $\lim_{t\rightarrow\kappa_1\gamma_1}\Gamma(\kappa_1-t/\gamma_1) = \infty$,
$\lim_{t\rightarrow\kappa_1\gamma_1}\lambda_1^{t} = \lambda_1^{\kappa_1\gamma_1}$,
and $\lim_{t\rightarrow\kappa_1\gamma_1}\Gamma(1+t/\gamma_1) = \Gamma(1 + \kappa_1)>0$. On the other hand, we have that
\begin{equation}
\begin{aligned}
\lim_{t\rightarrow t_1}\phi_2(t) & = 
\lim_{t\rightarrow \kappa_1\gamma_1}\lambda_2^{t}\kappa_2\,B(\kappa_2 - t/\gamma_2, 
1 + t/\gamma_2)\\
& = \lambda_2^{\kappa_1\gamma_1}\,
\Gamma(\kappa_2 - \kappa_1\gamma_1/\gamma_2)\,
\Gamma(1 + \kappa_1\gamma_1/\gamma_2) > 0.
\end{aligned}
\end{equation}
It follows that $\lim_{t\rightarrow t_1}\phi_2(t)/\phi_1(t) = 0$
and Theorem 2.4 in \citesm{chandra1977mixtures} applies.
\end{proof}

Since a finite mixture of three-parameter Burr-Type-\RomanNumeralCaps{7}
distributions is identifiable, based on Definition \ref{def: mtdpp_iden}, 
we have that
$L = L'$, $w_l = w'_l$, $\kappa = \kappa'$, $\gamma = \gamma'$,
and $(\lambda^{\gamma} + x_{i-l}^{\gamma})^{1/\gamma} = 
(\lambda'^{\gamma'} + x_{i-l}^{\gamma'})^{1/\gamma'}$, for each $l$ 
and for each $i$.
It follows that $\lambda = \lambda'$ for each $i$.
Thus, the class of Burr MTDPPs is identifiable based on Definition
\ref{def: mtdpp_iden}.

\paragraph{Lomax and Scaled-Lomax MTDPPs}
The scaled-Lomax distribution can be treated as a reparameterized Lomax 
distribution, and thus it suffices to prove the identifiability for the Lomax 
MTDPP. Note that \citesm{ahmad1988identifiability} has verified that a finite mixture 
of Pareto-Type-\RomanNumeralCaps{1} distributions is identifiable,
and that the Lomax distribution is a shifted version of the 
Pareto-Type-\RomanNumeralCaps{1} distribution. 
For a Pareto-Type-I distribution with c.d.f. 
$F(x) = 1 - (x/\la)^{-\alpha}$, the c.d.f. of the corresponding Lomax distribution
is $F(x) = 1  - ((x+\la)/\la)^{-\alpha} =  1  - (1 + x/\la)^{-\alpha}$.
It follows that a finite mixture of Lomax distributions is identifiable.
Based on Definition \ref{def: mtdpp_iden}, we have that
$L = L'$, $w_l = w'_l$, $\alpha_l = \alpha'_l$, 
and $\phi_l + x_{i-l} = \phi_l' + x_{i-l}$, for each $l$ and for each $i$.
It follows that $\phi_l = \phi'_l$ for each $l$ and for each $i$.
Thus, the class of Lomax MTDPPs is identifiable, and so is the class
of scaled-Lomax MTDPPs.

\section{Bivariate Burr distribution}

Let $X$ be a random variable, and its 
cumulative distribution function (c.d.f.) is
$F(x) = 1 - \left(1 + (x/\la)^{\ga}\right)^{-\psi}$.
We say that $X$ follows a three-parameter Burr distribution \citepsm{tadikamalla1980look}, denoted as 
$\mathrm{Burr}(x\,|\,\ga,\la,\psi)$.
We will use such notation throughout to indicate 
either the distribution or its density for a Burr random variable, 
depending on the context (we follow the same notation approach for other distributions).

Consider a bivariate random vector $(X,Y)$, with marginal c.d.f.s for $X$ and $Y$ 
given by $F(x) = 1 - \left(1 + (x/\la)^{\ga}\right)^{-\psi}$
and $F(y) = 1 - \left(1 + (y/\la)^{\ga}\right)^{-\psi}$, respectively.
The joint c.d.f. $F(x,y)$ is specified by the heavy right tail (HRT) copula
given by 
\begin{equation}\label{eq:hrt}
C(u,v) = u +  v - 1 + \left[(1-u)^{-1/a} + (1-v)^{-1/a} -1\right]^{-a},
\end{equation}
where $0\leq u\leq 1$, $0\leq v\leq 1$, and $a > 0$
\citepsm{frees1998understanding}.

We set the copula parameter to be the same as the second shape parameter of 
the Burr distribution, that is, $a = \psi$. 
Replace $u$ and $v$ with $F(x)$ and $F(y)$, respectively, in
\eqref{eq:hrt}. Then, the joint c.d.f. of the random vector $(X,Y)$ is 
given by 
\begin{equation}
\begin{aligned}
F(x, y) & = F(x) + F(y) - 1 + \left[(1 - F(x))^{-1/\psi} + (1-F(y))^{-1/\psi} - 1\right]^{-\psi}\\
& = 1 - \left(1 + \left(\frac{x}{\la}\right)^{\ga}\right)^{-\psi} -
\left(1 + \left(\frac{y}{\la}\right)^{\ga}\right)^{-\psi}  + 
\left[1 + \left(\frac{x}{\la}\right)^{\ga} + \left(\frac{y}{\la}\right)^{\ga}\right]^{-\psi}.
\end{aligned}
\end{equation}
The conditional c.d.f. of $Y$ given $X=x$ is $F(y\,|\,x) = \partial C(F(x),F(y))/\partial F(x)$.
Note that
$\partial C(u,v)/\partial u = 1 - 
\left[(1-u)^{-1/\psi}\right]^{\psi+1}\left[(1-u)^{-1/\psi} + (1-v)^{-1/\psi} -1\right]^{-(\psi+1)}$.
It follows that
\begin{equation}\label{eq:cond_burr}
\begin{aligned}
F(y\,|\,x) & = 1 - \left[1 + \left(\frac{x}{\la}\right)^{\ga}\right]^{\psi+1}
\left[1 + \left(\frac{x}{\la}\right)^{\ga} + 1 + \left(\frac{y}{\la}\right)^{\ga} - 1\right]^{-(\psi + 1)}\\
& = 1 - \left[1 + \frac{\left(\frac{y}{\la}\right)^{\ga}}{1 + \left(\frac{x}{\la}\right)^{\ga}}\right]^{-(\psi+1)} = 1 - \left[1 + \frac{y^\ga}{\la^{\ga} + x^{\ga}}\right]^{-(\psi+1)}\\
& = 1 - \left[1 + \left(\frac{y}{\tilde\la(x)}\right)^{\ga}\right]^{-(\psi+1)},
\end{aligned}
\end{equation}
where $\tilde\la(x) = (\la^{\ga} + x^{\ga})^{1/\ga}$. Therefore, the conditional 
distribution of $Y$ given $X = x$ is a Burr distribution,
$\mathrm{Burr}(\ga,\tilde\la(x),\psi+1)$. 
Since the HRT copula is symmetric in its arguments, the conditional 
distribution of $X$ given $Y = y$ is also a Burr distribution.

We note that the bivariate Burr distribution defined through the HRT copula and 
Burr marginals was considered in \citesm{venter2002tails}. However, the expressions 
for the conditional c.d.f.s reported in \citesm{venter2002tails} include an error. 
Equation \eqref{eq:cond_burr} provides the corrected expression for the 
conditional c.d.f. of $Y$ given $X$.

\section{Additional details for Bayesian implementation}

In Section \ref{sec: prediction}, we provide details for posterior prediction 
using MTDPPs. Posterior inference and prediction for MTDCPPs are introduced 
in Section \ref{sec: inference-mtdcpp}.

The observed point pattern comprises event times $0 = t_0 <t_1 <\ldots < t_n < T$,
with corresponding observed durations $x_i = t_i - t_{i-1} > 0$, for $i = 1,\dots,n$.
Let $D_n = \{t_1,\dots,t_n, t_{n+1} > T\}$, for $n > L$, represent the information 
from the observed point pattern. Note that, as its description highlights, $D_n$ 
includes the information that the (unobserved) event time $t_{n+1}$ is greater than 
the upper bound $T$ of the time observation window, i.e., that the (unobserved) 
duration $x_{n+1}$ is greater than $T - t_{n}$.

\subsection{Posterior prediction for MTDPPs}
\label{sec: prediction}

To obtain the posterior predictive density for the next duration, $p(x_{n+1} \,|\, D_n)$,
we first derive the conditional density for the next duration given the model parameters,
$\{ \bth,\bm w \}$, and $D_n$. As discussed above, $D_n$ implies conditioning on event 
$x_{n+1} > T - t_{n}$. Therefore, using Equation (4) in the main paper, we obtain:
\begin{equation}\label{eq: cond_out_1}
\begin{aligned}
p(x_{n+1} \,|\, D_n, \bth,\bm w) 
& = f^*(x_{n+1})\,/\,S^*(T-t_n)\\
& = \frac{\sum_{l=1}^L w_l \, f_l(x_{n+1}\,|\,x_{n+1-l},\bth_l)}
{\int_{T-t_n}^{\infty}\sum_{l=1}^L w_l \, f_l(x_{n+1}\,|\,x_{n+1-l},\bth_l)\,dx_{n+1}}\\
& = \frac{\sum_{l=1}^L w_l \, S_l(T-t_n\,|\,x_{n+1-l},\bth_l) \,
\tilde{f}_l(x_{n+1} \,|\,x_{n+1-l},\bth_l)}
{\sum_{l=1}^L w_l \, S_l(T-t_n\,|\,x_{n+1-l},\bth_l)}\\
& = \sum_{l=1}^Lw_l^*(T) \, \tilde f_l(x_{n+1} \,|\,x_{n+1-l},\bth_l),
\,\,\,\,\, x_{n+1} \in (T-t_n,\infty).
\end{aligned}
\end{equation}
Here, the weights $w_l^*(T) =
w_l \, S_l(T-t_n\,|\,x_{n+1-l},\bth_l)/
\{ \sum_{l=1}^L w_l \, S_l(T-t_n\,|\,x_{n+1-l},\bth_l) \}$, and
$\tilde f_l(x_{n+1} \,|\,x_{n+1-l},\btheta_l)
= f_l(x_{n+1}\,|\,x_{n+1-l},\btheta_l)/S_l(T-t_n\,|\,x_{n+1-l},\btheta_l)$, 
for $x_{n+1} \in (T-t_n,\infty)$, is the $l$th component density for $X_{n+1}$
truncated below at $T - t_{n}$.

Hence, the posterior predictive density for the next duration is given by
\begin{equation}
\label{eq: pred_out_1}   
p(x_{n+1} \,|\, D_n) = \int\int 
\left\{ \sum_{l=1}^L w_l^*(T) \, \tilde f_l(x_{n+1} \,|\,x_{n+1-l},\btheta_l) \right\}
p(\btheta,\bm w \,|\,D_n) \, d\btheta\,d\bm w,
\end{equation}
for $x_{n+1} \in (T-t_n,\infty)$, where $p(\btheta,\bm w\,|\,D_n)$ is the 
posterior distribution of $\{\btheta,\bm w\}$.

Then, for $k\geq 2$, the $k$-step-ahead posterior predictive density 
of duration $x_{n+k}$, 
\begin{equation}\label{eq: pred_out_k}
\begin{aligned}
\int\int
& \Big\{\int\dots\int
\Big\{\prod_{j = n+2}^k\sum_{l=1}^L w_l \, f_l(x_j\,|\,x_{j-l},\bth_l)\Big\}\,
p(x_{n+1} \,|\, D_n, \bth,\bm w) \, \\
&\;\;\; dx_{n+1}\dots dx_{n+k-1}\Big\}\
p(\btheta,\bm w\,|\,D_n)\,d\btheta\,d\bm w.
\end{aligned}
\end{equation}

For in-sample prediction of $x_i$, for $i = L+1,\dots, n$, given the observed point 
pattern $0<t_1<\ldots <t_n<T$, the posterior predictive density for $x_i$ is given by
\begin{equation}\label{eq: pred_in}
\int\int 
\left\{ \sum_{l=1}^L w_l \, f_l(x_i\,|\,x_{i-l},\bth_l) \right\} \,
p(\bth,\bm w\,|\,D_n) \, d\bth\,d\bm w.
\end{equation}

\subsection{Posterior inference and prediction for MTDCPPs}
\label{sec: inference-mtdcpp}

Consider an MTDCPP for durations $x_1,\dots, x_n$, and 
take $\tx_{n+1} = T - t_n$. The likelihood conditional on $(x_1,\dots,x_L)$ is 
\begin{equation}\label{eq:mtdcpp_con_likhod}
\begin{aligned}
&\;\;\;\;\;\; p(x_1, \dots, x_n,\tx_{n+1} \,;\, \pi_0,\bm w, \bphi, \btheta)\\ 
& \propto
\prod_{i=L+1}^n\left\{\pi_0f_I(x_i\,|\,\bphi) + 
(1-\pi_0)\sum_{l=1}^Lw_lf_l(x_i\,|\,x_{i-l},\btheta_l)\right\}\\
&\;\;\;\;\;\;\times \left(1 - \int_0^{\tx_{n+1}}\left\{\pi_0f_I(u\,|\,\bphi) + 
(1-\pi_0)\sum_{l=1}^Lw_lf_l(u\,|\,x_{n+1-l}, \btheta_l)\right\}du\right)
\end{aligned}
\end{equation}
where $\bm w = (w_1,\dots, w_L)^\top$. The vectors $\bphi$ and 
$\btheta = \{\btheta_l\}_{l=1}^L$, respectively, collect the parameters of the independent 
duration density $f_I$ and the MTDPP component densities $f_l$, $l = 1,\dots, L$. 
A Bayesian model formulation involves priors for 
parameters $\{\pi_0, \bw, \bphi, \bth\}$.
The priors for $\bphi$ and $\btheta$, respectively, 
depend on particular choices of the densities $f_I$ and 
$f_l$, $l = 1,\dots, L$. For $\pi_0$, we consider a beta prior,
denoted as $\mathrm{Beta}(u_0,v_0)$.
For the weight vector $\bm w$, we use the same prior
as that for the MTDPP, which can be found in
Section 3.1 of the main paper. In particular, 
the vector $\bm w$ follows a Dirichlet distribution with shape parameter vector
$\alpha_0 (a_1, \dots, a_L)^\top$.

We outline an MCMC posterior simulation method, Metropolis-within-Gibbs, 
for the model parameters of MTDCPP. 
For more efficient notation, we rewrite the MTDCPP transition density as
\begin{equation}
f^*_C(x_i) = \sum_{l=0}^L\pi_lf^c_l(x_i\,|\,\bphi,\btheta_l),
\end{equation}
where 
$f^c_0 \equiv f_I$, $f^c_l \equiv f_l$,
$\pi_l = (1-\pi_0)w_l$, for $l = 1,\dots, L$, and $\sum_{l=0}^L\pi_l = 1$.

We augment the model with configuration variables $\ell_i$, taking values in 
$\{0, 1, \dots, L\}$, with discrete distribution 
$\sum_{l=0}^L \pi_l \, \delta_l(\ell_i)$, where $\delta_l(\ell_i) = 1$
if $\ell_i = l$ and $0$ otherwise, for $i = L+1,\dots, n$. 
Therefore, $\ell_i = 0$ indicates that
the duration $x_i$ is generated from $f_I$, and $\ell_i = l$ indicates that $x_i$ 
is generated from the $l$th component of the MTDPP, for $l = 1,\dots, L$.
Note that the likelihood normalizing term in \eqref{eq:mtdcpp_con_likhod} 
can be written as $\sum_{l=0}^L \pi_l \, S_l^c(\tx_{n+1}\,|\,\bphi,\btheta_l)$,
where $S_0^c \equiv S_I$ and $S_l^c \equiv S_l$, for $l = 1,\dots, L$. 
Similarly with the observed durations, we can introduce a configuration
variable $\ell_{n+1}$ to identify the component of the mixture for $\tx_{n+1}$.
The posterior distribution of the augmented model is proportional to
\begin{equation}
\begin{aligned}
& p(\bphi) \times \prod_{l=1}^Lp(\btheta_l) \times 
\mathrm{Dir}(\bm w\,|\,\alpha_0a_1,\dots,\alpha_0a_L) 
\times \mathrm{Beta}(\pi_0\,|\,u_0,v_0) \\
& \times \prod_{i=L+1}^n \left\{ f_{\ell_i}^c(x_i \,|\,\bphi,\btheta_{\ell_i})
\sum_{l=0}^L \pi_l \, \delta_l(\ell_i)\right\}
\left\{ S^c_{\ell_{n+1}}(\tx_{n+1}\,|\,\bphi,\btheta_{\ell_{n+1}})
\sum_{l=0}^L \pi_l \, \delta_l(\ell_{n+1})\right\}.
\end{aligned}
\end{equation}

The posterior full conditional distribution 
of $\ell_i$ is a discrete distribution on $\{ 0,...,L \}$ with probabilities 
proportional to $\pi_l f^c_l(x_i\,|\,\bphi,\btheta_l)$, for $i = L+1,\dots,n$,
and with probabilities proportional to $\pi_lS^c_l(\tx_{n+1} \,|\,\bphi,\btheta_l)$, 
for $i = n+1$. 
% Let $M_l = |\{i:\ell_i = l\}|$, 
$M_l = |\{i:\ell_i = l,\, L+1\leq i\leq n+1\}|$,
for $l=0,...,L$, where $|\{\cdot\}|$ returns the size of set $\{\cdot\}$.
Given the configuration variables, we update the weights $\bm{w}$ with a 
Dirichlet posterior full conditional distribution 
with parameter vector $(\alpha_0 a_1 + M_1, \dots, \alpha_0 a_L + M_L)^\top$. 
The beta prior for $\pi_0$ yields a conjugate posterior full conditional 
distribution, $\mathrm{Beta}(\pi_0\,|\,u_0 + M_0, v_0 + \sum_{l=1}^LM_l)$.
Posterior updates for parameters $\bphi$ and $\btheta_l$, respectively,
depend on $f_I$ and $f_l$, $l = 1,\dots, L$.
Implementation details for the MTDCPP model in Section 4.3 of the main 
paper are provided in Section \ref{supp-sec: mcmc}.

We now turn to posterior prediction for MTDCPPs. 
The conditional duration density for $X_{n+1}$,
denoted as $p_C(x_{n+1}\,|\,D_n, \bth, \bphi, \bm w, \pi_0)$,
can be obtained similarly by replacing
$f^*(x_{n+1})$ and $S^*(T-t_n)$ 
in \eqref{eq: cond_out_1}, respectively, with
$f_C^*(x_{n+1})$ and $S_C^*(T-t_n)$ (both $f_C^*$ and 
$S_C^*$ are available in Section 2.4 of the main paper).
Then, the posterior predictive density of $X_{n+1}$ can 
be obtained by marginalizing $p_C(x_{n+1}\,|\, D_n, \bth, \bphi, \bm w, \pi_0)$ 
with respect to
the model parameters' posterior distribution $p(\bth,\bphi,\bm w,\pi_0\,|\,D_n)$.
For $k\geq 2$, the posterior predictive density of $X_{n+k}$ 
is obtained by replacing the MTDPP conditional duration density
$\sum_{l=1}^Lw_lf_l(x_j\,|\,x_{j-l},\bth_l)$,
$p(x_{n+1}\,|\, D_n, \bth,\bm w)$,
and $p(\btheta,\bm w\,|\,D_n)$ in \eqref{eq: pred_out_k},
respectively, with the MTDCPP
conditional duration density in (9) of the main paper, 
$p_C(x_{n+1}\,|\, D_n, \bth, \bphi, \bm w, \pi_0)$,
and $p(\bth,\bphi,\bm w,\pi_0\,|\,D_n)$. Finally,
for in-sample predictions, the posterior predictive density
of $X_i$, for $i = L+1,\dots,n$, can be obtained by
replacing the MTDPP conditional duration density and 
$p(\btheta,\bm w\,|\,D_n)$ in \eqref{eq: pred_in}, respectively,
with the MTDCPP conditional duration density in (9) of the main paper 
and the posterior distribution $p(\bth,\bphi,\bm w,\pi_0\,|\,D_n)$.

\section{MCMC algorithms}
\label{supp-sec: mcmc}

We outline the posterior simulation steps for the Burr MTDPP,
the extended scaled-Lomax MTDPP,
and the Lomax MTDCPP models illustrated in Section 4. 
Given an observed point pattern $0 = t_0 < t_1 < \dots < t_n < T$, 
we have that $x_i = t_i - t_{i-1}$ for $i = 1,\dots,n$,
and we take $\tilde x_{n+1} = T - t_n$.
Our posterior inference is based on a likelihood,
conditional on $(x_1,\dots,x_L)$.
Posterior samples of model parameters and latent variables
are obtained with Metropolis-within-Gibbs updates, by iteratively sampling from
their posterior full conditional distributions.
Throughout the remainder of this section, 
for a generic parameter or latent variable $\psi$,
we denote $p(\psi\,|\, -)$ as its posterior full conditional 
distribution or density, depending on the context.

\subsection{Burr MTDPP}

We associate each $x_i$ with a latent discrete variable $\ell_i$ such that 
$P(\ell_i=l) = \sum_{l=1}^Lw_l\delta_l(\ell_i)$, 
$i = 1,\dots, n$, and similarly, consider a latent discrete variable 
$\ell_{n+1}$ for $\tx_{n+1}$ such that
$P(\ell_{n+1}=l) = \sum_{l=1}^Lw_l\delta_l(\ell_{n+1})$.
We consider independent priors 
$\mathrm{Ga}(\la\,|\,u_{\la},v_{\la})\,
\mathrm{Ga}(\ga\,|\,u_{\ga},v_{\ga})\,$\\
$\mathrm{Ga}(\kappa\,|\,u_{\kappa},v_{\kappa})\mathbbm{1}(\kappa > 1)$
for the Burr-distribution parameters $(\ga,\la,\kappa)$.
Then the joint posterior distribution of 
the model parameters and latent variables,
$\{\ga,\la,\kappa,\bm w,\ell_{L+1},\dots,\ell_{n+1}\}$, is proportional to
$$
\begin{aligned}
&\mathrm{Ga}(\la\,|\,u_{\la},v_{\la})\times
\mathrm{Ga}(\ga\,|\,u_{\ga},v_{\ga})\times
\mathrm{Ga}(\kappa\,|\,u_{\kappa},v_{\kappa})\mathbbm{1}(\kappa > 1)
\times \mathrm{Dir}(\bm w\,|\,\alpha_0a_1,\dots,\alpha_0a_L)\\
&\times\left\{\prod_{i=L+1}^n\mathrm{Burr}\big(x_i\,|\,\ga, 
\tilde{\la}(x_{i-\ell_i}),\kappa\big)
\sum_{l=1}^Lw_l\delta_l(\ell_i)\right\}
\left\{S_{\mathrm{Burr}}\big(\tx_{n+1}\,|\,\ga, \tilde{\la}(x_{n+1-\ell_{n+1}}),\kappa\big)
\sum_{l=1}^Lw_l\delta_l(\ell_{n+1})\right\},
\end{aligned}
$$
where $\tilde{\la}(v) = (\la^\ga + v^\ga)^{1/\ga}$, and
$S_{\mathrm{Burr}}(x\,|\,\ga,\la,\kappa) = (1 + (x/\la)^\gamma)^{-\kappa}$
is the survival function associated with the Burr distribution,
$\mathrm{Burr}(x\,|\,\ga,\la,\kappa)$.

Take $p(\x,\btheta) = 
\left\{\prod_{i=L+1}^n\mathrm{Burr}(x_i\,|\,\ga,
\tilde{\la}(x_{i-\ell_i}),\kappa)\right\}
S_{\mathrm{Burr}}(\tx_{n+1}\,|\,\ga,\tilde{\la}(\x_{n+1-\ell_{n+1}}),\kappa)$,
where $\x = (x_1,\dots,x_n,\tx_{n+1})^\top$ and 
$\btheta = \{\la,\ga,\kappa,\ell_{L+1},\dots,\ell_{n+1}\}$.
Then we can obtain the posterior samples of $\{\ga,\la,\kappa,\bm w,\ell_{L+1},\dots,\ell_{n+1}\}$
by iterating the following steps.

\begin{enumerate}[label=(\roman*)]
    \item Update $\ga$ with target distribution
    $\mathrm{Ga}(\ga\,|\,u_{\ga},v_{\ga})\,p(\x,\btheta)$,
    using a random walk Metropolis step implemented on the 
    log scale with a Gaussian proposal distribution.

    \item Update $\la$ with target distribution
    $\mathrm{Ga}(\la\,|\,u_{\la},v_{\la})\,p(\x,\btheta)$,
    using a random walk Metropolis step implemented on the 
    log scale with a Gaussian proposal distribution.    

    \item Sample $\kappa$ from a gamma distribution
    with shape parameter $u_\kappa$ and rate parameter $v_\kappa$
    truncated at the interval $(1,\infty)$, 
    denoted as $\mathrm{Ga}(\kappa\,|\, \tilde u_\kappa, \tilde v_\kappa; 1,\infty)$,
    where $\tilde u_\kappa = u_{\kappa}+n-L$
    and $\tilde v_\kappa = v_{\kappa} + 
    \sum_{i=L+1}^{n}\log(1 + \{x_i/\tilde{\la}(x_{i-\ell_i})\}^{\ga})
    + \log(1 + \{\tx_{n+1}/\tilde{\la}(x_{n+1-\ell_{n+1}})\}^{\ga})$.
    
    \item Sample $\ell_i$, $i = L+1, \dots, n$, from
    $$
    p(\ell_i = l\,|\,-) = 
    \frac{w_l\mathrm{Burr}\big(x_i\,|\,\ga, \tilde{\la}(x_{i-l}),\kappa\big)}
    {\sum_{r=1}^Lw_r\mathrm{Burr}\big(x_i\,|\,\ga, \tilde{\la}(x_{i-r}),\kappa\big)},
    $$
    and sample $\ell_{n+1}$ from
    $$
    p(\ell_{n+1} = l\,|\,-) = 
    \frac{w_l S_{\mathrm{Burr}}\big(\tx_{n+1}\,|\,\ga, \tilde{\la}(x_{n+1-l})}
    {\sum_{r=1}^Lw_r S_{\mathrm{Burr}}\big(\tx_{n+1}\,|\,\ga, \tilde{\la}(x_{n+1-l})}.
    $$    

    \item Sample $\bm w$ from a Dirichlet distribution
    $\mathrm{Dir}(\bm w\,|\,\alpha_0\alpha_1+M_1,\dots,
    \alpha_0\alpha_L + M_L)^\top$, 
    where $M_l = |\{i:\ell_i=l,\,1\leq i\leq n\}|$, for $l = 1,\dots, L$.    

\end{enumerate}

\subsection{Extended scaled-Lomax MTDPP}

Let $x_i = \mu(t_i)z_i$, with 
$\log\mu(t_i) = \sum_{j=1}^J\{\beta_{1j}\sin(j\omega t_i)+
\beta_{2j}\cos(j\omega t_i)\}
$.
The conditional duration density is 
$f^*(x_i) = \mu(t_i)^{-1}\sum_{l=1}^Lw_lP\big(\mu(t_i)^{-1}x_i\,|\,
\alpha\phi+\mu(t_{i-l})^{-1}x_{i-l},\alpha\big)$, 
for $i > L$.

Denote $\bbeta = (\beta_{11},\dots,\beta_{1J},\beta_{21},\dots,\beta_{2J})^\top$,
and let $\tilde{\beta}_k$ be the $k$th component of $\bbeta$, for $k = 1,\dots, 2J$.
We introduce a collection of configuration variables $\{\ell_i\}_{i=L+1}^{n+1}$
such that $P(\ell_i=l) = \sum_{l=1}^Lw_l\delta_l(\ell_i)$.
We consider independent priors 
$\prod_{k=1}^{2J}N(\tilde{\beta}_k\,|\,\mu_{\tilde{\beta}_k}, \sigma^2_{\tilde{\beta}_k})\mathrm{Ga}(\phi\,|\,u_{\phi},v_{\phi})$\\
$\mathrm{Ga}(\alpha\,|\,u_{\alpha},v_{\alpha})\mathbbm{1}(\alpha > 1)$
for parameters $\{\bbeta,\phi,\alpha\}$. 
Then the joint posterior distribution of the model parameters and
latent variables, 
$\{\bbeta, \phi, \alpha, \bm w,\ell_{L+1},\dots, \ell_{n+1}\}$, is 
proportional to
$$
\begin{aligned}
&
\prod_{k=1}^{2J}N(\tilde{\beta}_k\,|\,
\mu_{\tilde{\beta}_k}, \sigma^2_{\tilde{\beta}_k})
\times\mathrm{Ga}(\phi\,|\,u_{\phi},v_{\phi})\times
\mathrm{Ga}(\alpha\,|\,u_{\alpha},v_{\alpha})\mathbbm{1}(\alpha > 1)
\times \mathrm{Dir}(\bm w\,|\,\alpha_0a_1,\dots,\alpha_0a_L)\\
&\times\left\{\prod_{i=L+1}^n\mu(t_i)^{-1}P\left(\mu(t_i)^{-1}x_i\,|\,
\alpha\phi+\mu(t_{i-\ell_i})^{-1}x_{i-\ell_i},\alpha\right)
\sum_{l=1}^Lw_l\delta_l(\ell_i)\right\}\\
&\times\left\{S_{\mathrm{Lo}}\big(\mu(T)^{-1}\tx_{n+1}\,|\,\alpha\phi + \mu(t_{n+1-\ell_{n+1}})^{-1}x_{n+1-\ell_{n+1}},\alpha\big)
\sum_{l=1}^Lw_l\delta_l(\ell_{n+1})\right\},
\end{aligned}
$$
where $S_{\mathrm{Lo}}(x\,|\,\psi,\alpha) = (1 + (x/\psi))^{-\alpha}$ 
is the survival function associated with the distribution
$P(x\,|\,\psi, \alpha)$. 

Let $\bm t = (t_1,\dots,t_{n}, T)^\top$, $\x = (x_1,\dots,x_n,
\tx_{n+1})^\top$, and 
$\btheta = \{\bbeta, \phi, \alpha, \ell_{L+1},\dots,\ell_{n+1}\}$. Take
$$
\begin{aligned}
p(\bm t, \x,\btheta) = & 
\left\{\prod_{i=L+1}^nP\left(\mu(t_i)^{-1}x_i\,|\,
\alpha\phi+\mu(t_{i-\ell_i})^{-1}x_{i-\ell_i},\alpha\right)\right\}\\
& \times S_{\mathrm{Lo}}\big(\mu(T)^{-1}\tx_{n+1}\,|\,\alpha\phi + \mu(t_{n+1-\ell_{n+1}})^{-1}x_{n+1-\ell_{n+1}},\alpha\big).
\end{aligned}
$$
Then we can obtain the posterior samples of $\{\bbeta,\phi,\alpha,
\bm w,\ell_{L+1},\dots,\ell_{n+1}\}$
by iterating the following steps.

\begin{enumerate}[label=(\roman*)]
    \item Update $\tilde\beta_k$ with 
    target distribution $N(\tilde{\beta}_k\,|\,\mu_{\tilde{\beta}_k},\sigma^2_{\tilde{\beta}_k})
    \,p(\bm t, \x,\btheta)\prod_{i=L+1}^{n}\mu(t_i)^{-1}$,
    using a random walk Metropolis step with a Gaussian proposal distribution,
    for $k = 1,\dots, 2J$.

    \item Update $\phi$ with target distribution
    $\mathrm{Ga}(\phi\,|\,u_{\phi},v_{\phi})\,p(\bm t, \x,\btheta)$,
    using a random walk Metropolis step implemented on the log
    scale with a Gaussian proposal distribution.

    \item Update $\alpha$ with target distribution
    $\mathrm{Ga}(\alpha\,|\,u_{\alpha},v_{\alpha})\mathbbm{1}(\alpha > 1)
    \,p(\bm t, \x,\btheta)$,
    using a random walk Metropolis step implemented on the log
    scale with a truncated Gaussian proposal distribution.

    \item Sample $\ell_i$, $i = L+1, \dots, n$, from
    $$
    p(\ell_i = l\,|\,-) = 
    \frac{w_lP\left(\mu(t_i)^{-1}x_i\,|\,
    \alpha\phi+\mu(t_{i-l})^{-1}x_{i-l},\alpha\right)}
    {\sum_{r=1}^Lw_rP\left(\mu(t_i)^{-1}x_i\,|\,
    \alpha\phi+\mu(t_{i-r})^{-1}x_{i-r},\alpha\right)},\,l = 1,\dots,L,
    $$
    and sample $\ell_{n+1}$ from
    $$
    p(\ell_{n+1} = l\,|\,-) = 
    \frac{w_l S_{\mathrm{Lo}}\big(\mu(T)^{-1}\tx_{n+1}\,|\,\alpha\phi + 
    \mu(t_{n+1-l})^{-1}x_{n+1-l},\alpha\big)}
    {\sum_{r=1}^Lw_r S_{\mathrm{Lo}}\big(\mu(T)^{-1}\tx_{n+1}\,|\,\alpha\phi + 
    \mu(t_{n+1-r})^{-1}x_{n+1-r},\alpha\big)},\,l = 1,\dots, L.
    $$    

    \item Sample $\bm w$ from a Dirichlet distribution
    $\mathrm{Dir}(\bm w\,|\,\alpha_0\alpha_1+M_1,\dots,
    \alpha_0\alpha_L + M_L)^\top$, 
    where $M_l = |\{i:\ell_i=l,\,1\leq i\leq n\}|$, for $l = 1,\dots, L$.

\end{enumerate}

\subsection{Lomax MTDCPP}

The Lomax MTDCPP conditional duration density, for $i > L$, can be written as
$f_C^*(x_i) = \sum_{l=0}^L\pi_lf_l^c(x_i\,|\,\mu,\phi,\alpha)$, where
$f_0^c(x_i\,|\,\mu,\phi,\alpha) = \mu\exp(-\mu x_i)$, 
$f_l^c(x_i\,|\,\mu,\phi,\alpha) = P(x_i\,|\,\phi + x_{i-l},\alpha)$,
and $\pi_l = (1 - \pi_0)w_l$, for $l = 1,\dots,L$.
Let $S_0^c$ and $S_l^c$ be the survival functions associated with $f_0^c$ and $f_l^c$,
respectively. 

We augment the model with latent variables $\ell_i$, 
with discrete distribution 
$\sum_{l=0}^L \pi_l \, \delta_l(\ell_i)$, 
for $i = L+1,\dots, n + 1$. 
For parameters $(\mu,\phi,\alpha, \pi_0)$, 
we consider independent priors 
$\mathrm{Ga}(\mu\,|\,u_{\mu},v_{\mu})\,
\mathrm{Ga}(\phi\,|\,u_{\phi},v_{\phi})\,
\mathrm{Ga}(\alpha\,|\,u_{\alpha},v_{\alpha})\mathbbm{1}(\alpha > 1)\,
\mathrm{Beta}(\pi_0\,|\,u_0,v_0)$.
Then the joint posterior distribution of the model parameters and
latent variables, 
$\{\mu, \phi, \alpha, \pi_0, \bm w,\ell_{L+1},\dots, \ell_{n+1}\}$, is 
proportional to
$$
\begin{aligned}
&\mathrm{Ga}(\mu\,|\,u_{\mu},v_{\mu})\times 
\mathrm{Ga}(\phi\,|\,u_{\phi},v_{\phi})\times 
\mathrm{Ga}(\alpha\,|\,u_{\alpha},v_{\alpha})\mathbbm{1}(\alpha > 1)
\times \mathrm{Dir}(\bm w\,|\,\alpha_0a_1,\dots,\alpha_0a_L)\\
&\times \mathrm{Beta}(\pi_0\,|\,u_0,v_0)
\times \left\{\prod_{i=L+1}^nf_{\ell_i}^c(x_i\,|\,\mu,\phi,\alpha)
\sum_{l=0}^L\pi_l\delta_l(\ell_i)\right\}
\left\{S^c_{\ell_{n+1}}(\tx_{n+1}\,|\,\mu,\phi,\alpha)
\sum_{l=0}^L\pi_l\delta_l(\ell_{n+1})\right\}.
\end{aligned}
$$

Let $M_l = |\{i:\,\ell_i = l,\,L+1\leq i\leq n+1\}|$, for $l=0,...,L$.
Take 
$$
p(\x,\bth) = 
\left\{\prod_{i=1}^n\left[f_{\ell_i}^c(x_i\,|\,\mu,\phi,\alpha)\right]
^{1-\delta_0(\ell_i)}\right\}
\left\{S_{\ell_{n+1}}^c(\tx_{n+1}\,|\,\mu,\phi,\alpha)\right\}^
{1-\delta_0(\ell_{n+1})},
$$
where $\x = (x_1,\dots,x_n,
\tx_{n+1})^\top$, and 
$\btheta = \{\mu, \phi, \alpha, \ell_{L+1},\dots, \ell_{n+1}\}$.

We can obtain the posterior samples of $\{\mu, \phi, \alpha, \pi_0, \bm w,\ell_{L+1},\dots, \ell_{n+1}\}$
by iterating the following steps.

\begin{enumerate}[label=(\roman*)]
    \item Sample $\mu$ from a gamma distribution with shape 
    parameter $u_{\mu}+M_0 - \delta_0(\ell_{n+1})$
    and rate parameter $v_{\mu} + \sum_{i=1}^nx_i\delta_0(\ell_i)
    + \tx_{n+1}\delta_0(\ell_{n+1})$.

    \item Sample $\alpha$ from a truncated gamma distribution
    $\mathrm{Ga}(\alpha\,|\,, \tilde u_\alpha, \tilde v_\alpha;1,\infty)$,
with $\tilde u_\alpha = u_{\alpha} + \sum_{l=1}^LM_l-1+\delta_0(\ell_{n+1})$,
and $\tilde v_\alpha = v_{\alpha} + 
\sum_{i=1}^n(1-\delta_0(\ell_i))\log(1+x_i/(\phi+x_{i-\ell_i})) + 
(1-\delta_0(\ell_{n+1}))\log(1+\tx_{n+1}/(\phi+x_{n+1-\ell_{n+1}}))$.

    \item Update $\phi$  with target distribution 
    $\mathrm{Ga}(\phi\,|\,u_{\phi},v_{\phi})\,p(\x,\bth)$,
    using a random walk Metropolis step implemented on the log
    scale with a Gaussian proposal distribution.

    \item Sample $\ell_i$, $i = L+1, \dots, n$, from
    $$
    p(\ell_i = l\,|\,-) = 
    \frac{\pi_lf_l^c(x_i\,|\,\mu,\phi,\alpha)}
    {\sum_{r=0}^L\pi_rf_r^c(x_i\,|\,\mu,\phi,\alpha)},\,l = 0,1,\dots, L,
    $$
    and sample $\ell_{n+1}$ from
    $$
    p(\ell_{n+1} = l\,|\,-) = 
    \frac{\pi_lS_l^c(\tx_{n+1}\,|\,\mu,\phi,\alpha)}
    {\sum_{r=0}^L\pi_rS_r^c(\tx_{n+1}\,|\,\mu,\phi,\alpha)},\,l = 0,1,\dots, L,
    $$    
    where $S_0^c(\tx_{n+1}\,|\,\mu,\phi,\alpha) = \exp(-\mu\tx_{n+1})$,
    and, for $l = 1,\dots, L$, 
    $S_l^c(\tx\,|\,\mu,\phi,\alpha) = S_{\mathrm{Lo}}(\tx\,|\,
    \phi + x_{n+1-l}, \alpha)$.

    \item Sample $\bm w$ from a Dirichlet distribution
    $\mathrm{Dir}(\bm w\,|\,\alpha_0\alpha_1+M_1,\dots,
    \alpha_0\alpha_L + M_L)^\top$.

    \item Sample $\pi_0$ from a beta distribution
    $\mathrm{Beta}(\pi_0\,|\, u_0 + M_0, v_0 + \sum_{l=1}^LM_l)$.
\end{enumerate}

\section{Additional simulation studies}

\subsection{First simulation study: Comparison with ACD models}
\label{sec: comp-acd}

MTDPPs are duration-based models for point processes with memory.
In this section, we compare MTDDPs with alternative duration-based models,
the autoregressive conditional duration (ACD) models, via simulation studies.
Consider an ordered sequence of event times $0 = t_0 < t_1 < \dots < t_n < T$,
and durations $x_i = t_i - t_{i-1} > 0$, $i\geq1$.
We consider the following Burr ACD model 
\begin{equation}\label{eq: burr-acd}
\begin{aligned}
    x_i & = \psi_i \, \eps_i,\\
    \psi_i & = a_0 + \sum_{l=1}^p a_l \, x_{i-l},
\end{aligned}
\end{equation}
for $i = p + 1,\dots, n$, where 
the innovations $\eps_i$ are independent and identically distributed
as the Burr distribution, denoted
as $\mathrm{Burr}_{GM}(\theta, \gamma,\sigma^2)$,
with density $\theta\gamma\eps_i^{\gamma-1}(1 + \sigma^2\theta\eps_i^{\gamma})^{-(1/\sigma^2+1)}$, where $\theta >0$ and $0 < \sigma^2 < \gamma$;
for more details, see \cite{engle1998autoregressive}, \cite{grammig2000non},
and \citesm{belfrage2022acdm}.
The parameter $\theta$ is taken as a function of $\gamma$ and $\sigma^2$ 
so that $E(\eps_i) = 1$, for all $i$,
to ensure identifiability of the Burr ACD model in \eqref{eq: burr-acd}.
 Specifically, 
\begin{equation}\label{eq: theta}
\theta \equiv \theta(\gamma,\sigma^2) = 
\left(\frac{\Gamma(1 + 1/\gamma)\Gamma(1/\sigma^2 - 1/\gamma}
{\sigma^{2(1+1/\gamma)}\Gamma(1/\sigma^2 + 1)}\right)^{\gamma}.
\end{equation}

For the simulation study, we consider two scenarios:
(i) $(\gamma,\sigma^2) = (1, 0.8)$;
(ii) $(\gamma,\sigma^2) = (1.5, 0.8)$.
Note that the associated hazard
function is monotonic if $0<\gamma\leq 1$ and hump-shaped if $\gamma > 1$,
and thus the two scenarios result in different types of conditional
intensity functions.
We generate data using the Burr ACD model in \eqref{eq: burr-acd}
with $p = 3$ and $(a_0,a_1,a_2,a_3) = (0.1, 0.3, 0.2, 0.1)$.
For each scenario, we choose observation window $(0,T)$ so that
the resulting number of event times is between $500$ and $600$.
Take $\x_{1:n} \equiv (x_1,\dots, x_n)^\top$ as the vector 
of simulated durations.

We compare the Burr MTDPP model (Section 2.3 of the main paper)
with the Burr ACD model used to generate the synthetic data (details for 
implementation of the two models are given below). Note therefore that the 
simulation setting favors the Burr ACD model, whereas the Burr MTDPP is a 
misspecified model. We focus on predictive performance of the two models,
based on the following criteria: 
median absolute deviation (MAD), root mean squared error (RMSE), 
continuous ranked probability score (CRPS; \citealtsm{gneiting2007strictly}),
and interval score (IS; \citealtsm{gneiting2007strictly}) based on
$95\%$ interval estimations.  Models are compared under the
following two settings:
\begin{itemize}
    \item[(i)] \textbf{One-step-ahead in-sample prediction.}
    We fit models to data $\x_{1:n}$, and evaluate models by comparing
    $x_i$ and its predictions from the two models, for $i = p+1,\dots, n$, 
    where $p = 3$ is the order of the ACD model.
    
    \item[(ii)]\textbf{One-step-ahead out-of-sample prediction.}
    We consider an expanding observation window $(0, T_m)$, where 
    $T_m\sim\mathrm{Unif}(t_m,t_{m+1})$, for $m = n - M,\dots, n - 1$,
    where $M$ is the number of observation windows.
    For each $m$, we fit models to $\x_{1:m}$, and evaluate models 
    by comparing $x_{m+1}$ and its predictions from the two models,
    for $m = n - M,\dots, n - 1$. We chose $M = 50$ for the simulation study.
\end{itemize}

\paragraph{Burr ACD model}
We fitted the Burr ACD model 
under the setting corresponding to the one in \eqref{eq: burr-acd}
with $p = 3$.
For in-sample prediction, a prediction of $X_i$ conditional on 
$\x_{1:(i-1)}$ was available from
the ACD-model-fitting output, denoted as $\hat{\psi}_i$, 
for $i = p+1,\dots, n$.
We constructed $95\%$ prediction intervals using bootstrap samples
that were also output of the ACD-model-fitting function. 
Denote by $\bm a^{(b)} = (a_0^{(b)}, a_1^{(b)}, a_2^{(b)}, a_3^{(b)})^\top$
the $b$th bootstrap sample of $\bm a = (a_0,a_1,a_2,a_3)^\top$,
for $b = 1,\dots, B$.
Take $\psi_i^{(b)} = a_0^{(b)} + \sum_{l=1}^pa_l^{(b)}x_{i-l}$,
% Denote by $\psi_i^{(b)}$ the $b$th bootstrap sample of $\psi_i$, 
and let $\eps_i^{*(b)}$ be a random draw from $\mathrm{Burr}_{GM}(\theta^{(b)},\gamma^{(b)},
\sigma^{2(b)})$, where $(\gamma^{(b)},\sigma^{2(b)})$ is the 
$b$th bootstrap sample and calculate 
$\theta^{(b)}\equiv \theta(\gamma^{(b)},\sigma^{2(b)})$ 
according to \eqref{eq: theta},
for $b = 1,\dots, B$.
The collection
$\{X_i^{*(b)} = \psi_i^{(b)}\eps_i^{*(b)}:\, b = 1,\dots, B\}$
was used to construct prediction intervals for $X_i$ given 
$\x_{1:(i-1)}$, for $i =  p+1,\dots, n$.

For one-step-ahead out-of-sample prediction given 
data $\x_{1:m}$, $m = n-M,\dots, n - 1$, note that the conditional
expectation of $X_{m+1}$, incorporating the condition that 
$X_{m+1} > T_m-t_m$, is given by
\begin{equation}\label{eq: acd_pred_out}
\begin{aligned}
& E(X_{m+1}\,|\,X_{m+1}>T_m-t_m,X_m = x_m,\dots,X_1 = x_1)\\
= & \;\psi_{m+1}
E(\eps_{m+1}\,|\,\eps_{m+1} > (T_m-t_m)/\psi_{m+1}).
\end{aligned}
\end{equation}
Thus, a prediction of $X_{m+1}$ is given by
$\hat\psi_{m+1}\bar\eps_{m+1}$,
where $\hat\psi_{m+1} = \hat a_0 + \sum_{l=1}^p\hat a_lx_{m+1-l}$,
and $\bar\eps_{m+1} = E(\eps_{m+1}\,|\,\eps_{m+1} > (T-t_m)/\psi_{m+1})$
can be approximated by samples generated from the corresponding truncated Burr 
distribution with parameter estimates $(\hat{\gamma},\hat{\sigma}^2, \theta(\hat{\gamma},
\hat{\sigma}^2))$. Similar to the in-sample prediction, we constructed
prediction intervals from the collection
$\{X_{m+1}^{*(b)} = \psi_{m+1}^{(b)}\eps_{m+1}^{*(b)}:\, b = 1,\dots, B\}$,
where $\psi_{m+1}^{(b)} = a_0^{(b)} + \sum_{l=1}^pa_l^{(b)}x_{m+1-l}$,
and $\eps_{m+1}^{*(b)}$ is a random draw from 
$\mathrm{Burr}_{GM}(\theta^{(b)},\gamma^{(b)},
\sigma^{2(b)})$, truncated at $((T_m-t_m)/\psi_{m+1}^{(b)}, \infty)$.
For each scenario, we fitted the Burr ACD with 5000 bootstrap iterations
using the package \textbf{ACDm} \citepsm{belfrage2022acdm}.

\begin{figure*}[t!]
    \centering
    \captionsetup[subfigure]{justification=centering, font=footnotesize}
    \begin{subfigure}[b]{0.48\textwidth}
         \centering
         \includegraphics[width=\textwidth]{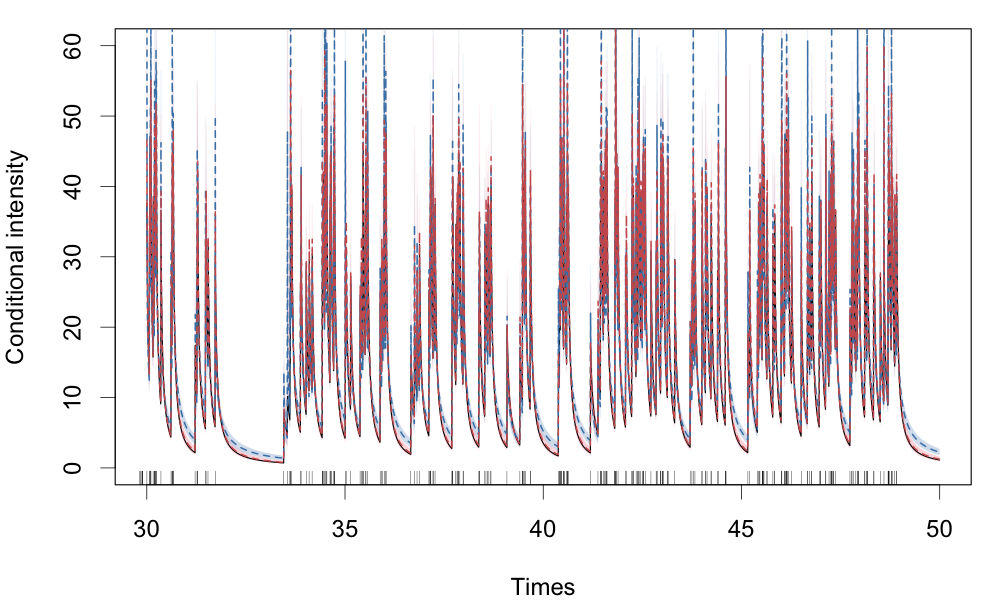}
    \end{subfigure}
    % \hspace{20pt}
    \begin{subfigure}[b]{0.48\textwidth}
         \centering
         \includegraphics[width=\textwidth]{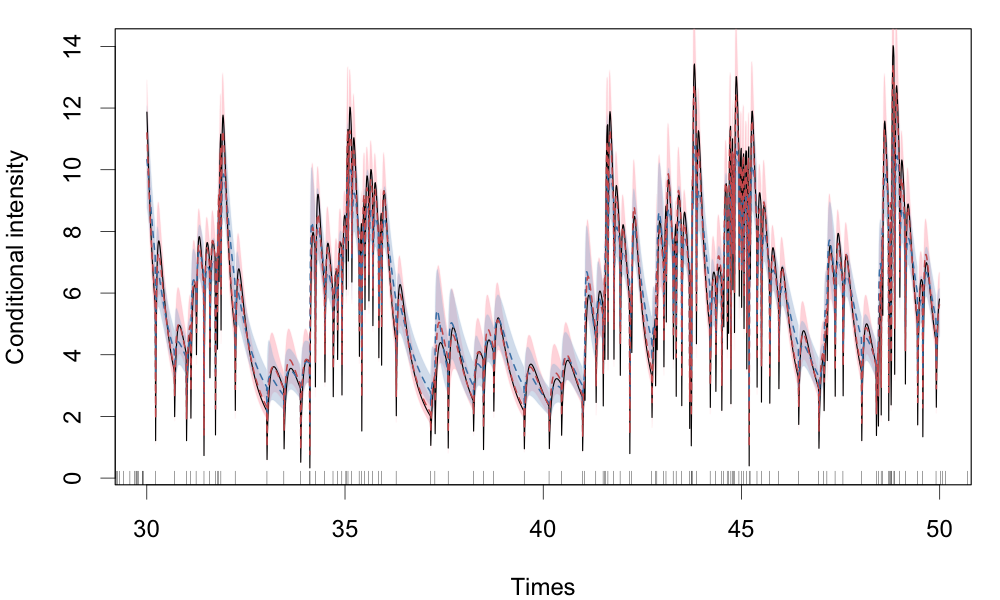}
    \end{subfigure}
    \caption{Comparison of Burr MTDPP and Burr ACD models regarding
             conditional intensity functions evaluated at the time
             interval $(30, 50)$, under the first simulation scenario 
             (left; $(\gamma,\sigma^2) = (1, 0.8))$
             and the second simulation scenario 
             (right; $(\gamma,\sigma^2) = (1.5, 0.8))$.
             The black solid line corresponds to the true conditional 
             intensity of the Burr ACD model.
             The blue dashed line and blue polygon, respectively, 
             correspond to the posterior mean and the 
             pointwise $95\%$ interval estimates of the conditional
             intensity for the Burr MTDPP.
             The red dashed line and red polygon, respectively, correspond to 
             the point and pointwise $95\%$ interval estimates 
             of the conditional intensity for the Burr ACD model.
             }
    \label{fig:sim1}
\end{figure*}

\paragraph{Burr MTDPP model}
We applied the stationary Burr MTDPP model defined in Equation (8) of the main paper, 
with $L = 3$. We assigned $\mathrm{Ga}(1,1)$ priors to $\ga$ and $\la$,
respectively, and independently of $(\ga,\la)$, we placed a truncated gamma 
prior, $\mathrm{Ga}(6,1)\mathbbm{1}(\kappa > 1)$, which implies the 
first five moments of the marginal duration distribution exist. 
The vector of weights received a CDP$(5, 1, 2)$ prior.
For each scenario, we obtained 5000 posterior samples, collected
every tenth iteration from a Markov chain, after discarding the 
first 5000 samples. We generated posterior predictive samples
for in-sample and out-of-sample predictions, respectively, 
according to \eqref{eq: pred_in} and \eqref{eq: pred_out_1}.

\subsubsection*{Results}

The results of in-sample predictions are shown in Table \ref{tbl:comp-acd1}.
The in-sample predictive performance of the MTDPP model is almost identical to 
that of the ACD model, indicating a good model fit. 
Figure \ref{fig:sim1}
shows the point and interval estimates of the conditional intensity
functions for the two models, under the two scenarios. Although
the Burr MTDPP's interval estimates miss some true intensity values,
overall, the model is able to produce intensity estimates that
resemble the true pattern. Again, note that for this simulation
study, the Burr MTDPP is a misspecified model, while we fitted the
same Burr ACD used to generate synthetic data.

We next turn to the results on out-of-sample predictive performance.
Note that for the ACD model, although we incorporated the condition
that the new duration $X_{m+1}> T_m - t_m$ 
for out-of-sample predictions (Equation \eqref{eq: acd_pred_out}), 
this condition is not considered in the 
approach for inference of ACD models \citep{engle1998autoregressive},
i.e., the point process likelihood normalizing term 
(the second term in Equation (1) of the main paper) is ignored. 
Besides, the interval estimates for the ACD model
using bootstrap samples (available from the package's 
ACD-model-fitting output) can be conservative \citepsm{diciccio1996bootstrap}. 
On the other hand, the MTDPP's model inference takes into consideration
the likelihood normalizing term, and the model bases its predictions on 
its posterior predictive distribution (main paper Section 3.3, and
Supplementary Material Section S3).
Thus, although the Burr ACD model was correctly specified, the Burr MTDPP 
outperformed the Burr ACD model in both scenarios (see Table \ref{tbl:comp-acd2}).

\begin{table}[t!]
\renewcommand{\arraystretch}{1.25}
\captionsetup{font=footnotesize}
\caption{$\;$Comparison of Burr MTDPP and Burr ACD models regarding
one-step-ahead in-sample prediction, under each 
of the two simulation scenarios.}
\vspace{-5pt}
\small
    \centering
    \begin{tabular*}{\hsize}{@{\extracolsep{\fill}}lcccccccccc}
    \hline
\multicolumn{1}{l}{} & 
\multicolumn{4}{c}{$\gamma = 1, \,\sigma^2 = 0.8$} & \multicolumn{1}{c}{} & 
\multicolumn{4}{c}{$\gamma = 1.5, \,\sigma^2 = 0.8$}\\
\cline{2-5} \cline{7-10} 
\multicolumn{1}{l}{} & \multicolumn{1}{c}{MAD} & \multicolumn{1}{c}{RMSE} & \multicolumn{1}{c}{CRPS} & \multicolumn{1}{c}{IS} & \multicolumn{1}{c}{} &
\multicolumn{1}{c}{MAD} & \multicolumn{1}{c}{RMSE} & \multicolumn{1}{c}{CRPS} & \multicolumn{1}{c}{IS}\\
\hline
ACD & 0.07 & 0.18 & 0.06 & 0.96 & 
         & 0.11 & 0.23 & 0.09 & 1.19 \\
MTDPP & 0.06 & 0.18 & 0.06 & 0.97 & 
         & 0.10 & 0.23 & 0.10 & 1.15 \\         
\hline
    \end{tabular*}
    \label{tbl:comp-acd1}
\end{table}

\begin{table}[t!]
\renewcommand{\arraystretch}{1.25}
\captionsetup{font=footnotesize}
\caption{$\;$Comparison of Burr MTDPP and Burr ACD models regarding
one-step-ahead out-of-sample prediction, under each 
of the two simulation scenarios.}
\vspace{-5pt}
\small
    \centering
    \begin{tabular*}{\hsize}{@{\extracolsep{\fill}}lcccccccccc}
    \hline
\multicolumn{1}{l}{} & 
\multicolumn{4}{c}{$\gamma = 1, \,\sigma^2 = 0.8$} & \multicolumn{1}{c}{} & 
\multicolumn{4}{c}{$\gamma = 1.5, \,\sigma^2 = 0.8$}\\
\cline{2-5} \cline{7-10} 
\multicolumn{1}{l}{} & \multicolumn{1}{c}{MAD} & \multicolumn{1}{c}{RMSE} & \multicolumn{1}{c}{CRPS} & \multicolumn{1}{c}{IS} & \multicolumn{1}{c}{} &
\multicolumn{1}{c}{MAD} & \multicolumn{1}{c}{RMSE} & \multicolumn{1}{c}{CRPS} & \multicolumn{1}{c}{IS}\\
\hline
ACD & 0.09 & 0.11 & 0.03 & 0.84 & 
         & 0.16 & 0.19 & 0.07 & 1.06 \\
MTDPP & 0.07 & 0.08 & 0.03 & 0.55 & 
         & 0.14 & 0.16 & 0.06 & 0.80 \\         
\hline
    \end{tabular*}
    \label{tbl:comp-acd2}
\end{table}

\begin{table}[t!]
\renewcommand{\arraystretch}{1.25}
\captionsetup{font=footnotesize}
\caption{MCMC diagnostics of the Burr MTDPP and effective sample sizes.}
\small
    \centering
    \begin{tabular*}{\hsize}{@{\extracolsep{\fill}}lcccccccccc}
\hline
  & $\kappa$ & $\gamma$ & $\lambda$ &  $w_1$ & $w_2$ & $w_3$\\
\hline
$\hat{R}$ & 1.01 & 1.00 & 1.01 & 1.01 & 1.01 & 1.02\\
\hline
$\hat{n}_{\text{eff}}$ & 476.67 & 930.94 & 465.13 & 661.79 & 588.55 & 371.87\\
\hline
$\hat{n}_{\text{eff}}$ per second & 4.46 & 8.71 & 4.35 & 6.19 & 5.51 & 3.48\\
\hline
    \end{tabular*}
    \label{tbl:comp-acd3}
\end{table}

\begin{figure*}[t!]
    \centering
    \captionsetup[subfigure]{justification=centering, font=footnotesize}
    \includegraphics[width=0.85\textwidth]{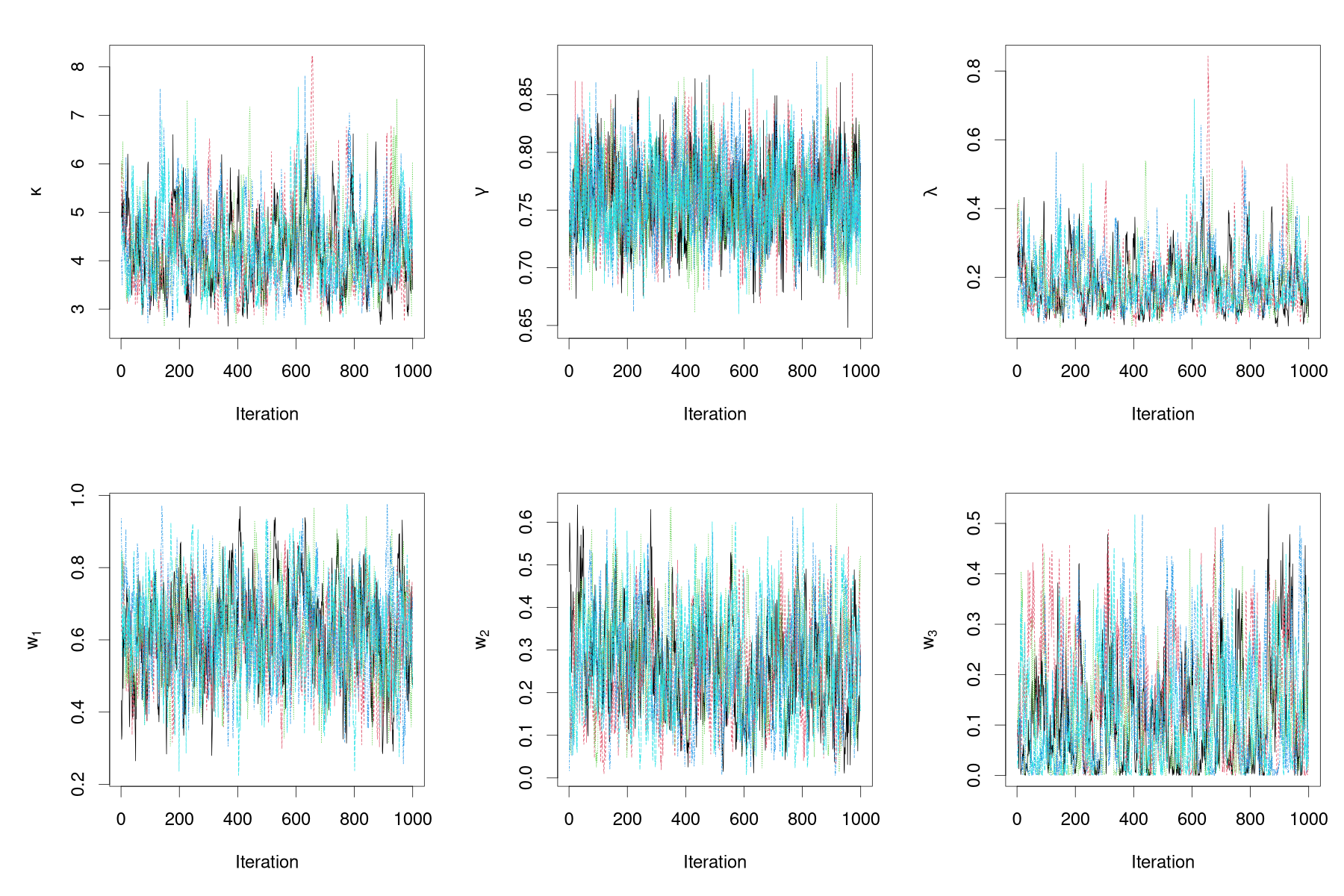}
    \caption{MCMC convergence diagnostics for Section \ref{sec: comp-acd}: 
    trace plots in each panel correspond to 5 independent chains of a parameter.}
    \label{fig:sim1-traceplots}
\end{figure*}

Finally, we assess MCMC convergence 
and compare the computational costs of the Burr MTDPP and Burr
ACD models using data from the first scenario as an example.
Following Chapter 11.5 in \cite{gelman2013bayesian}, 
we assessed MCMC convergence by computing the potential scale reduction factor
$\hat R$ and effective sample size $\hat{n}_{\text{eff}}$.
Specifically, we first ran 5 independent 
Markov chains, each with 1000 posterior samples obtained 
from a total of 15000 iterations,, discarding the first 5000 as 
burn-in samples and retaining samples every $10$ iterations.
We then used the R package \textbf{coda} to compute $\hat R$,
$\hat{n}_{\text{eff}}$, and $\hat{n}_{\text{eff}}$ per second,
as shown in Table \ref{tbl:comp-acd3}.
The factors $\hat R$ near 1 for all parameters indicate convergence of the 
chains. As suggested in Chapter 11.5 in \cite{gelman2013bayesian}
$100$ independent simulation draws are typically sufficient for many purposes,
and thus the effective sample sizes $\hat{n}_{\text{eff}}$ 
in Table \ref{tbl:comp-acd3} are adequate, 
which is further supported by the comparison 
results in Tables \ref{tbl:comp-acd1} and \ref{tbl:comp-acd2}.
Figure \ref{fig:sim1-traceplots} shows trace plots of the five independent
chains for each parameter.

Thus, we concluded that a total of $55000$ MCMC iterations are sufficient 
for the particular simulated data set.
Fitting the Burr MTDPP model to the data (572 observations) 
took around 78 seconds to complete $55000$ iterations (resulting in 5000 
posterior samples after burn-in and thinning), corresponding to an average of 
$5.5$ independent samples per second. On the other hand,
it took less than a second to run the Burr ACD model for a 
single fit; however, its computation cost increased substantially when 
bootstrap sampling (available from the \textbf{ACDm} package)
was employed to compute interval estimates, 
which took around 121 seconds for 5000 bootstrap iterations. 
Both models were fitted in R on a Linux server with 512 GB of RAM and 
two Intel Xeon Gold 6348 processors.

\subsection{Second simulation study: Sensitivity analyses}

\begin{figure*}[t!]
    \centering
    \captionsetup[subfigure]{justification=centering, font=footnotesize}
    \begin{subfigure}[b]{0.32\textwidth}
         \centering
         \includegraphics[width=\textwidth]{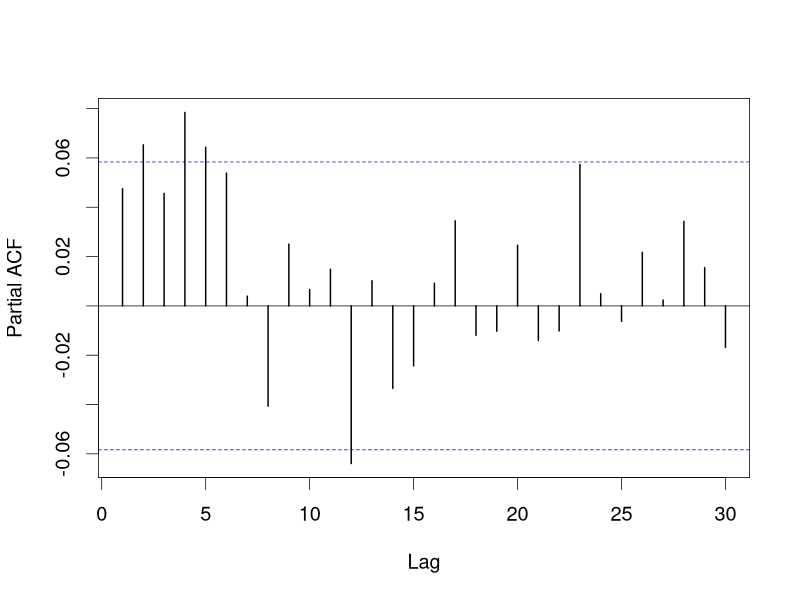}
         \caption{PACF ($n = 1128$)}
    \end{subfigure}
    \begin{subfigure}[b]{0.32\textwidth}
         \centering
         \includegraphics[width=\textwidth]{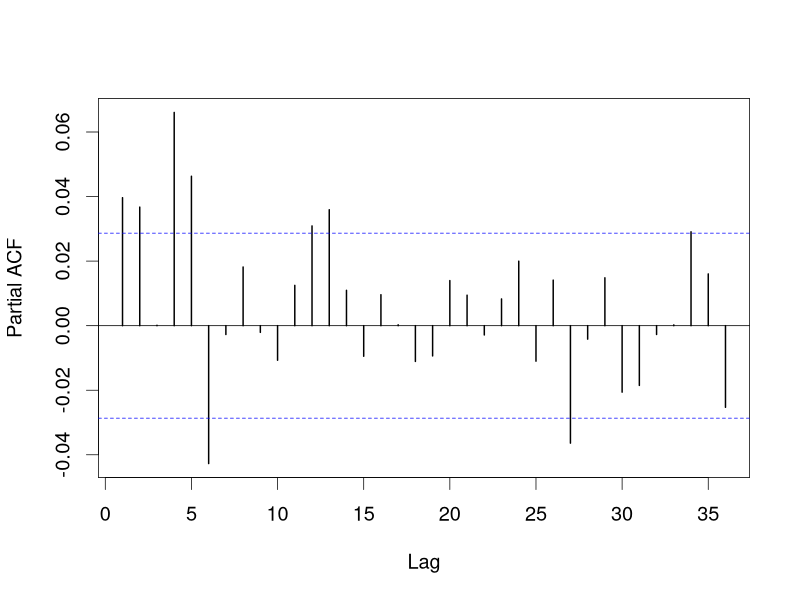}
         \caption{PACF ($n = 4675$)}
    \end{subfigure}
    \begin{subfigure}[b]{0.32\textwidth}
         \centering
         \includegraphics[width=\textwidth]{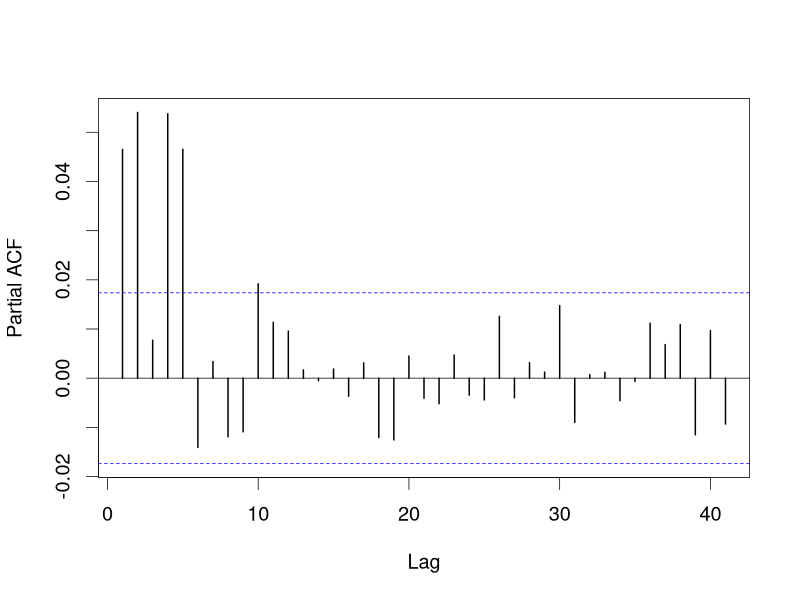}
         \caption{PACF ($n = 12473$)}
    \end{subfigure}    
    \caption{PACFs of the three simulated data sets used in 
    the sensitivity analysis for $L$, the number of mixture components.}
    \label{fig:burr-sa-pacfs}
\end{figure*}

We conducted a sensitivity analysis for
$L$, the number of mixture components,
and a prior sensitivity analysis for the mixture weights
through a simulation study. Specifically,
we generated data from the stationary Burr MTDPP models,
with component parameters $(\la,\ga,\kappa) = (1, 2, 6)$
and weights $\bm w = (0.3, 0.2, 0.01, 0.29, 0.2)$.
We chose observation windows $(0, 500)$, $(0, 2000)$, $(0, 5500)$,
resulting in $n = 1128$, $4675$, $12743$ event times, respectively.
Figure \ref{fig:burr-sa-pacfs} shows the partial
autocorrelation functions (PACFs) of the simulated data sets.

\paragraph{Sensitivity analysis for the number of mixture components}\mbox{}

Since the PACFs of the simulated data sets cut off after lag $5$ or lag $6$, 
we chose $L =$ $10$, $15$, and $20$ for the sensitivity analysis.
For each one of the observation windows, we fitted the Burr MTDPP model
with $L =$ $10$, $15$, and $20$, respectively, with
priors for the weights, $\mathrm{CDP}(5, 1, 5)$, $\mathrm{CDP}(5, 1, 6)$, $\mathrm{CDP}(5, 1, 8)$.
We assigned $\mathrm{Ga}(1,1)$ prior to $\ga$, and independently of $\ga$, we placed a truncated gamma prior, 
$\mathrm{Ga}(6,1)\mathbbm{1}(\kappa > 1)$, on $\kappa$.
The scale parameter $\la$ was assigned a $\mathrm{Ga}(1,1)$ prior.

\begin{figure*}[t]
    \centering
    \captionsetup[subfigure]{justification=centering, font=footnotesize}
    \begin{subfigure}[b]{0.3\textwidth}
         \centering
         \includegraphics[width=\textwidth]{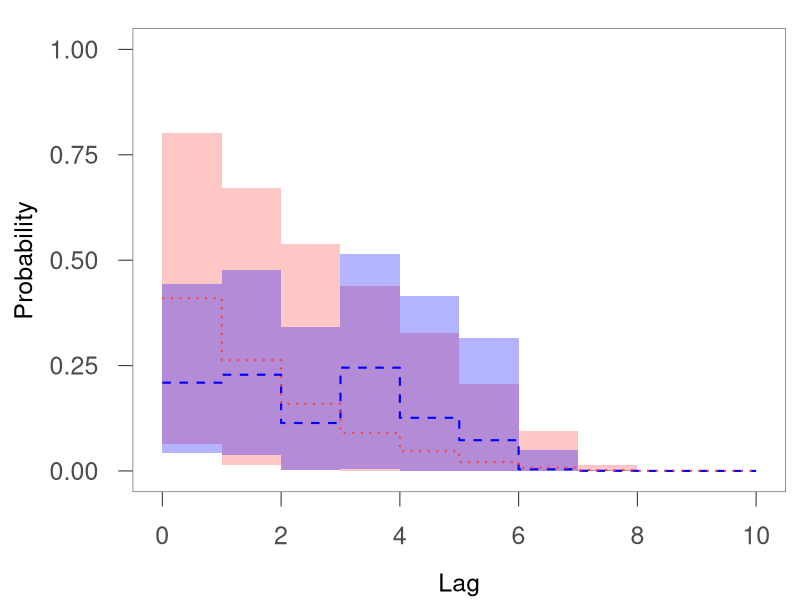}
         \caption{$L = 10$, $n = 1128$}
    \end{subfigure}
    \begin{subfigure}[b]{0.3\textwidth}
         \centering
         \includegraphics[width=\textwidth]{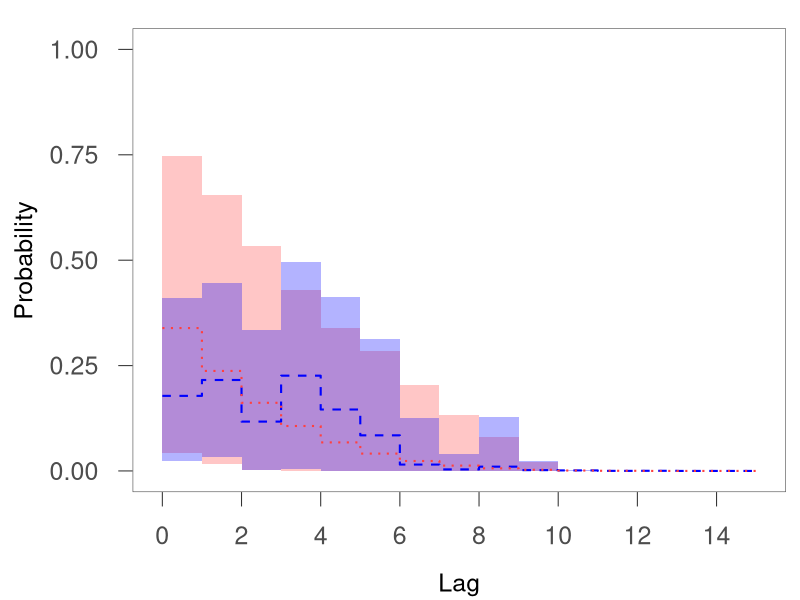}
         \caption{$L = 15$, $n = 1128$}
    \end{subfigure}
    \begin{subfigure}[b]{0.3\textwidth}
         \centering
         \includegraphics[width=\textwidth]{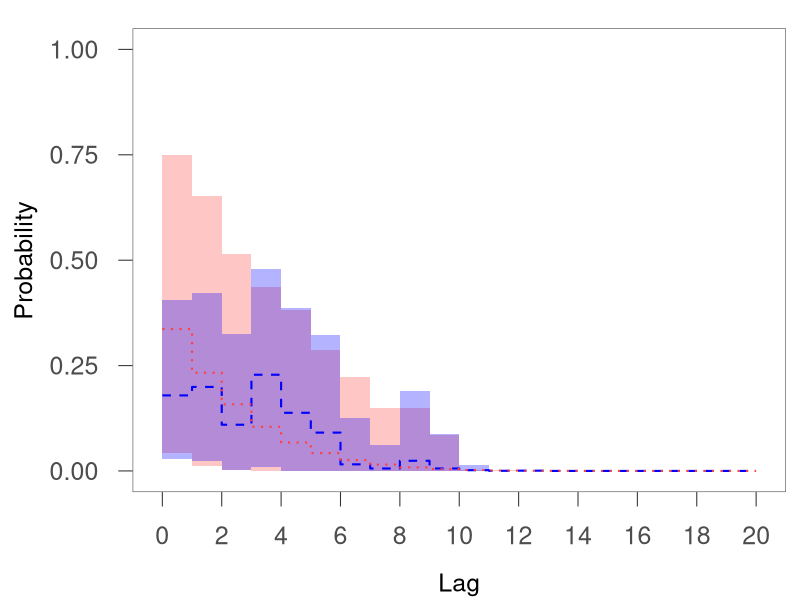}
         \caption{$L = 20$, $n = 1128$}
    \end{subfigure}\\
    \vspace{5pt}
    \begin{subfigure}[b]{0.3\textwidth}
         \centering
         \includegraphics[width=\textwidth]{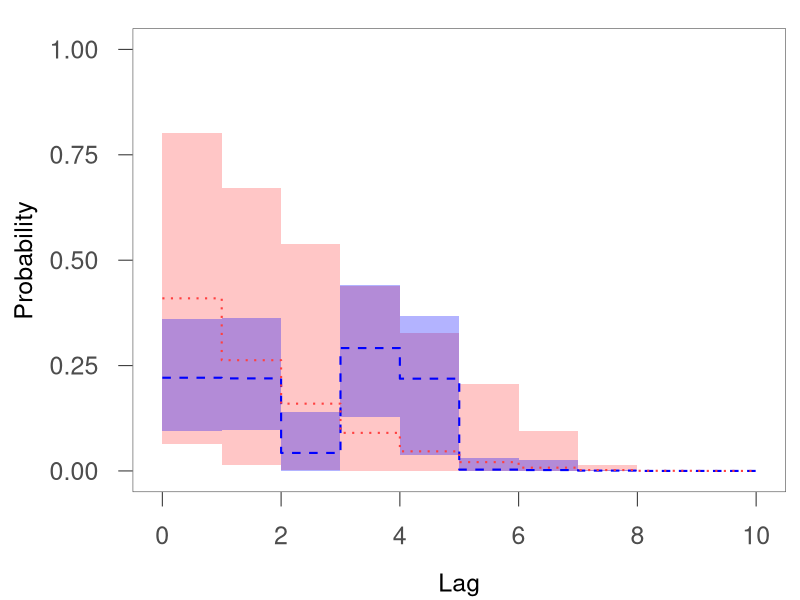}
         \caption{$L = 10$, $n = 4675$}
    \end{subfigure}
    \begin{subfigure}[b]{0.3\textwidth}
         \centering
         \includegraphics[width=\textwidth]{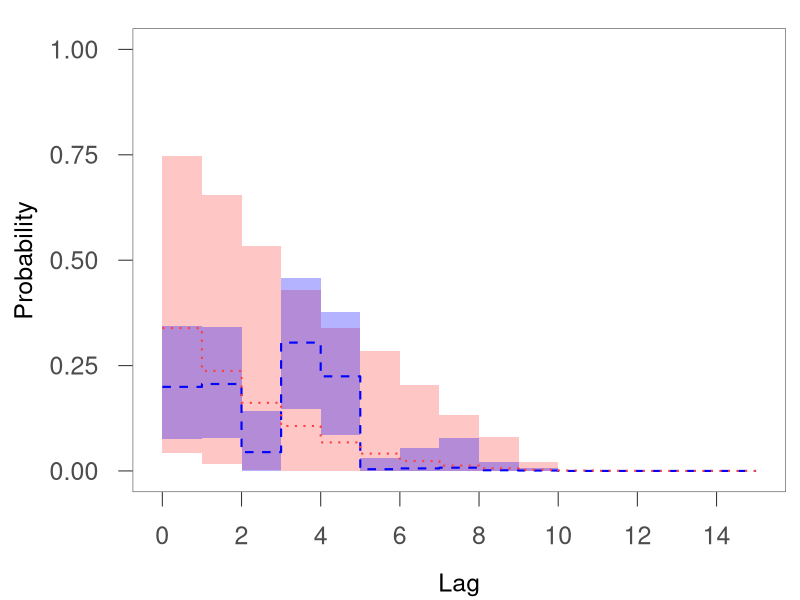}
         \caption{$L = 15$, $n = 4675$}
    \end{subfigure}
    \begin{subfigure}[b]{0.3\textwidth}
         \centering
         \includegraphics[width=\textwidth]{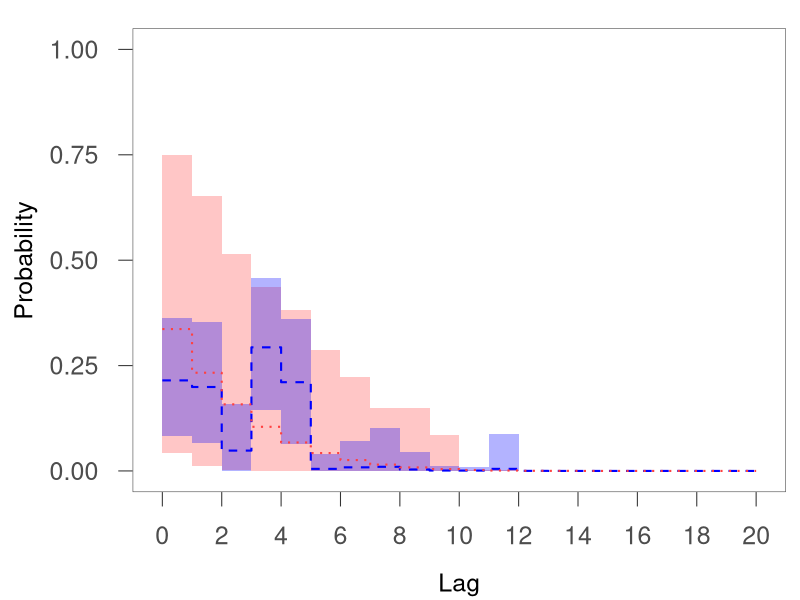}
         \caption{$L = 20$, $n = 4675$}
    \end{subfigure}\\
    \vspace{5pt}
    \begin{subfigure}[b]{0.3\textwidth}
         \centering
         \includegraphics[width=\textwidth]{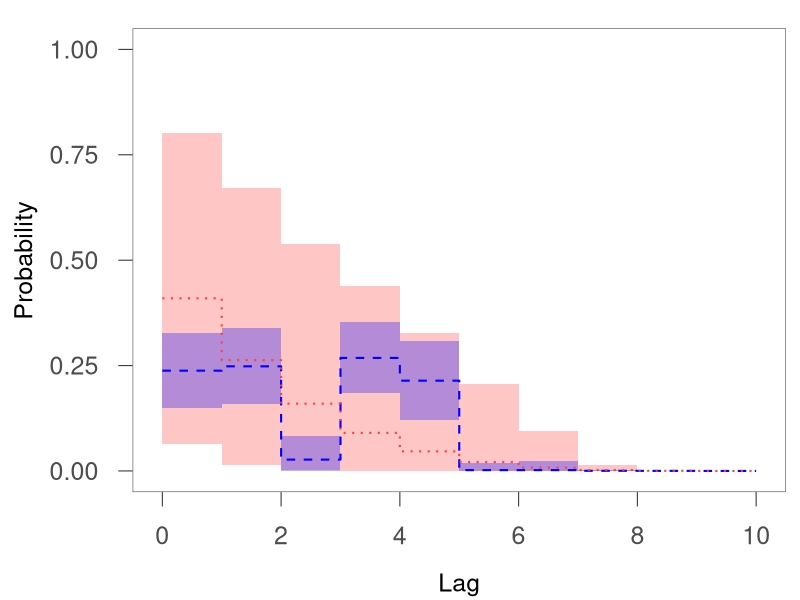}
         \caption{$L = 10$, $n = 12743$}
    \end{subfigure}
    \begin{subfigure}[b]{0.3\textwidth}
         \centering
         \includegraphics[width=\textwidth]{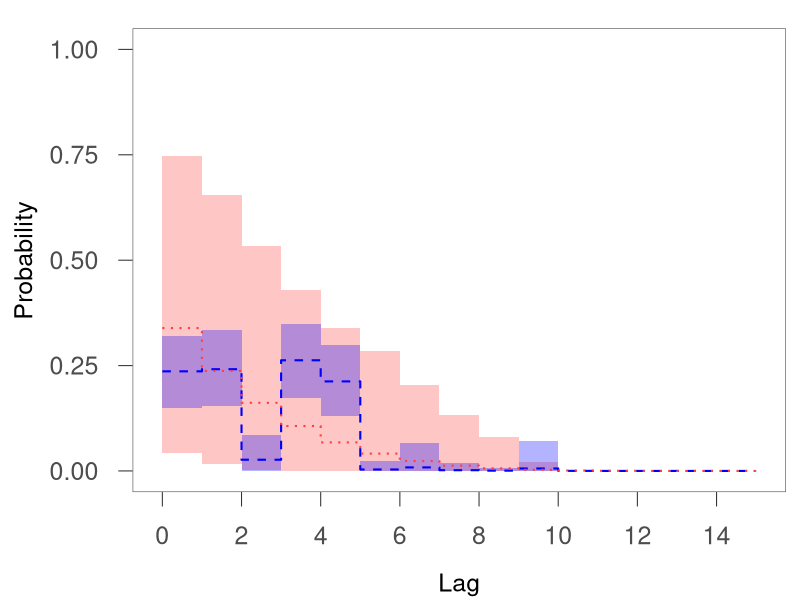}
         \caption{$L = 15$, $n = 12743$}
    \end{subfigure}
    \begin{subfigure}[b]{0.3\textwidth}
         \centering
         \includegraphics[width=\textwidth]{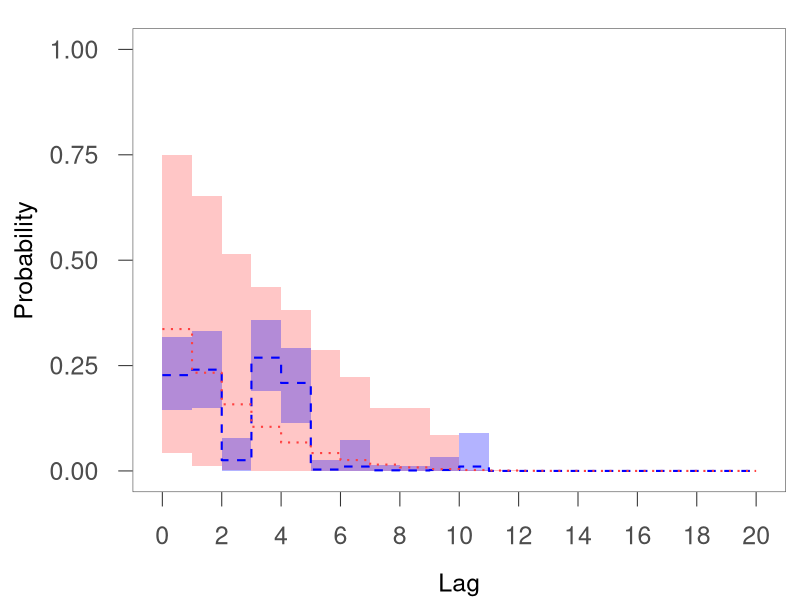}
         \caption{$L = 20$, $n = 12743$}
    \end{subfigure}     
    \caption{
    Posterior mean and $95\%$ credible interval estimates of 
    the mixture weights, under $L = 10$ (left column),
    $L = 15$ (middle column), and $L = 20$ (right column), 
    for observation windows: 
    $(0, 500)$ ($n = 1128$; first row), 
    $(0, 2000)$ ($n = 4675$; second row),
    and $(0, 5500)$ ($n = 12743$; third row).
    Blue lines and blue polygons correspond to posterior mean
    and interval estimates, while red lines and red polygons
    correspond to prior mean and interval estimates.
    }
    \label{fig:burr-sa-LL}
\end{figure*}

Figure \ref{fig:burr-sa-LL} and Table \ref{tbl:burr-sa-LL} 
illustrate the posterior mean and $95\%$ credible interval estimates, respectively, 
for the weights and for the component-density parameters.
All the results were based on posterior samples collected every sixteenth
iteration from a Markov chain of 85000 iterations with a burn-in of 5000 samples.
We observed that the estimates were quite consistent
for different values of $L$, and the model was able to penalize
non-influential lags by assigning them small probabilities.

\begin{table}[t!]
\renewcommand{\arraystretch}{1.25}
\captionsetup{font=footnotesize}
\caption{Posterior mean and $95\%$ credible interval estimates of 
the parameters $\la$, $\ga$, and $\kappa$, 
with $L =$ $10$, $15$, and $20$,  for observation windows: 
$(0, 500)$ ($n = 1128$), $(0, 2000)$ ($n = 4675$),
and $(0, 5500)$ ($n = 12743$).}
\small
    \centering
    \begin{tabular*}{\hsize}{@{\extracolsep{\fill}}lcccccccccc}
    % \hline
\hline
  & $L = 10$ & $L = 15$ & $L = 20$\\
\hline
 &  & $n = 1128$ & \\
\hline
$\lambda$ & 0.96 (0.72, 1.32) & 0.95 (0.72, 1.29) & 0.96 (0.74, 1.30)\\
\hline
$\gamma$ & 1.94 (1.81, 2.07) & 1.94 (1.82, 2.07) & 1.94 (1.82, 2.07)\\
\hline
$\kappa$ & 5.31 (3.76, 7.69) & 5.23 (3.76, 7.57) & 5.30 (3.85, 7.62)\\
\hline
 &  & $n = 4675$ & \\
\hline
$\lambda$ & 1.02 (0.87, 1.19) & 1.02 (0.87, 1.20) & 1.03 (0.88, 1.22)\\
\hline
$\gamma$ & 1.97 (1.91, 2.04) & 1.97 (1.91, 2.03) & 1.97 (1.90, 2.03)\\
\hline
$\kappa$ & 6.12 (4.96, 7.52) & 6.12 (5.03, 7.51) & 6.20 (5.07, 7.67)\\
\hline
 &  & $n = 12743$ & \\
\hline
$\lambda$ & 1.02 (0.93, 1.13) & 1.01 (0.92, 1.11) & 1.01 (0.92, 1.11)\\
\hline
$\gamma$ & 1.98 (1.94, 2.02) & 1.98 (1.94, 2.02) & 1.98 (1.94, 2.02)\\
\hline
$\kappa$ & 6.06 (5.38, 6.91) & 6.03 (5.31, 6.83) & 6.00 (5.33, 6.76)\\
\hline
    \end{tabular*}
    \label{tbl:burr-sa-LL}
\end{table}

\begin{figure*}[t!]
    \centering
    \captionsetup[subfigure]{justification=centering, font=footnotesize}
    \begin{subfigure}[b]{0.3\textwidth}
         \centering
         \includegraphics[width=\textwidth]{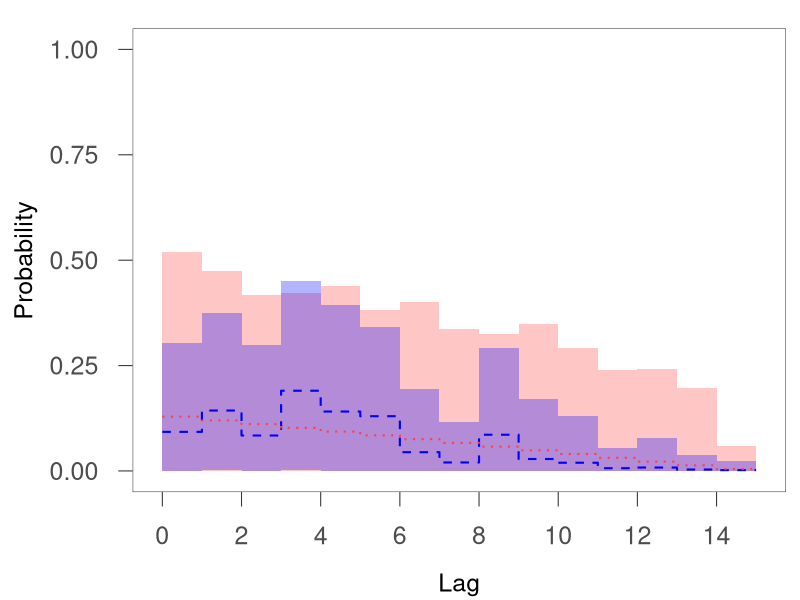}
         \caption{$b_0 = 2$, $n = 1128$}
    \end{subfigure}
    \begin{subfigure}[b]{0.3\textwidth}
         \centering
         \includegraphics[width=\textwidth]{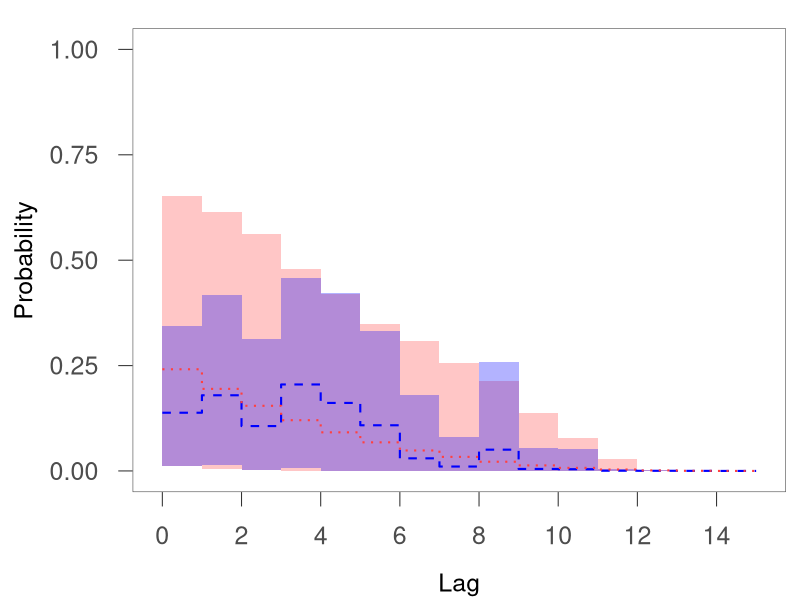}
         \caption{$b_0 = 4$, $n = 1128$}
    \end{subfigure}
    \begin{subfigure}[b]{0.3\textwidth}
         \centering
         \includegraphics[width=\textwidth]{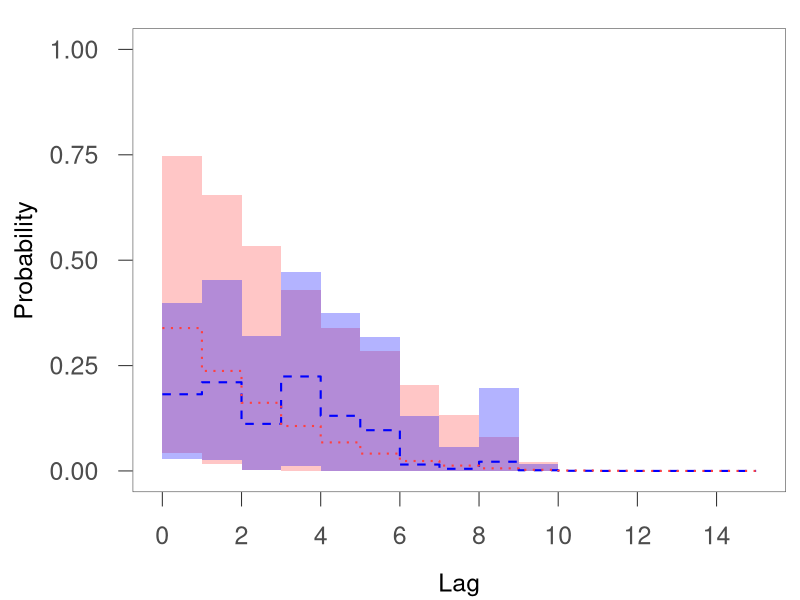}
         \caption{$b_0 = 6$, $n = 1128$}
    \end{subfigure}\\
    \vspace{5pt}
    \begin{subfigure}[b]{0.3\textwidth}
         \centering
         \includegraphics[width=\textwidth]{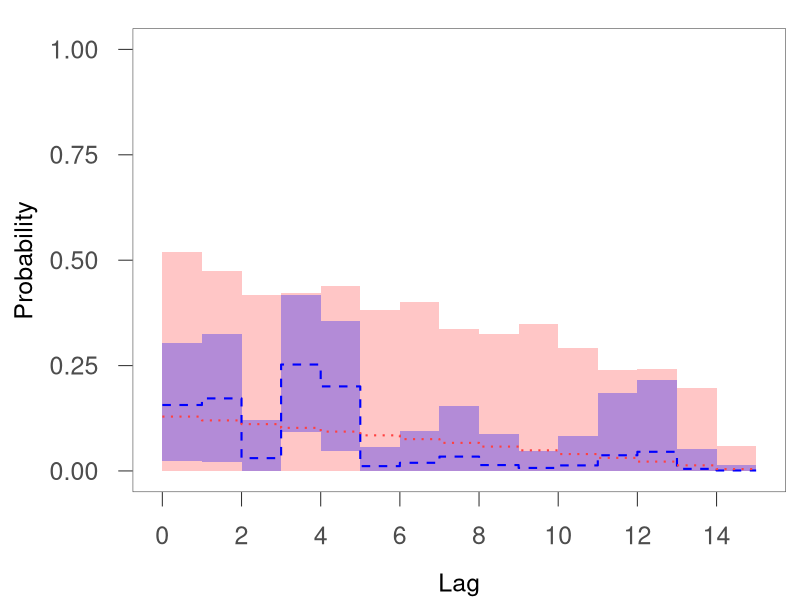}
         \caption{$b_0 = 2$, $n = 4675$}
    \end{subfigure}
    \begin{subfigure}[b]{0.3\textwidth}
         \centering
         \includegraphics[width=\textwidth]{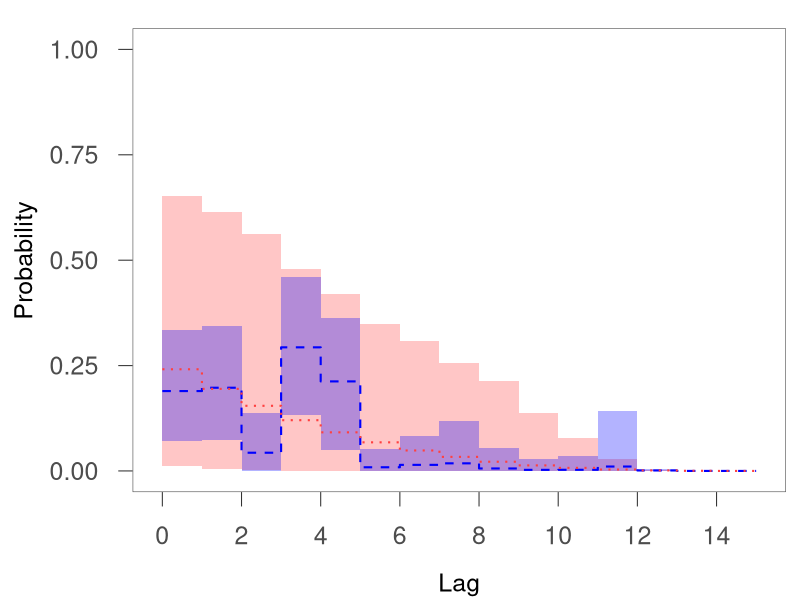}
         \caption{$b_0 = 4$, $n = 4675$}
    \end{subfigure}
    \begin{subfigure}[b]{0.3\textwidth}
         \centering
         \includegraphics[width=\textwidth]{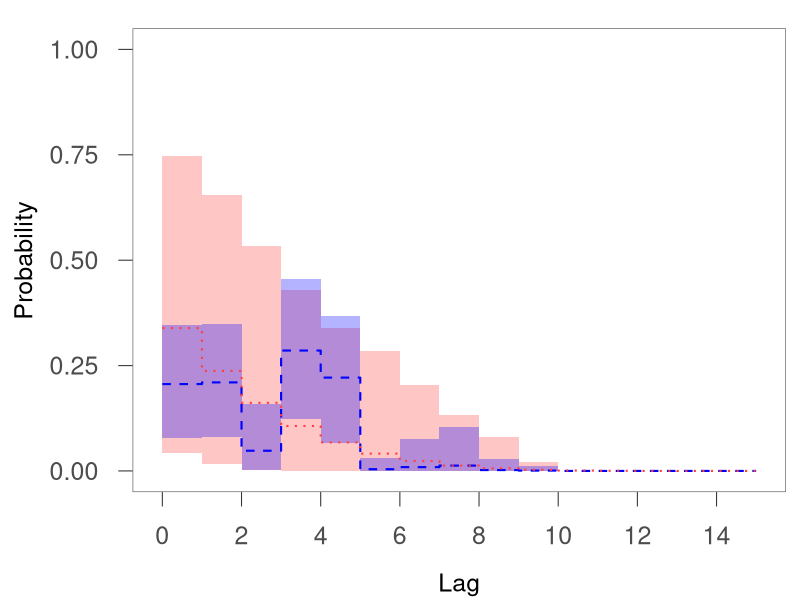}
         \caption{$b_0 = 6$, $n = 4675$}
    \end{subfigure}\\
    \vspace{5pt}
    \begin{subfigure}[b]{0.3\textwidth}
         \centering
         \includegraphics[width=\textwidth]{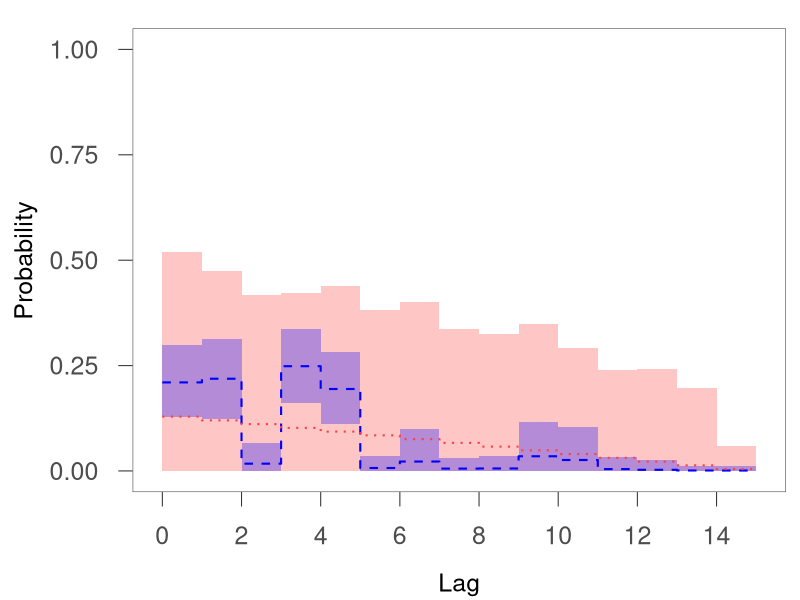}
         \caption{$b_0 = 2$, $n = 12743$}
    \end{subfigure}
    \begin{subfigure}[b]{0.3\textwidth}
         \centering
         \includegraphics[width=\textwidth]{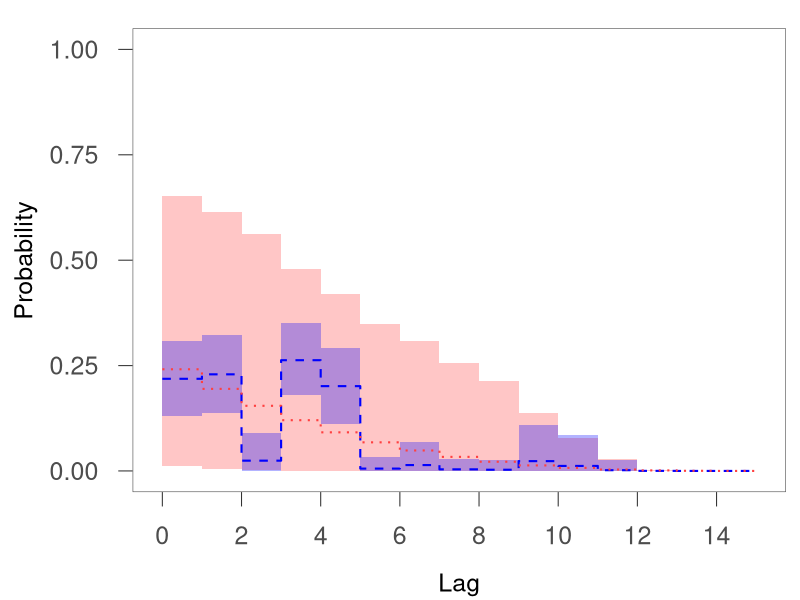}
         \caption{$b_0 = 4$, $n = 12743$}
    \end{subfigure}
    \begin{subfigure}[b]{0.3\textwidth}
         \centering
         \includegraphics[width=\textwidth]{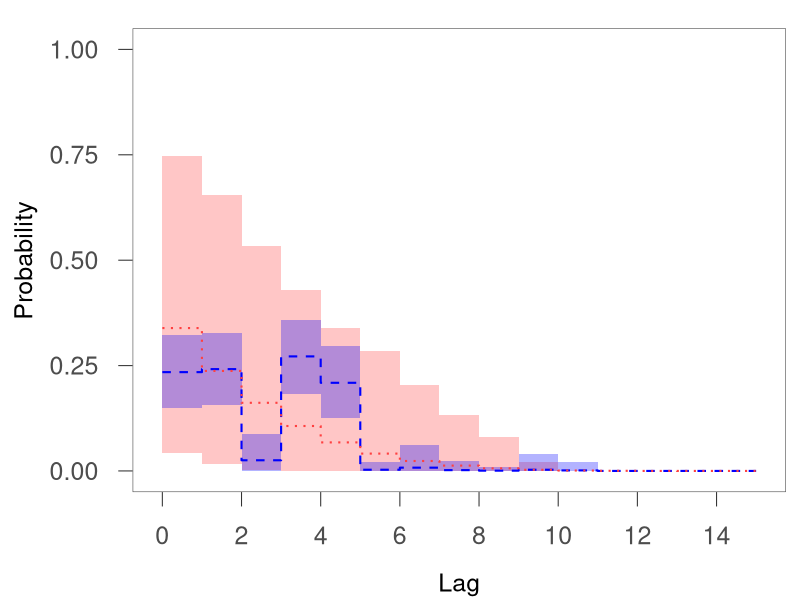}
         \caption{$b_0 = 6$, $n = 12743$}
    \end{subfigure}     
    \caption{Posterior mean and $95\%$ credible interval estimates of 
    the mixture weights, under priors $\mathrm{CDP}(5, 1, b_0)$, 
    with $b_0 = 2$ (left column), $b_0 = 4$ (middle column),
    and $b_0 = 6$ (right column), 
    for observation windows: 
    $(0, 500)$ ($n = 1128$; first row), 
    $(0, 2000)$ ($n = 4675$; second row),
    and $(0, 5500)$ ($n = 12743$; third row).
    Blue lines and blue polygons correspond to posterior mean
    and interval estimates, while red lines and red polygons
    correspond to prior mean and interval estimates.  
    }
    \label{fig:burr-sa-b0}
\end{figure*}

\begin{figure*}[t!]
    \centering
    \captionsetup[subfigure]{justification=centering, font=footnotesize}
    \begin{subfigure}[b]{0.3\textwidth}
         \centering
         \includegraphics[width=\textwidth]{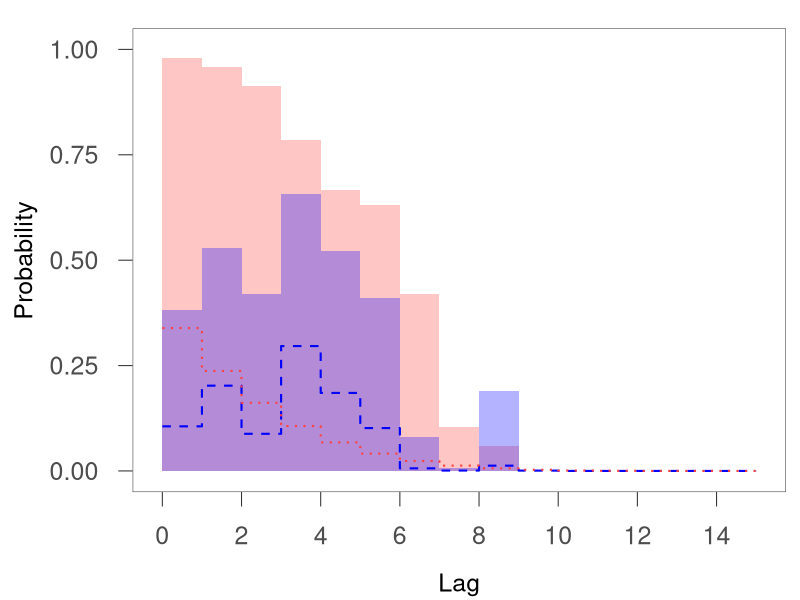}
         \caption{$\alpha_0 = 1$, $n = 1128$}
    \end{subfigure}
    \begin{subfigure}[b]{0.3\textwidth}
         \centering
         \includegraphics[width=\textwidth]{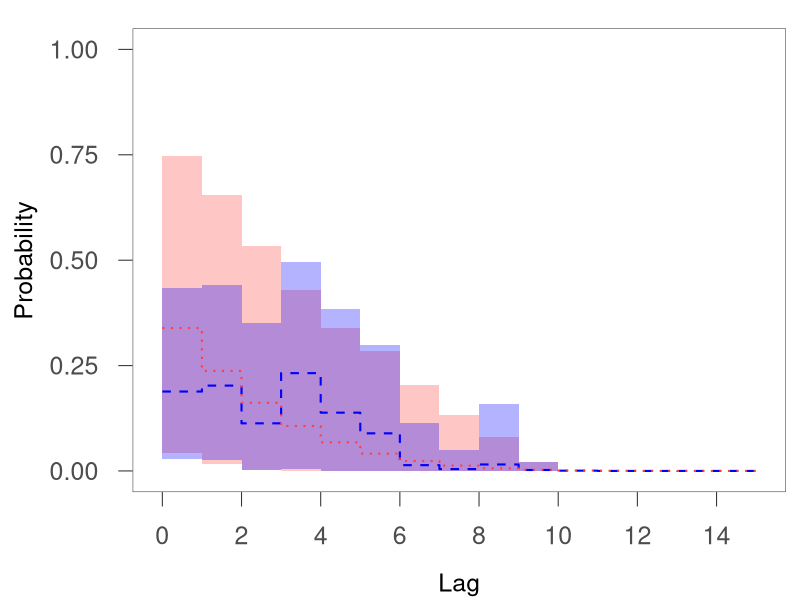}
         \caption{$\alpha_0 = 5$, $n = 1128$}
    \end{subfigure}
    \begin{subfigure}[b]{0.3\textwidth}
         \centering
         \includegraphics[width=\textwidth]{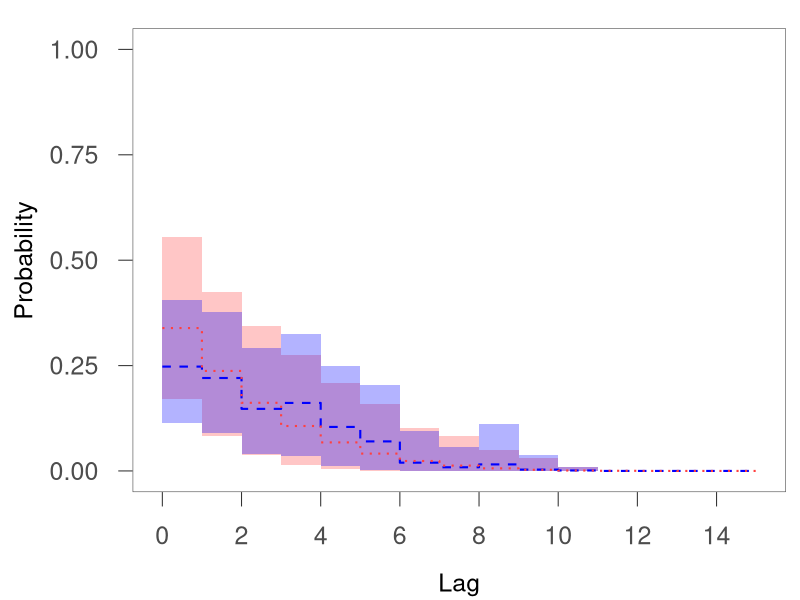}
         \caption{$\alpha_0 = 20$, $n = 1128$}
    \end{subfigure}\\
    \vspace{5pt}
    \begin{subfigure}[b]{0.3\textwidth}
         \centering
         \includegraphics[width=\textwidth]{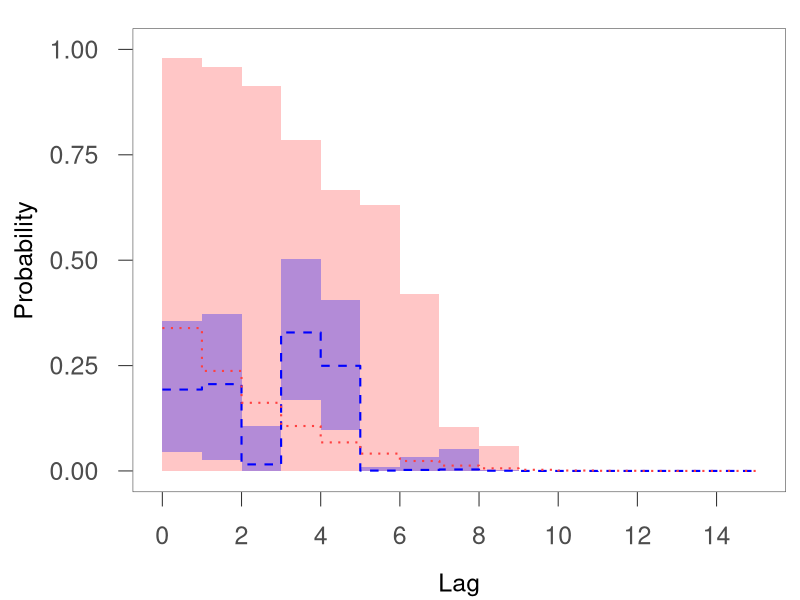}
         \caption{$\alpha_0 = 1$, $n = 4675$}
    \end{subfigure}
    \begin{subfigure}[b]{0.3\textwidth}
         \centering
         \includegraphics[width=\textwidth]{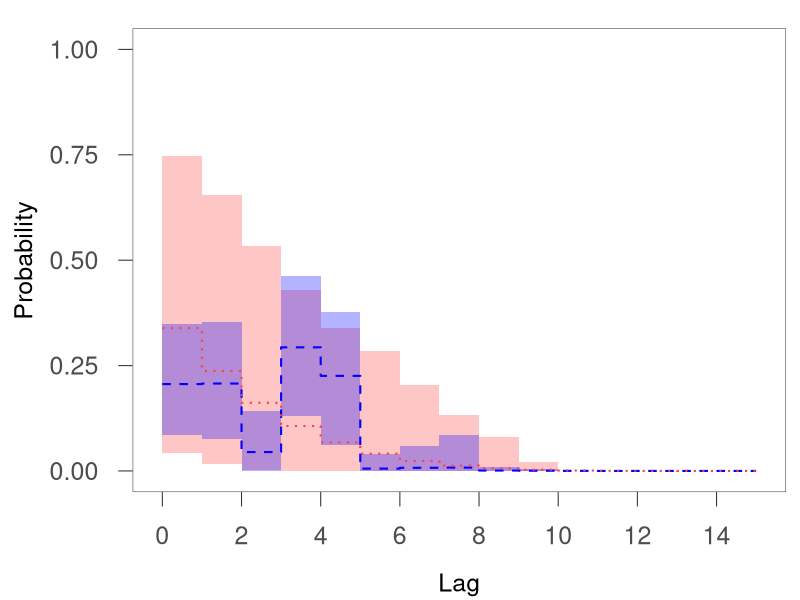}
         \caption{$\alpha_0 = 5$, $n = 4675$}
    \end{subfigure}
    \begin{subfigure}[b]{0.3\textwidth}
         \centering
         \includegraphics[width=\textwidth]{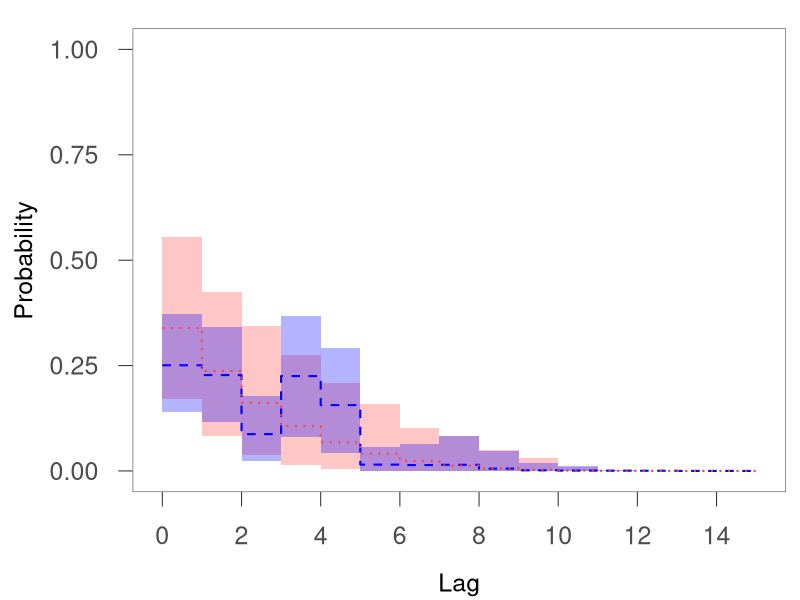}
         \caption{$\alpha_0 = 20$, $n = 4675$}
    \end{subfigure}\\
    \vspace{5pt}
    \begin{subfigure}[b]{0.3\textwidth}
         \centering
         \includegraphics[width=\textwidth]{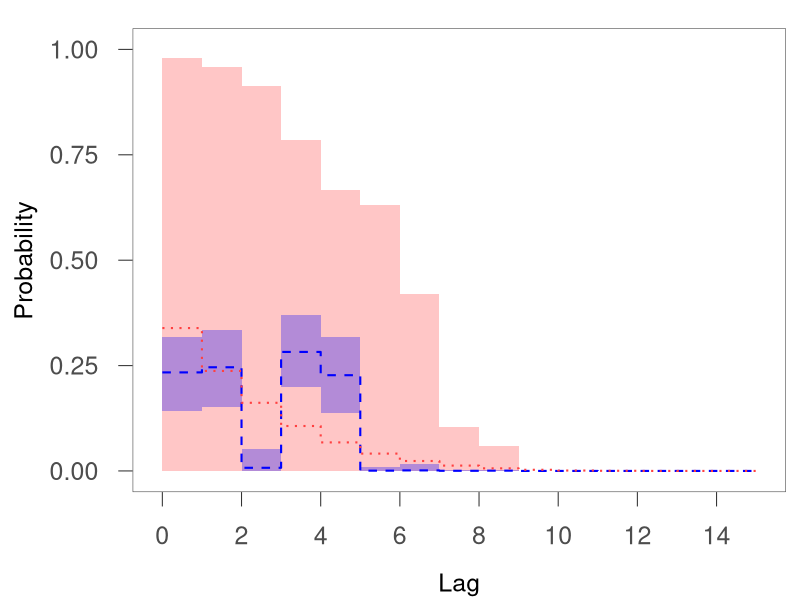}
         \caption{$\alpha_0 = 1$, $n = 12743$}
    \end{subfigure}
    \begin{subfigure}[b]{0.3\textwidth}
         \centering
         \includegraphics[width=\textwidth]{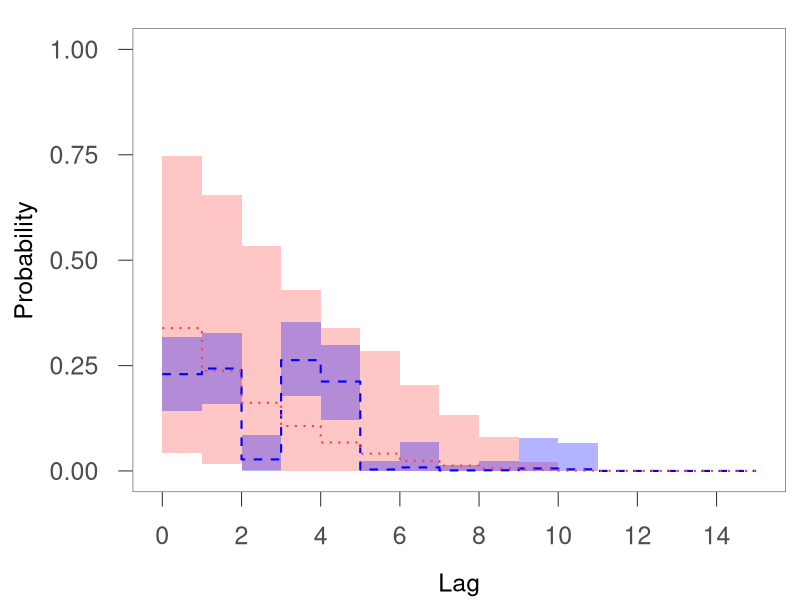}
         \caption{$\alpha_0 = 5$, $n = 12743$}
    \end{subfigure}
    \begin{subfigure}[b]{0.3\textwidth}
         \centering
         \includegraphics[width=\textwidth]{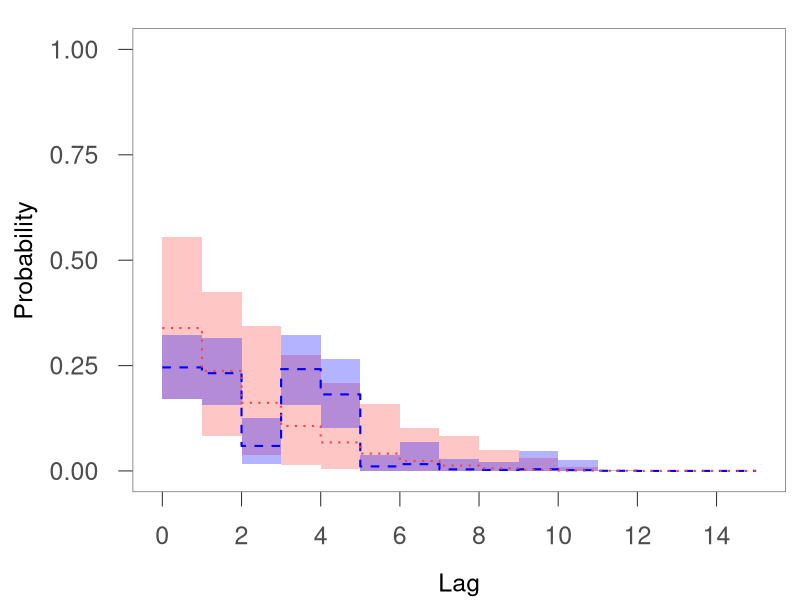}
         \caption{$\alpha_0 = 20$, $n = 12743$}
    \end{subfigure}     
    \caption{
    Posterior mean and $95\%$ credible interval estimates of 
    the mixture weights, under priors $\mathrm{CDP}(\alpha_0, 1, 6)$, 
    with $\alpha_0 = 1$ (left column), $\alpha_0 = 5$ (middle column),
    and $\alpha_0 = 20$ (right column), 
    for observation windows: 
    $(0, 500)$ ($n = 1128$; first row), 
    $(0, 2000)$ ($n = 4675$; second row),
    and $(0, 5500)$ ($n = 12743$; third row).
    Blue lines and blue polygons correspond to posterior mean
    and interval estimates, while red lines and red polygons
    correspond to prior mean and interval estimates. 
    }
    \label{fig:burr-sa-alpha0}
\end{figure*}

\paragraph{Prior sensitivity analysis for the mixture weights}\mbox{}

The sensitivity analysis for $L$ suggested that $L = 15$ worked as 
a reasonable upper bound. Thus, for the following prior sensitivity analysis, 
we chose $L = 15$ when fitting models.
For each one of the observation windows, we first fitted the Burr MTDPP model, 
using priors for the weights, $\mathrm{CDP}(5, 1, b_0)$, with
$b_0 = 2, 4, 6$; then we fitted the Burr model 
using $\mathrm{CDP}(\alpha_0, 1, 6)$
with $\alpha_0 = 1, 5, 20$. Thus, in total, we fitted the model six times
for each observation window. The posterior estimates
of the weights are summarized
in Figures \ref{fig:burr-sa-b0} and \ref{fig:burr-sa-alpha0}.
All the results were based on posterior samples collected every sixteenth
iteration from a Markov chain of 85000 iterations with a burn-in of 5000 samples.

Figure \ref{fig:burr-sa-b0} shows the posterior mean and interval estimates
of the weights under priors, $\mathrm{CDP}(5, 1, 2)$, $\mathrm{CDP}(5, 1, 4)$,
and $\mathrm{CDP}(5, 1, 6)$. As $b_0$ increases, 
the model imposes a greater penalty on distant lags, resulting in less uncertainty.
Regardless of the choice of $b_0$,
the posterior estimates of the weights generally recovered the 
pattern of the true weights, i.e., there is a gap between 
the first two and the last two influential lags. 
Moreover, the posterior estimates become closer to the true weights
as $n$ increases.

\begin{table}[t!]
\renewcommand{\arraystretch}{1.25}
\captionsetup{font=footnotesize}
\caption{$\;$Posterior mean and $95\%$ credible interval estimates of 
the component parameter estimates $\la$, $\ga$, and $\kappa$, 
under priors $\mathrm{CDP}(5, 1, b_0)$, with $b_0 = 2, 4, 6$.}
\small
    \centering
    \begin{tabular*}{\hsize}{@{\extracolsep{\fill}}lcccccccccc}
    % \hline
\hline
  & $b_0 = 2$ & $b_0 = 4$ & $b_0 = 6$\\
\hline
 &  & $n = 1128$ & \\
\hline
$\lambda$ & 0.93 (0.71, 1.21) & 0.95 (0.71, 1.31) & 0.93 (0.71, 1.20)\\
\hline
$\gamma$ & 1.96 (1.84, 2.09) & 1.95 (1.82, 2.09) & 1.95 (1.84, 2.08)\\
\hline
$\kappa$ & 5.09 (3.71, 7.01) & 5.23 (3.70, 7.71) & 5.07 (3.71, 6.90)\\
\hline
 &  & $n = 4675$ & \\
\hline
$\lambda$ & 1.00 (0.83, 1.21) & 1.02 (0.86, 1.20) & 1.02 (0.87, 1.21)\\
\hline
$\gamma$ & 1.98 (1.91, 2.05) & 1.97 (1.91, 2.03) & 1.97 (1.91, 2.03)\\
\hline
$\kappa$ & 5.95 (4.75, 7.59) & 6.10 (4.94, 7.58) & 6.15 (5.00, 7.70)\\
\hline
 &  & $n = 12743$ & \\
\hline
$\lambda$ & 0.99 (0.89, 1.10) & 1.00 (0.90, 1.11) & 1.01 (0.92, 1.12)\\
\hline
$\gamma$ & 1.99 (1.95, 2.03) & 1.99 (1.94, 2.03) & 1.98 (1.94, 2.02)\\
\hline
$\kappa$ & 5.86 (5.13, 6.70) & 5.93 (5.20, 6.79) & 5.98 (5.30, 6.83)\\
\hline
    \end{tabular*}
    \label{tbl:burr-sa-b0}
\end{table}

\begin{table}[t!]
\renewcommand{\arraystretch}{1.25}
\captionsetup{font=footnotesize}
\caption{$\;$Posterior mean and $95\%$ credible interval estimates of 
the component-density parameters $\la$, $\ga$, and $\kappa$, 
under priors $\mathrm{CDP}(\alpha_0, 1, 6)$, with $\alpha_0 = 1, 5, 10$.}
\small
    \centering
    \begin{tabular*}{\hsize}{@{\extracolsep{\fill}}lcccccccccc}
    % \hline
\hline
  & $\alpha_0 = 1$ & $\alpha_0 = 5$ & $\alpha_0 = 20$\\
\hline
 &  & $n = 1128$ & \\
\hline
$\lambda$ & 1.00 (0.75, 1.38) & 0.96 (0.73, 1.29) & 0.93 (0.71, 1.24)\\
\hline
$\gamma$ & 1.93 (1.81, 2.06) & 1.94 (1.82, 2.07) & 1.95 (1.83, 2.08)\\
\hline
$\kappa$ & 5.54 (3.91, 8.22) & 5.27 (3.82, 7.51) & 5.08 (3.70, 7.23)\\
\hline
 &  & $n = 4675$ & \\
\hline
$\lambda$ & 1.02 (0.86, 1.20) & 1.02 (0.88, 1.19) & 1.04 (0.88, 1.22)\\
\hline
$\gamma$ & 1.97 (1.91, 2.04) & 1.97 (1.91, 2.03) & 1.97 (1.90, 2.03)\\
\hline
$\kappa$ & 6.14 (4.93, 7.56) & 6.13 (5.01, 7.50) & 6.25 (5.08, 7.72)\\
\hline
 &  & $n = 12743$ & \\
\hline
$\lambda$ & 1.02 (0.93, 1.13) & 1.01 (0.92, 1.12) & 1.00 (0.90, 1.10)\\
\hline
$\gamma$ & 1.98 (1.94, 2.02) & 1.98 (1.94, 2.02) & 1.99 (1.95, 2.02)\\
\hline
$\kappa$ & 6.11 (5.40, 6.91) & 6.01 (5.30, 6.84) & 5.93 (5.21, 6.72)\\
\hline
    \end{tabular*}
    \label{tbl:burr-sa-alpha0}
\end{table}

Figure \ref{fig:burr-sa-alpha0} shows the posterior mean and interval estimates
of the weights under priors, $\mathrm{CDP}(1, 1, 6)$, $\mathrm{CDP}(5, 1, 6)$,
and $\mathrm{CDP}(20, 1, 6)$. Note that 
for the DP that defines the weights, $\alpha_0$ is the precision parameter.
That is, for large $\alpha_0$, there is small variability in the DP realizations,
and thus the prior realizations
of the weights are less variable as $\alpha_0$ increases, as shown 
in Figure \ref{fig:burr-sa-alpha0}. 
When $\alpha = 20$ and the sample size is small
(e.g., $n = 1128$), the results were sensitive to the prior
(Figure \ref{fig:burr-sa-alpha0}(c)). However, in other cases,
the posterior estimates of the weights were able to recover the true pattern. 
In all scenarios, the posterior estimates approach the true weights as
$n$ increases.

Overall, the prior for the mixture weights, with careful choices of 
$\alpha_0$ and $b_0$, is quite robust, in the sense that given the data, the
model can assign
large weights to influential lags and small probabilities to non-important lags.
Tables \ref{tbl:burr-sa-b0} and \ref{tbl:burr-sa-alpha0} show
the posterior mean and $95\%$ credible interval estimates of the
component-parameter estimates.

\subsection{Third simulation study: The MTDCPP model for event clustering}
\label{sec: mtdcpp-sim}

The goal of this study is to examine the ability 
of the MTDCPP to recover various clustering behaviors attributed to two
different factors. 
We generate data from a Lomax MTDCPP, that is, with $f_I$ corresponding to an 
exponential distribution with rate parameter $\mu$
and $f^*(t-\tN)$ a stationary Lomax MTDPP.

In the study, we consider four scenarios, with $\pi_0$ taking one of the following values, 
$(0.2$, $0.5$, $0.8$, $1)$. 
The first three values indicate the proportion
of the duration process affected by a factor through $f_I$. When
$\pi_0 = 1$, the data are equivalently generated by a Poisson process with
rate $\mu$.
For all scenarios, we take $\mu = 0.2$, $\alpha = 5$, $\phi = 0.1$ 
and decaying weights $\bm w = (0.35, 0.25, 0.2, 0.1, 0.1)^\top$.

We applied the Lomax MTDCPP model with $L = 5$ to the synthetic data. 
We specified a beta prior
$\mathrm{Beta}(\pi_0\,|\,1,1)$ for the probability $\pi_0$ and a gamma prior
$\mathrm{Ga}(\mu\,|\,1,1)$ for the rate parameter $\mu$. 
For the stationary Lomax MTDPP, the shape and scale parameters 
received priors $\mathrm{Ga}(\alpha\,|\,6,1)\mathbbm{1}(\alpha>1)$ and 
$\mathrm{Ga}(\phi\,|\,1, 1)$, respectively. 
In particular, we chose prior for $\alpha$
with the expectation that the first four moments exist with respect to the 
component and marginal Lomax distributions. 
The vector $\bm w$ was assigned CDP$(\bm w\,|\,5, 1, 3)$,
which elicits a decreasing pattern in the weights. 

\begin{table}[t!]
    \renewcommand{\arraystretch}{1.3}
    \captionsetup{font=footnotesize}
    \caption{Simulation study for event clustering.
    Posterior mean and $95\%$ credible interval estimates of the MTDCPP 
    model parameters under different scenarios.}
    \small
    \centering
    \begin{threeparttable}
    \begin{tabular*}{\hsize}{@{\extracolsep{\fill}}lcccc}
\hline
  & $\pi_0 = 0.2$ & $\pi_0 = 0.5$ & $\pi_0 = 0.8$ & $\pi_0 = 1$\\
\hline
$\pi_0$ & 0.22 (0.20, 0.25) & 0.52 (0.48, 0.56) & 0.81 (0.76, 0.86) & 0.99 (0.96, 1.00)\\
\hline
$\mu$ & 0.22 (0.19, 0.24) & 0.19 (0.17, 0.20) & 0.20 (0.19, 0.21) & 0.19 (0.18, 0.20)\\
\hline
$\phi$ & 0.12 (0.09, 0.15) & 0.13 (0.09, 0.19) & 0.12 (0.02, 0.33) & 1.33 (0.04, 4.70)\\
\hline
$\alpha$ & 5.46 (4.63, 6.43) & 6.37 (4.91, 8.15) & 4.78 (2.99, 8.21) & 5.39 (2.19, 10.82)\\
\hline
    \end{tabular*}
    \begin{tablenotes}[para,flushleft]
        \small 
    \end{tablenotes}
    \end{threeparttable}
    \label{tbl:mtdcpp} 
\end{table}

We focus on the inference on the two-component mixture probability $\pi_0$
and the component density parameters ($\mu, \phi,\alpha$). The posterior
mean and $95\%$ credible interval estimates of the parameters are presented in 
Table \ref{tbl:mtdcpp}. 
The posterior estimates of $\pi_0$
suggest that the model was able to recover the proportion of the
point process driven by $f_I$, even in the extreme case when $\pi_0 = 1$.
For other parameters, the model produced estimates close to the 
true values for all scenarios.

\section{Additional data-example results}
\subsection{IVT recurrence interval analysis}

\subsubsection*{Sensitivity analysis of \texorpdfstring{$L$}{Lg}}

\begin{figure*}[t!]
    \centering
    \captionsetup[subfigure]{justification=centering, font=footnotesize}
    \begin{subfigure}[b]{0.3\textwidth}
         \centering
         \includegraphics[width=\textwidth]{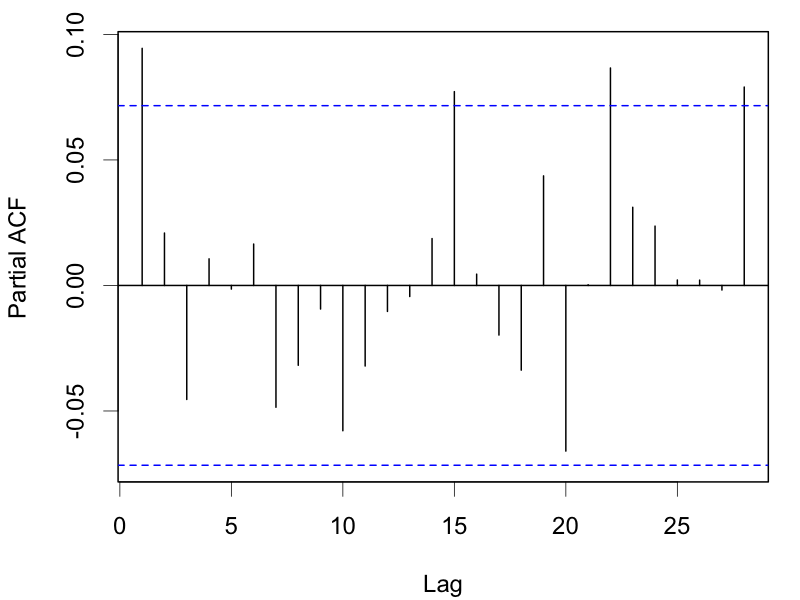}
         \caption{PACF (original data)}
    \end{subfigure}    
    \begin{subfigure}[b]{0.3\textwidth}
         \centering
         \includegraphics[width=\textwidth]{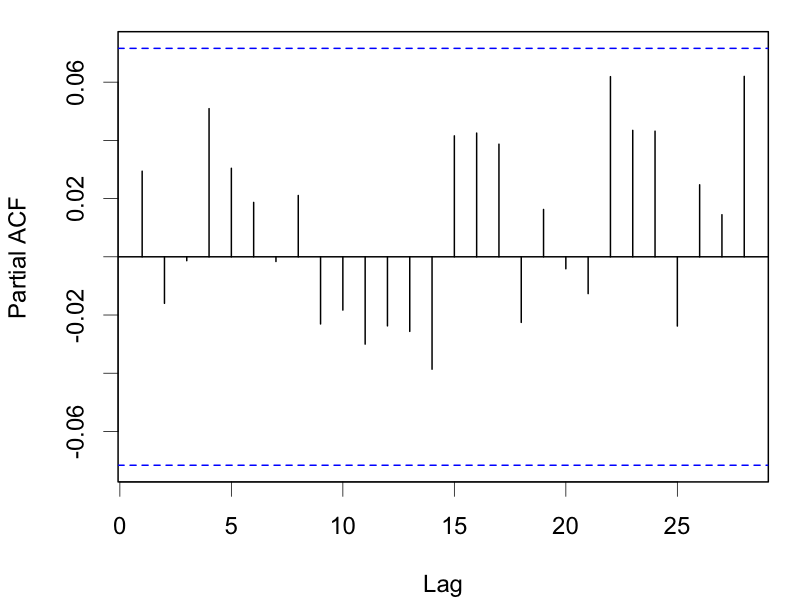}    
         \caption{PACF (detrended data)}
    \end{subfigure}    
    \begin{subfigure}[b]{0.3\textwidth}
         \centering
         \includegraphics[width=\textwidth]{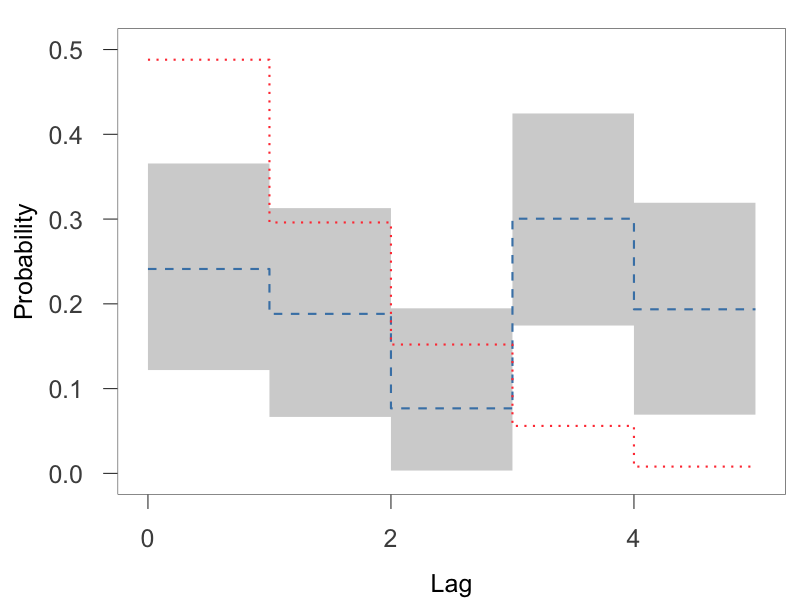}
         \caption{$L = 5$}
    \end{subfigure}\\
    \medskip
    \vspace{5pt}
    \begin{subfigure}[b]{0.3\textwidth}
         \centering
         \includegraphics[width=\textwidth]{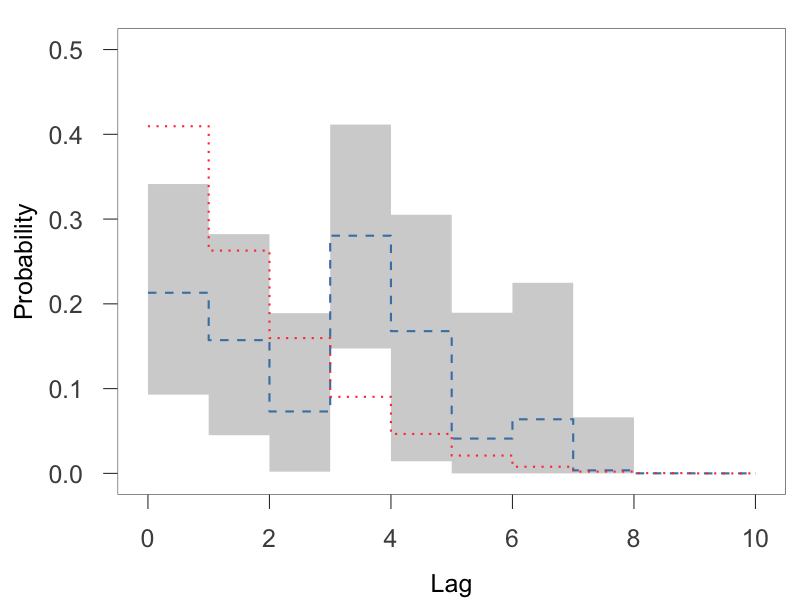}
         \caption{$L = 10$}
    \end{subfigure}    
    \begin{subfigure}[b]{0.3\textwidth}
         \centering
         \includegraphics[width=\textwidth]{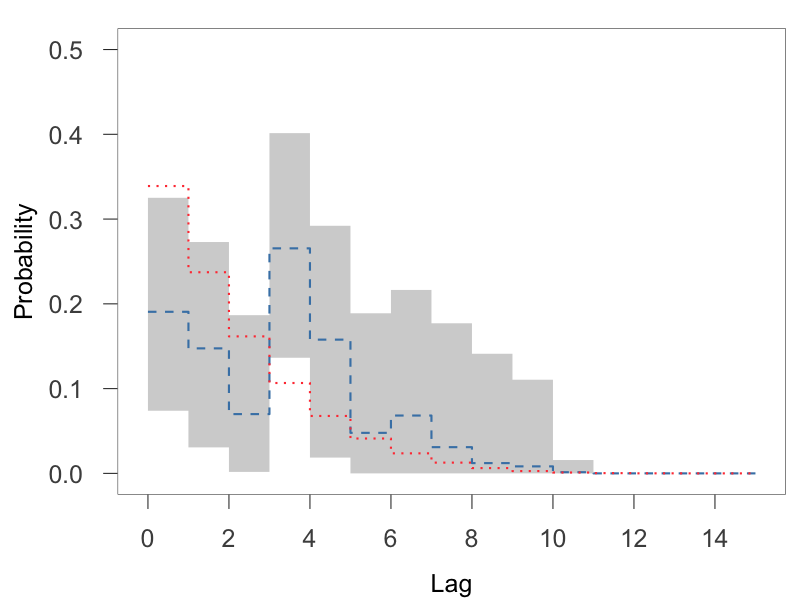}
         \caption{$L = 15$}
    \end{subfigure}    
    \begin{subfigure}[b]{0.3\textwidth}
         \centering
         \includegraphics[width=\textwidth]{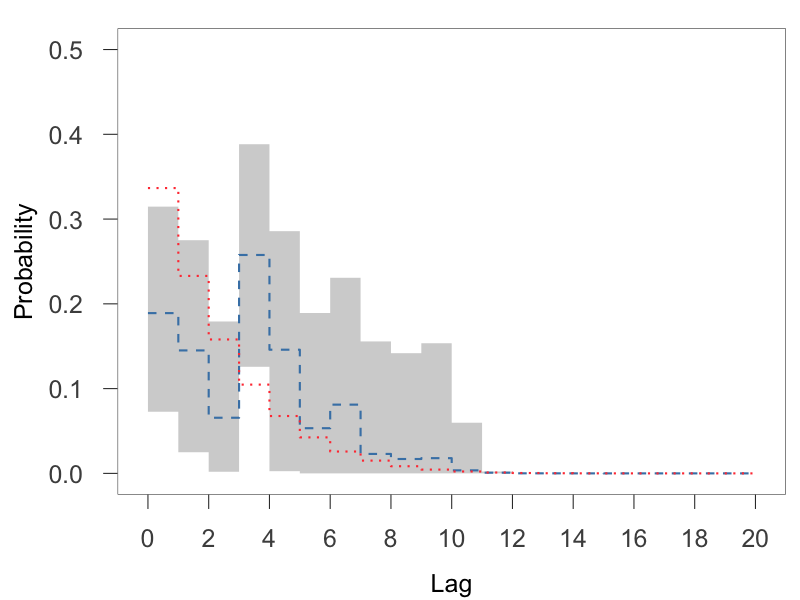}
         \caption{$L = 20$}
    \end{subfigure}    
    \caption{
    IVT recurrence interval analysis:
    Partial autocorrelation functions for
    the original durations (Panel (a)) and for the detrended durations
    (Panel (b)) based on a harmonic regression;
    posterior mean and 95\% credible
    interval estimates for the weights, under the MTDPP multiplicative 
    model with $L = 5$ (Panel (c)), $L = 10$ (Panel (d)), 
    $L = 15$ (Panel (e)), and $L = 20$ (Panel (f)).
    Blue dashed lines, red dotted lines, and grey polygons are, respectively, 
    posterior mean, prior mean, are $95\%$ credible interval estimates.}
    \label{fig:ivt-L-sa-weight}
\end{figure*}

We first examined the PACF of the 
durations. As shown in Figure \ref{fig:ivt-L-sa-weight}(a),
the PACF cuts off after lag 1. We then examined the PACF of 
the detrended durations based on a harmonic regression.
That is, we regressed the natural logarithm of duration $x_i$
on covariates, $\sin(j\omega t_i)$ and $\cos(j\omega t_i)$,
$j = 1,\dots, J$, where $t_i$ is the event time associated
with $x_i$, $J = 5$, $\omega = 2\pi/T_0$, and $T_0 = 365$.
Then, we obtained the detrended durations by taking 
the exponential of the residuals
of the harmonic regression. The PACF of the detrended durations
is illustrated in Figure \ref{fig:ivt-L-sa-weight}(b).
Overall, the PACFs indicate the possibility of temporal dependence
in the durations.

According to the PACFs, 
we fitted the multiplicative model (see Section 4.2 of the main paper)
with $L = 5, 10, 15, 20$, respectively, with
priors for the weights, $\mathrm{CDP}(5, 1, 3)$, $\mathrm{CDP}(5, 1, 5)$, 
$\mathrm{CDP}(5, 1, 6)$, $\mathrm{CDP}(5, 1, 8)$.
For all models, we assigned a normal distribution $N(0, 10)$ 
to each regression parameter.
The shape and scale parameters $\alpha$ and $\phi$, respectively, received 
$\mathrm{Ga}(6,1)\mathbbm{1}(\alpha>1)$ and $\mathrm{Ga}(1,1)$ priors.
We examined model performance on
parameter estimates, which are demonstrated in 
Figure \ref{fig:ivt-L-sa-weight}
and Table \ref{tbl:ivt-sa-param}. Results are based on 5000 posterior samples,
obtained after discarding the first 5000 iterations
of the MCMC and then retaining one every tenth iterations.
Computing times for running the models are also reported in
Table \ref{tbl:ivt-sa-param}.

\begin{table}[t!]
\renewcommand{\arraystretch}{1.35}
\captionsetup{font=footnotesize}
\caption{IVT recurrence interval analysis (sensitivity analysis of $L$): 
posterior mean and 95\% credible interval estimates for the component-density and harmonic-regression 
parameters, and computing time (minutes), 
under the MTDPP multiplicative model with different values of $L$.}
\small
    \centering
    \begin{tabular*}{\hsize}{@{\extracolsep{\fill}}lcccc}
    % \\[-5pt]
\hline
  & $L = 5$ & $L = 10$ & $L = 15$ & $L = 20$\\
\hline
$\alpha$ & 2.04 (1.75, 2.37) & 2.03 (1.75, 2.36) & 2.01 (1.73, 2.36) & 2.02 (1.73, 2.37)\\
\hline
$\phi$ & 5.16 (3.55, 7.09) & 5.05 (3.46, 7.05) & 4.95 (3.38, 6.90) & 4.95 (3.35, 6.92)\\
\hline
$\beta_{11}$ & -0.58 (-0.85, -0.30) & -0.57 (-0.85, -0.30) & -0.59 (-0.86, -0.30) & -0.59 (-0.87, -0.31)\\
\hline
$\beta_{12}$ & -0.67 (-1.03, -0.33) & -0.68 (-1.07, -0.34) & -0.69 (-1.06, -0.35) & -0.70 (-1.07, -0.35)\\
\hline
$\beta_{21}$ & 0.21 (-0.11, 0.55) & 0.20 (-0.12, 0.56) & 0.20 (-0.13, 0.52) & 0.19 (-0.14, 0.53)\\
\hline
$\beta_{22}$ & -0.53 (-0.82, -0.25) & -0.53 (-0.82, -0.24) & -0.53 (-0.82, -0.22) & -0.53 (-0.83, -0.23)\\
\hline
$\beta_{31}$ & 0.16 (-0.19, 0.49) & 0.16 (-0.23, 0.51) & 0.16 (-0.19, 0.49) & 0.15 (-0.21, 0.50)\\
\hline
$\beta_{32}$ & 0.04 (-0.20, 0.29) & 0.04 (-0.21, 0.29) & 0.04 (-0.22, 0.29) & 0.04 (-0.21, 0.29)\\
\hline
$\beta_{41}$ & -0.06 (-0.34, 0.21) & -0.07 (-0.34, 0.22) & -0.08 (-0.35, 0.20) & -0.08 (-0.37, 0.21)\\
\hline
$\beta_{42}$ & 0.23 (-0.04, 0.50) & 0.24 (-0.03, 0.50) & 0.24 (-0.03, 0.52) & 0.23 (-0.04, 0.51)\\
\hline
$\beta_{51}$ & 0.09 (-0.11, 0.29) & 0.09 (-0.12, 0.30) & 0.09 (-0.12, 0.30) & 0.10 (-0.11, 0.31)\\
\hline
$\beta_{52}$ & -0.12 (-0.34, 0.09) & -0.13 (-0.34, 0.10) & -0.13 (-0.34, 0.10) & -0.13 (-0.35, 0.09)\\
\hline
Time & 17.22 & 18.95 & 20.36 & 21.87\\
\hline
    \end{tabular*}
    \label{tbl:ivt-sa-param}
\end{table}

\begin{figure*}[t!]
    \centering
    \captionsetup[subfigure]{justification=centering, font=footnotesize}
    \begin{subfigure}[b]{0.24\textwidth}
         \centering
         \includegraphics[width=\textwidth]{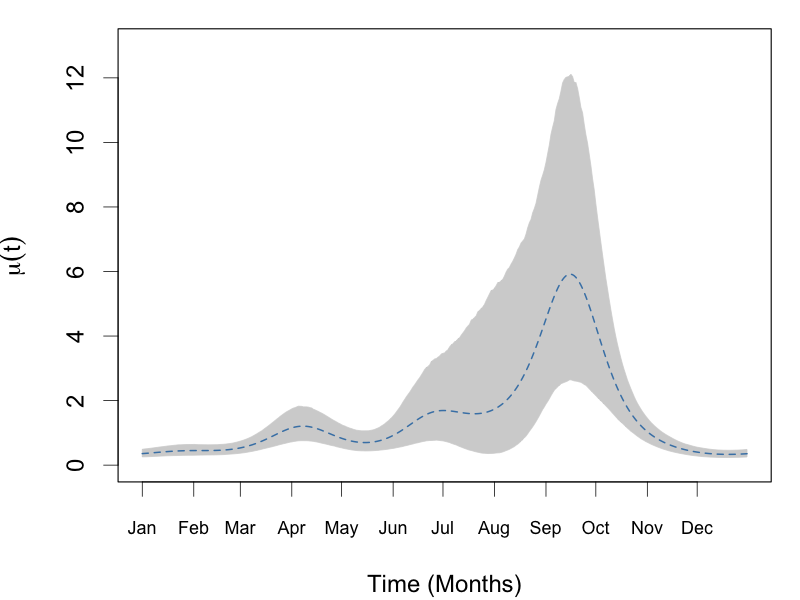}
         \caption{$L = 5$}
    \end{subfigure}    
    \begin{subfigure}[b]{0.24\textwidth}
         \centering
         \includegraphics[width=\textwidth]{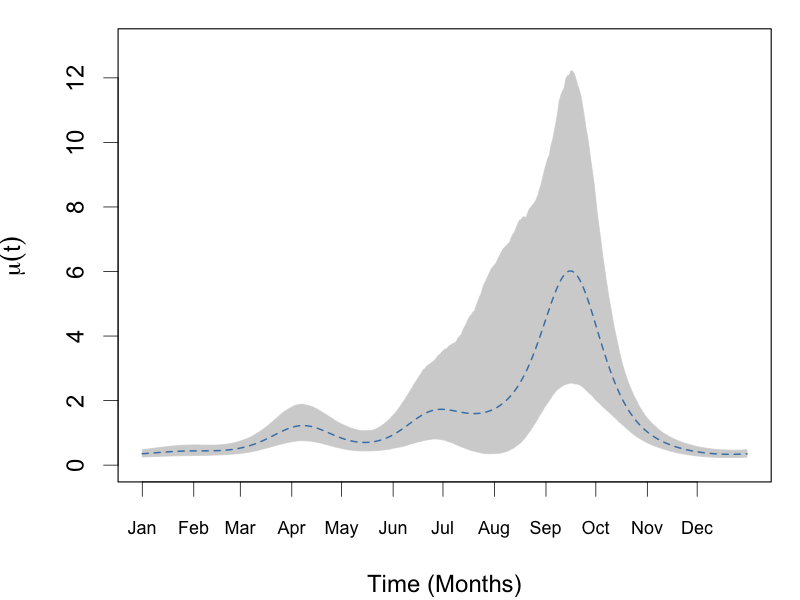}
         \caption{$L = 10$}
    \end{subfigure}    
    \begin{subfigure}[b]{0.24\textwidth}
         \centering
         \includegraphics[width=\textwidth]{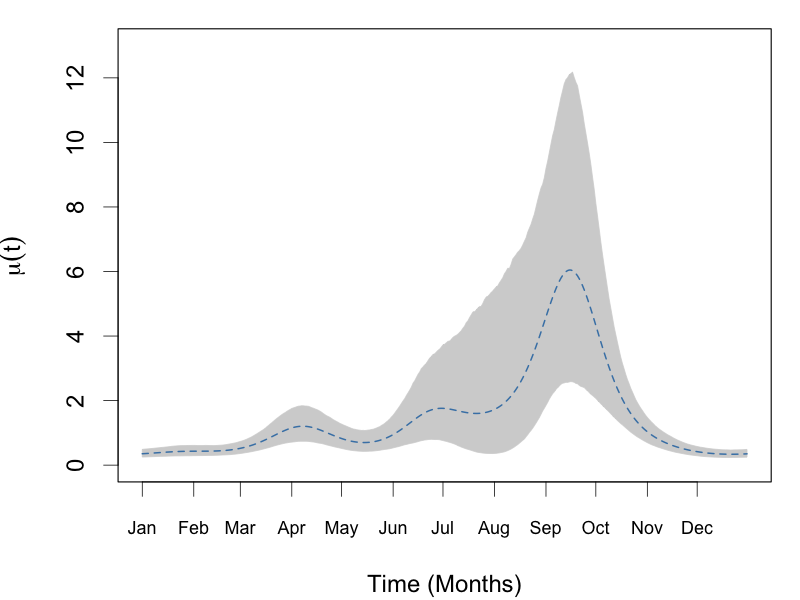}
         \caption{$L = 15$}
    \end{subfigure}
    \begin{subfigure}[b]{0.24\textwidth}
         \centering
         \includegraphics[width=\textwidth]{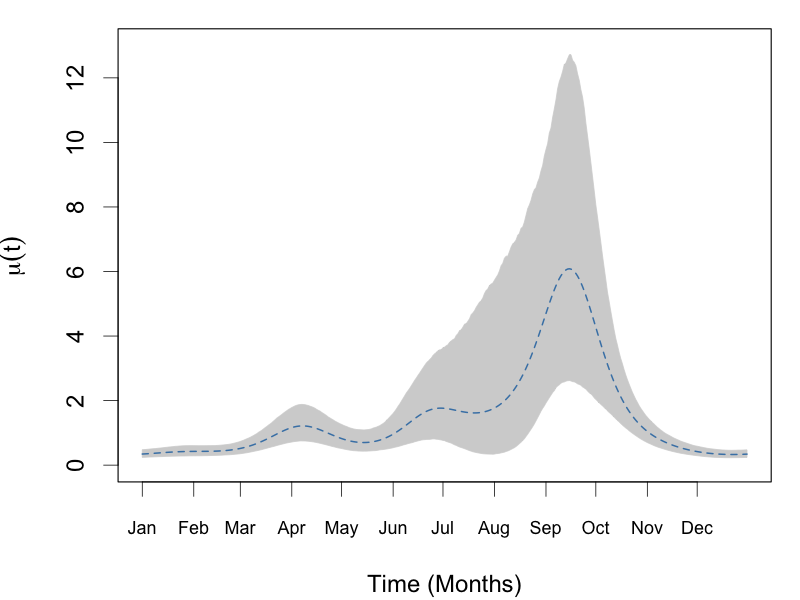}
         \caption{$L = 20$}
    \end{subfigure}     
    \caption{
    IVT recurrence interval analysis (sensitivity analysis of $L$): 
    harmonic function $\mu(t)$ (for an
    one-year window) of the MTDPP multiplicative model with different
    values of $L$.
    }
    \label{fig:ivt-L-sa-harmonic}
\end{figure*}

Figure \ref{fig:ivt-L-sa-weight}(c)-(f) shows 
the posterior mean and $95\%$ credible interval estimates
of the weights, 
under different values of $L$. We observed that, $L = 5$ seems not large
enough to work as an upper bound, but when $L$ ranged from $10$ to $20$,
the posterior estimates of the weights were quite consistent. 
Available in Table \ref{tbl:ivt-sa-param} are the posterior mean and 
$95\%$ credible interval estimates of the parameters 
$\{\alpha,\phi,\bbeta\}$, where $\bbeta =
(\beta_{11},\beta_{12},\beta_{21},\beta_{22},\dots,\beta_{51},\beta_{52})^\top$.
Overall, the estimates of $\{\alpha,\phi,\bbeta\}$ were quite robust
across different values of $L$. All the four models implied
the presence of annual and semiannual seasonality, and the posterior 
mean and pointwise $95\%$ credible interval estimates of the 
harmonic function $\mu(t)$ look similar across different $L$,
as shown in Figure \ref{fig:ivt-L-sa-harmonic}.
Overall, there were no discernible differences among the models
with $L$ between $10$ and $20$. Thus, we used $L = 15$
as the upper bound for the rest of the analyses for this particular
data example.

\subsubsection*{Comparison with the renewal process}

We also assessed model performance by comparison with a renewal process (RP)
model, which involves the simpler assumption of independent durations. 

The scaled-Lomax RP model is obtained by modeling the $z_i$ of the
multiplicative model in Section 4.2 of the main paper with an RP, such that
the $z_i$, $i = 1,\dots,n$, are independent and 
identically distributed as a scaled-Lomax distribution, 
$P(z\,|\,\alpha\phi,\alpha-1)$. Thus, the scaled-Lomax RP model 
corresponds to a simpler assumption of the scaled-Lomax MTDPP model 
in which the $z_i$ are identically distributed
as $P(z\,|\,\alpha\phi,\alpha-1)$, but are Markov dependent.
Our goal is to examine whether incorporating temporal dependence
in durations aligns better with the underlying data structure
and improves the prediction of future events.

We used the same prior specification for
both the scaled-Lomax MTDPP and scaled-Lomax RP models.
Specifically, the regression parameter vector was assigned
mean-zero dispersed normal priors, and the shape and scale parameters 
$\alpha$ and $\phi$ received $\mathrm{Ga}(6,1)\mathbbm{1}(\alpha>1)$
and $\mathrm{Ga}(1,1)$ priors, respectively.
Both models were fitted with MCMC, and we obtained
$5000$ posterior samples with appropriate burn-in and thinning.

To compare the goodness-of-fit of the two models, we used the
Bayesian information criterion (BIC; \citealtsm{schwarz1978estimating}),
deviance information criterion (DIC; \citealtsm{spiegelhalter2002bayesian}),
and negative log-likelihood (NLL).
Regarding predictive performance, we used the 
same criteria (i.e., MAD, RMSE, CRPS, IS) and same settings
(i.e., one-step-ahead in-sample and one-step-ahead out-of-sample
predictions) with Section \ref{sec: comp-acd}.

\begin{figure*}[t!]
    \centering
    \captionsetup[subfigure]{justification=centering, font=footnotesize}
    \begin{subfigure}[b]{0.35\textwidth}
         \centering
         \includegraphics[width=\textwidth]{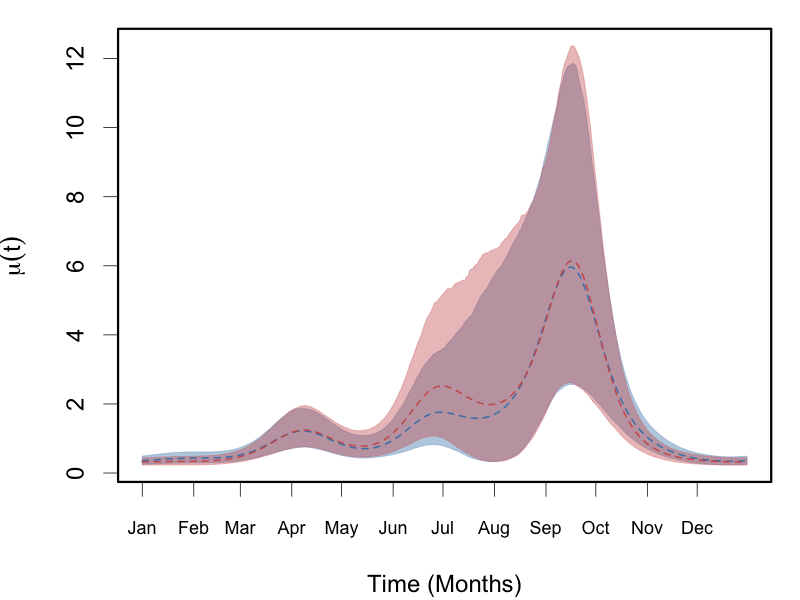}
         \caption{Harmonic functions}
    \end{subfigure}    
    \hspace{5pt}
    \begin{subfigure}[b]{0.35\textwidth}
         \centering
         \includegraphics[width=\textwidth]{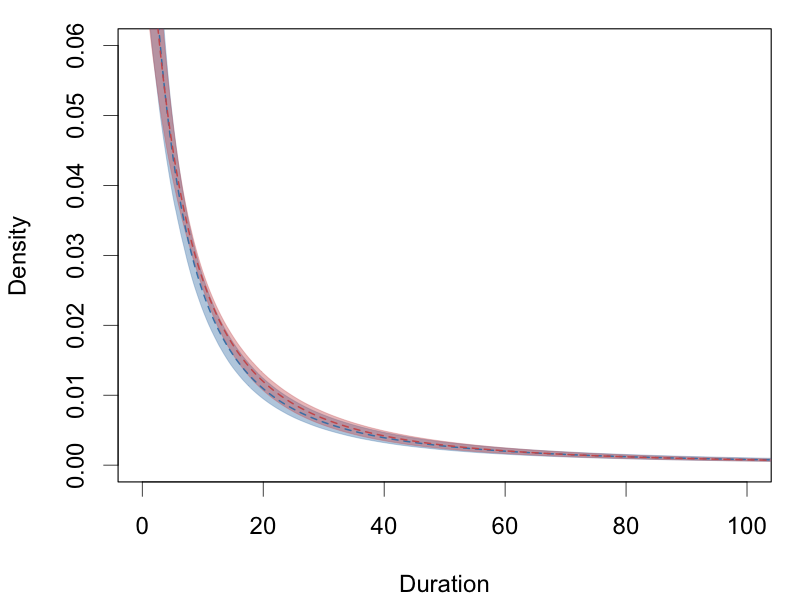}
         \caption{Stationary marginal estimates}
    \end{subfigure}      
    \caption{IVT recurrence interval analysis:
    comparison of scaled-Lomax MTDPP (blue) and scaled-Lomax RP (red)
    models regarding posterior estimates of the harmonic functions
    and the stationary marginal densities.
    Dashed lines and polygons correspond to posterior mean
    and pointwise $95\%$ credible interval estimates, respectively.
    }
    \label{fig:ivt-comp-RP-functional}
\end{figure*}

\begin{table}[t!]
\renewcommand{\arraystretch}{1.25}
\captionsetup{font=footnotesize}
\caption{IVT recurrence interval analysis:
    comparison of scaled-Lomax MTDPP and scaled-Lomax RP 
models regarding parameter estimates and goodness-of-fit.}
\small
    \centering
    \begin{tabular*}{\hsize}{@{\extracolsep{\fill}}lccccc}
    % \\[-5pt]
\hline
  & $\alpha$ & $\phi$ & DIC & BIC & NLL\\
\hline
MTDPP & 2.01 (1.72, 2.35) & 4.92 (3.35, 6.92) & 5981 & 6221 & 3022\\
\hline
RP & 2.26 (2.06, 2.51) & 6.07 (4.50, 7.98) & 6284 & 6339 & 3130\\
\hline
    \end{tabular*}
    \label{tbl:ivt-comp-est}
\end{table}

\begin{table}[t!]
\renewcommand{\arraystretch}{1.25}
\captionsetup{font=footnotesize}
\caption{IVT recurrence interval analysis:
    comparison of scaled-Lomax MTDPP and scaled-Lomax RP 
models regarding one-step-ahead in-sample and one-step-ahead out-of-sample predictions.}
\small
    \centering
    \begin{tabular*}{\hsize}{@{\extracolsep{\fill}}lcccccccccc}
    \hline
\multicolumn{1}{l}{} & 
\multicolumn{4}{c}{In-sample prediction} & \multicolumn{1}{c}{} & 
\multicolumn{4}{c}{Out-of-sample prediction}\\
\cline{2-5} \cline{7-10} 
\multicolumn{1}{l}{} & \multicolumn{1}{c}{MAD} & \multicolumn{1}{c}{RMSE} & \multicolumn{1}{c}{CRPS} & \multicolumn{1}{c}{IS} & \multicolumn{1}{c}{} &
\multicolumn{1}{c}{MAD} & \multicolumn{1}{c}{RMSE} & \multicolumn{1}{c}{CRPS} & \multicolumn{1}{c}{IS}\\
\hline
MTDPP & 14.40 & 39.91 & 13.33 & 202.30 & 
         & 22.08 & 59.45 & 11.87 & 254.96 \\
RP & 17.87 & 179.71 & 13.28 & 217.41 & 
         & 30.16 & 123.52 & 13.46 & 414.43 \\
\hline
    \end{tabular*}
    \label{tbl:ivt-comp-pred}
\end{table}

The results in Table \ref{tbl:ivt-comp-est} suggest that 
the RP model does not fit the data as well as the MTDPP model, 
which has smaller values of DIC, BIC, and NLL. Also available in 
Table \ref{tbl:ivt-comp-est} are posterior estimates
of the parameters for the stationary marginal distribution.
The MTDPP model suggests a heavier tail, 
after adjusting for seasonality. Figure \ref{fig:ivt-comp-RP-functional}
shows the tails of the marginal densities estimated by the 
two models, as well as the posterior estimates of the 
harmonic function, where the two models agree on most parts.
Turning to predictive performance (Table \ref{tbl:ivt-comp-pred}), the MTDPP 
model not only results in more accurate point predictions, as indicated by 
much lower RMSEs, but also provides more accurate and tighter prediction 
intervals, especially when it comes to out-of-sample predictions.

Overall, the comparison demonstrates the benefit of incorporating duration 
dependence. The scaled-Lomax MTDPP yields better goodness-of-fit, and it improves 
the prediction of future events for the particular data example.

\subsection{Mid-price changes of the AUD/USD exchange rate}

\begin{figure*}[t!]
    \centering
    \captionsetup[subfigure]{justification=centering, font=footnotesize}
    \begin{subfigure}[b]{0.3\textwidth}
         \centering
         \includegraphics[width=\textwidth]{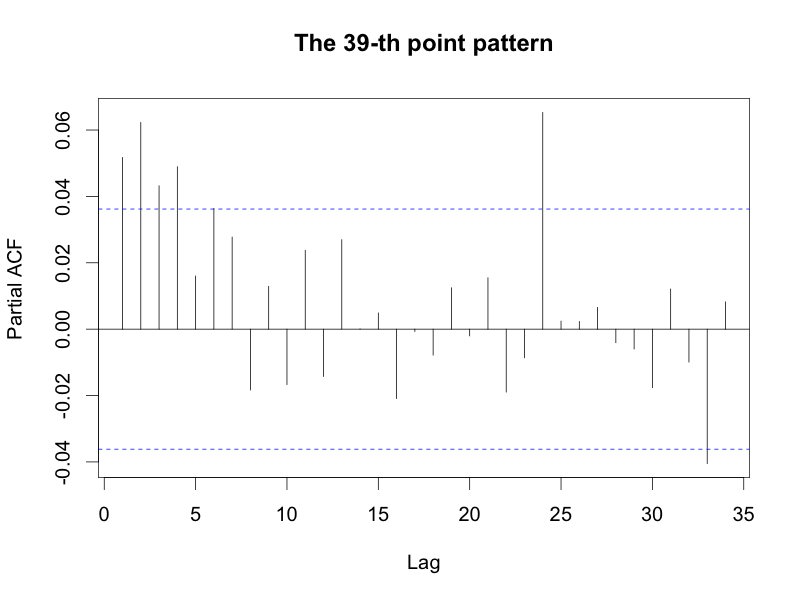}
         \caption{PACF}
    \end{subfigure}
    \begin{subfigure}[b]{0.3\textwidth}
         \centering
         \includegraphics[width=\textwidth]{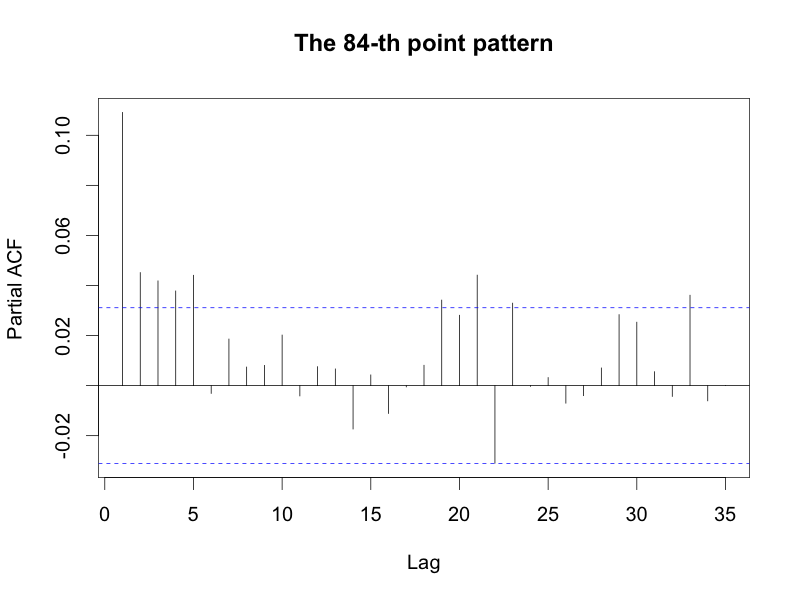}
         \caption{PACF}
    \end{subfigure}
    \begin{subfigure}[b]{0.3\textwidth}
         \centering
         \includegraphics[width=\textwidth]{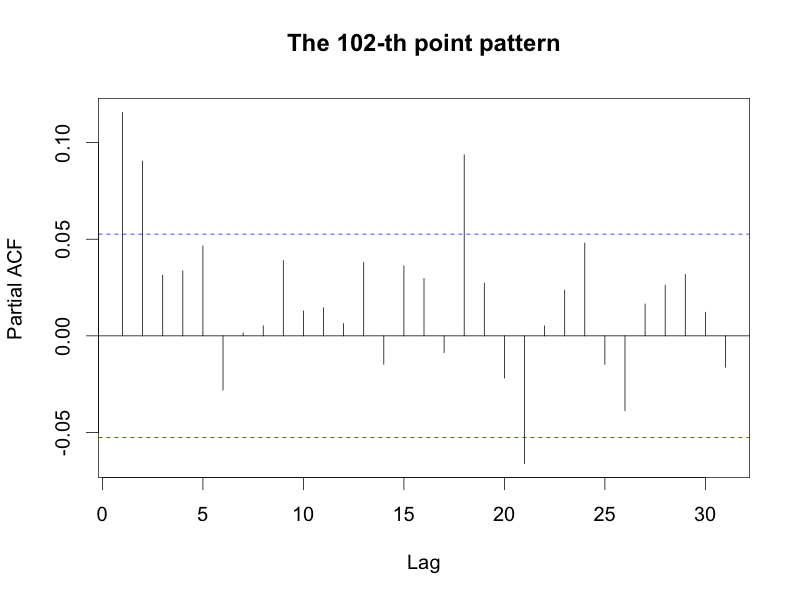}
         \caption{PACF}
    \end{subfigure} \\
    \vspace{5pt}
    \begin{subfigure}[b]{0.3\textwidth}
         \centering
         \includegraphics[width=\textwidth]{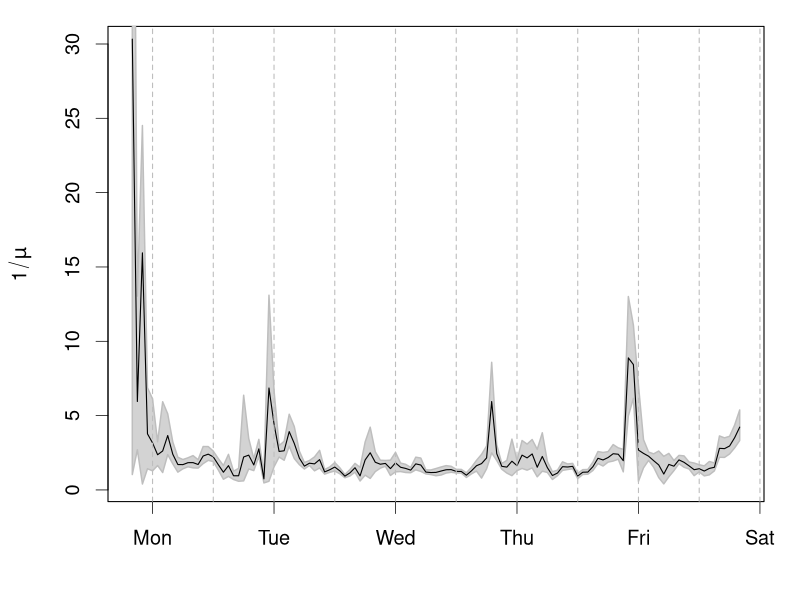}
         \caption{Exponential distribution waiting time $1/\mu$}
    \end{subfigure}
    \begin{subfigure}[b]{0.3\textwidth}
         \centering
         \includegraphics[width=\textwidth]{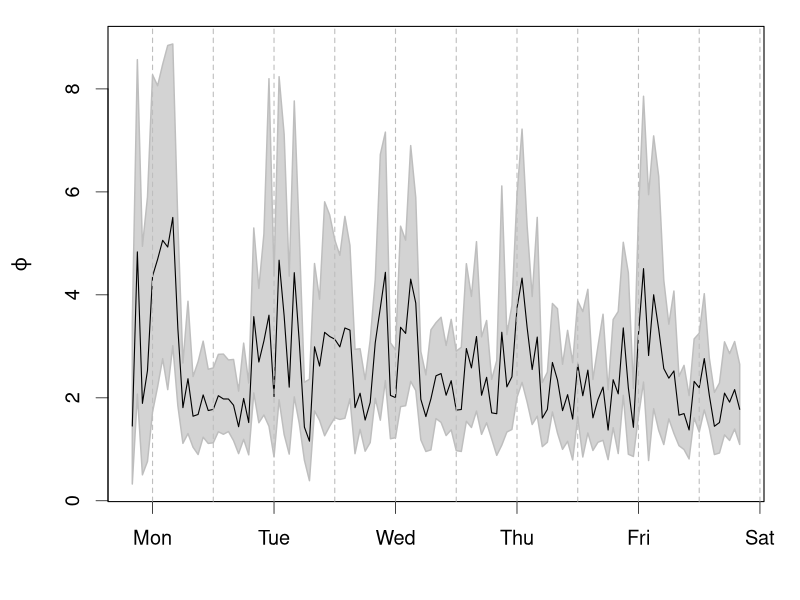}
         \caption{Lomax MTDPP scale parameter $\phi$}
    \end{subfigure}
    \begin{subfigure}[b]{0.3\textwidth}
         \centering
         \includegraphics[width=\textwidth]{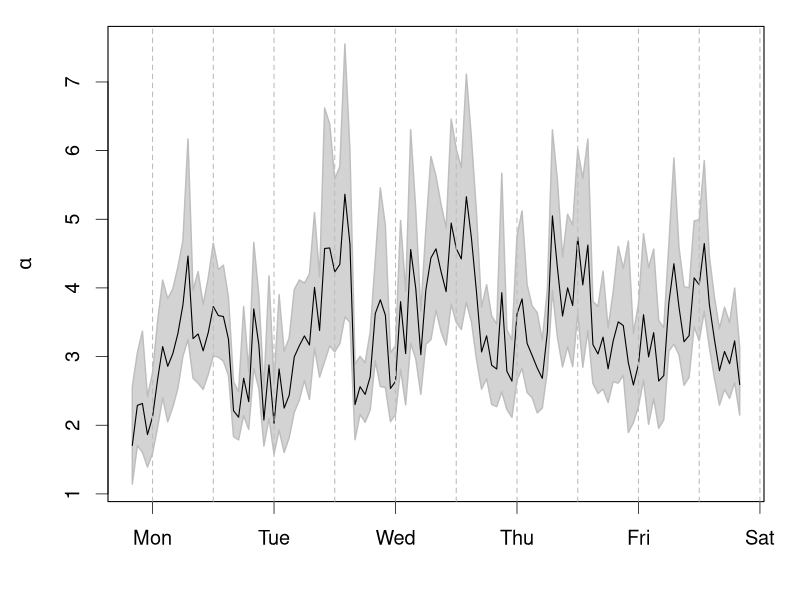}
         \caption{Lomax MTDPP shape parameter $\alpha$}
    \end{subfigure}    
    \caption{AUD/USD exchange rate data analysis.
    The top row shows the PACFs for three point patterns.
    The bottom row plots the 
    time series of the posterior mean (solid lines) and 
    pointwise $95\%$ credible interval (grey polygons) estimates for parameters
    $1/\mu$, $\phi$, and $\alpha$, based on the MTDCPP model.
    Vertical dashed lines correspond to midnight and midday GMT.}
    \label{fig: fx-pacf-params}
\end{figure*}

There are $121$ point patterns, each of which corresponds to 
an one-hour time window during the trading week between July 19 and 
July 24 in 2015.
Before fitting models, we examined the PACF of the durations for each
of the $121$ point patterns. Overall, the PACFs first cut off after one of 
the first five lags. Figure \ref{fig: fx-pacf-params}(a)-(c) shows
the PACFs of three point patterns.

Figure \ref{fig: fx-pacf-params}(d)-(f) 
illustrates the time series of posterior mean and interval estimates 
for three parameters: exponential distribution parameter $1/\mu$, Lomax MTDPP scale 
and shape parameters $\phi$ and $\alpha$.
Note that the exponential distribution and Lomax MTDPP are regarded as drivers of 
external and internal factors for waiting times between successive mid-price 
changes, respectively.
The estimates of the mean waiting time $1/\mu$ for external factors shows 
obvious diurnal pattern, with peaks and troughs appearing around midnight 
and midday GMT, respectively. 
The posterior estimates of $\phi$ for all point patterns seem more volatile,
with relatively high and low values occurring at 
midnight and midday GMT, whereas the posterior estimates of $\alpha$ reflect
an opposite pattern. The mean of the stationary marginal distribution of the 
Lomax MTDPP is $\phi/(\alpha-2)$, provided that $\alpha > 2$. Thus, given the 
patterns of estimated $\phi$ and $\alpha$, the estimates of the 
mean waiting time for internal factors appear high and low around midnight
and midday GMT, respectively. In addition, small values of $\alpha$ 
around midnight GMT suggest a heavy-tailed duration distribution of the Lomax MTDPP 
during that period. This indicates that mid-price changes tend to cluster around 
midnight GMT, which corresponds to the opening time of Asian markets (23:00-1:00 GMT).

\subsubsection*{Comparison with the RHawkes process models}
\label{sec: comp-rhp}

We compare the MTDCPP model and the RHawkes process model (RHP) 
of \cite{wheatley2016hawkes} 
for the foreign exchange data regarding
their computation time and out-of-sample predictive performance.
The RHP model
was fitted by \cite{chen2018direct} to the same data to study 
market endogeneity.

Let $M_i$ denote the unobservable event type indicator of the $i$th event,
such that $M_i = 0$ if event $i$ is an immigration and $M_i = 1$ otherwise.
Then $I(t) = \max\{i: t_i < t, M_i = 0\}$ is the index 
of the most recent immigration event time before time $t$.
The conditional intensity of the model is given by
\begin{equation}\label{eq: rhawkes}
\begin{aligned}
  \la^*(t) = \mu(t - t_{I(t)}) + \sum_{j:t_j<t}\eta h(t - t_j),
\end{aligned}
\end{equation}
where $\mu(\cdot)$ is interpreted as the hazard function of the i.i.d. waiting
times between the immigration events, which form a renewal process; 
$\eta$ is the branching ratio parameter; and $h(\cdot)$ is the offspring density.
Following \cite{chen2018direct}, we take 
$\mu(\omega) = \kappa\omega^{\kappa - 1}/\beta^{\kappa}$, $\omega\geq 0$,
which corresponds to a Weibull distribution with shape parameter 
$\kappa$ and scale parameter $\beta$, with density function 
$g(\omega) = (\kappa/\beta)(\omega/\beta)^{\kappa-1}\exp(-(\omega/\beta)^\kappa)$.
The offspring density $h(\cdot)$ corresponds to an exponential distribution.

Using the 121 point patterns, we compared the predictive performance of the 
MTDCPP with the RHP model, based on one-step-ahead out-of-sample prediction. 
In particular, 
let $0 = t_0^{(k)} < t_1^{(k)} < t_2^{(k)} < \dots < t_{n_k}^{(k)} < T$
be the $k$th observed point pattern, with durations 
$x_i^{(k)} = t_i^{(k)}-t_{i-1}^{(k)}$, $i = 1,\dots, n_k$, 
for $k = 1, \dots, 121$. 
For the $k$th point pattern, $k = 1, \dots, 120$, we 
fitted models and then generated predictions 
of $x^{(k)}_{n_k+1}$, and 
compare the predictions with the observed $x^{(k)}_{n_k+1}$ calculated as
$T - t^{(k)}_{n_k} + t_1^{(k+1)}$, where $T$ is the end time of the
$k$th observation window. In other words, $x^{(k)}_{n_k+1}$ is the duration between the 
last event time $t^{(k)}_{n_k}$ in the $k$th one-hour window and the first event time 
$t_1^{(k+1)}$in the $(k+1)$th one-hour window. 
Predictive performance was measured by MAD, RMSPE, and CRPS.

Both the MTDCPP and RHP models were fitted in R
on a Linux server with 512 GB of RAM and two Intel Xeon Gold 6348 processors.
For each point pattern, we obtained 10000 posterior samples for the MTDCPP, from 
155000 MCMC iterations, with 5000 samples as burn-in and retaining samples
every 15 iterations. We fitted the RHP model using the \textit{optim} 
function in R, with the negative log-likelihood function available from the
\textbf{RHawkes} package \citep{RHawkes}; all steps followed
the code available in the supplementary material of \cite{chen2018direct}.
The MAD, RMSPE, and CRPS from the MTDCPP model were 0.59, 4.57, and 0.98,
respectively. All the metrics are smaller than those (1.41, 7.22, 1.47) from the RHP model, indicating that the MTDCPP model had better predictive performance
than the RHP model.

\begin{figure*}[t!]
    \centering
    \captionsetup[subfigure]{justification=centering, font=footnotesize}
    \begin{subfigure}[b]{0.9\textwidth}
         \centering
         \includegraphics[width=\textwidth]{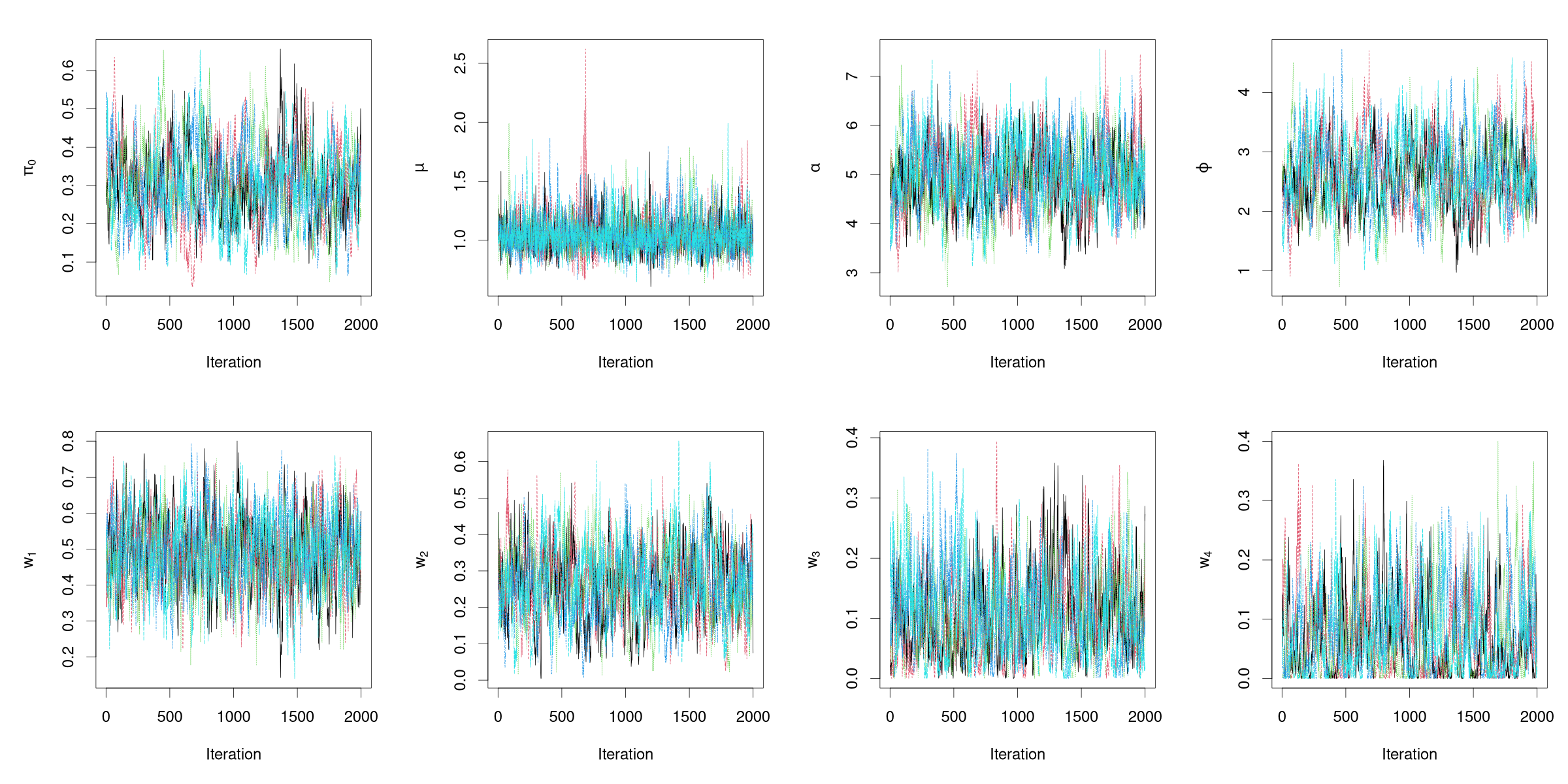}
    \end{subfigure}
    \caption{MCMC convergence diagnostics for Section \ref{sec: comp-rhp}: 
    trace plots in each panel correspond to 5 independent chains of a parameter.}
    \label{fig:comp-rhp}
\end{figure*}

We used the largest point pattern (3961 event times) to 
compare the computation times of the MTDPP and the RHP models.
Similar to Section \ref{sec: comp-acd}, before comparison, 
we assessed MCMC convergence of the MTDCPP by computing the potential scale 
reduction factor $\hat R$ and effective sample size $\hat{n}_{\text{eff}}$.
Specifically, we ran 5 independent 
Markov chains, each with 2000 posterior samples obtained 
from a total of 35000 iterations, discarding the first 5000 as 
burn-in samples and retaining samples every $15$ iterations.
Table \ref{tbl:comp-rhp} shows $\hat R$,
$\hat{n}_{\text{eff}}$, and $\hat{n}_{\text{eff}}$ per second,
computed using the R package \textbf{coda}.
The factors $\hat R$ near 1 indicate convergence of the 
chains. The effective sample sizes $\hat{n}_{\text{eff}} > 100$ 
\citep[][Chapter 11.5]{gelman2013bayesian},
along with the model comparison results regarding predictive performance,
suggest adequacy of the effective sample sizes.
Figure \ref{fig:comp-rhp} shows trace plots of the five independent
chains for each parameter listed in Table \ref{tbl:comp-rhp}.

\begin{table}[t!]
\renewcommand{\arraystretch}{1.25}
\captionsetup{font=footnotesize}
\caption{MCMC diagnostics of the MTDCPP and effective sample sizes.}
\small
    \centering
    \begin{tabular*}{\hsize}{@{\extracolsep{\fill}}lcccccccccc}
\hline
  & $\pi_0$ & $\mu$ & $\alpha$ & $\phi$ & $w_1$ & $w_2$ & $w_3$ & $w_4$\\
\hline
$\hat{R}$ & 1.01 & 1.01 & 1.00 & 1.01 & 1.01 & 1.01 & 1.02 & 1.03\\
\hline
$\hat{n}_{\text{eff}}$ & 341.76 & 2167.84 & 370.37 & 353.50 & 739.08 & 535.31 & 575.54 & 402.72\\
\hline
$\hat{n}_{\text{eff}}$ per second & 0.30 & 1.88 & 0.32 & 0.31 & 0.64 & 0.46 & 0.50 & 0.35\\
\hline
    \end{tabular*}
    \label{tbl:comp-rhp}
\end{table}

Thus, 10000 posterior samples from 
a total of $155000$ MCMC iterations are sufficient for this point pattern.
Fitting the MTDCPP model to the data 
took around $17$ minutes to complete $155000$ iterations, corresponding to 
a median of $0.6$ independent samples per second. On the other hand,
it took around $32$ minutes to complete a single fit of the RHP model,
since the time complexity of the algorithm to 
compute the point process likelihood is $O(n^2)$ \citep{chen2018direct}.

\section{MCMC diagnostics}

\begin{figure*}[t!]
    \centering
    \begin{subfigure}{0.98\textwidth}
         \centering
         \includegraphics[width=\textwidth]{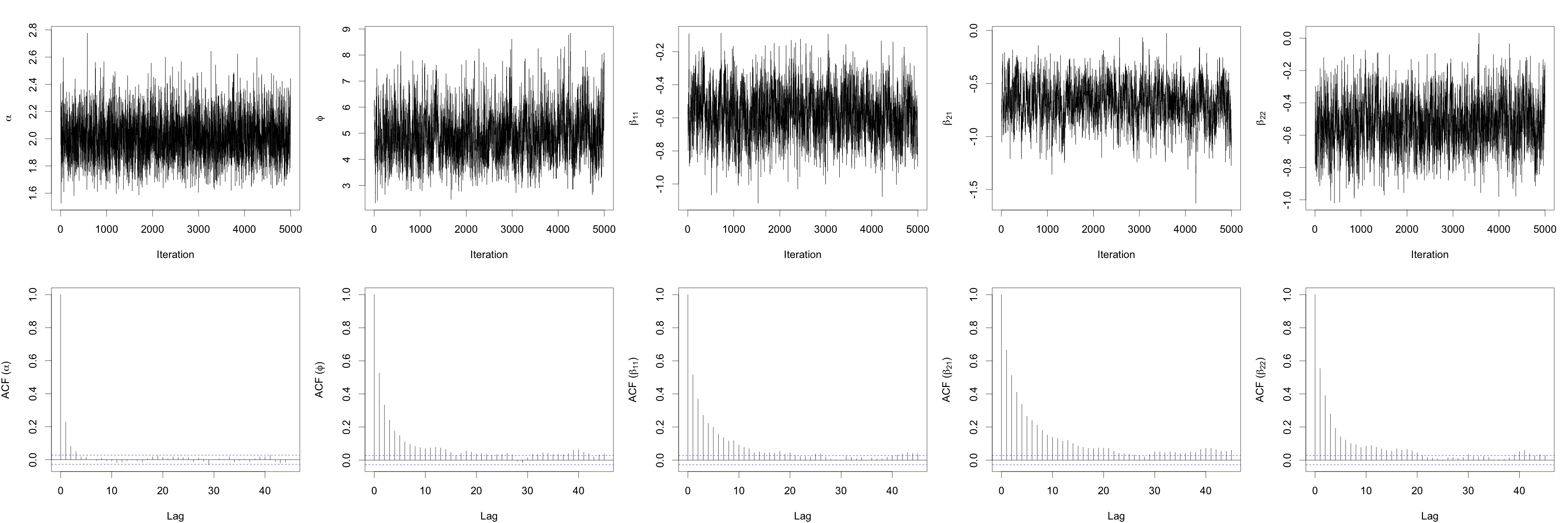}
         \caption{}
    \end{subfigure}\\
    \vspace{8pt}
    \begin{subfigure}{0.98\textwidth}
         \centering
         \includegraphics[width=\textwidth]{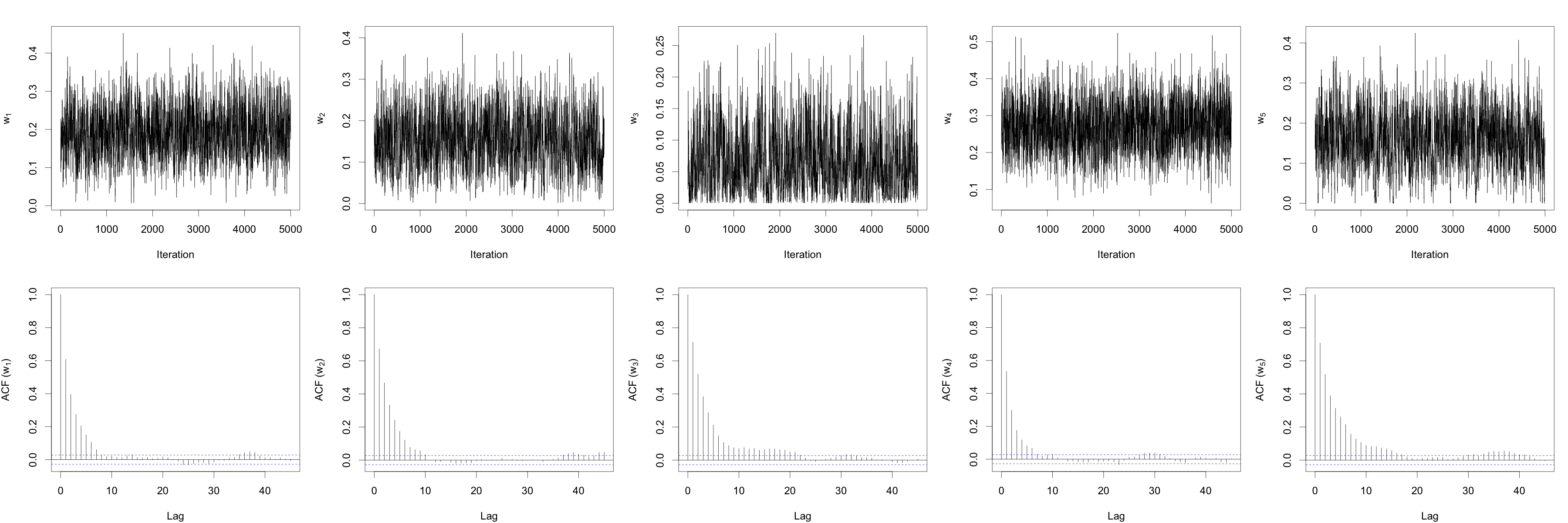}
         \caption{}
    \end{subfigure}\\
    \vspace{5pt}
    \caption{
    MCMC convergence diagnostics for Section 4.2 of the main paper: 
    trace plots and ACFs of the posterior samples
    of the parameters of the scaled-Lomax MTDPP.
    In Panel (a), columns from left to right correspond
    to $\alpha$, $\phi$, $\beta_{11}$, $\beta_{21}$, $\beta_{22}$,
    the last three of which are statistically significant
    coefficients.
    In Panel (b), columns from left to right correspond to 
    weights $w_1$, $w_2$, $w_3$, $w_4$, and $w_5$.
    }
    \label{fig:mtdpp-ivt-mcmc}
\end{figure*}

\begin{figure*}[t!]
    \centering
    \begin{subfigure}{0.98\textwidth}
         \centering
         \includegraphics[width=\textwidth]{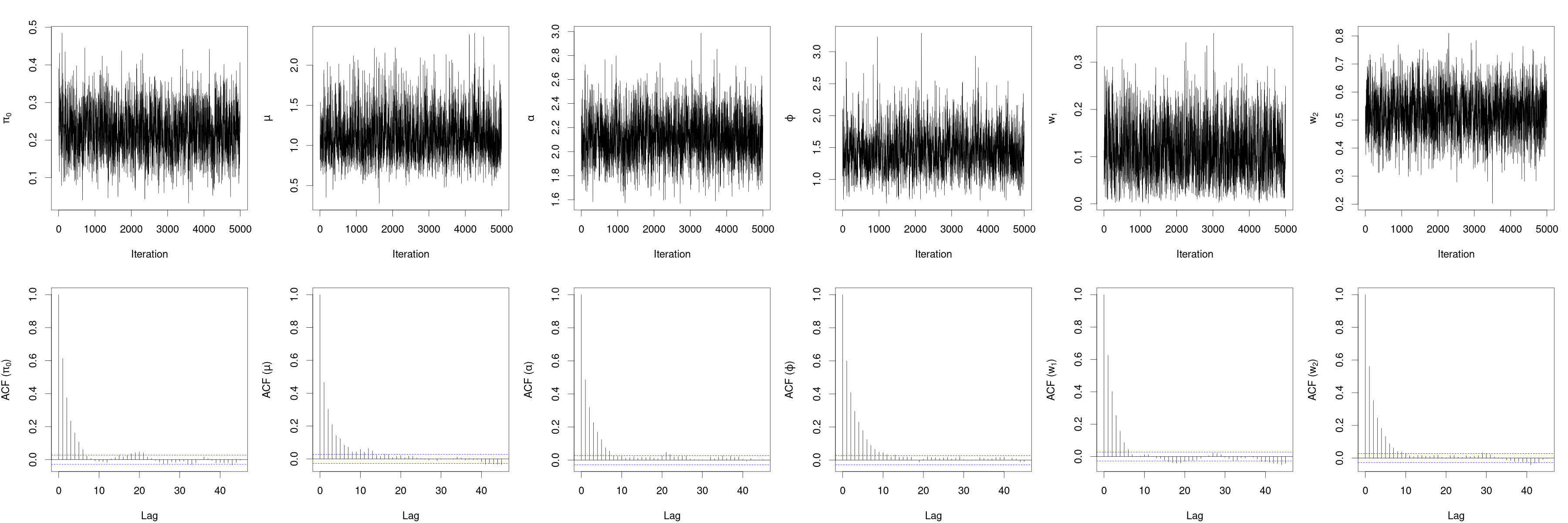}
         \caption{}
    \end{subfigure}\\
    \vspace{5pt}
    \begin{subfigure}{0.98\textwidth}
         \centering
         \includegraphics[width=\textwidth]{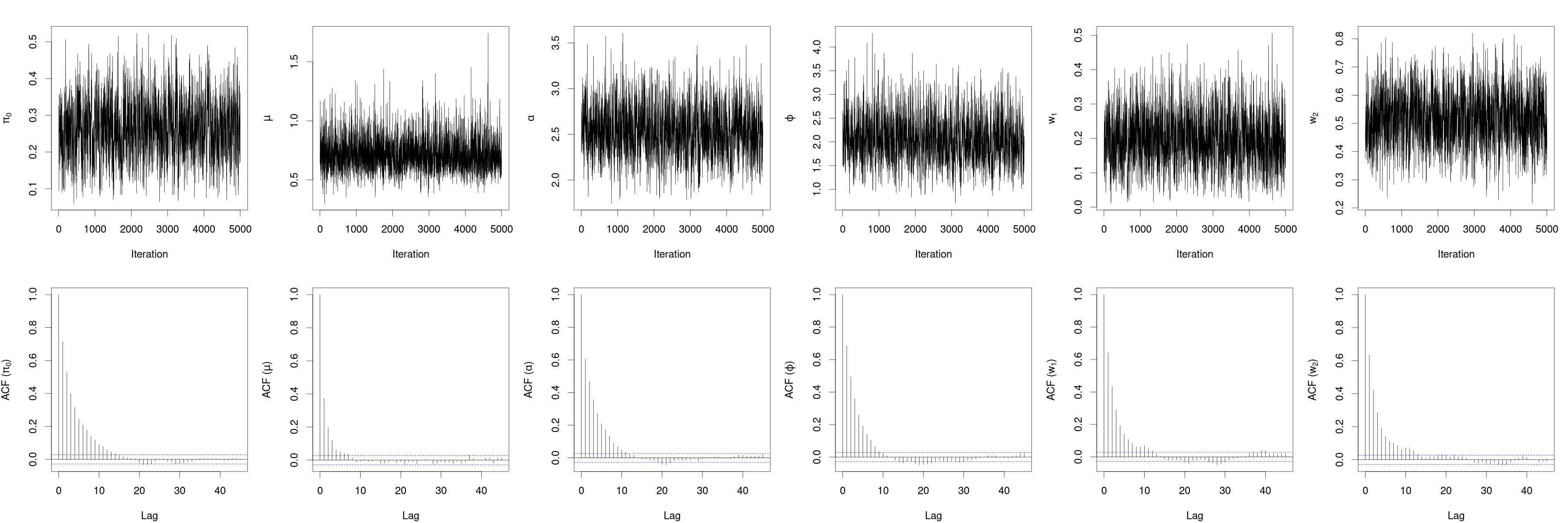}
         \caption{}
    \end{subfigure}\\
    \vspace{5pt}
    \begin{subfigure}{0.98\textwidth}
         \centering
         \includegraphics[width=\textwidth]{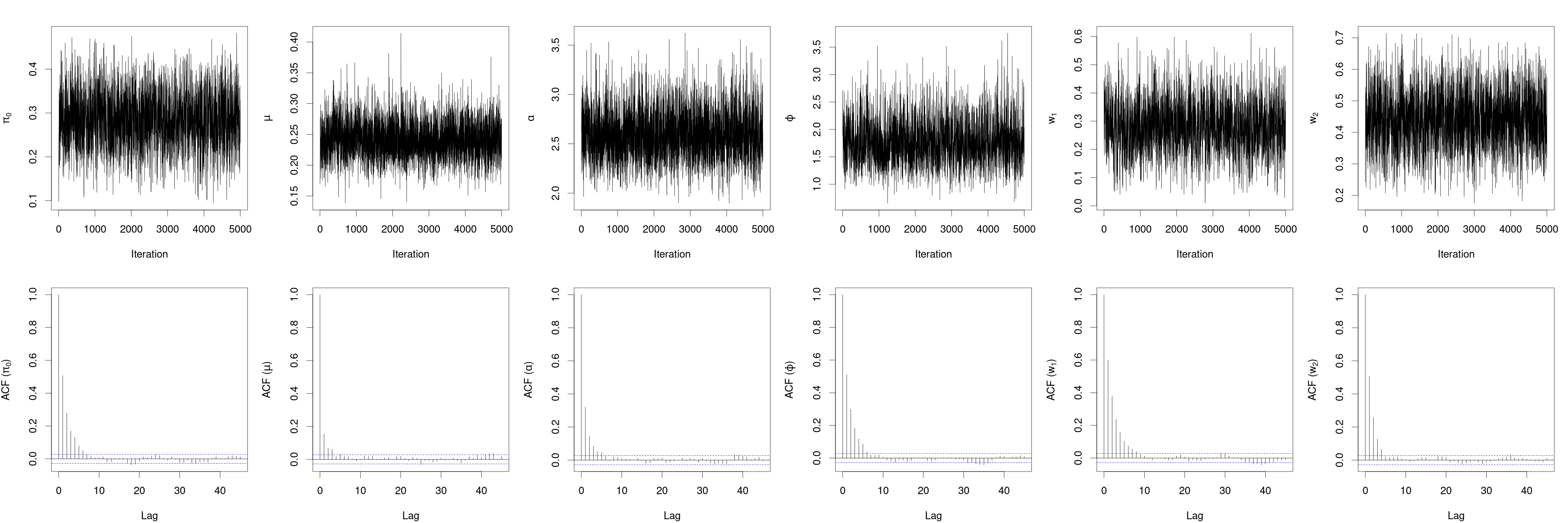}
         \caption{}
    \end{subfigure}\\
    \vspace{5pt}
    \caption{
    MCMC convergence diagnostics for Section 4.3 of the main paper: 
    trace plots and ACFs of the posterior samples
    of the parameters of the Lomax MTDCPP fitted to 
    three point patterns, each of which corresponds
    to a panel. In each panel, columns from left to right correspond 
    to $\pi_0$, $\mu$, $\alpha$, $\phi$, $w_1$, and $w_2$.
    }
    \label{fig:mtdcpp-fx-mcmc}
\end{figure*}

\begin{figure*}[t!]
    \centering
    \begin{subfigure}{0.98\textwidth}
         \centering
         \includegraphics[width=\textwidth]
         {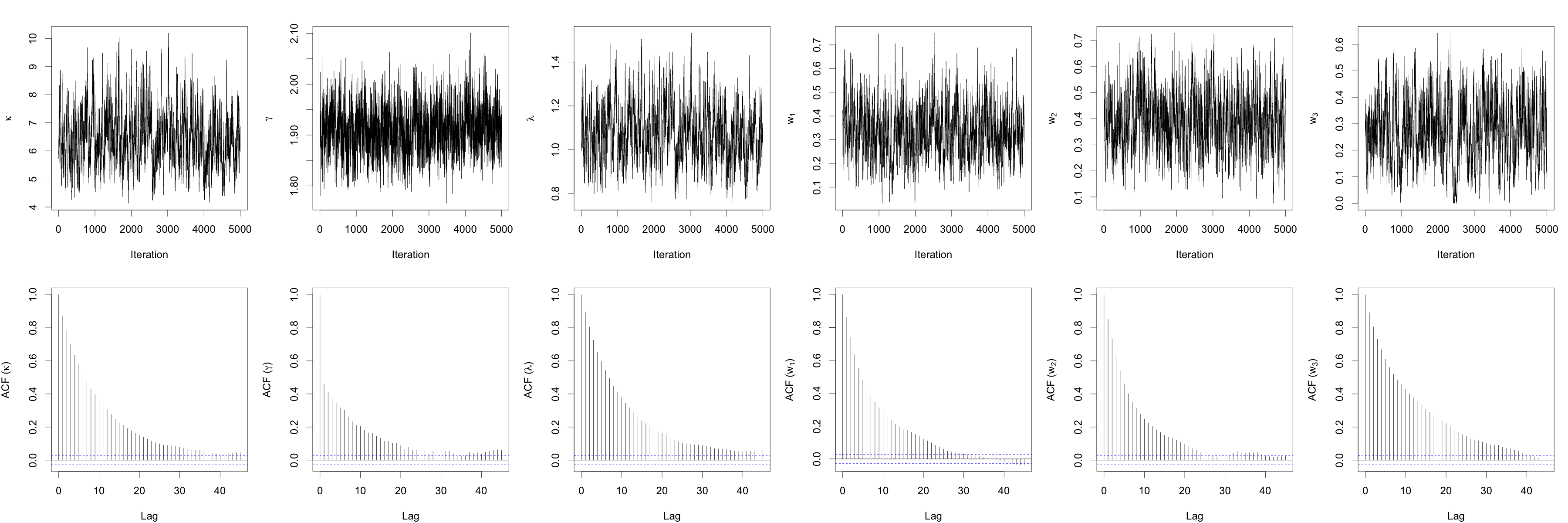}
         \caption{}
    \end{subfigure}\\
    \vspace{5pt}
    \begin{subfigure}{0.98\textwidth}
         \centering
         \includegraphics[width=\textwidth]
         {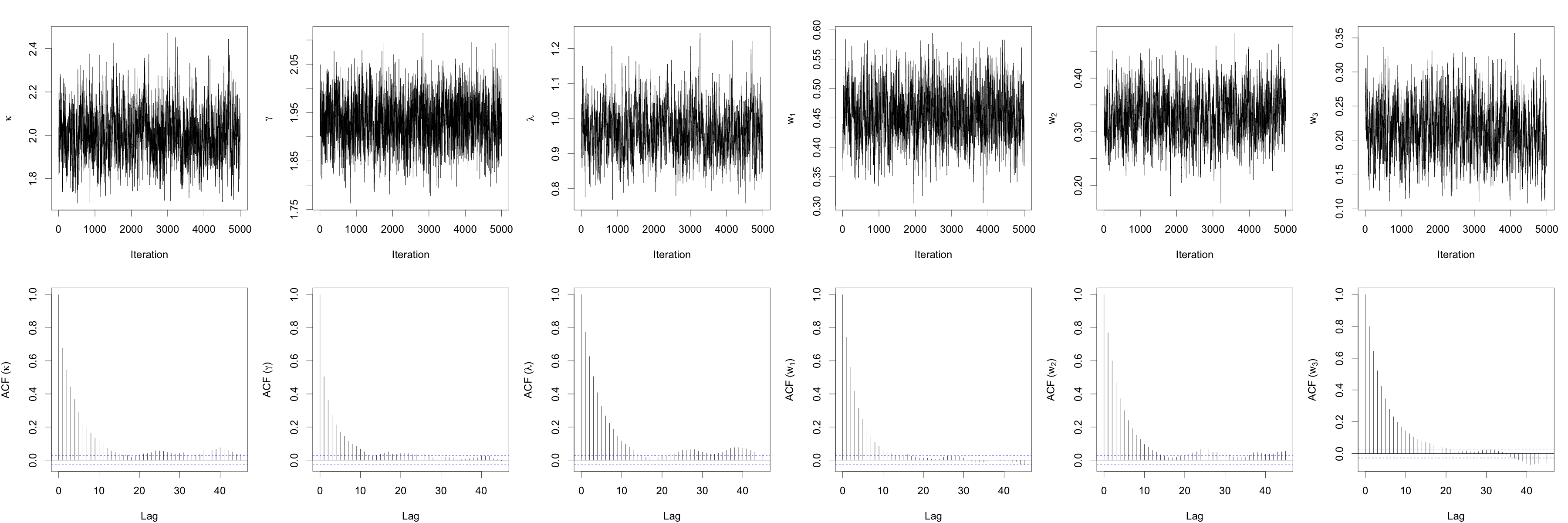}
         \caption{}
    \end{subfigure}\\
    \vspace{5pt}
    \begin{subfigure}{0.98\textwidth}
         \centering
         \includegraphics[width=\textwidth]
         {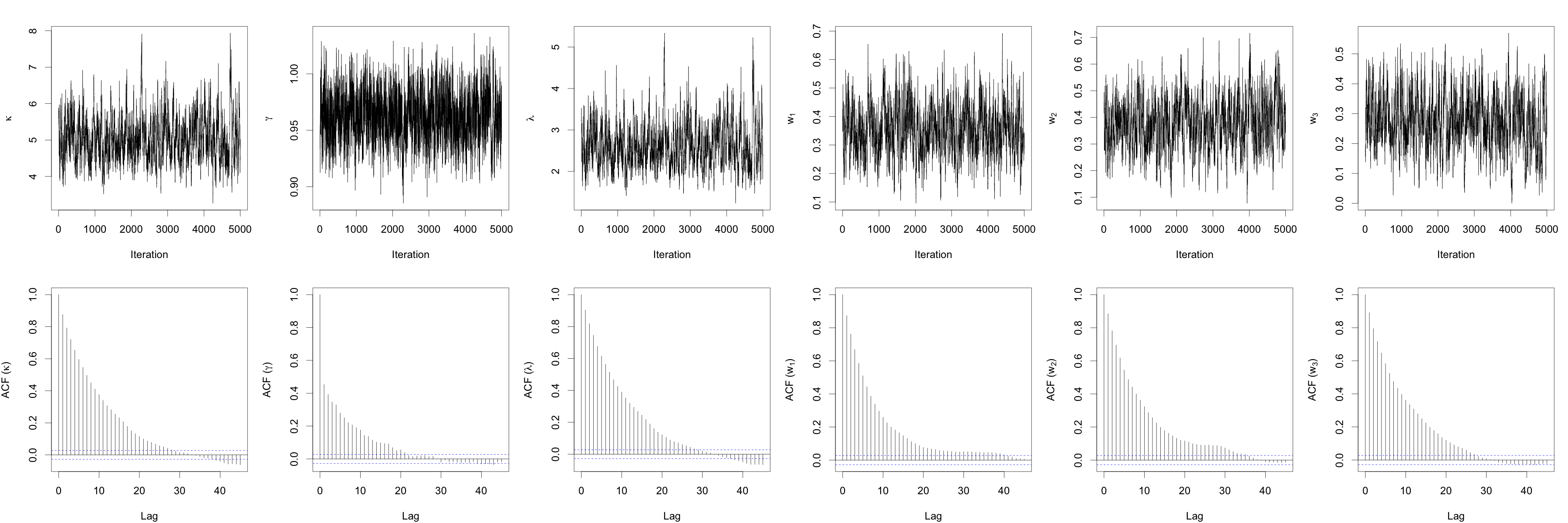}
         \caption{}
    \end{subfigure}\\
    \vspace{5pt}
    \caption{
    MCMC convergence diagnostics for Section 4.1 of the main paper: 
    trace plots and ACFs of the posterior samples
    of the parameters of the Burr MTDPP fitted to data simulated by a 
    a Burr MTDPP(a), a log-logistic MTDPP (b), and a scaled-Lomax MTDPP (c). Columns 1-3 correspond to 
    parameters $\kappa$, $\gamma$, and $\lambda$,
    and Columns 4-6 correspond to weights
    $w_1$, $w_2$, and $w_3$.
    }
    \label{fig:mtdpp-sim1-mcmc}
\end{figure*}

\begin{figure*}[t!]
    \centering
    \begin{subfigure}{0.98\textwidth}
         \centering
         \includegraphics[width=\textwidth]{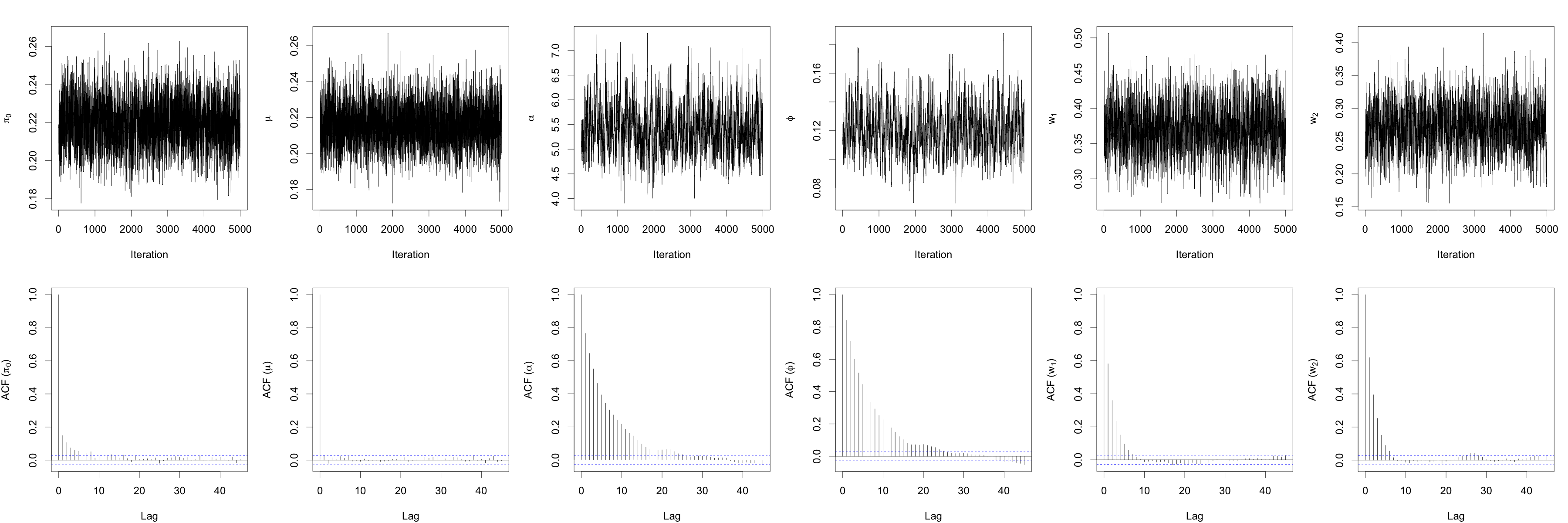}
         % \caption{}
    \end{subfigure}
    \begin{subfigure}{0.98\textwidth}
         \centering
         \includegraphics[width=\textwidth]{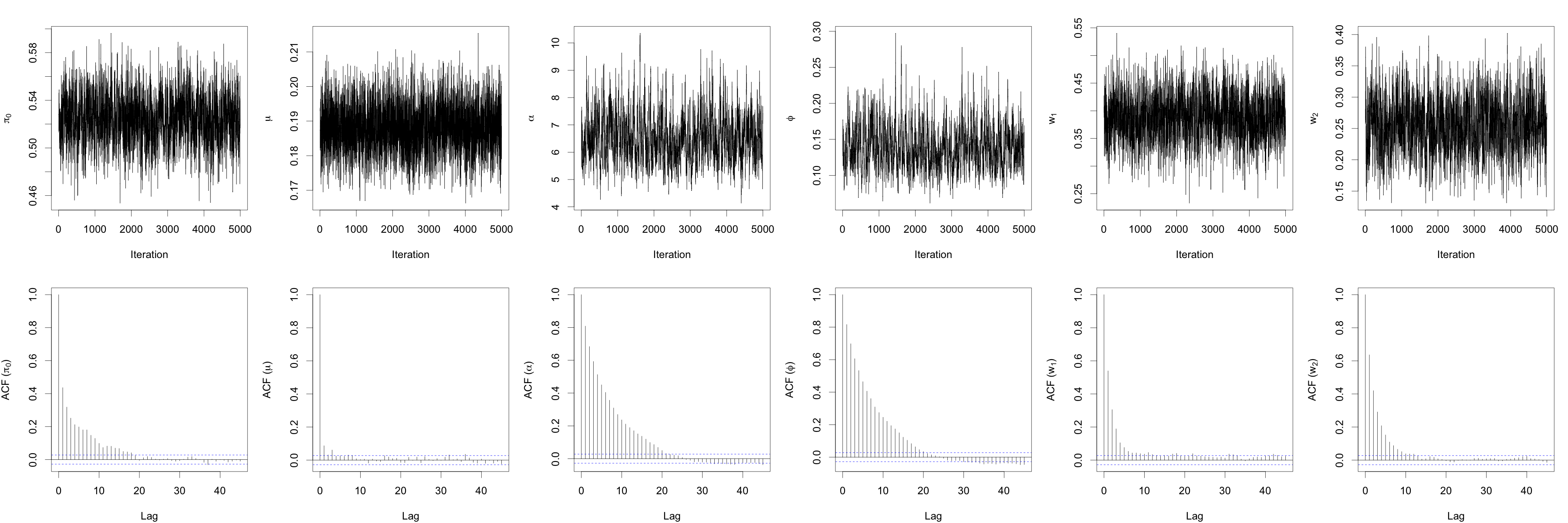}
         % \caption{}
    \end{subfigure}
    \begin{subfigure}{0.98\textwidth}
         \centering
         \includegraphics[width=\textwidth]{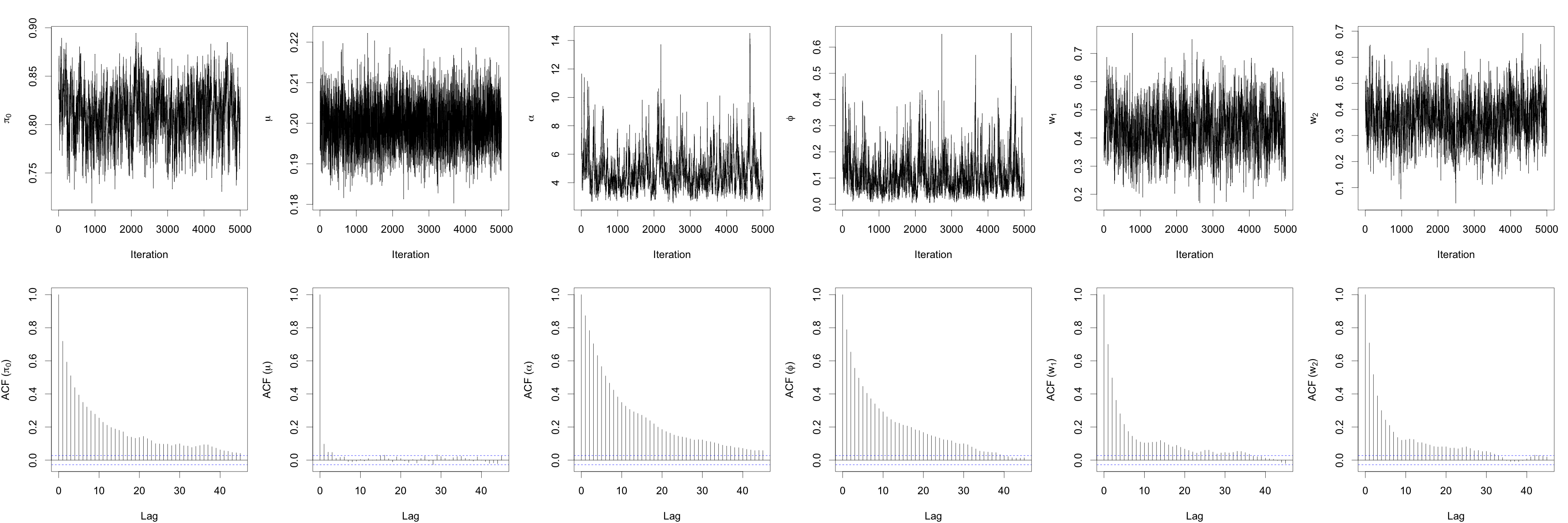}
         % \caption{}
    \end{subfigure}
    \begin{subfigure}{0.98\textwidth}
         \centering
         \includegraphics[width=\textwidth]{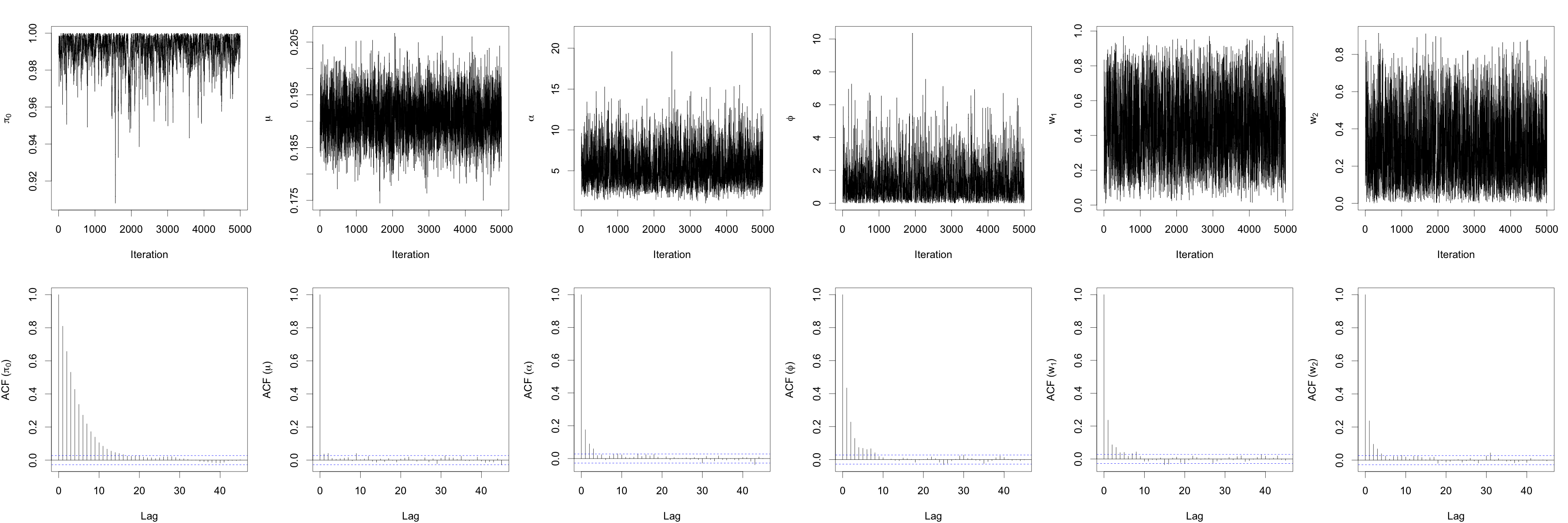}
         % \caption{}
    \end{subfigure}
    \caption{
    MCMC convergence diagnostics for Section \ref{sec: mtdcpp-sim}:
    From top to bottom are 
    trace plots and ACFs of the posterior samples
    of the parameters of the Lomax MTDCPP, corresponding to 
    scenarios where $\pi_0 = 0.2$, $0.5$, $0.8$, and $1$.
    }
    \label{fig:mtdcpp-sim-mcmc}
\end{figure*}

We assessed MCMC convergence through
trace plots and ACF plots.
Diagnostic results for real data examples (Sections 4.2 and 4.3)
are, respectively, shown in Figures \ref{fig:mtdpp-ivt-mcmc}
and \ref{fig:mtdcpp-fx-mcmc}. Since there were $121$ point patterns used to fit models in Section 4.3, we show diagnostic results for three point patterns 
in Figure \ref{fig:mtdcpp-fx-mcmc} as illustrations.
Diagnostic results for simulation studies (Sections 4.1 and \ref{sec: mtdcpp-sim})
are available in Figures \ref{fig:mtdpp-sim1-mcmc} and \ref{fig:mtdcpp-sim-mcmc},
respectively.

\section{Point process model checking results}

\begin{figure*}[t!]
    % \centering
    % \captionsetup[subfigure]{justification=centering, font=footnotesize}
    \includegraphics[width=.32\textwidth]{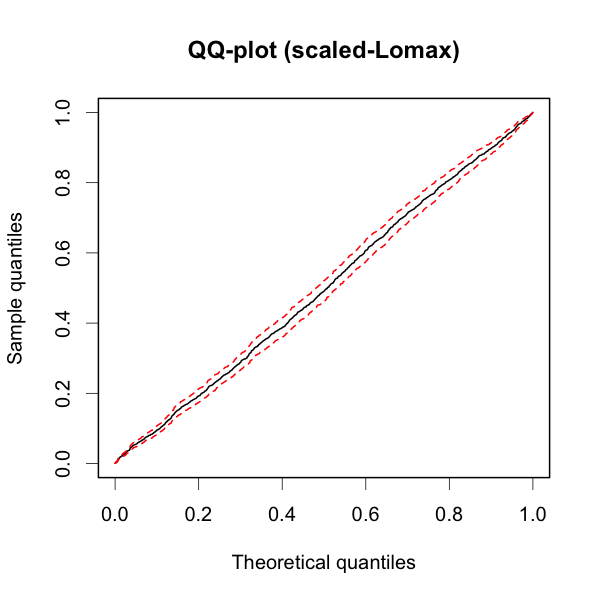}
    \includegraphics[width=.32\textwidth]{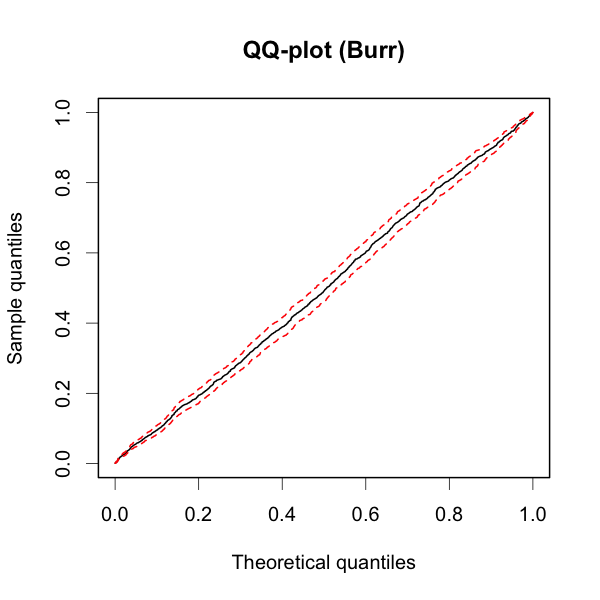}
    \includegraphics[width=.32\textwidth]{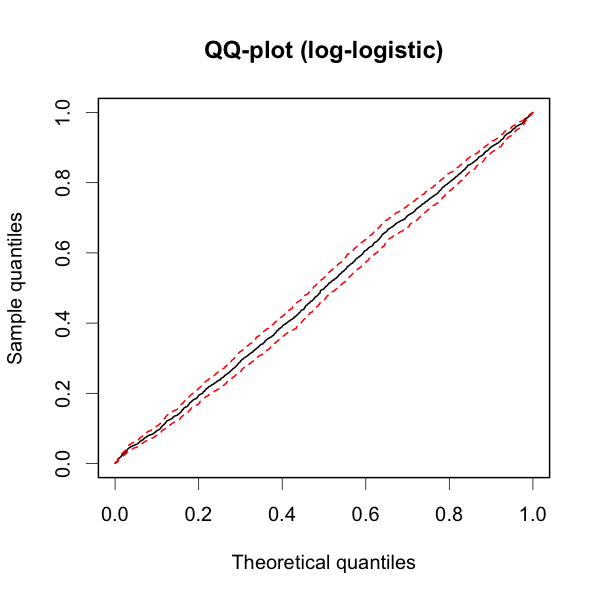}
    \medskip
    \includegraphics[width=.32\textwidth]{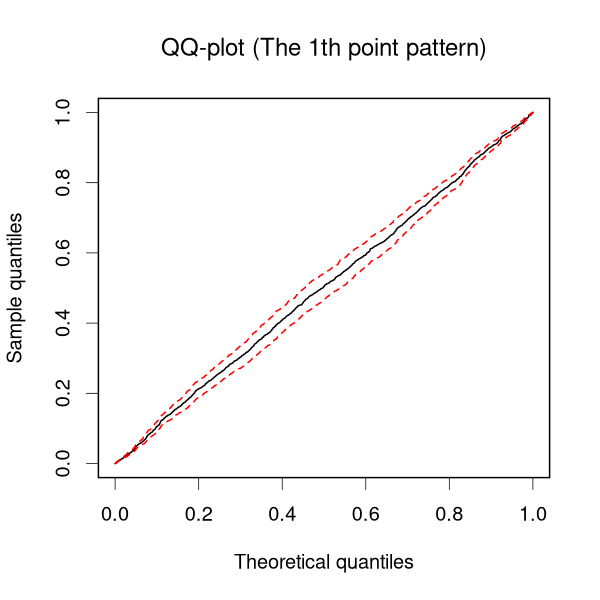}
    \includegraphics[width=.32\textwidth]{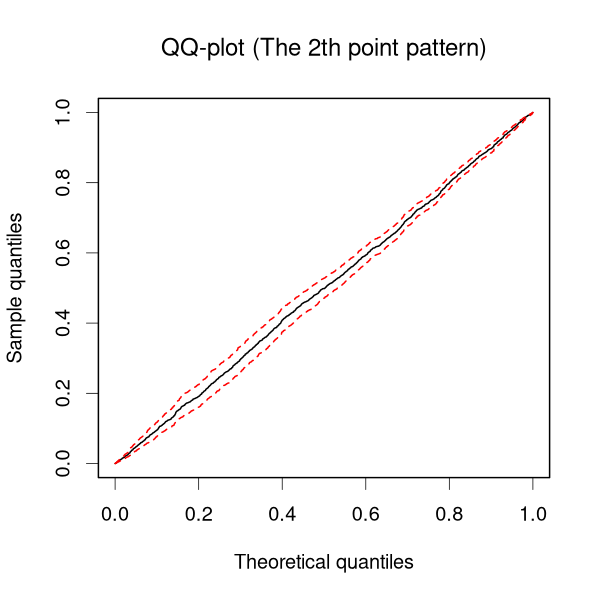}
    \includegraphics[width=.32\textwidth]{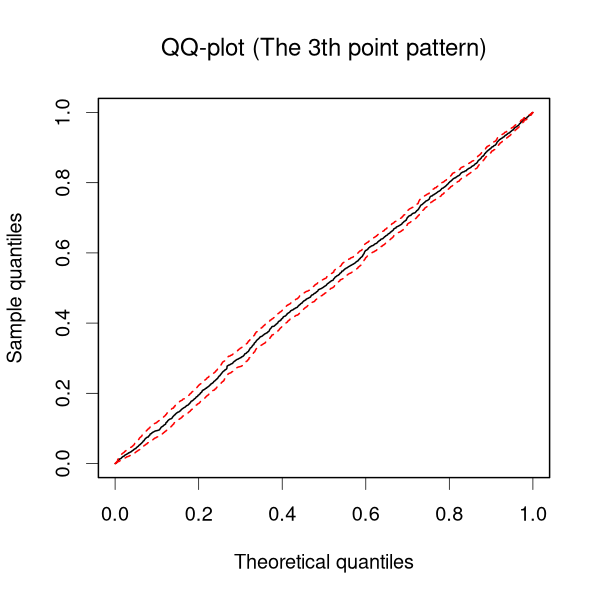}
    \medskip
    \includegraphics[width=.32\textwidth]{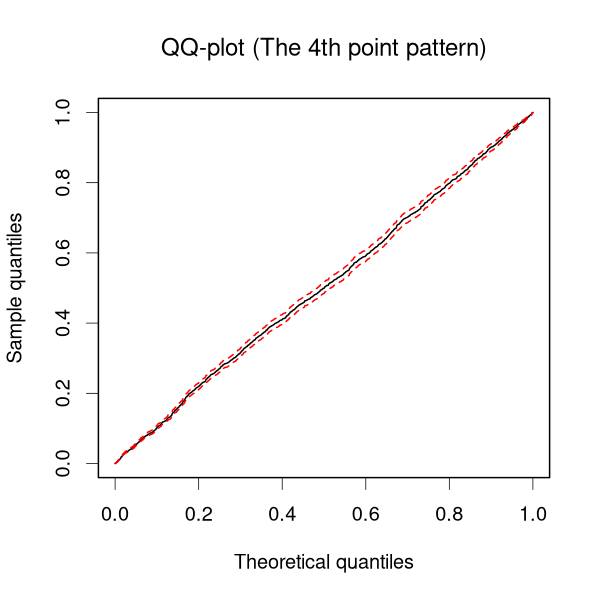}
    \includegraphics[width=.32\textwidth]{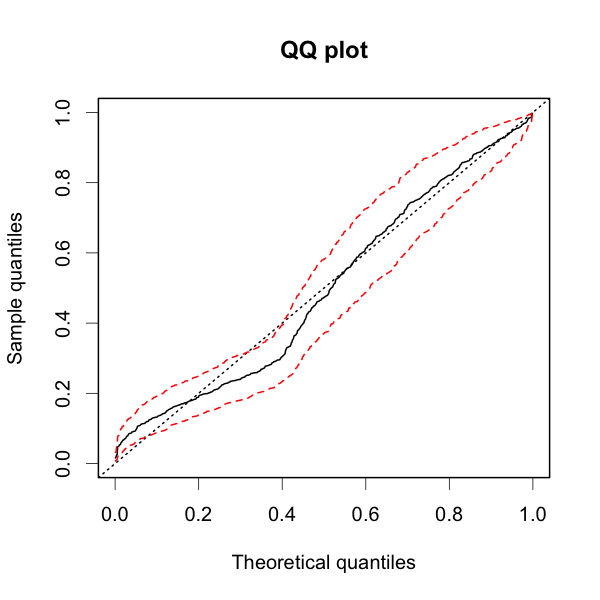}
    \caption{
    Model checking results. 
    The first row corresponds to the simulation study 
    in the main paper (Section 4.1).
    The second row and the first panel of the third row correspond to 
    the additional simulation study (Section \ref{sec: mtdcpp-sim}).
    The second panel of the third row corresponds to the IVT data example
    in the main paper (Section 4.2).
    Black solid lines and red dotted lines are posterior mean
    95\% credible interval estimates, respectively.}
    \label{fig:mtdpp-sim1}
\end{figure*}

\begin{figure*}[t!]
    \centering
    % \captionsetup[subfigure]{justification=centering, font=footnotesize}
    \includegraphics[width=.19\textwidth]{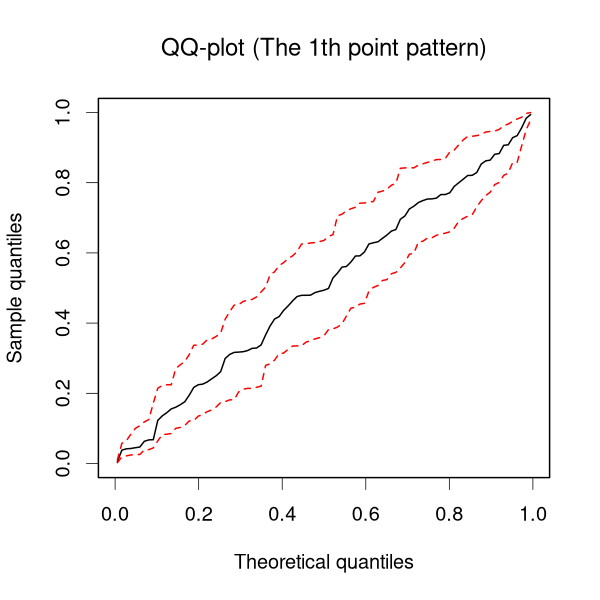}
    \includegraphics[width=.19\textwidth]{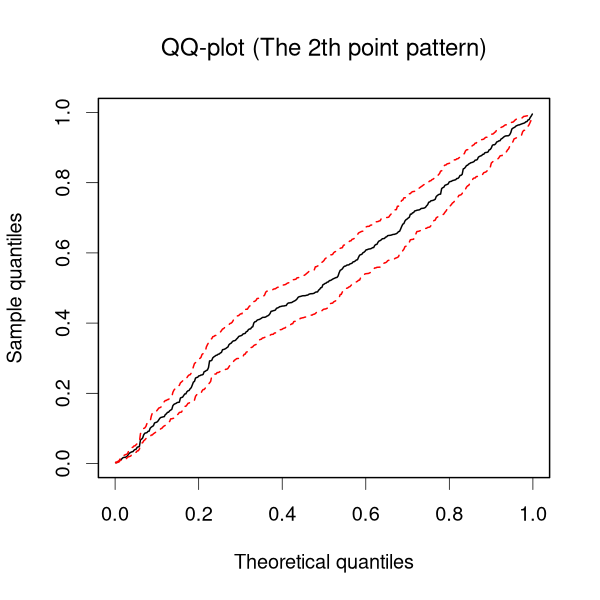}
    \includegraphics[width=.19\textwidth]{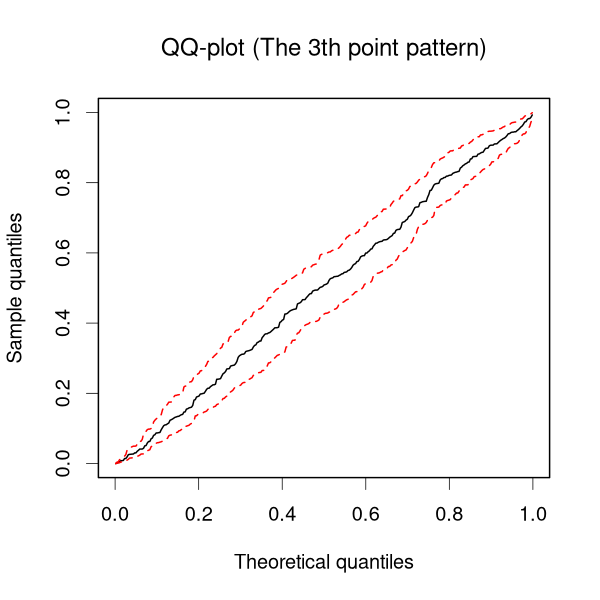}
    \includegraphics[width=.19\textwidth]{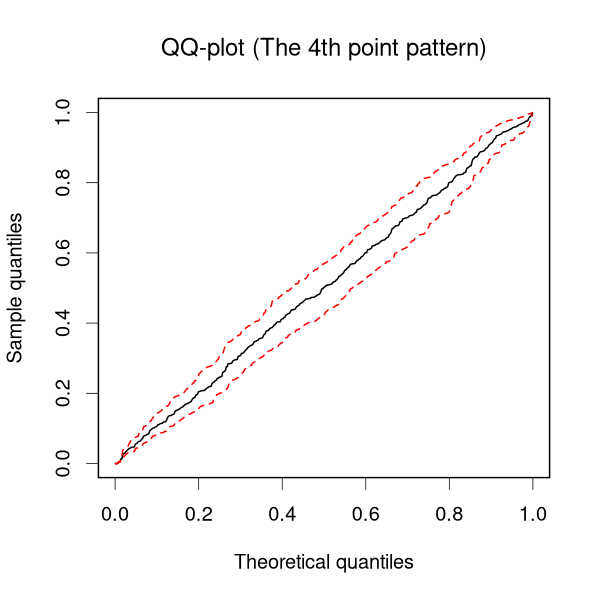}
    \includegraphics[width=.19\textwidth]{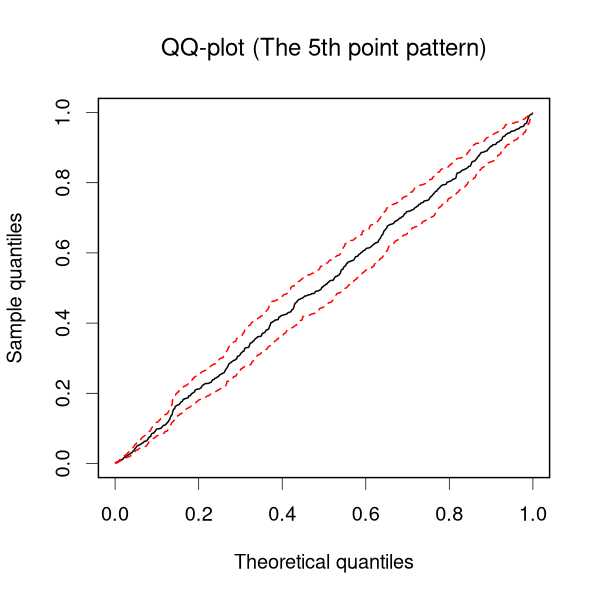}
    \includegraphics[width=.19\textwidth]{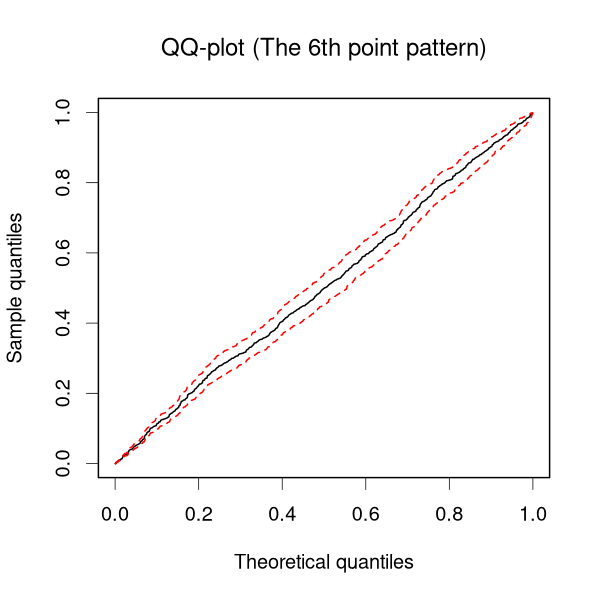}
    \includegraphics[width=.19\textwidth]{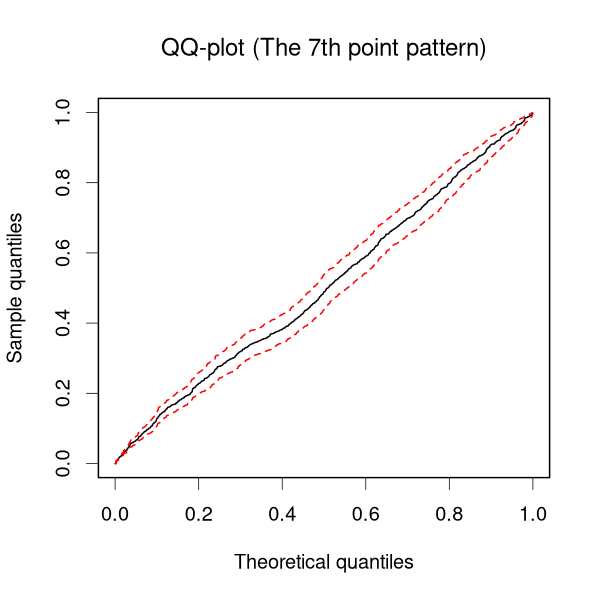}
    \includegraphics[width=.19\textwidth]{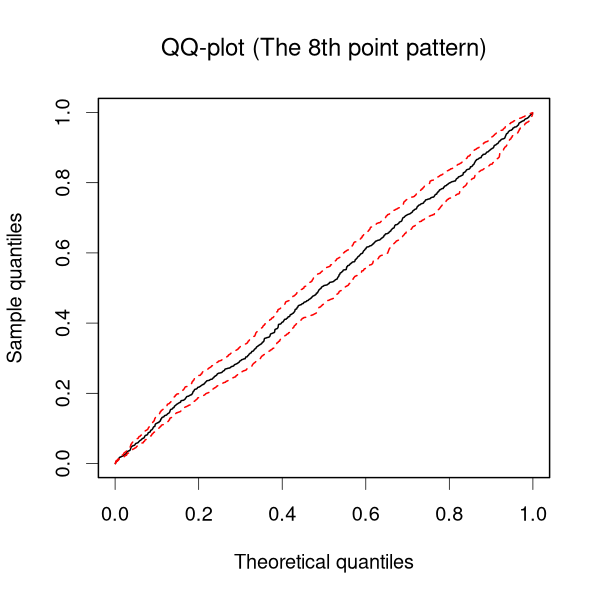}
    \includegraphics[width=.19\textwidth]{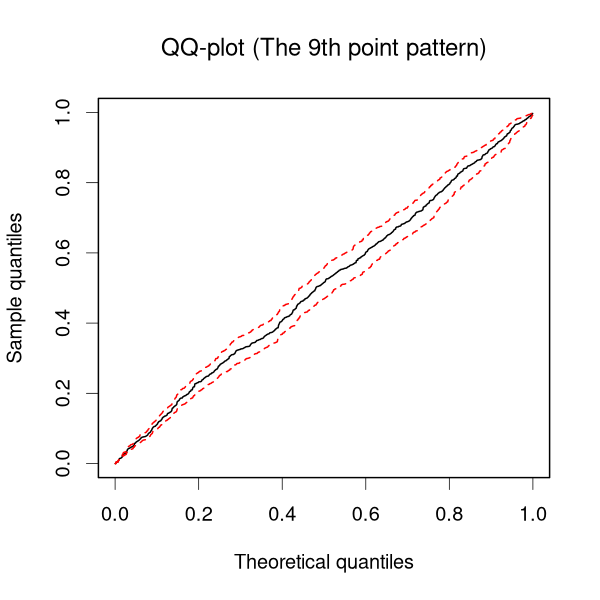}
    \includegraphics[width=.19\textwidth]{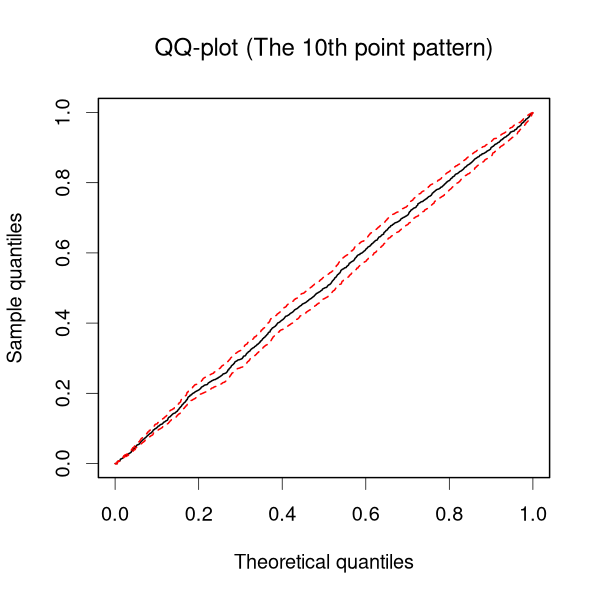}
    \includegraphics[width=.19\textwidth]{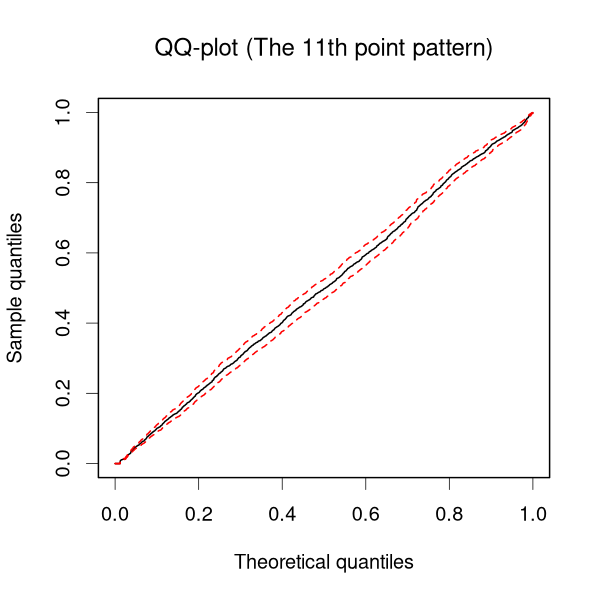}
    \includegraphics[width=.19\textwidth]{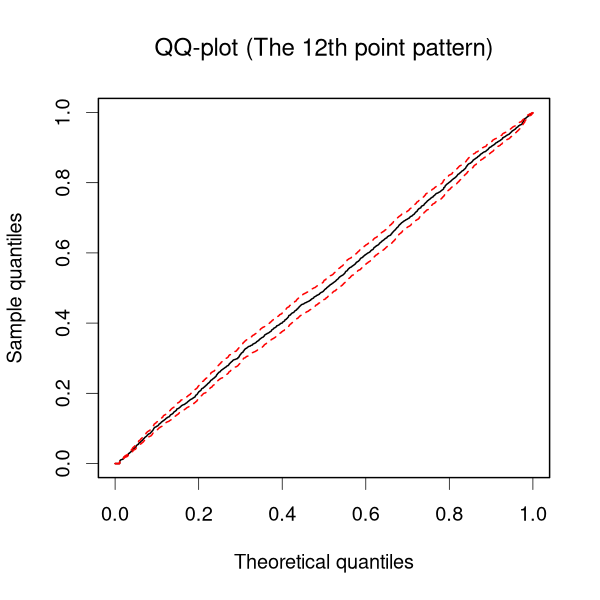}
    \includegraphics[width=.19\textwidth]{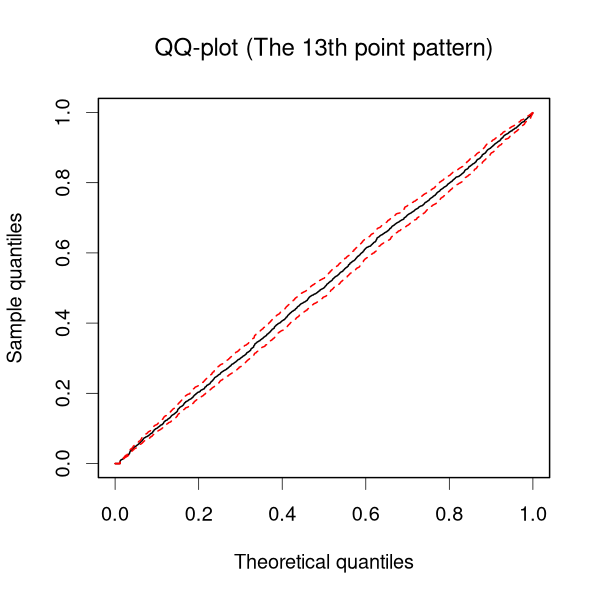}
    \includegraphics[width=.19\textwidth]{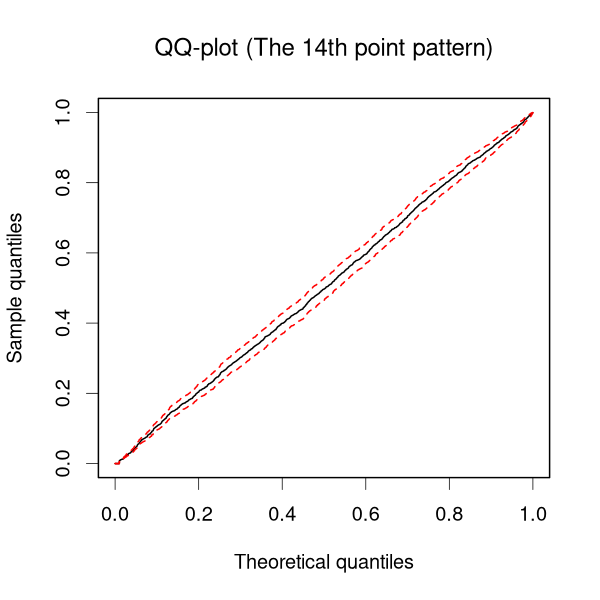}
    \includegraphics[width=.19\textwidth]{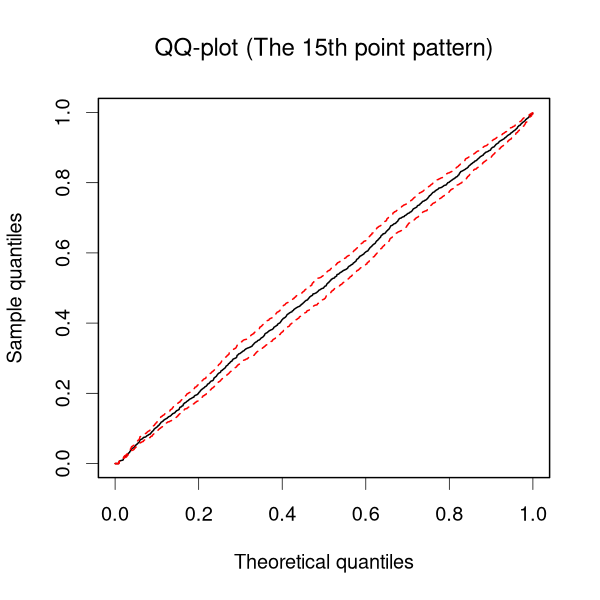}
    \includegraphics[width=.19\textwidth]{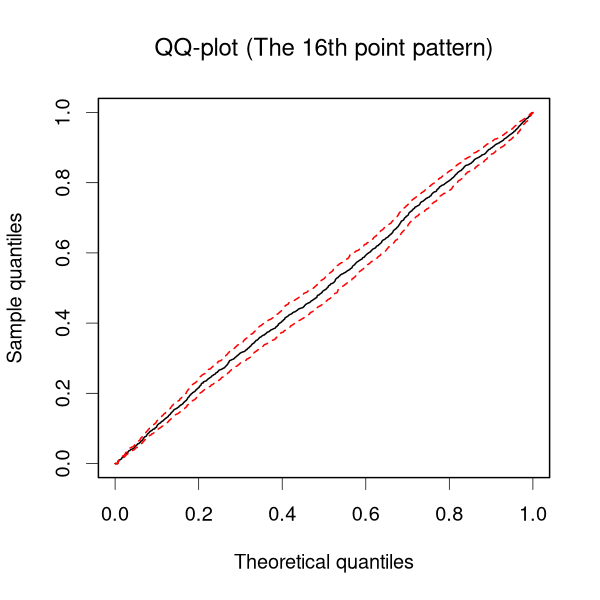}
    \includegraphics[width=.19\textwidth]{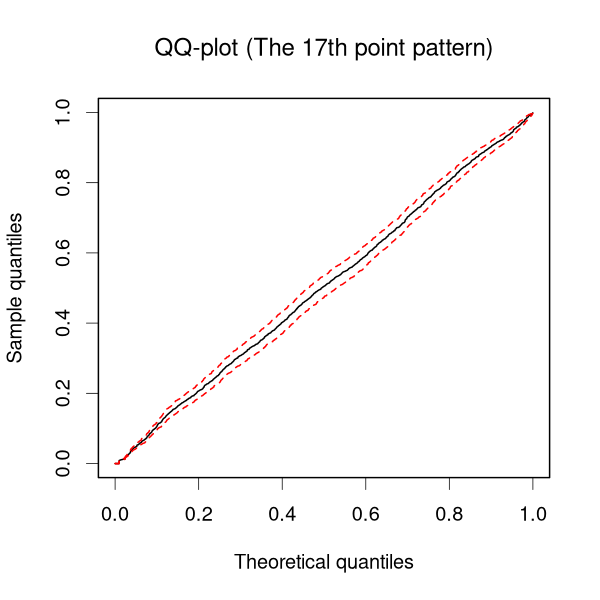}
    \includegraphics[width=.19\textwidth]{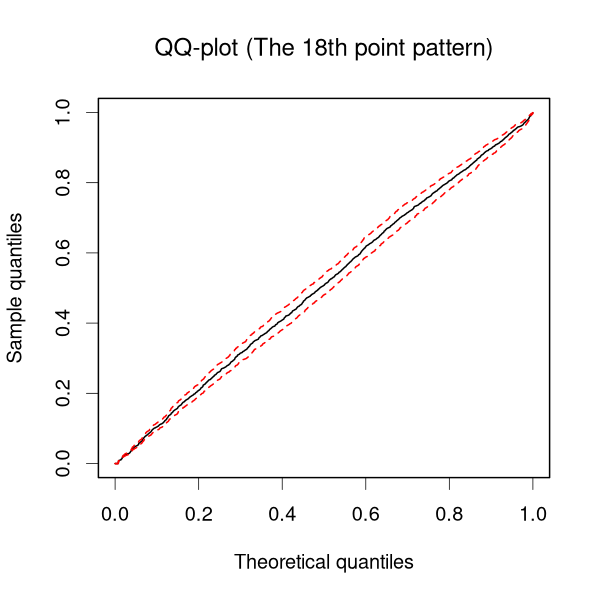}
    \includegraphics[width=.19\textwidth]{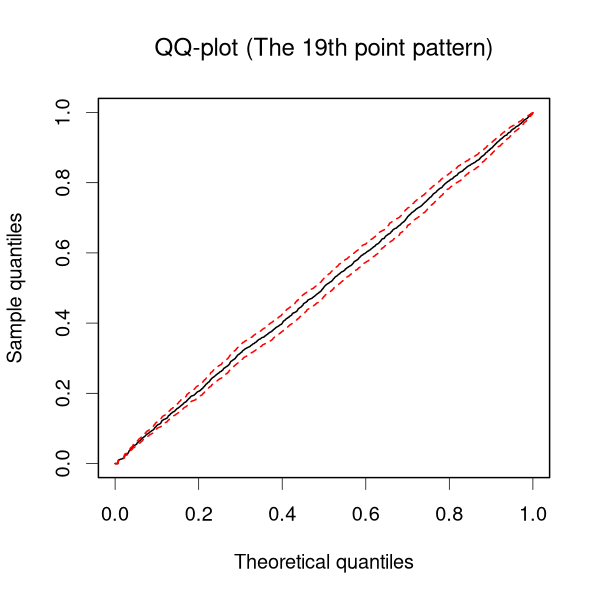}
    \includegraphics[width=.19\textwidth]{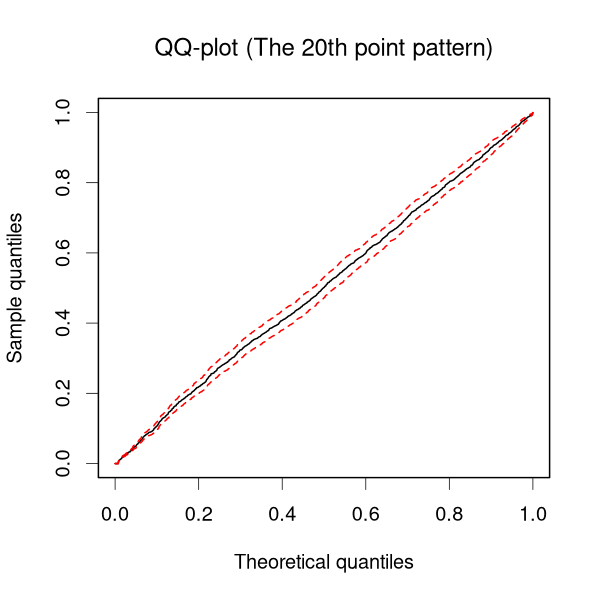}
    \includegraphics[width=.19\textwidth]{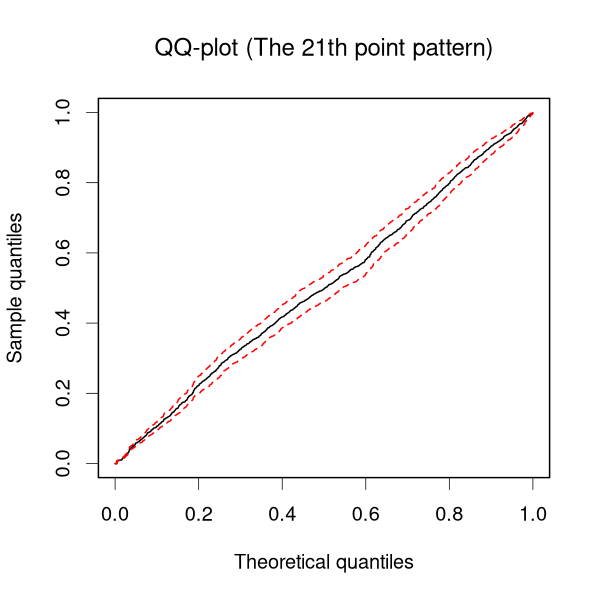}
    \includegraphics[width=.19\textwidth]{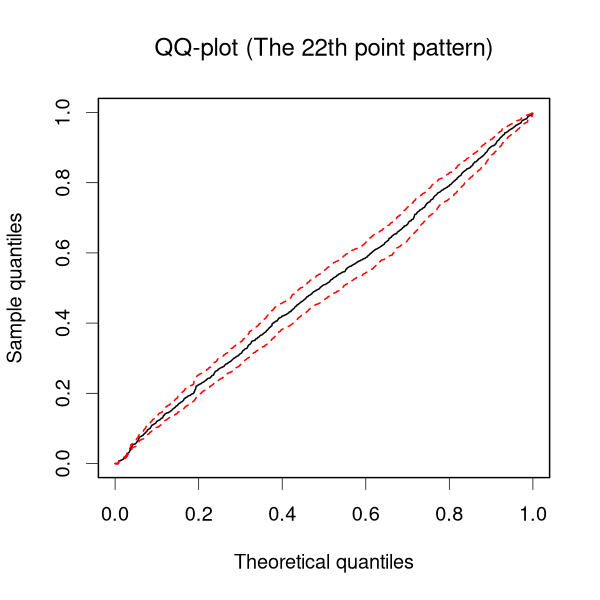}
    \includegraphics[width=.19\textwidth]{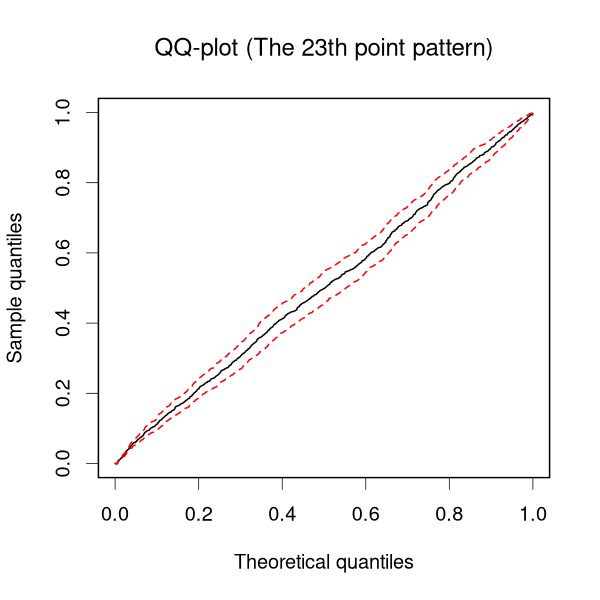}
    \includegraphics[width=.19\textwidth]{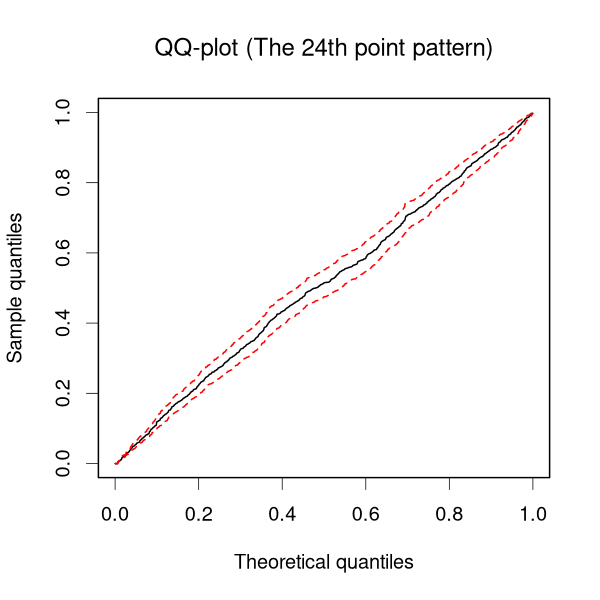}
    \includegraphics[width=.19\textwidth]{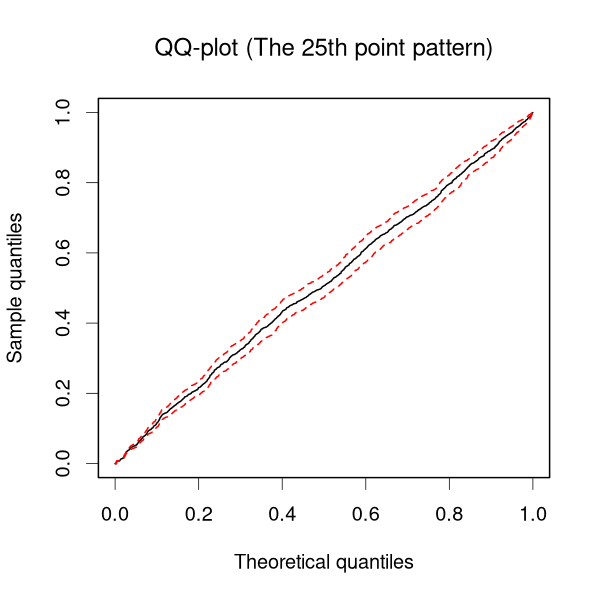}
    \includegraphics[width=.19\textwidth]{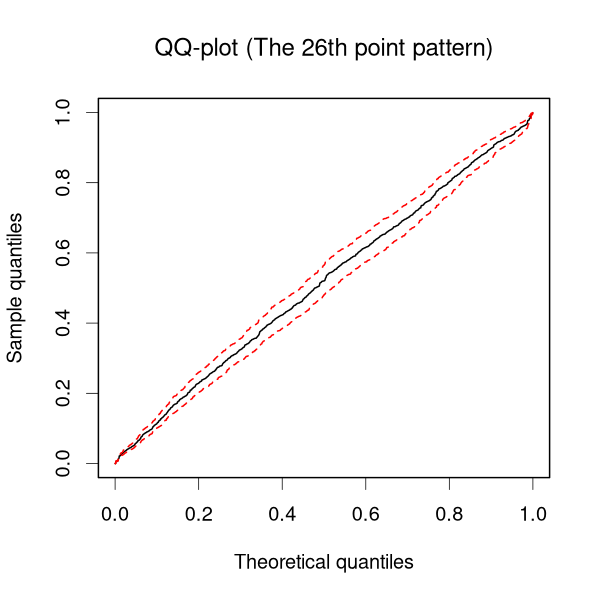}
    \includegraphics[width=.19\textwidth]{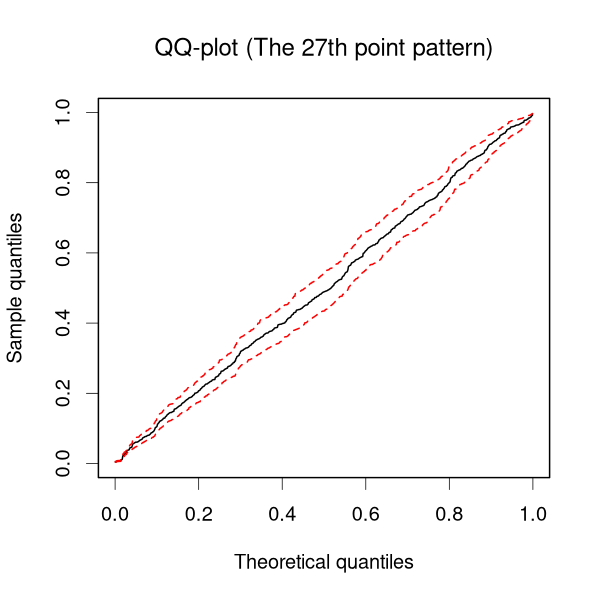}
    \includegraphics[width=.19\textwidth]{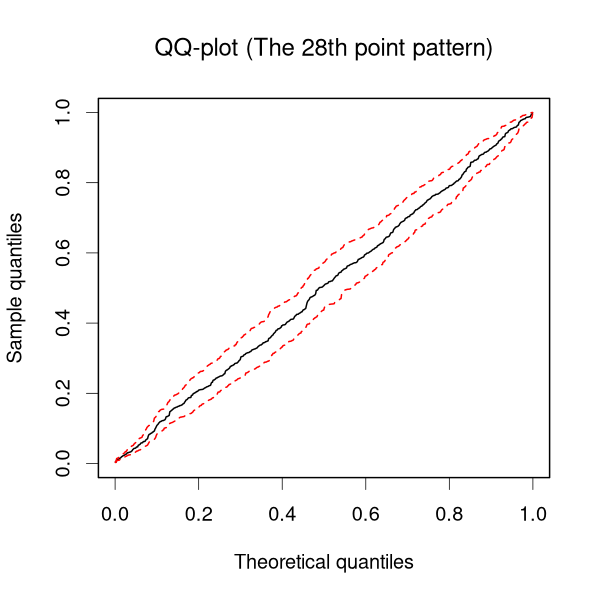}
    \includegraphics[width=.19\textwidth]{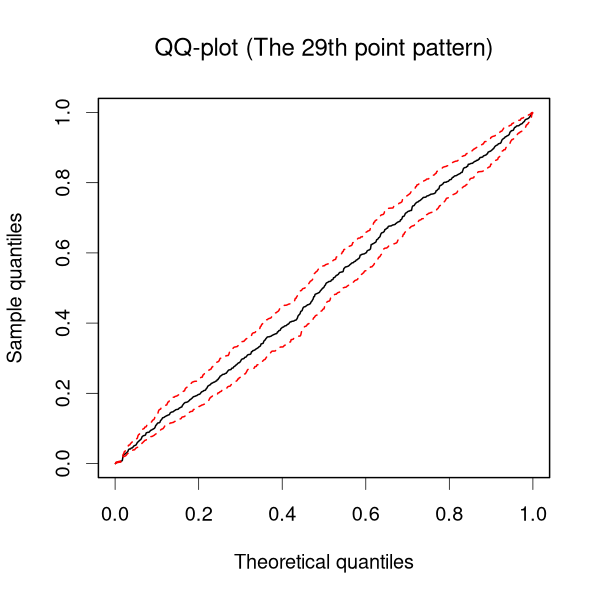}
    \includegraphics[width=.19\textwidth]{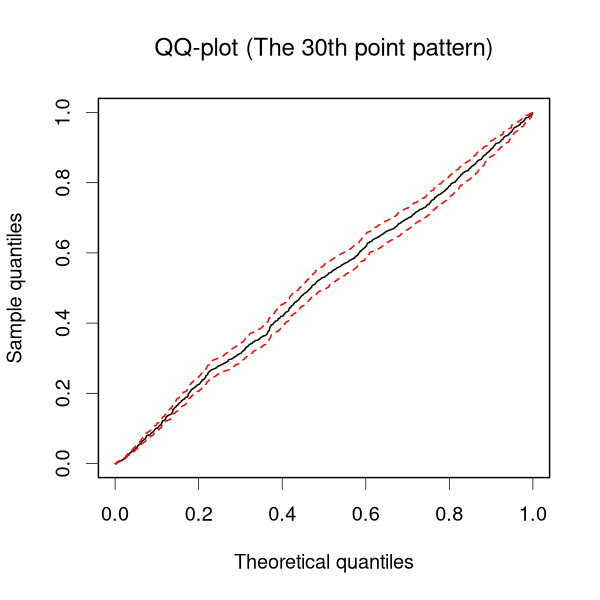}
    \includegraphics[width=.19\textwidth]{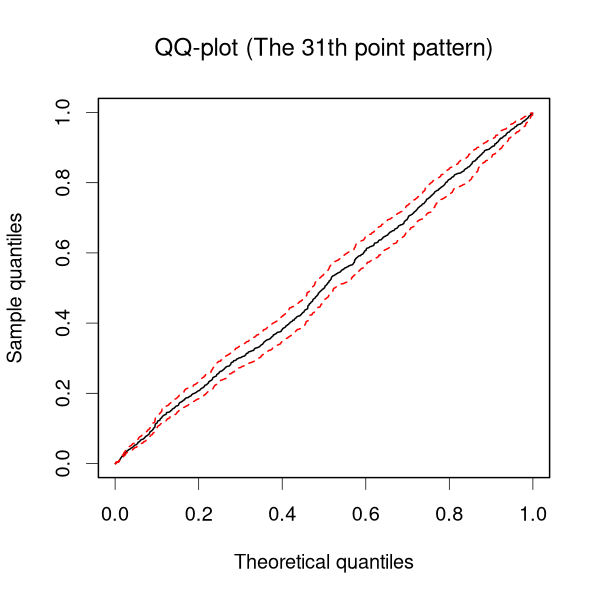}
    \includegraphics[width=.19\textwidth]{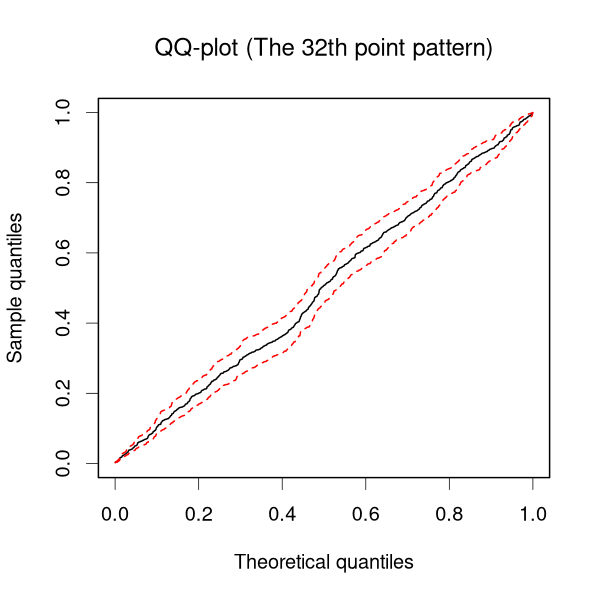}
    \includegraphics[width=.19\textwidth]{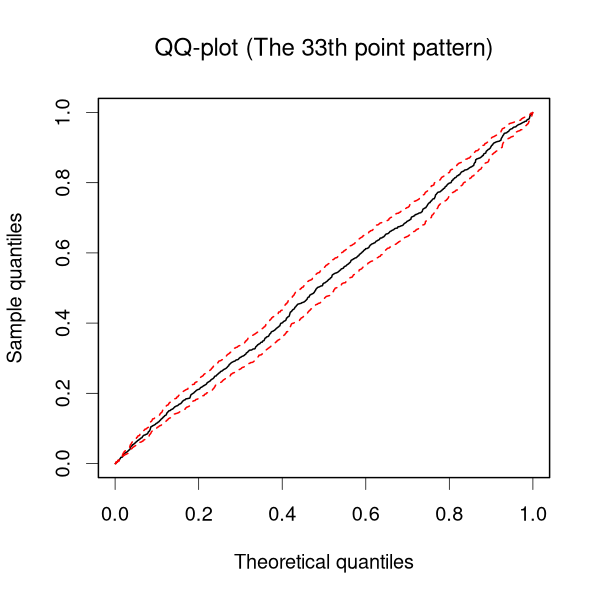}
    \includegraphics[width=.19\textwidth]{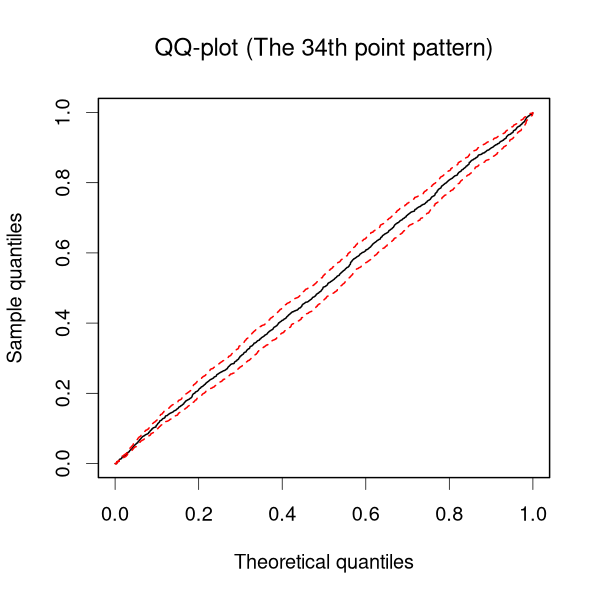}
    \includegraphics[width=.19\textwidth]{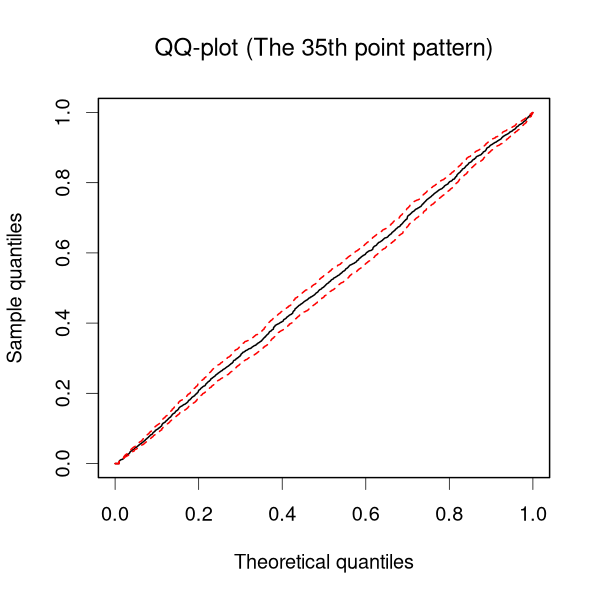}
    \caption{Model checking results for point patterns 1 - 35 
    in the second real data example in the main paper (Section 4.3).
    Black solid lines and red dotted lines are
    posterior mean and 95\% credible interval estimates,
    respectively.}
    \label{fig:mtdpp-fx1}
\end{figure*}

\begin{figure*}[t!]
    \centering
    % \captionsetup[subfigure]{justification=centering, font=footnotesize}
    \includegraphics[width=.19\textwidth]{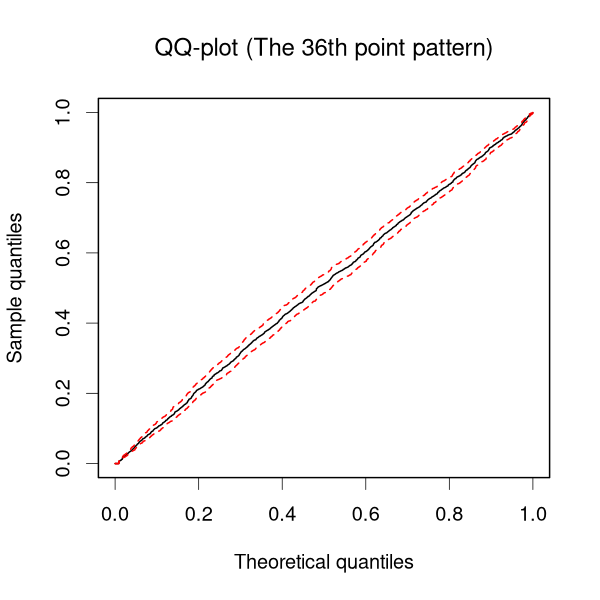}
    \includegraphics[width=.19\textwidth]{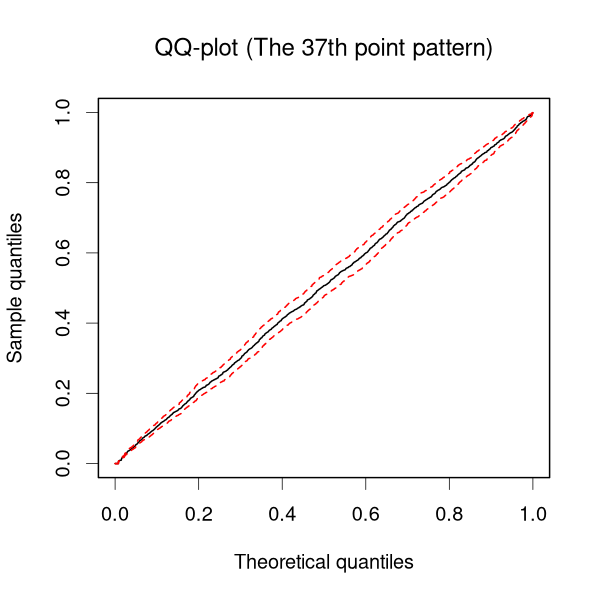}
    \includegraphics[width=.19\textwidth]{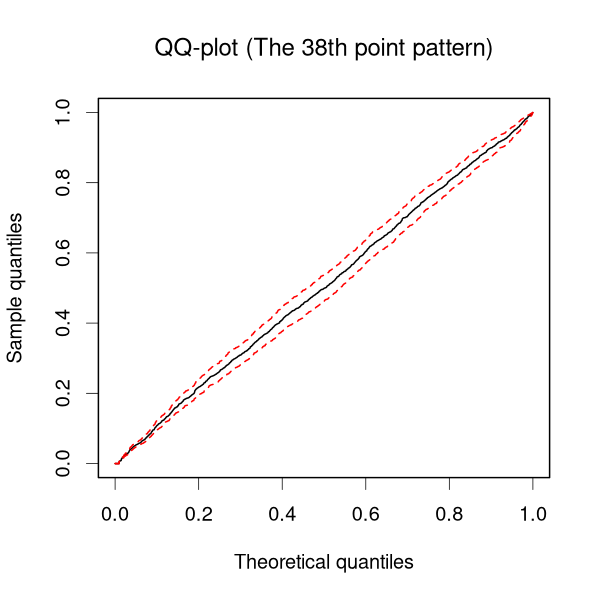}
    \includegraphics[width=.19\textwidth]{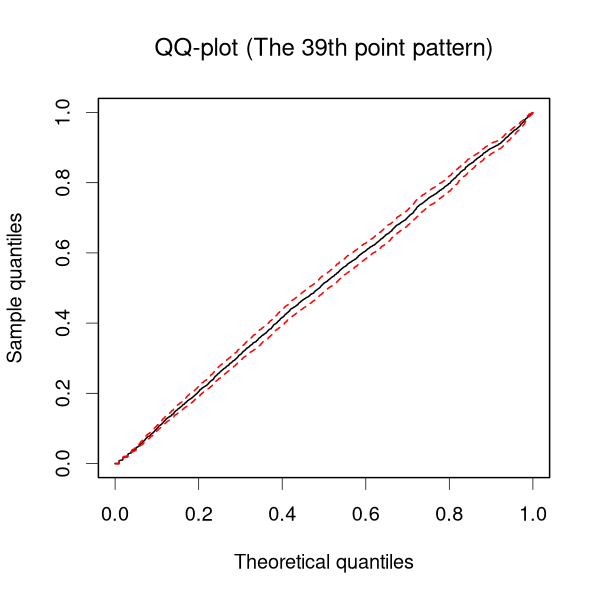}
    \includegraphics[width=.19\textwidth]{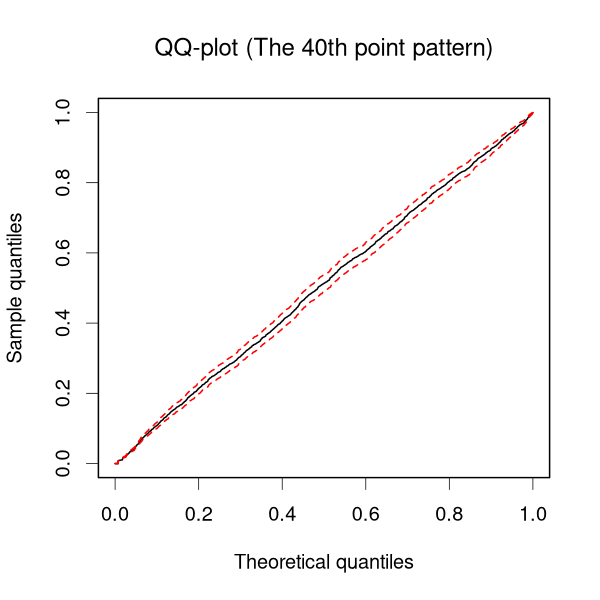}
    \includegraphics[width=.19\textwidth]{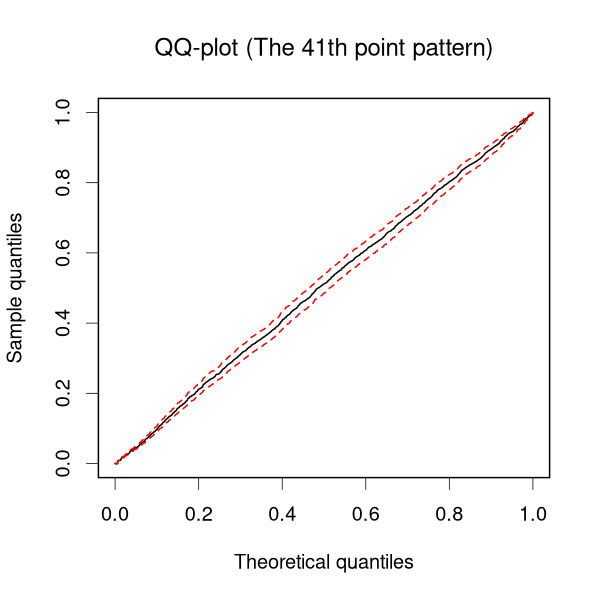}
    \includegraphics[width=.19\textwidth]{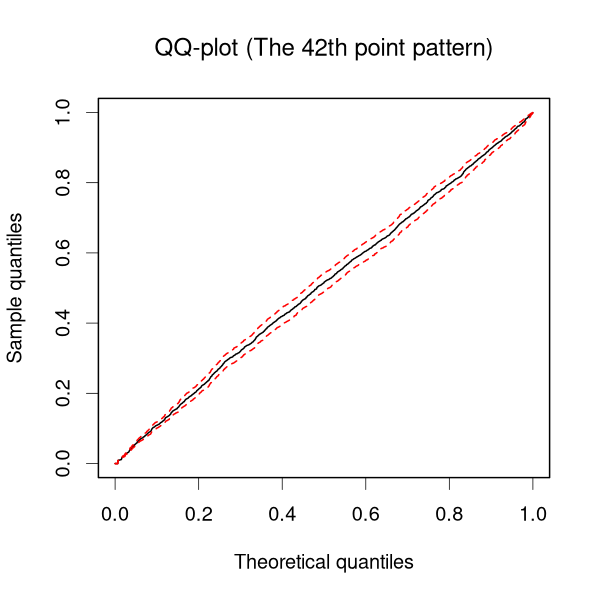}
    \includegraphics[width=.19\textwidth]{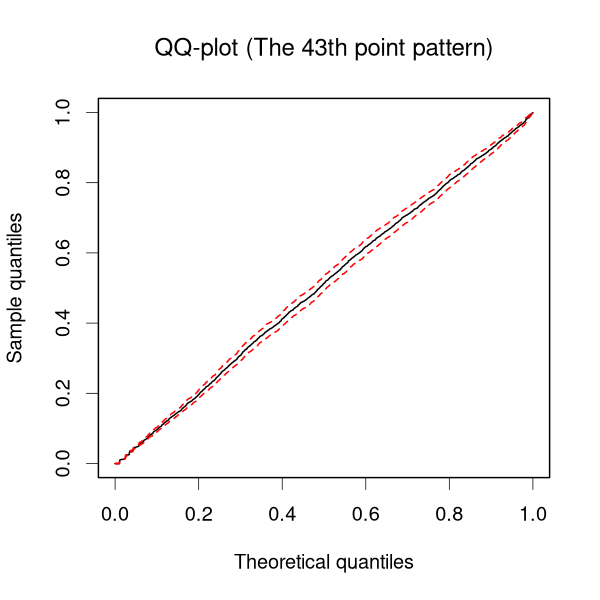}
    \includegraphics[width=.19\textwidth]{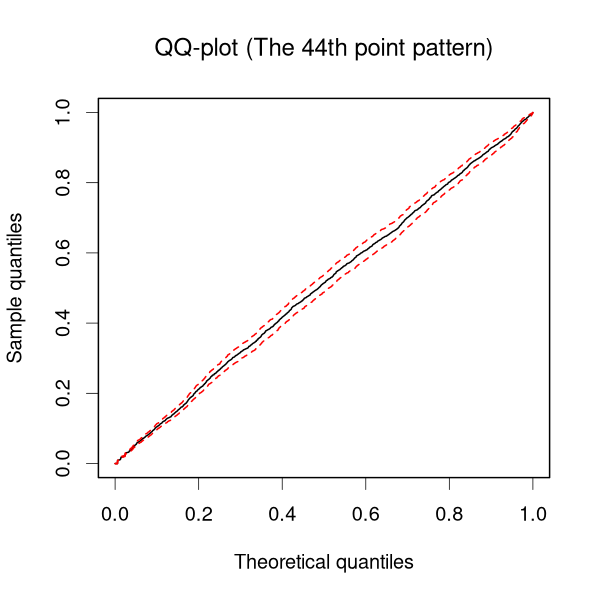}
    \includegraphics[width=.19\textwidth]{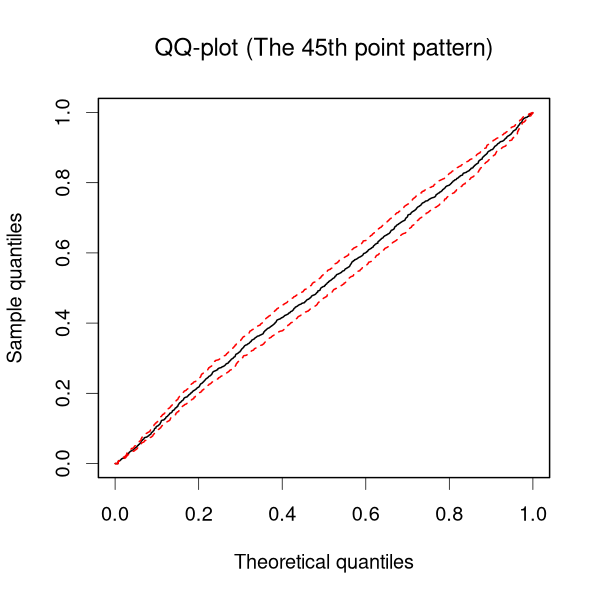}    
    \includegraphics[width=.19\textwidth]{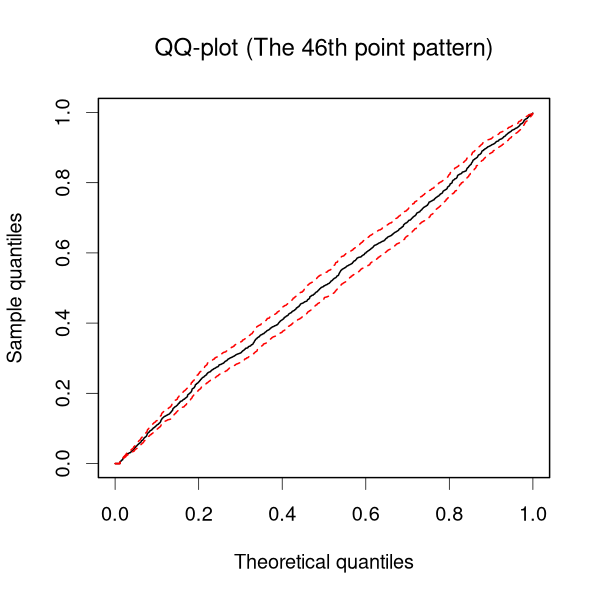}
    \includegraphics[width=.19\textwidth]{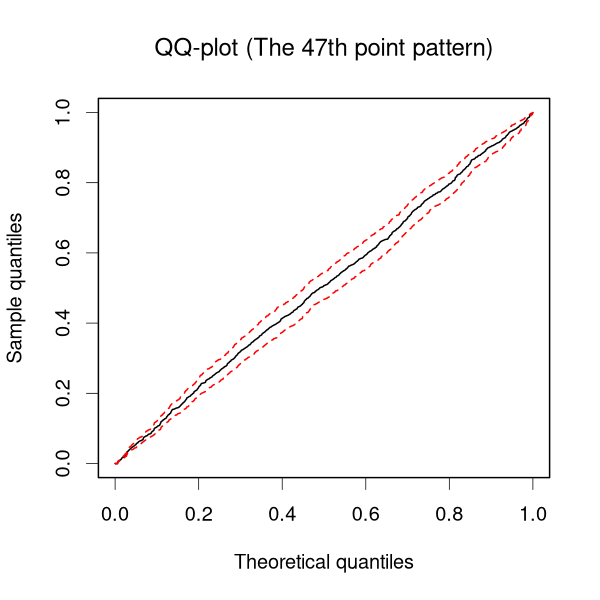}
    \includegraphics[width=.19\textwidth]{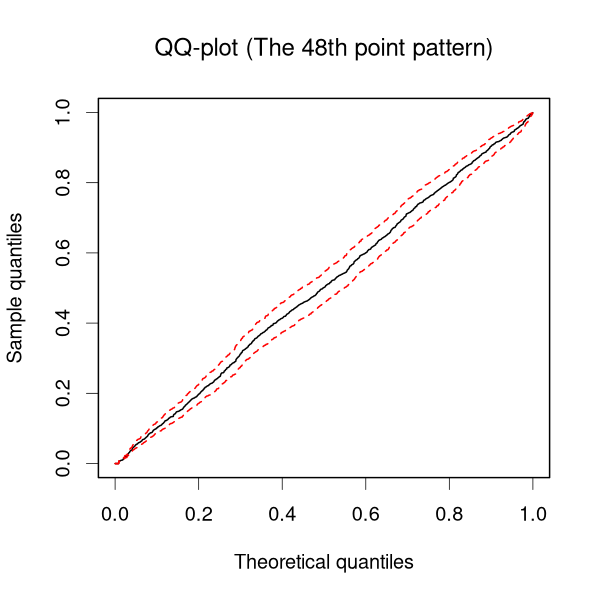}
    \includegraphics[width=.19\textwidth]{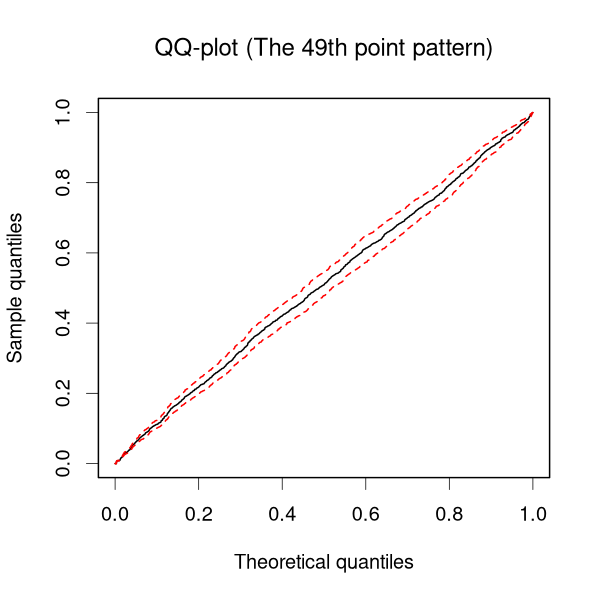}
    \includegraphics[width=.19\textwidth]{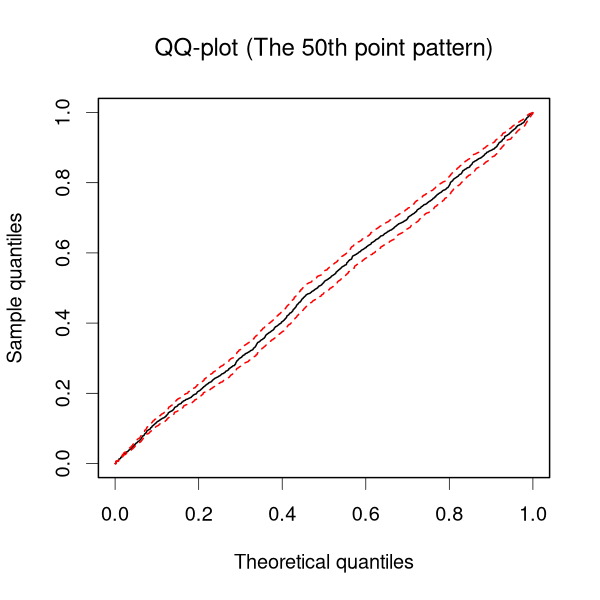}
    \includegraphics[width=.19\textwidth]{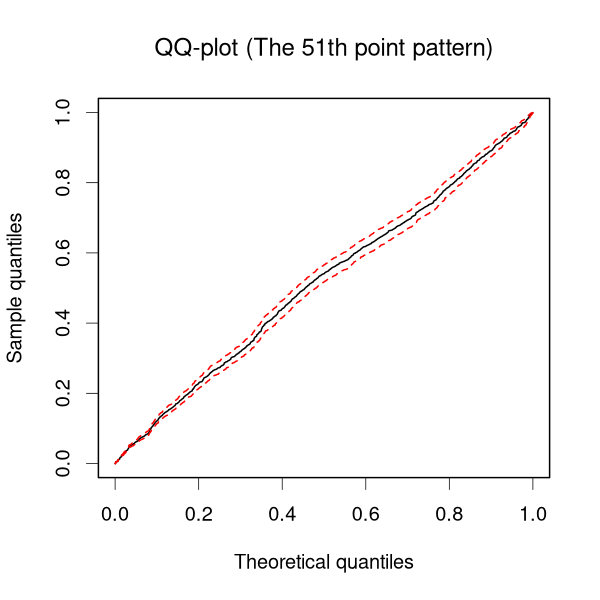}
    \includegraphics[width=.19\textwidth]{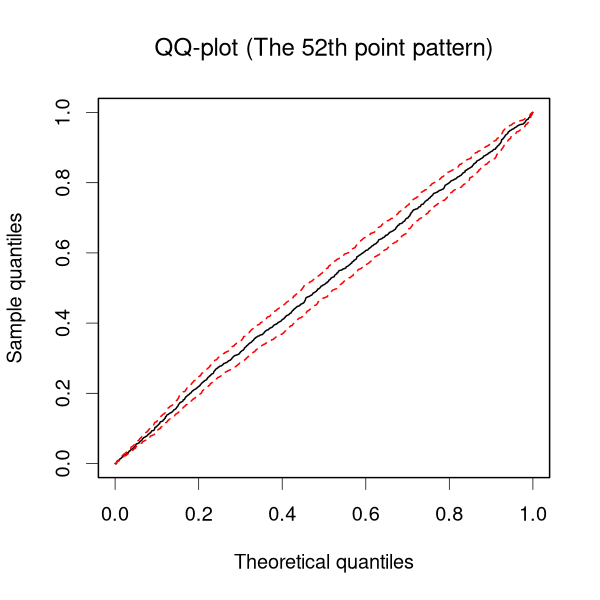}
    \includegraphics[width=.19\textwidth]{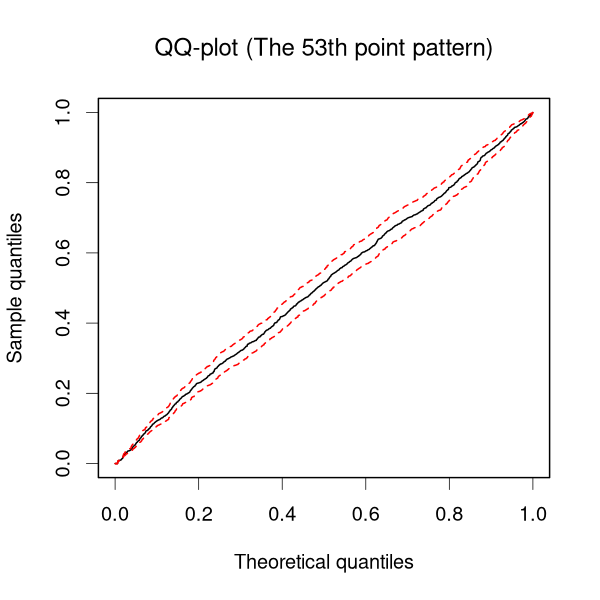}
    \includegraphics[width=.19\textwidth]{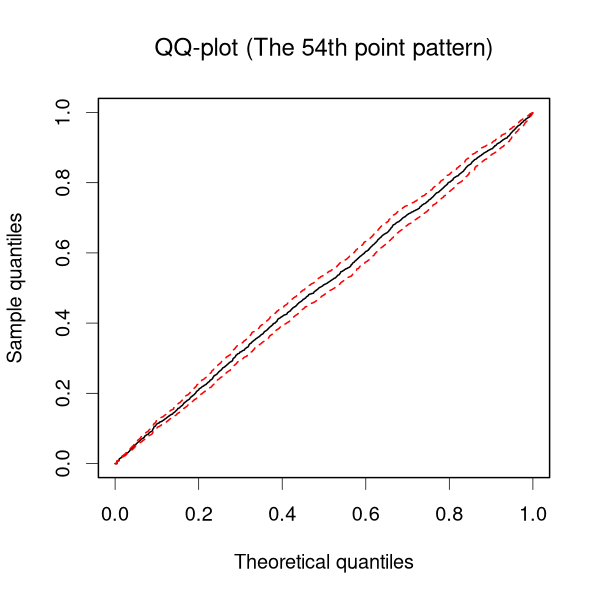}
    \includegraphics[width=.19\textwidth]{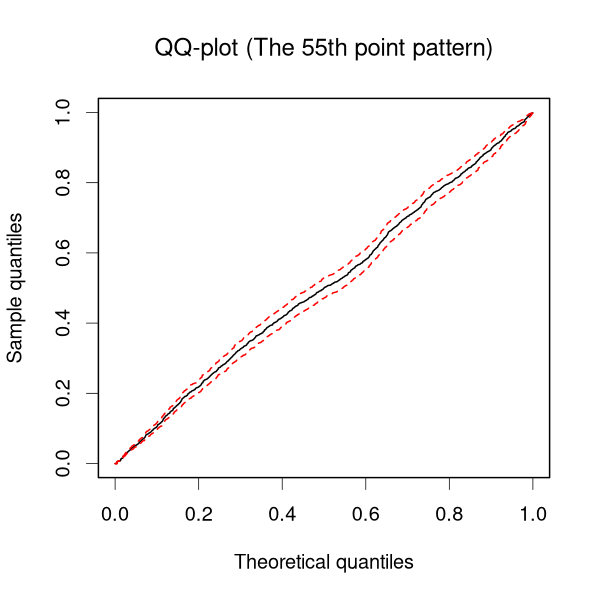}
    \includegraphics[width=.19\textwidth]{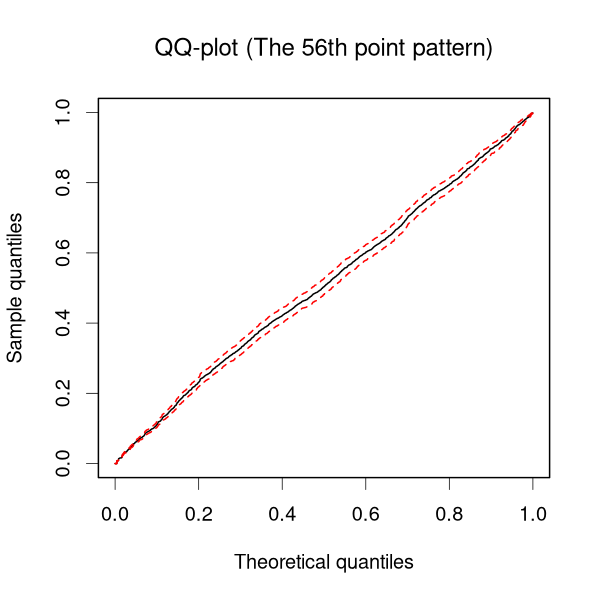}
    \includegraphics[width=.19\textwidth]{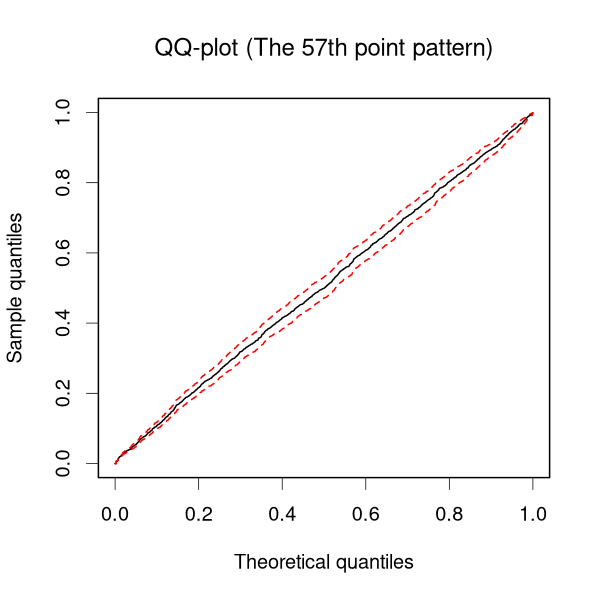}
    \includegraphics[width=.19\textwidth]{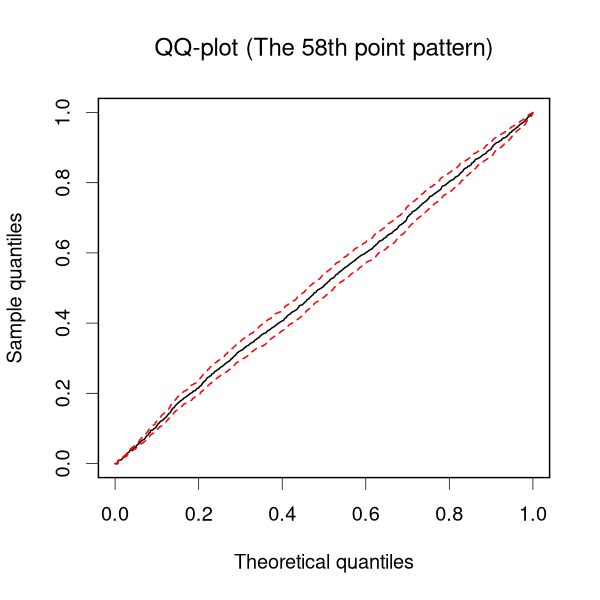}
    \includegraphics[width=.19\textwidth]{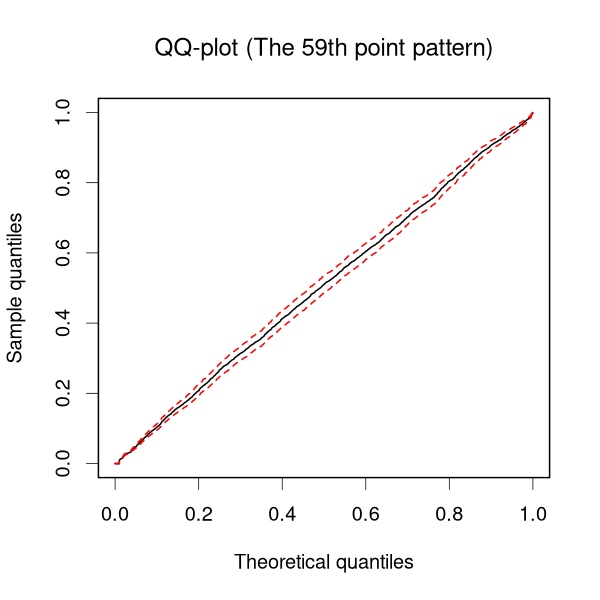}
    \includegraphics[width=.19\textwidth]{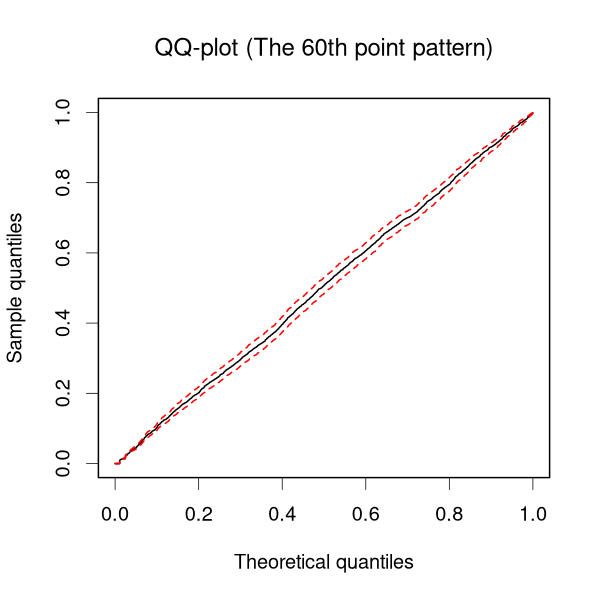}
    \includegraphics[width=.19\textwidth]{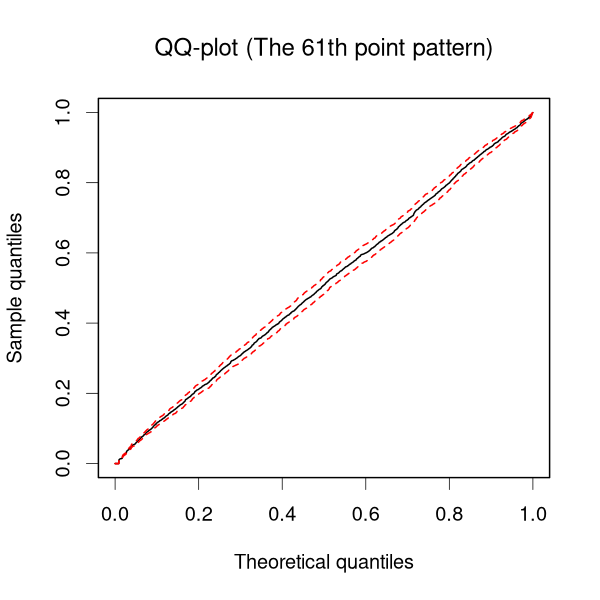}
    \includegraphics[width=.19\textwidth]{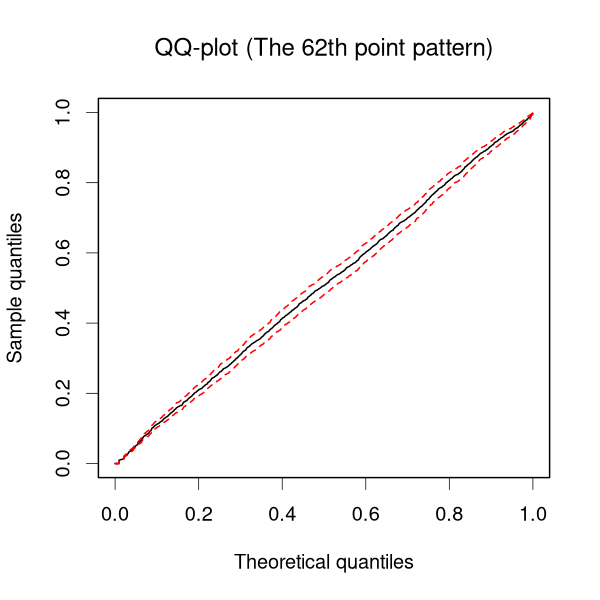}
    \includegraphics[width=.19\textwidth]{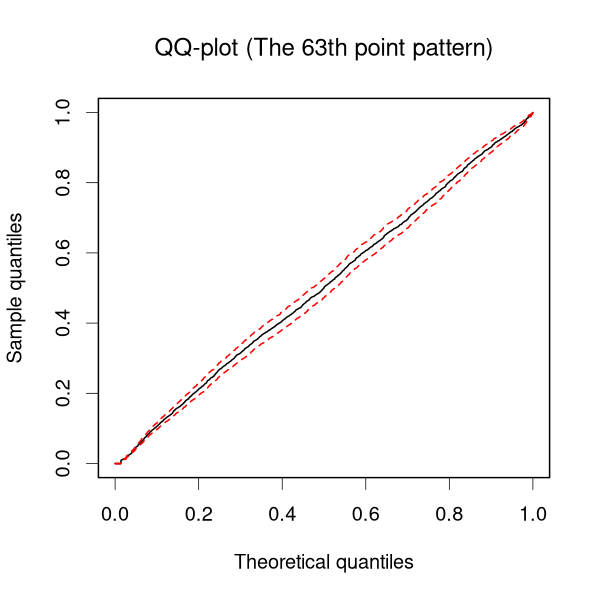}
    \includegraphics[width=.19\textwidth]{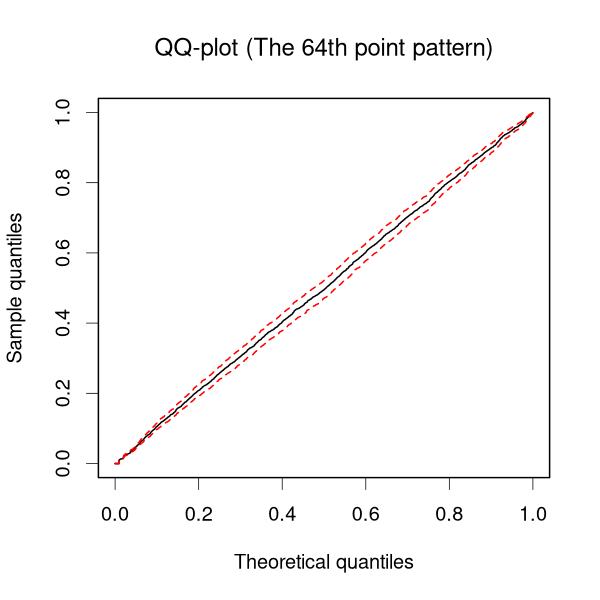}
    \includegraphics[width=.19\textwidth]{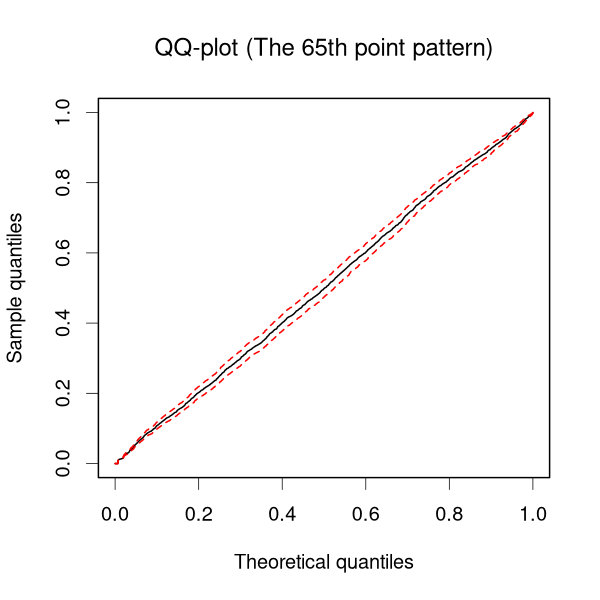}
    \includegraphics[width=.19\textwidth]{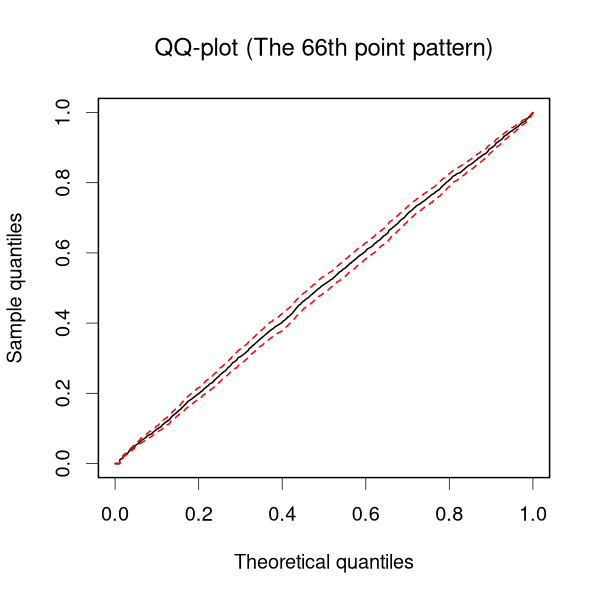}
    \includegraphics[width=.19\textwidth]{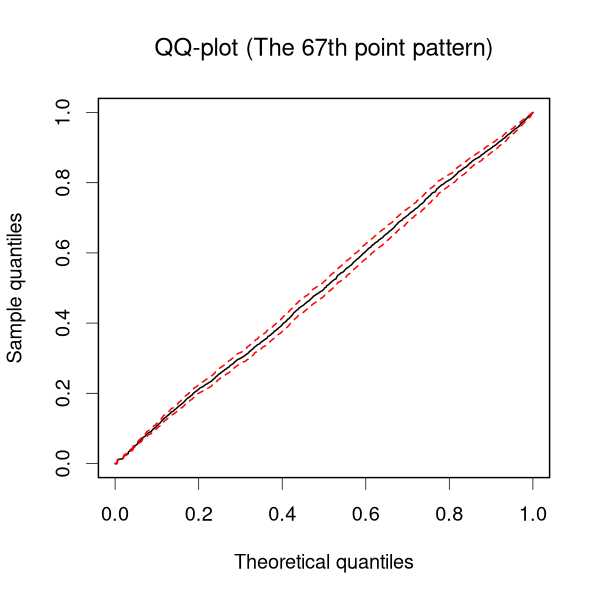}
    \includegraphics[width=.19\textwidth]{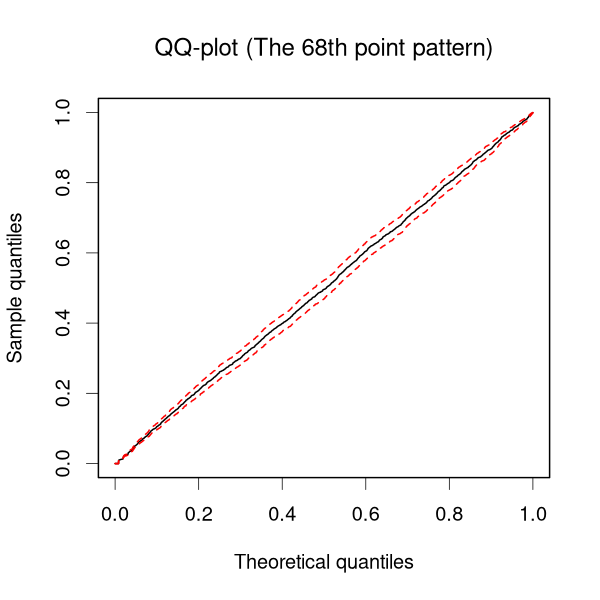}
    \includegraphics[width=.19\textwidth]{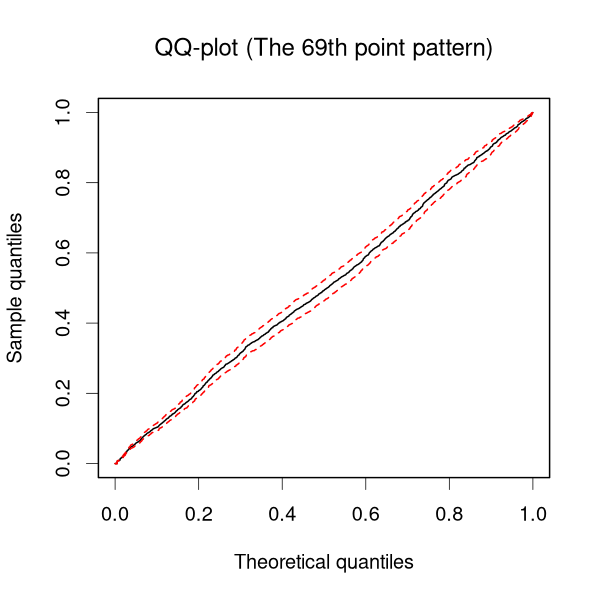}
    \includegraphics[width=.19\textwidth]{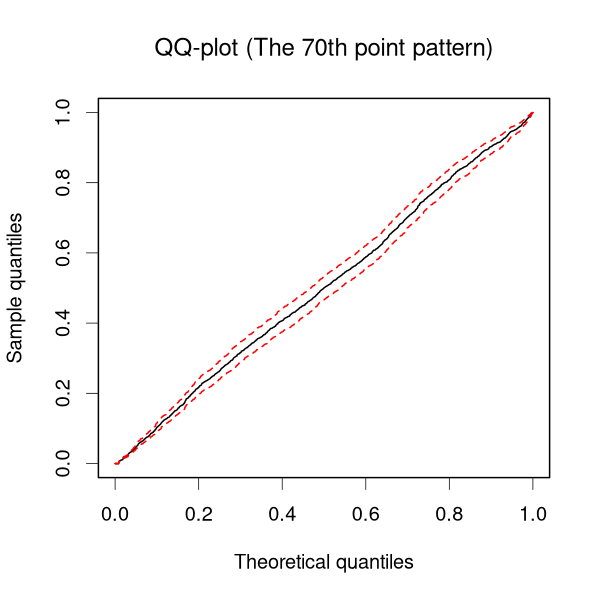}
    \caption{Model checking results for point patterns 36 - 70
    in the second real data example in the main paper (Section 4.3).
    Black solid lines and red dotted lines are
    posterior mean and 95\% credible interval estimates,
    respectively.}
    \label{fig:mtdpp-fx2}
\end{figure*}

\begin{figure*}[t!]
    % \centering
    % \captionsetup[subfigure]{justification=centering, font=footnotesize}
    \includegraphics[width=.19\textwidth]{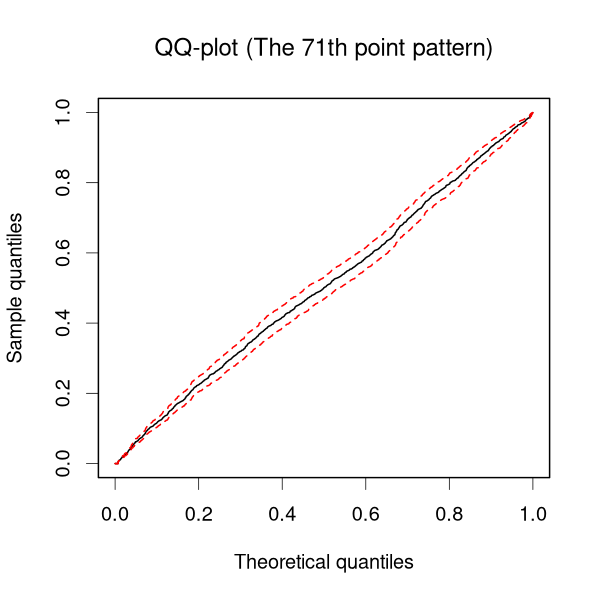}
    \includegraphics[width=.19\textwidth]{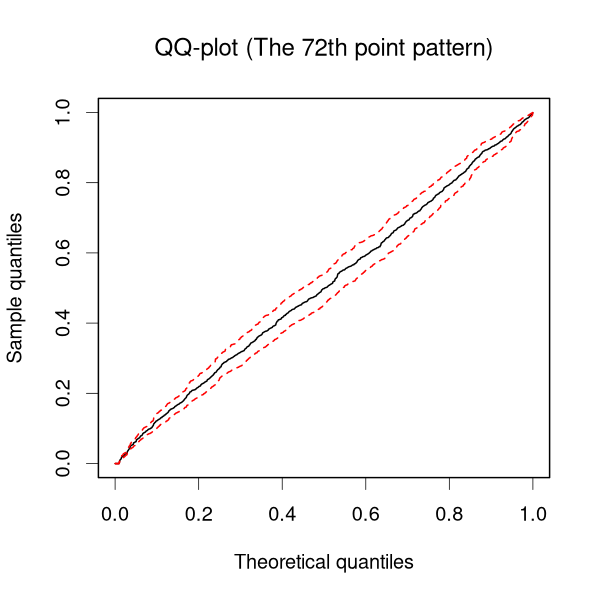}
    \includegraphics[width=.19\textwidth]{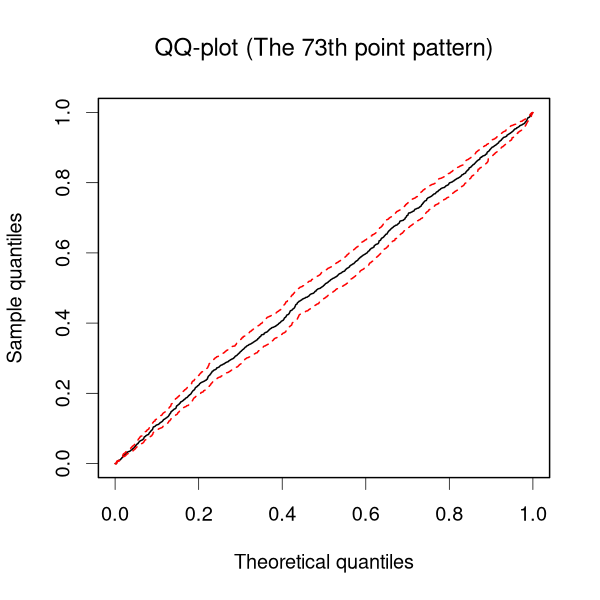}
    \includegraphics[width=.19\textwidth]{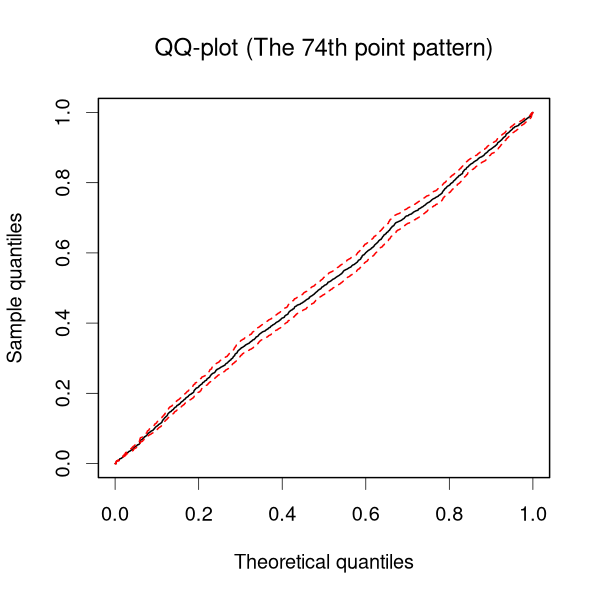}
    \includegraphics[width=.19\textwidth]{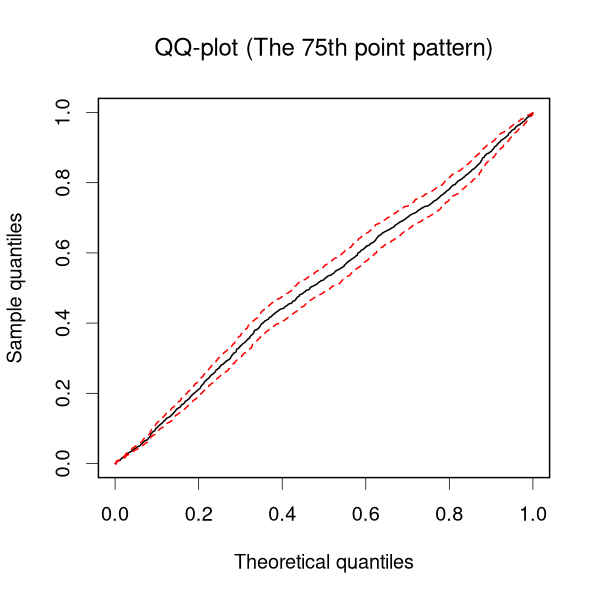}
    \includegraphics[width=.19\textwidth]{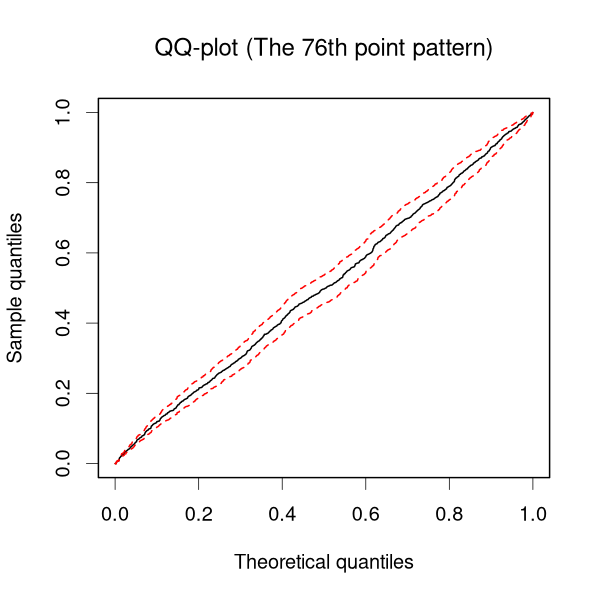}
    \includegraphics[width=.19\textwidth]{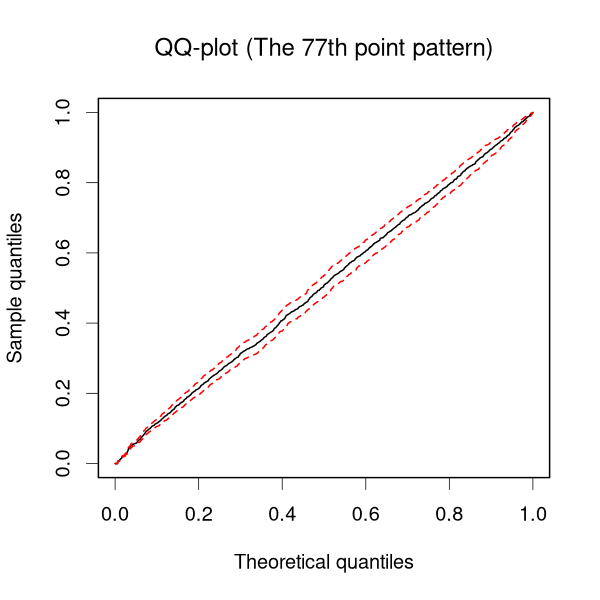}
    \includegraphics[width=.19\textwidth]{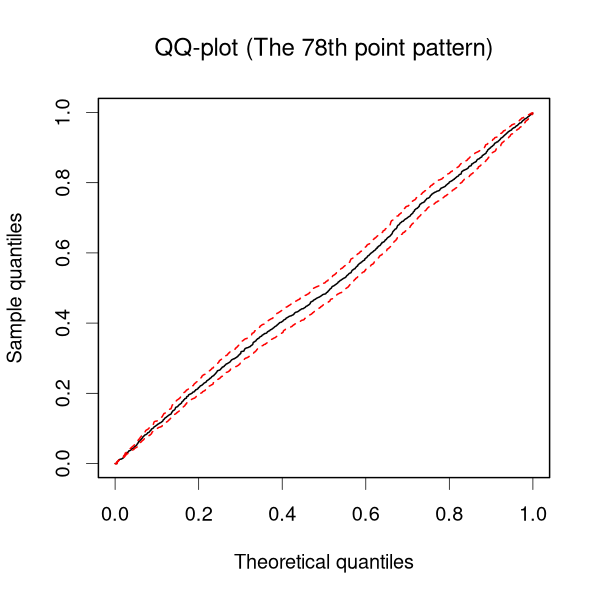}
    \includegraphics[width=.19\textwidth]{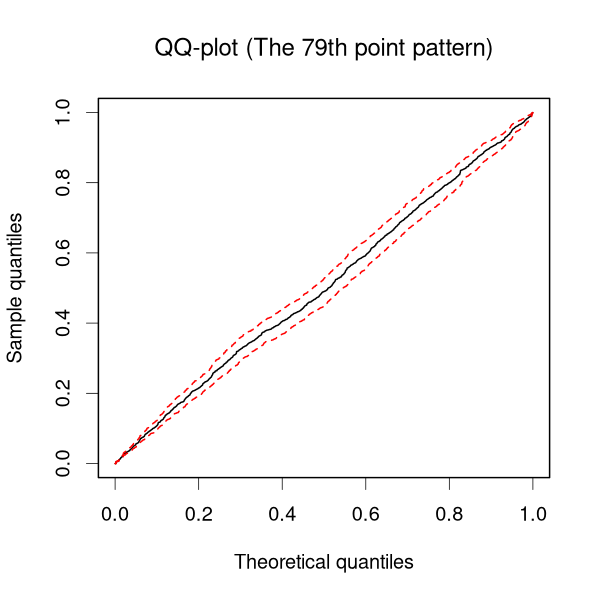}
    \includegraphics[width=.19\textwidth]{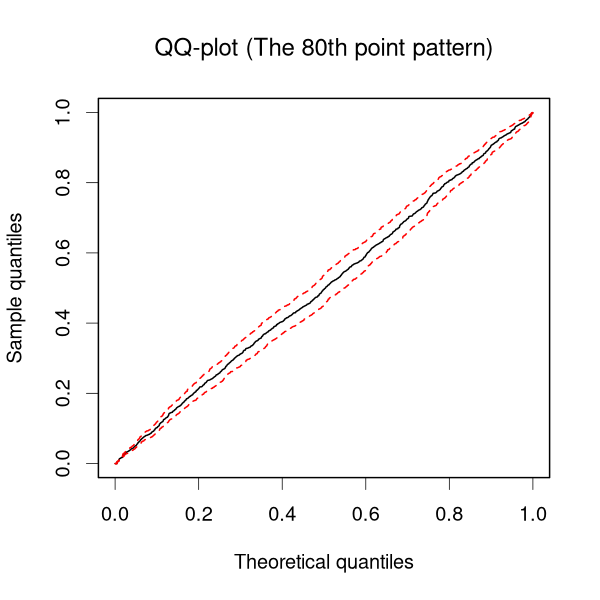}
    \includegraphics[width=.19\textwidth]{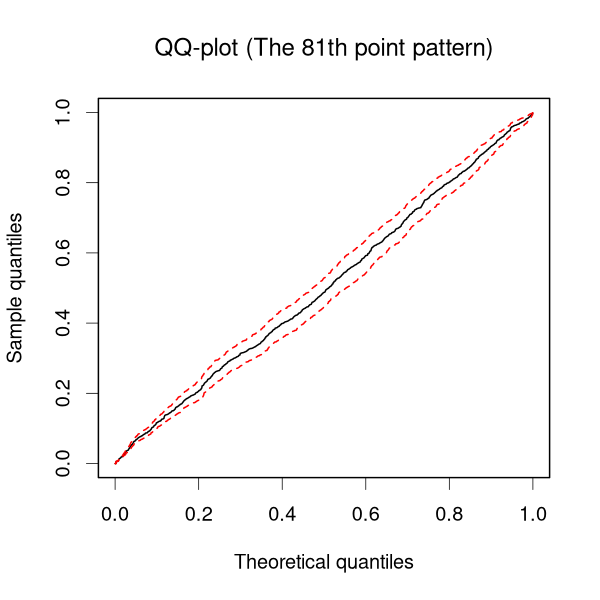}
    \includegraphics[width=.19\textwidth]{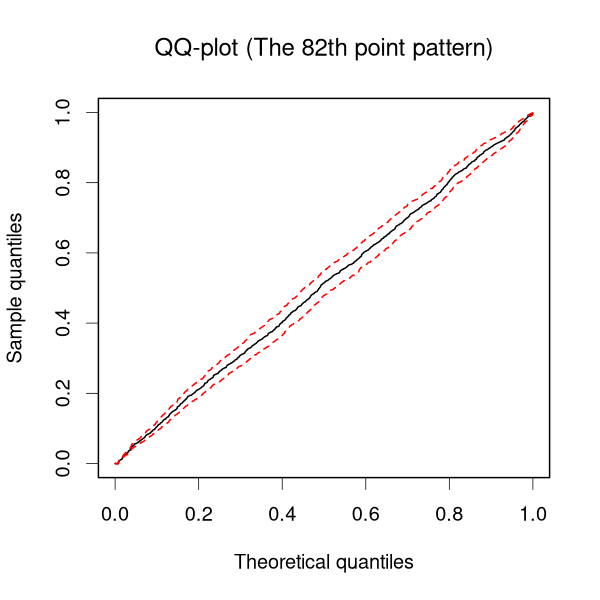}
    \includegraphics[width=.19\textwidth]{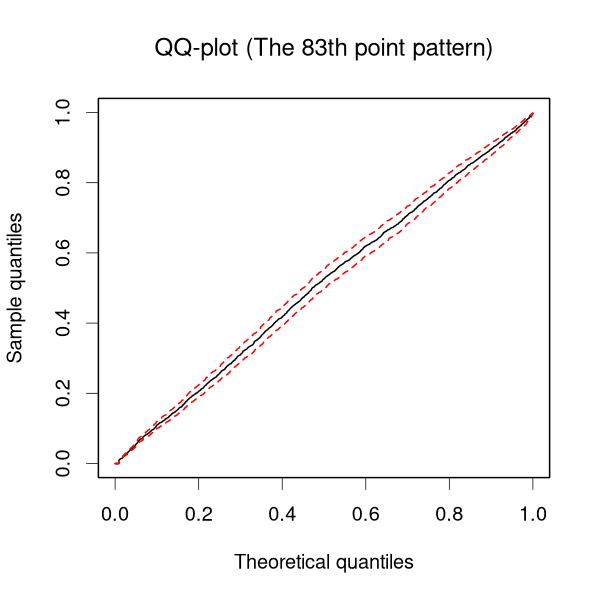}
    \includegraphics[width=.19\textwidth]{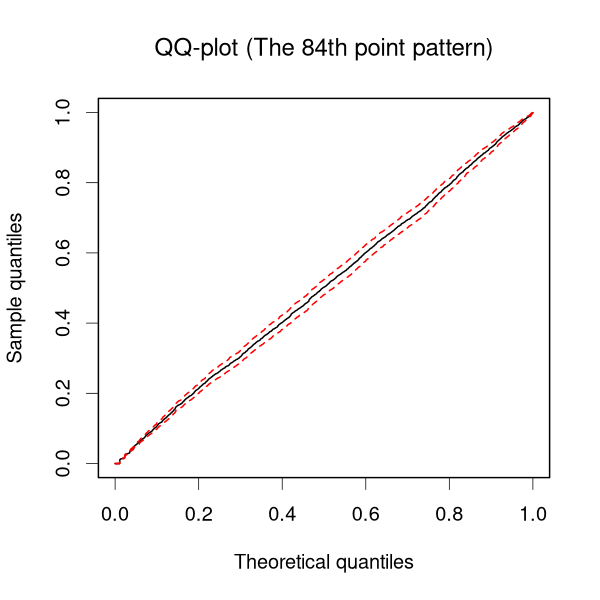}
    \includegraphics[width=.19\textwidth]{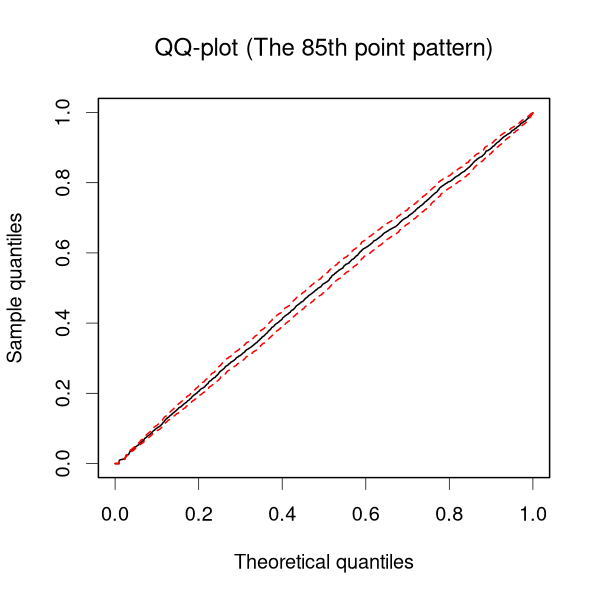}
    \includegraphics[width=.19\textwidth]{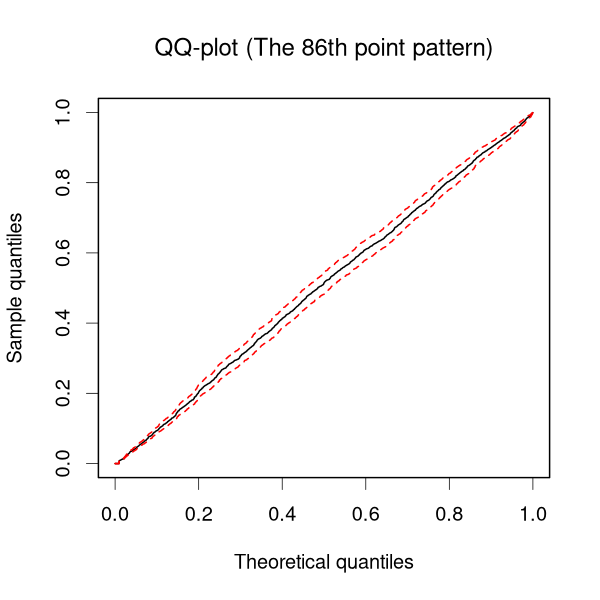}
    \includegraphics[width=.19\textwidth]{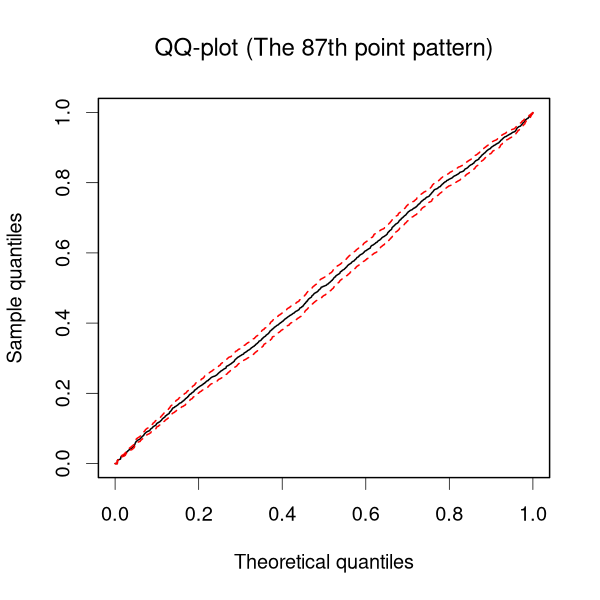}
    \includegraphics[width=.19\textwidth]{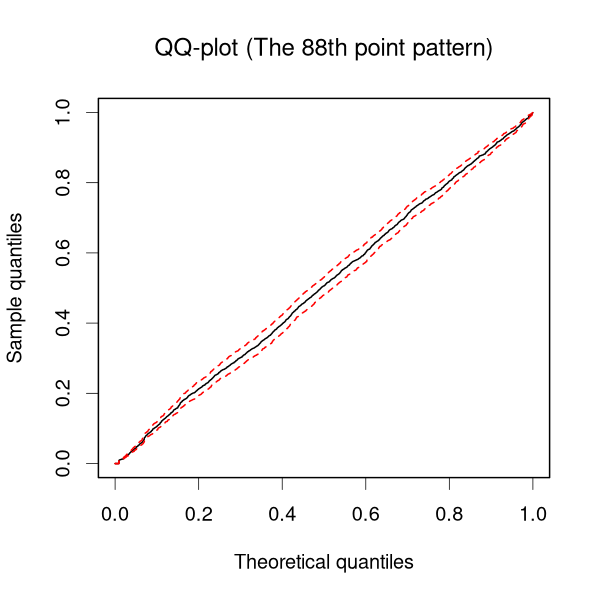}
    \includegraphics[width=.19\textwidth]{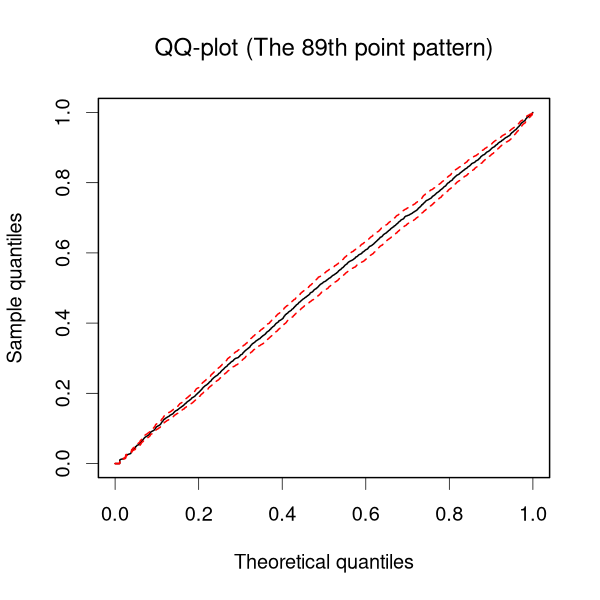}
    \includegraphics[width=.19\textwidth]{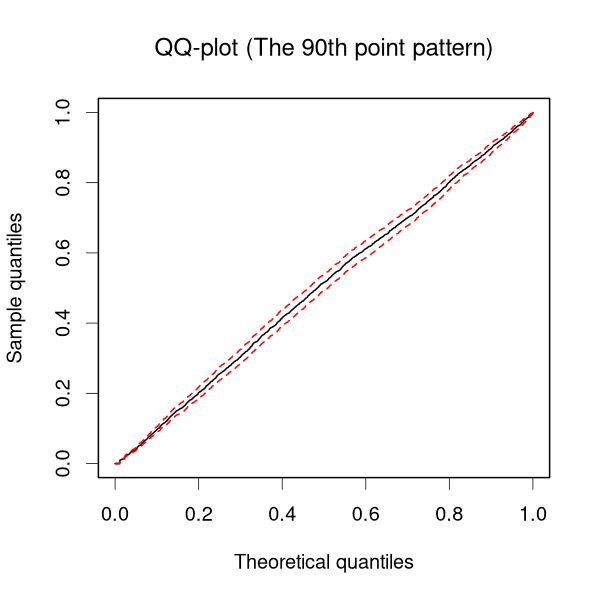}      
    \includegraphics[width=.19\textwidth]{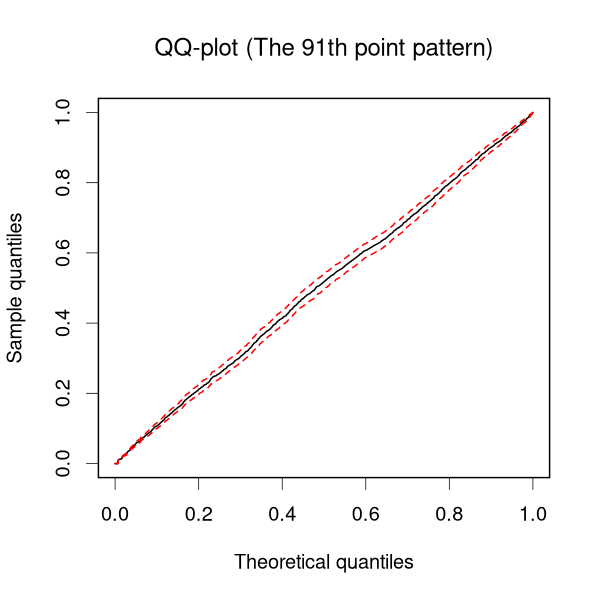}
    \includegraphics[width=.19\textwidth]{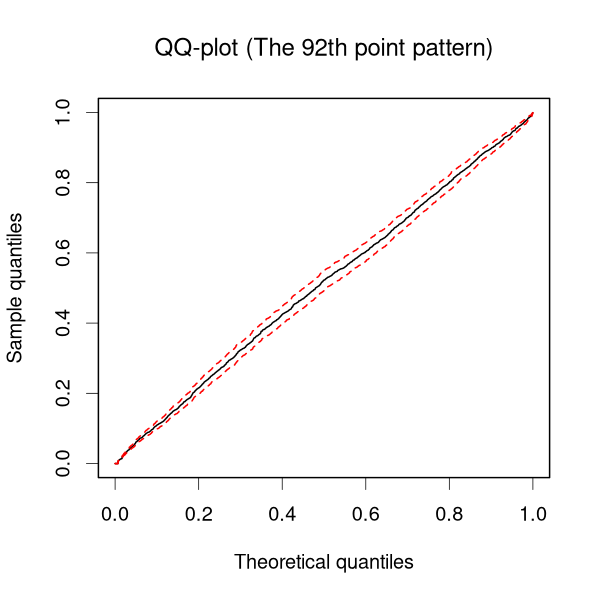}
    \includegraphics[width=.19\textwidth]{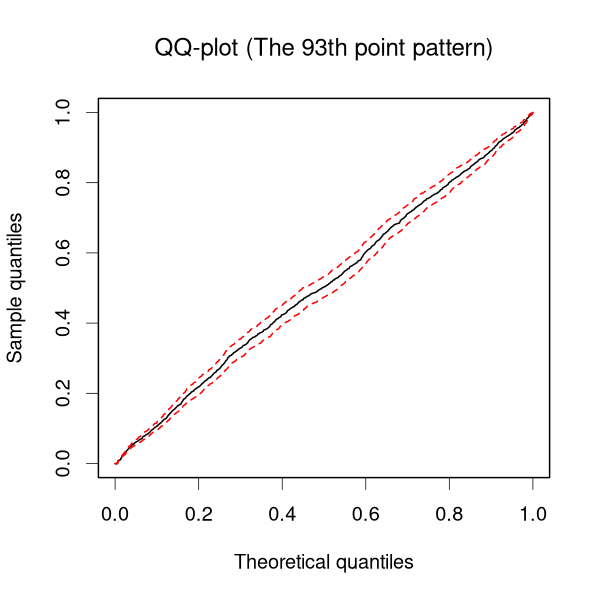}
    \includegraphics[width=.19\textwidth]{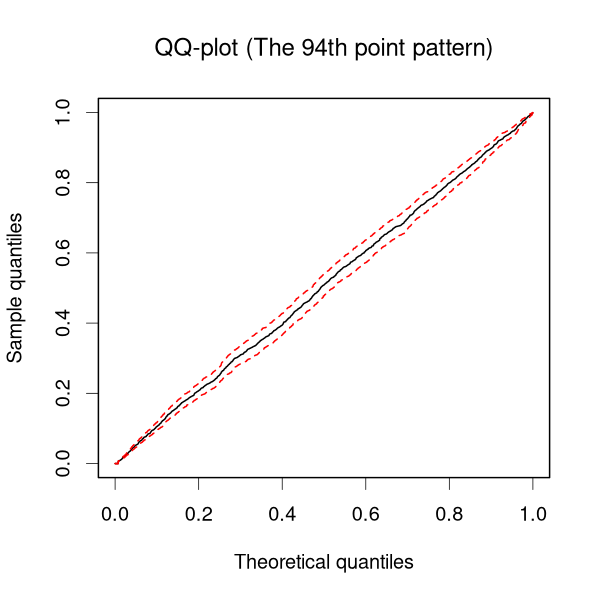}
    \includegraphics[width=.19\textwidth]{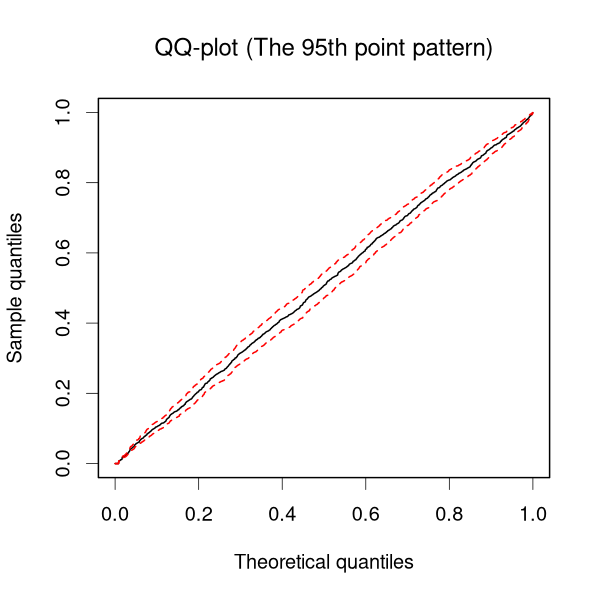}
    \includegraphics[width=.19\textwidth]{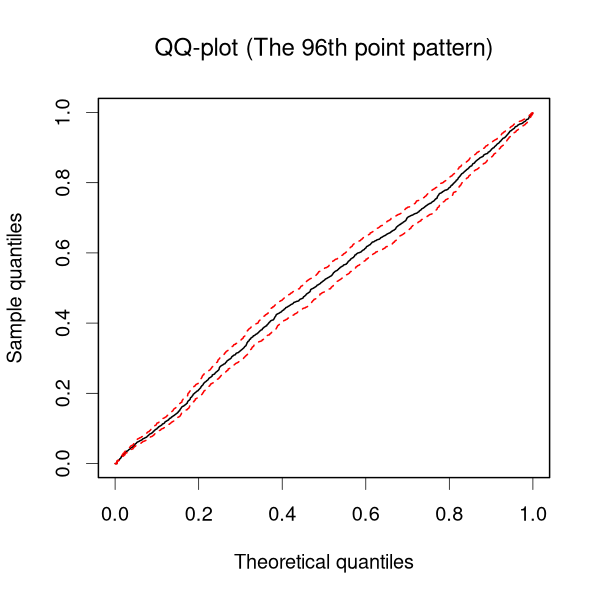}
    \includegraphics[width=.19\textwidth]{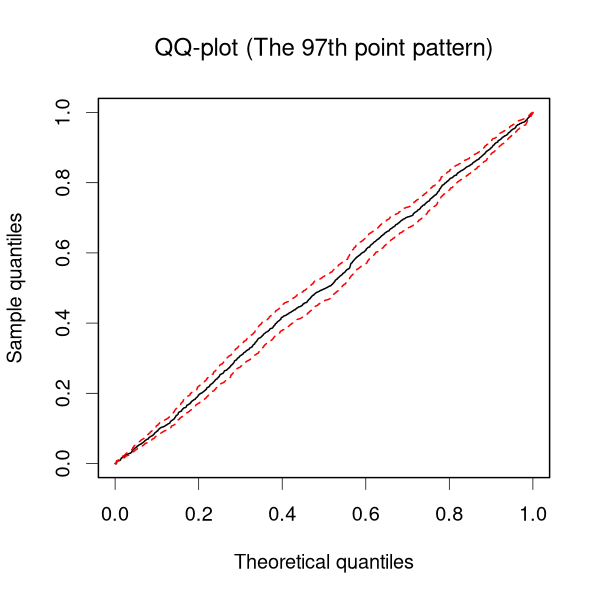}
    \includegraphics[width=.19\textwidth]{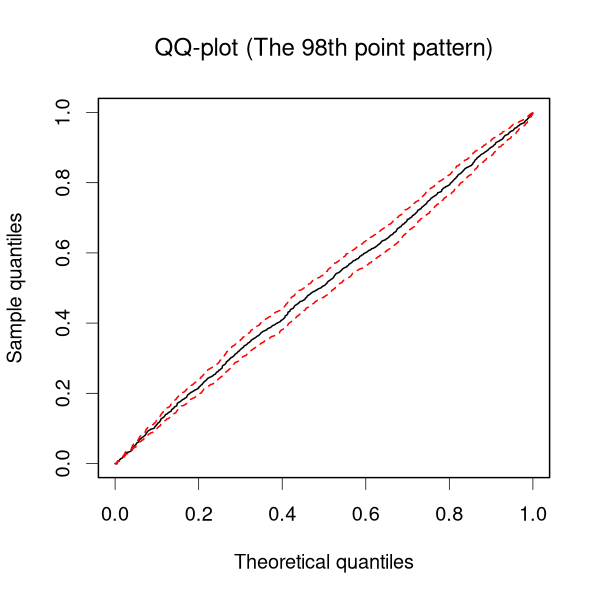}
    \includegraphics[width=.19\textwidth]{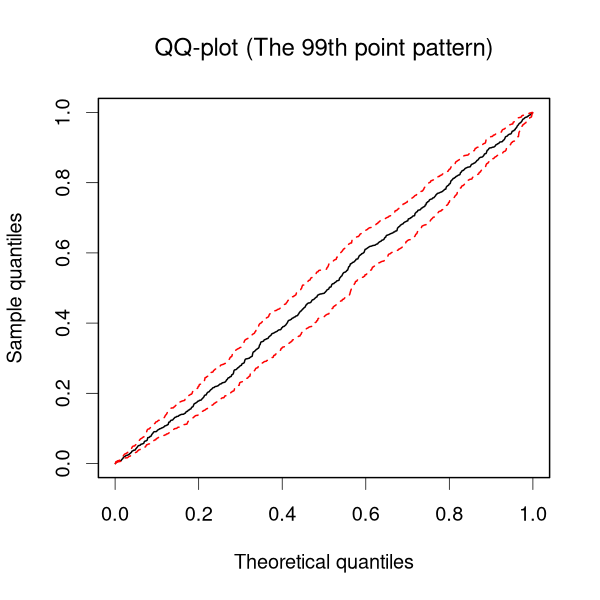}
    \includegraphics[width=.19\textwidth]{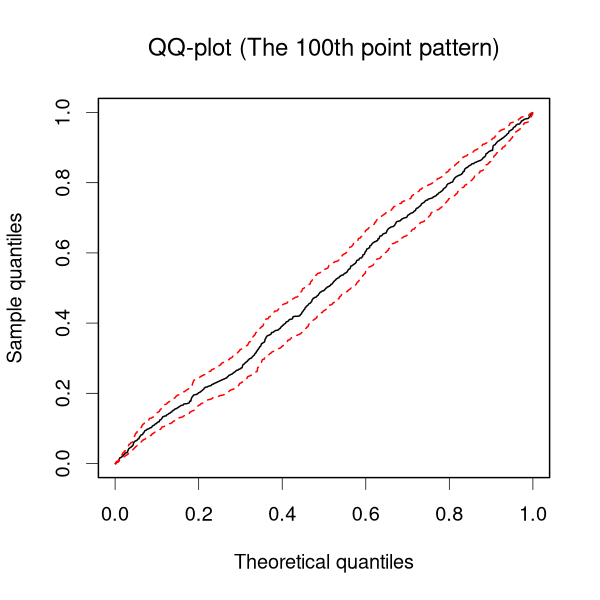}
    \includegraphics[width=.19\textwidth]{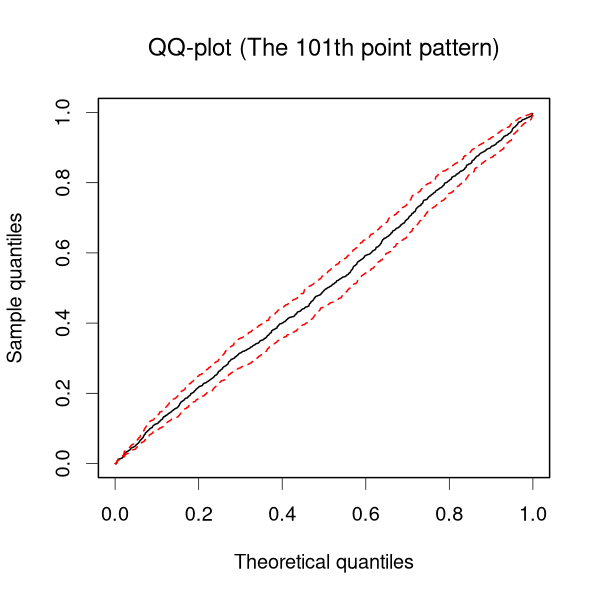}
    \includegraphics[width=.19\textwidth]{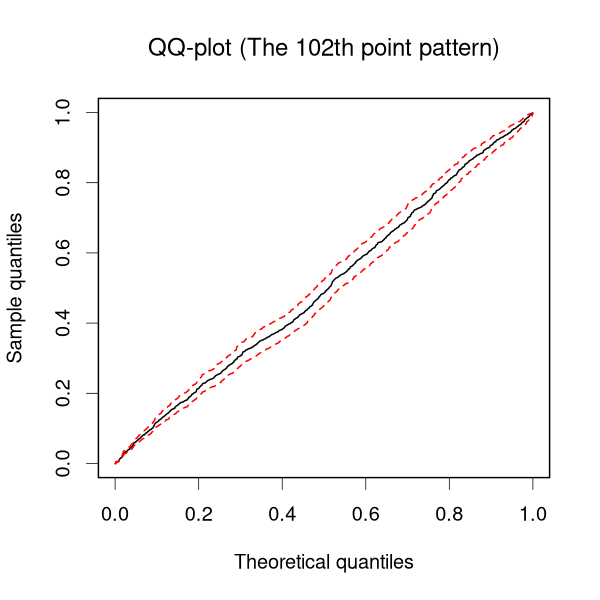}
    \includegraphics[width=.19\textwidth]{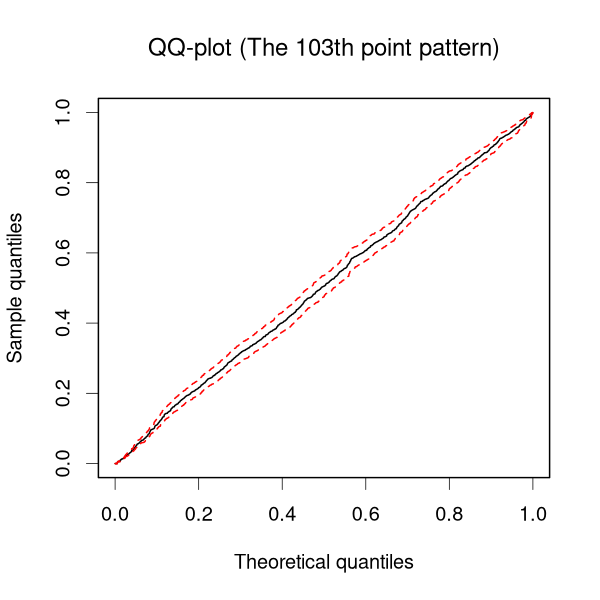}
    \includegraphics[width=.19\textwidth]{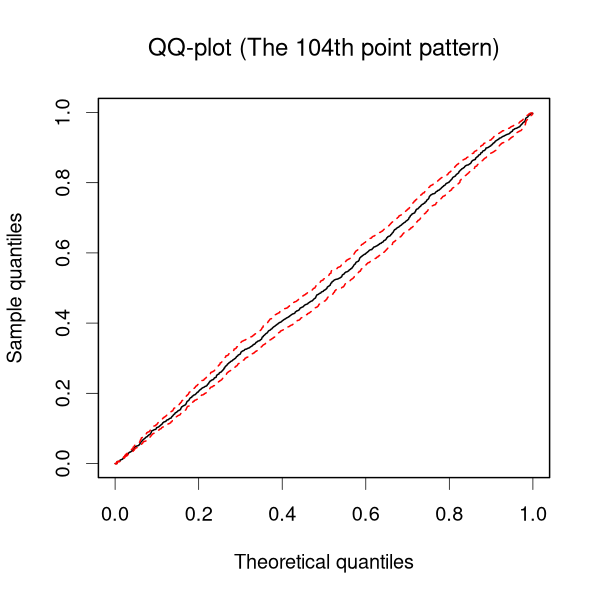}
    \includegraphics[width=.19\textwidth]{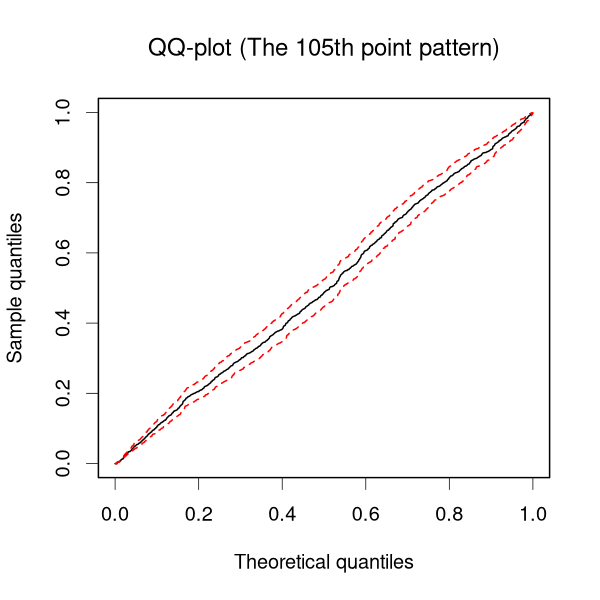}
    \caption{Model checking results for point patterns 71 - 105
    in the second real data example in the main paper (Section 4.3).
    Black solid lines and red dotted lines are
    posterior mean and 95\% credible interval estimates,
    respectively.}
    \label{fig:mtdpp-fx3}
\end{figure*}

\begin{figure*}[t!]
    % \centering
    % \captionsetup[subfigure]{justification=centering, font=footnotesize}
    \includegraphics[width=.19\textwidth]{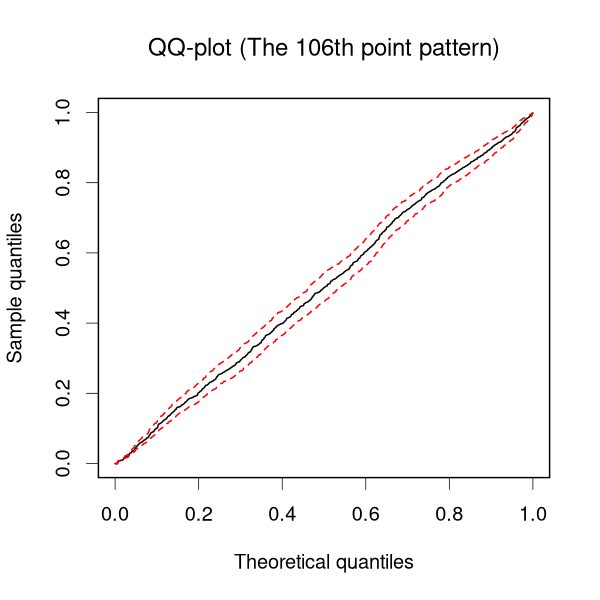}
    \includegraphics[width=.19\textwidth]{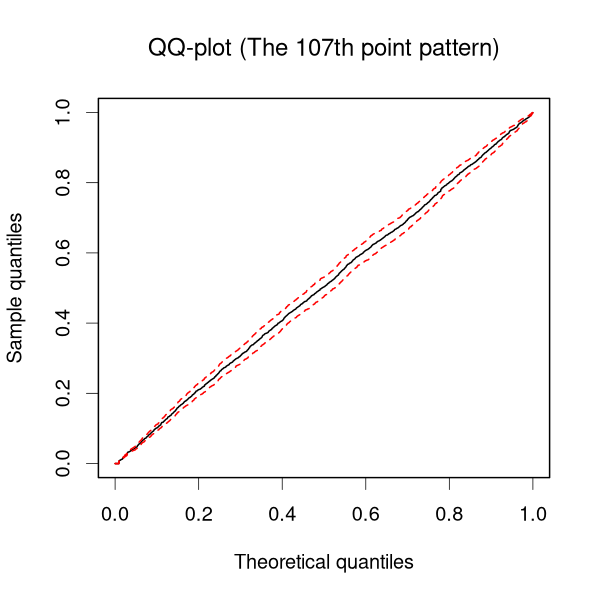}
    \includegraphics[width=.19\textwidth]{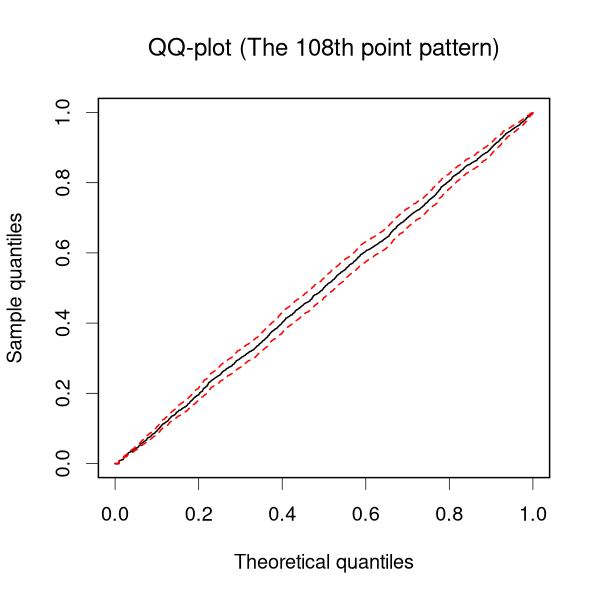}
    \includegraphics[width=.19\textwidth]{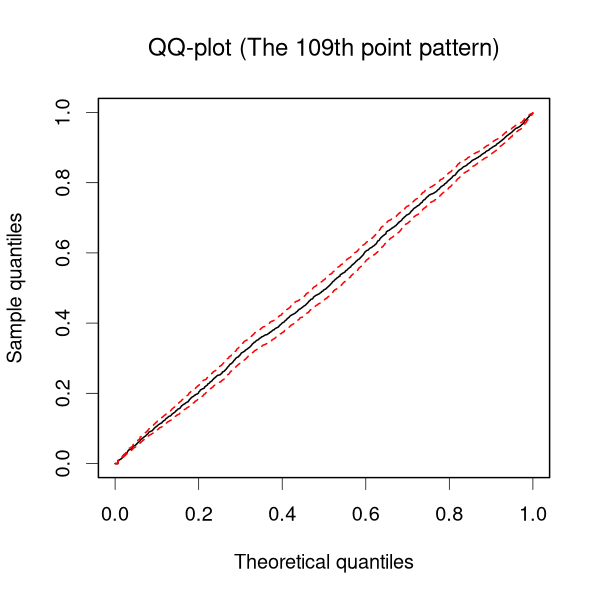}
    \includegraphics[width=.19\textwidth]{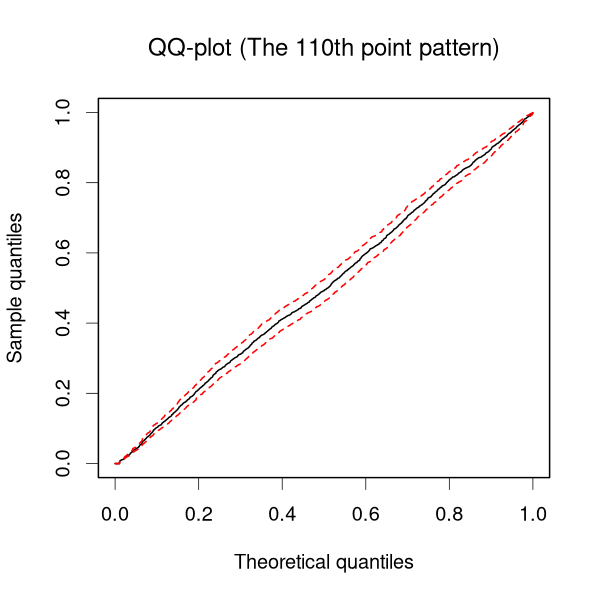}
    \includegraphics[width=.19\textwidth]{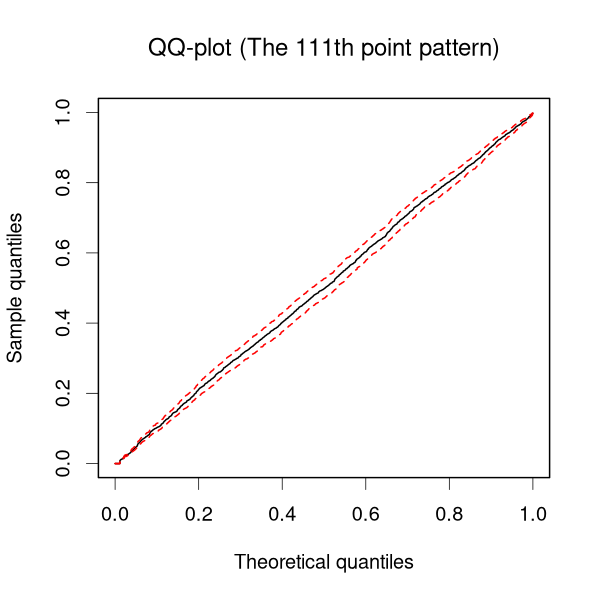}
    \includegraphics[width=.19\textwidth]{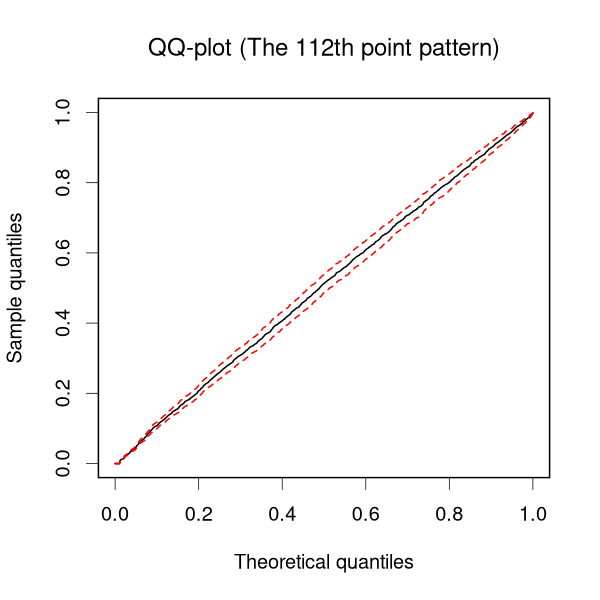}
    \includegraphics[width=.19\textwidth]{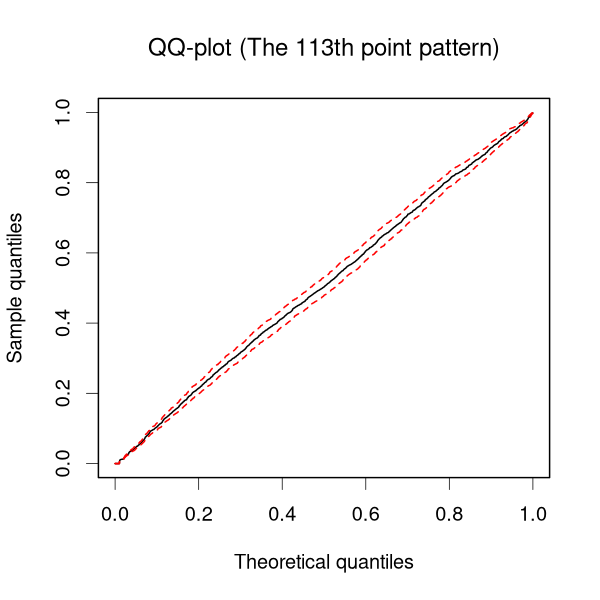}
    \includegraphics[width=.19\textwidth]{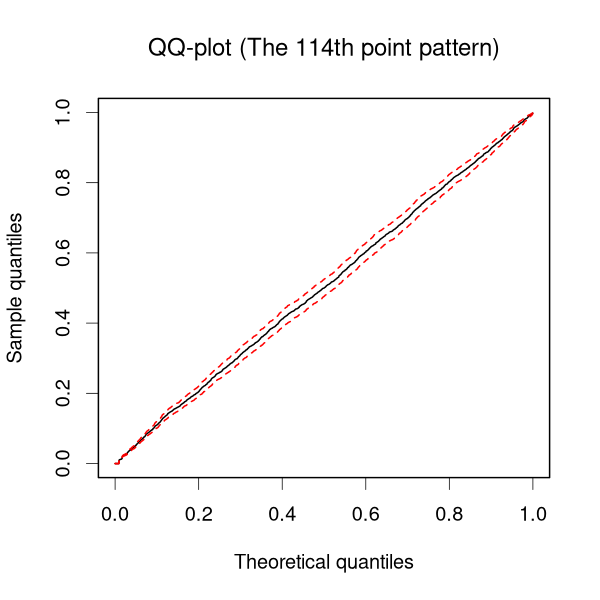}
    \includegraphics[width=.19\textwidth]{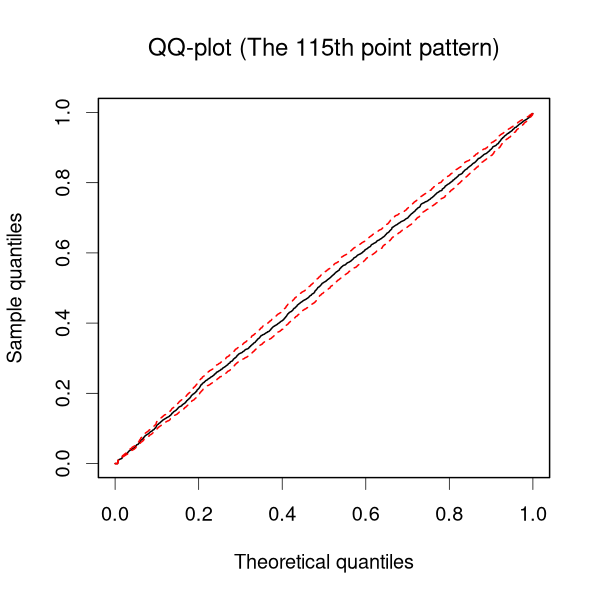}
    \includegraphics[width=.19\textwidth]{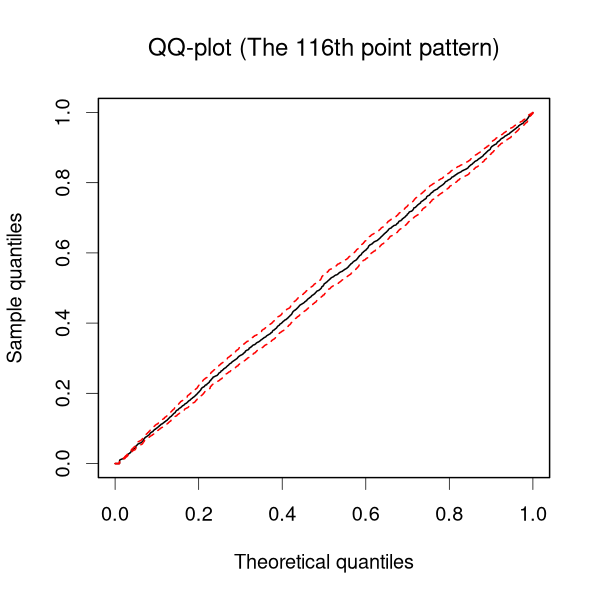}
    \includegraphics[width=.19\textwidth]{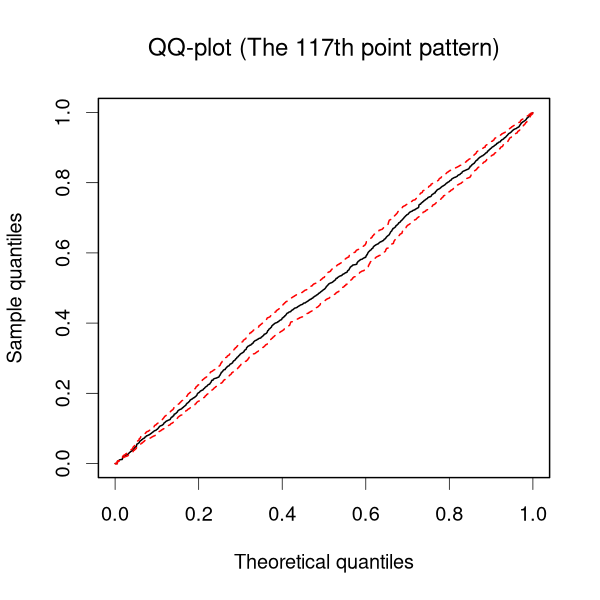}
    \includegraphics[width=.19\textwidth]{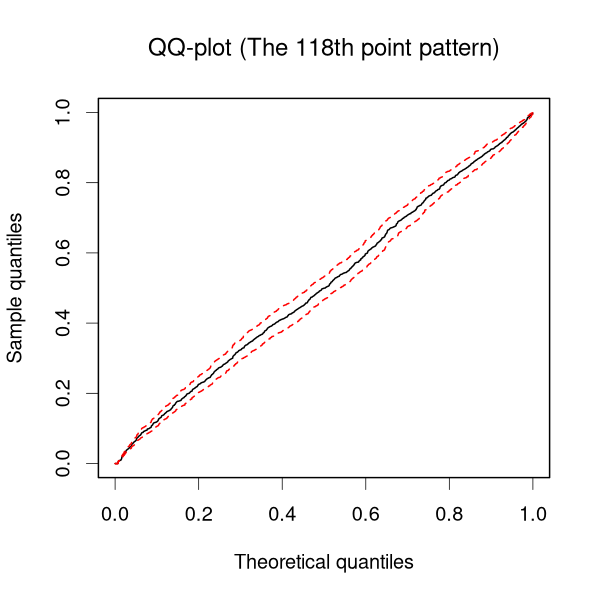}
    \includegraphics[width=.19\textwidth]{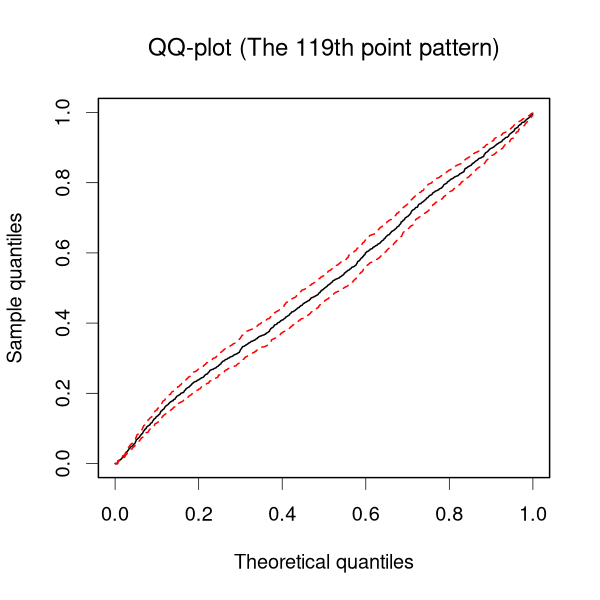}
    \includegraphics[width=.19\textwidth]{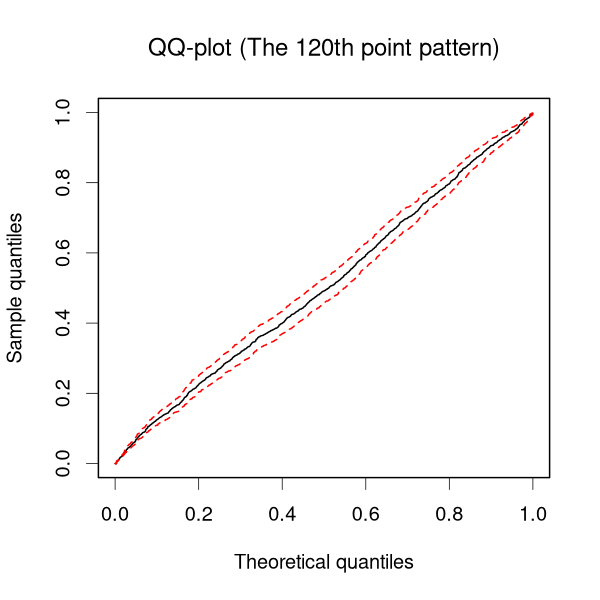}
    \includegraphics[width=.19\textwidth]{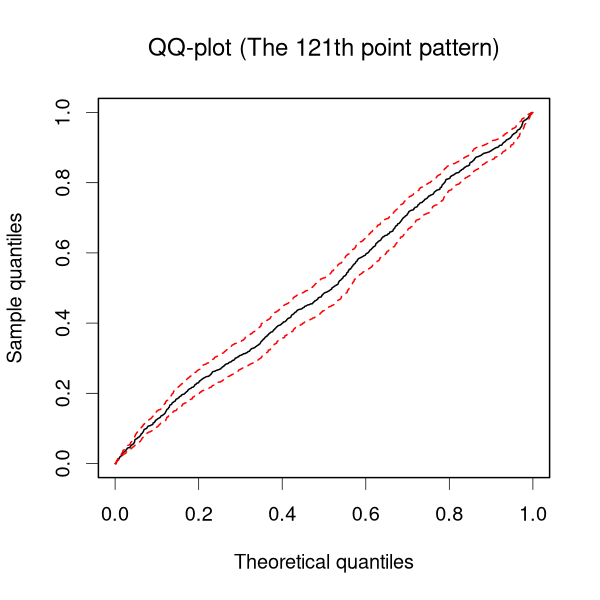}
    \caption{Model checking results for point patterns 106 - 121
    in the second real data example in the main paper (Section 4.3).
    Black solid lines and red dotted lines are
    posterior mean and 95\% credible interval estimates,
    respectively.}
    \label{fig:mtdpp-fx4}
\end{figure*}

Given observed points $0<t_1 < \dots < t_n < T$, consider random variables 
$U^*_i =$ $1 - \exp\{-(\Lambda^*(t_i) - \Lambda^*(t_{i-1}))\}
= F^*(t_i - t_{i-1})$, $i = L+1,\dots, n$, 
as described in the main paper. 
If the point process model is correctly specified, the estimates
of $(U^*_{L+1},\dots,U^*_n)$ will be independently and identically distributed 
as a standard uniform distribution. Figure \ref{fig:mtdpp-sim1} consists of 
quantile-quantile plots of the estimates of $(U^*_{L+1},\dots,U^*_n)$ for the 
simulation study and the first real data example in the main paper, as well 
as for the additional study in Section \ref{sec: mtdcpp-sim}.
Figures \ref{fig:mtdpp-fx1}-\ref{fig:mtdpp-fx4} contain quantile-quantile 
plots of the estimates of $(U^*_{L+1},\dots,U^*_n)$ for the second real data 
example in the main paper. The graphical model assessment results indicate 
good model fit for all data examples.

\bibliographystylesm{jasa3}
\bibliographysm{ref}

\newpage

\end{document}